\documentclass[preprint,12pt,sort&compress]{elsarticle}

\usepackage[dvipsnames]{xcolor}
\usepackage{tikz}
\usepackage{makecell}
\usepackage{threeparttable}
\usetikzlibrary{trees}
\usetikzlibrary{shapes}

\usepackage{amssymb}

\journal{Progress in Particle and Nuclear Physics}

\usepackage[T1]{fontenc}
\usepackage{lmodern}
\usepackage{calc}
\usepackage{graphicx}
\usepackage{booktabs}
\usepackage{textcomp}
\usepackage{xspace}
\usepackage{relsize}
\usepackage{amssymb}
\usepackage{amsmath}
\usepackage{listings}
\usepackage{microtype}
\usepackage{multirow}
\usepackage{tabularx}
\usepackage{array}
\usepackage{placeins}
\usepackage{cuted}
\usepackage{soul} 
\usepackage{fixltx2e}
\usepackage{slashed}
\usepackage{bm}
\usepackage[labelfont=bf,font=small]{caption}
\usepackage[skip=-2pt]{subcaption}
\usepackage[colorlinks,citecolor=blue,urlcolor=blue,linkcolor=blue,breaklinks=breakall]{hyperref}
\usepackage{breakurl}
\usepackage[clockwise,figuresright]{rotating}
\usepackage{siunitx}
\usepackage{tikz}
\usepackage[normalem]{ulem}
\usepackage[utf8]{inputenc}

\usepackage{etoolbox}
\AfterEndEnvironment{strip}{\leavevmode}

\allowdisplaybreaks

\newcolumntype{L}{>{\raggedright\let\newline\\\arraybackslash\hspace{0pt}}X}
\newcolumntype{R}{>{\raggedleft\let\newline\\\arraybackslash\hspace{0pt}}X}
\newcolumntype{C}{>{\centering\let\newline\\\arraybackslash\hspace{0pt}}X}

\newcommand{\prrange}[2]{$[\num{#1}, \, \num{#2}]$}
\newcommand{\otherpi}{\pi}
\newcommand{\mrm}[2]{#1_\text{#2}}
\newcommand{\lagrangian}{\pazocal{L}}
\DeclareMathAlphabet{\pazocal}{OMS}{zplm}{m}{n}
\newcommand{\lsim}{\mathrel{\rlap{\lower4pt\hbox{$\sim$}}\raise1pt\hbox{$<$}}}
\newcommand{\fa}{f_a}
\newcommand{\mazero}{m_{a,0}}
\newcommand{\ma}{m_a}
\newcommand{\Tcrit}{T_\chi}
\newcommand{\thetai}{\mrm{\theta}{i}}
\newcommand{\gagg}{g_{a\gamma\gamma}}

\newcommand{\caggtilde}{\widetilde{C}_{a\gamma\gamma}}
\newcommand{\gaee}{g_{aee}}
\newcommand{\caee}{C_{aee}}
\newcommand{\del}{\partial}
\newcommand{\highl}[1]{{\color{RedOrange}#1}}
\newcommand{\CMSSM}{\textsf{CMSSM}\xspace}
\newcommand{\NUHMone}{\textsf{NUHM1}\xspace}
\newcommand{\NUHMtwo}{\textsf{NUHM2}\xspace}
\newcommand{\MSSMseven}{\textsf{MSSM7}\xspace}
\newcommand{\EWMSSM}{\textsf{EWMSSM}\xspace}
\newcommand{\LambdaQCD}{\Lambda_\chi}
\newcommand{\genalp}{\textsf{GeneralALP}\xspace}

\newcommand{\qcdaxion}{\textsf{QCDAxion}\xspace}
\newcommand{\ksvz}{\textsf{KSVZAxion}\xspace}
\newcommand{\dfsz}{\textsf{DFSZAxion}\xspace}
\newcommand{\dfszI}{\textsf{DFSZAxion-I}\xspace}
\newcommand{\dfszII}{\textsf{DFSZAxion-II}\xspace}
\newcommand{\iuo}[2]{$#1 \, [\si{#2}]$}
\newcommand{\gsim}{\mathrel{\rlap{\lower4pt\hbox{$\sim$}}\raise1pt\hbox{$>$}}}
\newcommand{\dd}{\mathrm{d}}
\newcommand{\Neut}{\tilde{\chi}^0}
\newcommand{\NeutOne}{\tilde{\chi}^0_1}
\newcommand{\NeutTwo}{\tilde{\chi}^0_2}
\newcommand{\NeutThree}{\tilde{\chi}^0_3}
\newcommand{\NeutFour}{\tilde{\chi}^0_4}

\newcommand{\CharOne}{\tilde{\chi}^\pm_1}
\newcommand{\CharTwo}{\tilde{\chi}^\pm_2}

\newcommand{\gambitinstitute}[1]{\expandafter\csname #1\endcsname \label{#1}}

\newcommand{\imperial}{Department of Physics, Imperial College London, Blackett Laboratory, Prince Consort Road, London SW7 2AZ, UK}

\newcommand{\oslo}{Department of Physics, University of Oslo, N-0316 Oslo, Norway}
\newcommand{\adelaide}{Department of Physics, University of Adelaide, Adelaide, SA 5005, Australia}

\newcommand{\uq}{School of Mathematics and Physics, The University of Queensland, St.\ Lucia, Brisbane, QLD 4072, Australia}

\newcommand{\notimplies}{%
  \mathrel{{\ooalign{\hidewidth$\not\phantom{=}$\hidewidth\cr$\implies$}}}}

\makeatletter

\newcommand{\preprintnumber}[1]{\gdef\@preprintnumber{\begin{flushright}{#1}\end{flushright}}}

\g@addto@macro\bfseries{\boldmath}
\makeatother

\let\underscore\_
\renewcommand{\_}{\discretionary{\underscore}{}{\underscore}}

\makeatletter
\let\orgdescriptionlabel\descriptionlabel
\renewcommand*{\descriptionlabel}[1]{%
  \let\orglabel\label
  \let\label\@gobble
  \phantomsection
  \protected@edef\@currentlabel{#1}%
  \let\label\orglabel
  \orgdescriptionlabel{#1}%
}
\makeatother

\newcommand\postnewlinemarker{\hbox{\ensuremath{\hookrightarrow}}}
\newcommand\cpp[1]{{\lstinline!#1!}}  

\newcommand\yaml[1]{{\lstset{style=yaml}\lstinline!#1!\lstset{style=cpp}}}

\newcommand\term[1]{{\lstset{style=terminal}\lstinline!#1!\lstset{style=cpp}}}
\newcommand\fortran[1]{{\lstset{style=fortran}\lstinline!#1!\lstset{style=cpp}}}
\newcommand\py[1]{{\lstset{style=python}\lstinline!#1!\lstset{style=cpp}}}
\newcommand\customtilde{{\raisebox{0.2ex}{\scalebox{0.6}{\boldmath$\sim$}}}}
\newcommand\mathematica[1]{{\lstset{style=Mathematica}\lstinline!#1!\lstset{style=cpp}}}

\lstnewenvironment{lstlistingyaml}{\lstset{style=yaml}}{\lstset{style=cpp}}
\lstnewenvironment{lstlistingterm}{\lstset{style=terminal}}{\lstset{style=cpp}}
\lstnewenvironment{lstlistingfortran}{\lstset{style=fortran}}{\lstset{style=cpp}}
\lstnewenvironment{lstcpp}{\lstset{style=cpp}}{\lstset{style=cpp}}
\lstnewenvironment{lstcppalt}{\lstset{style=cppalt}}{\lstset{style=cpp}}
\lstnewenvironment{lstcppnum}{\lstset{style=cppnum}}{\lstset{style=cpp}}
\lstnewenvironment{lstyaml}{\lstset{style=yaml}}{\lstset{style=cpp}}
\lstnewenvironment{lstterm}{\lstset{style=terminal}}{\lstset{style=cpp}}
\lstnewenvironment{lsttermalt}{\lstset{style=terminalalt}}{\lstset{style=cpp}}
\lstnewenvironment{lsttext}{\lstset{style=text}}{\lstset{style=cpp}}
\lstnewenvironment{lstfortran}{\lstset{style=fortran}}{\lstset{style=cpp}}
\lstnewenvironment{lstpy}{\lstset{style=python}}{\lstset{style=cpp}}
\lstnewenvironment{lstmathematica}{\lstset{style=mathematica}}{\lstset{style=cpp}}

\newcommand{\tmpname}{}
\newcommand{\tmplistingname}{}
\makeatletter
\newif\ifATOlabelname
\lst@Key{labelname}{Listing}{\def\ATOlabelname{#1}\global\ATOlabelnametrue}
\makeatother
\lstnewenvironment{lstcpplabel}[1][]{
  \lstset{style=cpp,#1} 
  \ifATOlabelname
    \renewcommand{\tmpname}{\lstlistingname}
    \renewcommand{\tmplistingname}{\lstlistlistingname}
    \renewcommand{\lstlistingname}{\ATOlabelname}
    \renewcommand{\lstlistlistingname}{List of \lstlistingname s}
  \fi
}{
  \renewcommand{\lstlistingname}{\tmpname}
  \renewcommand{\lstlistlistingname}{\tmplistingname}
  \lstset{style=cpp}
}
\definecolor{solarized@base03}{HTML}{002B36}
\definecolor{solarized@base02}{HTML}{073642}
\definecolor{solarized@base01}{HTML}{586e75}
\definecolor{solarized@base00}{HTML}{657b83}
\definecolor{solarized@base0}{HTML}{839496}
\definecolor{solarized@base1}{HTML}{93a1a1}
\definecolor{solarized@base2}{HTML}{EEE8D5}
\definecolor{solarized@base3}{HTML}{FDF6E3}
\definecolor{solarized@yellow}{HTML}{B58900}
\definecolor{solarized@orange}{HTML}{CB4B16}
\definecolor{solarized@red}{HTML}{DC322F}
\definecolor{solarized@magenta}{HTML}{D33682}
\definecolor{solarized@violet}{HTML}{6C71C4}
\definecolor{solarized@blue}{HTML}{268BD2}
\definecolor{solarized@cyan}{HTML}{2AA198}
\definecolor{solarized@green}{HTML}{859900}
\definecolor{darkred}{HTML}{550003}
\definecolor{darkgreen}{HTML}{00AA00}

\newcommand\YAMLstringstyle{\footnotesize\color{solarized@green}\mdseries}
\newcommand\YAMLkeystyle{\footnotesize\color{solarized@blue}\ttfamily}
\newcommand\YAMLvaluestyle{\footnotesize\color{blue}\mdseries}
\newcommand\ProcessThreeDashes{\llap{\color{cyan}\mdseries-{-}-}}

\newcommand\CPPcommentstyle{\color{solarized@violet}\footnotesize\ttfamily}
\newcommand\CPPdirectivestyle{\color{solarized@magenta}\footnotesize\ttfamily}
\newcommand\termplainstyle{\footnotesize\ttfamily}

\newcommand\processLongMacroDelimiter
{%
\CPPdirectivestyle%
\#define%
}

\lstdefinestyle{cpp}
{
  language=C++,
  basicstyle=\footnotesize\ttfamily,
  basewidth={0.53em,0.44em}, 
  numbers=none,
  tabsize=2,
  breaklines=true,
  escapeinside={@}{@},
  showstringspaces=false,
  numberstyle=\tiny\color{solarized@base01},
  keywordstyle=\color{solarized@orange},
  stringstyle=\color{solarized@red}\ttfamily,
  identifierstyle=\color{solarized@blue},
  commentstyle=\CPPcommentstyle,
  directivestyle=\CPPdirectivestyle,
  emphstyle=\color{solarized@green},
  frame=single,
  rulecolor=\color{solarized@base2},
  rulesepcolor=\color{solarized@base2},
  literate={~} {\customtilde}1,
  moredelim=*[directive]\ \ \#,
  moredelim=*[directive]\ \ \ \ \#
}

\lstdefinestyle{cppalt}
{
  language=C++,
  basicstyle=\footnotesize\ttfamily,
  basewidth={0.53em,0.44em}, 
  numbers=none,
  tabsize=2,
  breaklines=true,
  escapeinside={*@}{@*},
  showstringspaces=false,
  numberstyle=\tiny\color{solarized@base01},
  keywordstyle=\color{solarized@orange},
  stringstyle=\color{solarized@red}\ttfamily,
  identifierstyle=\color{solarized@blue},
  commentstyle=\CPPcommentstyle,
  directivestyle=\CPPdirectivestyle,
  emphstyle=\color{solarized@green},
  frame=single,
  rulecolor=\color{solarized@base2},
  rulesepcolor=\color{solarized@base2},
  literate={~}{\customtilde}1,
  moredelim=**[is][\processLongMacroDelimiter]{BeginLongMacro}{EndLongMacro} 
}

\lstdefinestyle{cppnum}
{
  language=C++,
  basicstyle=\footnotesize\ttfamily,
  basewidth={0.53em,0.44em}, 
  numbers=none,
  tabsize=2,
  breaklines=true,
  escapeinside={@}{@},
  numberstyle=\tiny\color{solarized@base01},
  showstringspaces=false,
  numberstyle=\tiny\color{solarized@base01},
  keywordstyle=\color{solarized@orange},
  stringstyle=\color{solarized@red}\ttfamily,
  identifierstyle=\color{solarized@blue},
  commentstyle=\CPPcommentstyle,
  directivestyle=\CPPdirectivestyle,
  emphstyle=\color{solarized@green},
  frame=single,
  rulecolor=\color{solarized@base2},
  rulesepcolor=\color{solarized@base2},
  literate={~} {\customtilde}1,
  moredelim=*[directive]\ \ \#,
  moredelim=*[directive]\ \ \ \ \#
}

\lstdefinestyle{python}
{
  language=Python,
  basicstyle=\footnotesize\ttfamily,
  basewidth={0.53em,0.44em},
  numbers=none,
  tabsize=2,
  breaklines=true,
  escapeinside={@}{@},
  showstringspaces=false,
  numberstyle=\tiny\color{solarized@base01},
  keywordstyle=\color{blue},
  stringstyle=\color{orange}\ttfamily,
  identifierstyle=\color{darkred},
  commentstyle=\color{purple},
  emphstyle=\color{green},
  frame=single,
  rulecolor=\color{solarized@base2},
  rulesepcolor=\color{solarized@base2},
  literate = {~}{\customtilde}1
             {\ as\ }{{\color{blue}\ as\ \color{black}}}3
}

\lstdefinestyle{fortran}
{
  language=Fortran,
  basicstyle=\footnotesize\ttfamily,
  basewidth={0.53em,0.44em},
  numbers=none,
  tabsize=2,
  breaklines=true,
  escapeinside={@}{@},
  showstringspaces=false,
  numberstyle=\tiny\color{solarized@base01},
  keywordstyle=\color{blue},
  stringstyle=\color{orange}\ttfamily,
  identifierstyle=\color{Periwinkle},
  commentstyle=\color{purple},
  emphstyle=\color{green},
  morekeywords={and, or, true, false},
  frame=single,
  rulecolor=\color{solarized@base2},
  rulesepcolor=\color{solarized@base2},
  literate={~}{\customtilde}1
}

\lstdefinestyle{terminal}
{
  language=bash,
  basicstyle=\termplainstyle,
  numbers=none,
  tabsize=2,
  breaklines=true,
  escapeinside={@}{@},
  frame=single,
  showstringspaces=false,
  numberstyle=\tiny\color{solarized@base01},
  keywordstyle=\color{solarized@orange},
  stringstyle=\color{solarized@red}\ttfamily,
  identifierstyle=\color{black},
  commentstyle=\color{solarized@violet},
  emphstyle=\color{solarized@green},
  frame=single,
  rulecolor=\color{solarized@base2},
  rulesepcolor=\color{solarized@base2},
  morekeywords={gambit, cmake, make, mkdir, gum, python},
  deletekeywords={test},
  literate = {\ gambit}{{\ }{\color{black}}gambit}7
             {/gambit}{{/}{\color{black}}gambit}6
             {gambit/}{{\color{black}}gambit{/}}6
             {/include}{{/}{\color{black}}include}8
             {cmake/}{{\color{black}}cmake/}6
             {.cmake}{{.}{\color{black}}cmake}6
             {.gum}{{.}{\color{black}}gum}6
             {~}{\customtilde}1
}

\lstdefinestyle{terminalalt}
{
  language=bash,
  basicstyle=\footnotesize\ttfamily,
  numbers=none,
  tabsize=2,
  breaklines=true,
  escapeinside={*@}{@*},
  frame=single,
  showstringspaces=false,
  numberstyle=\tiny\color{solarized@base01},
  keywordstyle=\color{solarized@orange},
  stringstyle=\color{solarized@red}\ttfamily,
  identifierstyle=\color{black},
  commentstyle=\color{solarized@violet},
  emphstyle=\color{solarized@green},
  frame=single,
  rulecolor=\color{solarized@base2},
  rulesepcolor=\color{solarized@base2},
  morekeywords={gambit, cmake, make, mkdir},
  deletekeywords={test},
  literate = {\ gambit}{{\ }{\color{black}}gambit}7
             {/gambit}{{/}{\color{black}}gambit}6
             {gambit/}{{\color{black}}gambit{/}}6
             {/include}{{/}{\color{black}}include}8
             {cmake/}{{\color{black}}cmake/}6
             {.cmake}{{.}{\color{black}}cmake}6
             {~}{\customtilde}1
}

\lstdefinestyle{text}
{
  language={},
  basicstyle=\footnotesize\ttfamily,
  identifierstyle=\color{black},
  numbers=none,
  tabsize=2,
  breaklines=true,
  escapeinside={*@}{@*},
  showstringspaces=false,
  frame=single,
  rulecolor=\color{solarized@base2},
  rulesepcolor=\color{solarized@base2},
  literate={~}{\customtilde}1
}

\lstdefinestyle{yaml}
{
  language=bash,
  escapeinside={@}{@},
  keywords={true,false,null},
  otherkeywords={},
  keywordstyle=\color{solarized@base0}\bfseries,
  basicstyle=\footnotesize\color{black}\ttfamily,
  identifierstyle=\YAMLkeystyle,
  sensitive=false,
  commentstyle=\color{solarized@orange}\ttfamily,
  morecomment=[l]{\#},
  morecomment=[s]{/*}{*/},
  stringstyle=\YAMLstringstyle\ttfamily,
  moredelim=**[s][\YAMLkeystyle]{,}{:},   
  moredelim=**[l][\YAMLvaluestyle]{:},    
  morestring=[b]',
  morestring=[b]",
  literate =    {---}{{\ProcessThreeDashes}}3
                {>}{{\textcolor{solarized@red}\textgreater}}1
                {|}{{\textcolor{solarized@red}\textbar}}1
                {\ -\ }{{\mdseries\color{black}\ -\ \negmedspace}}3
                {\}}{{{\color{black} \}}}}1
                {\{}{{{\color{black} \{}}}1
                {[}{{{\color{black} [}}}1
                {]}{{{\color{black} ]}}}1
                {~}{\customtilde}1,
  breakindent=0pt,
  breakatwhitespace,
  columns=fullflexible
}

\lstdefinestyle{mathematica}
{
  language={Mathematica},
  basicstyle=\footnotesize\ttfamily,
  basewidth={0.53em,0.44em},
  numbers=none,
  tabsize=2,
  breaklines=true,
  escapeinside={@}{@},
  numberstyle=\tiny\color{black},
  showstringspaces=false,
  numberstyle=\tiny\color{solarized@base01},
  keywordstyle=\color{solarized@orange},
  stringstyle=\color{solarized@red}\ttfamily,
  identifierstyle=\color{solarized@orange}\ttfamily,
  commentstyle=\color{solarized@gray}\ttfamily,
  directivestyle=\color{solarized@orange}\ttfamily,
  emphstyle=\color{solarized@green},
  frame=single,
  rulecolor=\color{solarized@base2},
  rulesepcolor=\color{solarized@base2},
  literate={~} {\customtilde}1,
  moredelim=*[directive]\ \ \#,
  moredelim=*[directive]\ \ \ \ \#,
  mathescape=true
}

\newcommand{\doublecross}[2]{\hyperref[#2]{\textbf{#1}}}
\newcommand{\doublecrosssf}[2]{\hyperref[#2]{\textbf{\textsf{#1}}}}

\newcommand{\startglossary}{\section{Glossary}\label{glossary}Here we explain some terms that have specific technical definitions in \GB.\begin{description}}
\newcommand{\finishglossary}{\end{description}}

\newcommand{\bcode}{\begin{lstlisting}}
\newcommand{\ecode}{\end{lstlisting}}

\newcommand{\eV}{\ensuremath{\text{e}\mspace{-0.8mu}\text{V}}\xspace}
\newcommand{\MeV}{\text{M\eV}\xspace}
\newcommand{\GeV}{\text{G\eV}\xspace}

\newcommand{\sss}{\scriptscriptstyle}
\newcommand{\ms}{m_{\sss S}}

\newcommand{\lhs}{\lambda_{h\sss S}}
\newcommand{\ls}{\lambda_{\sss S}}

\newcommand{\mh}{m_h}

\newcommand{\DR}{$\overline{DR}$\xspace}
\newcommand{\DRbar}{\DR}
\newcommand{\MSbar}{$\MSBar$\xspace}
\newcommand{\MSBar}{\overline{MS}}

\newcommand{\gambit}{\textsf{GAMBIT}\xspace}

\newcommand{\darkbit}{\textsf{DarkBit}\xspace}
\newcommand{\colliderbit}{\textsf{ColliderBit}\xspace}
\newcommand{\flavbit}{\textsf{FlavBit}\xspace}
\newcommand{\specbit}{\textsf{SpecBit}\xspace}
\newcommand{\decaybit}{\textsf{DecayBit}\xspace}
\newcommand{\precisionbit}{\textsf{PrecisionBit}\xspace}
\newcommand{\scannerbit}{\textsf{ScannerBit}\xspace}

\newcommand{\neutrinobit}{\textsf{NeutrinoBit}\xspace}

\newcommand{\GB}{\gambit}

\newcommand{\pythia}{\textsf{Pythia}\xspace}
\newcommand{\pythiaeight}{\textsf{Pythia\,8}\xspace}

\newcommand{\higgsbounds}{\textsf{HiggsBounds}\xspace}

\newcommand{\higgssignals}{\textsf{HiggsSignals}\xspace}
\newcommand{\ds}{\textsf{DarkSUSY}\xspace}
\newcommand{\darksusy}{\ds}

\newcommand{\micromegas}{\textsf{micrOMEGAs}\xspace}

\newcommand{\feynhiggs}{\textsf{FeynHiggs}\xspace}

\newcommand\FlexibleSUSY{\textsf{FlexibleSUSY}\xspace}

\newcommand\HDECAY{\textsf{HDECAY}\xspace}

\newcommand\SDECAY{\textsf{SDECAY}\xspace}
\newcommand\SUSYHIT{\textsf{SUSY-HIT}\xspace}
\newcommand\susyhd{\textsf{SUSYHD}\xspace}

\newcommand\susyhit{\SUSYHIT}
\newcommand\gmtwocalc{\textsf{GM2Calc}\xspace}

\newcommand\SPheno{\textsf{SPheno}\xspace}
\newcommand\spheno{\SPheno}
\newcommand\superiso{\textsf{SuperIso}\xspace}
\newcommand\superisofour{\textsf{SuperIso 4}\xspace}
\newcommand\heplike{\textsf{HEPLike}\xspace}

\newcommand\FeynHiggs{\textsf{FeynHiggs}\xspace}
\newcommand\Mathematica{\textsf{Mathematica}\xspace}

\newcommand\nulike{\textsf{nulike}\xspace}
\newcommand\gamLike{\textsf{gamLike}\xspace}
\newcommand\gamlike{\gamLike}

\newcommand\pippi{\textsf{pippi}\xspace}
\newcommand\MultiNest{\textsf{MultiNest}\xspace}
\newcommand\multinest{\MultiNest}
\newcommand\great{\textsf{GreAT}\xspace}
\newcommand\twalk{\textsf{T-Walk}\xspace}
\newcommand\diver{\textsf{Diver}\xspace}
\newcommand\ddcalc{\textsf{DDCalc}\xspace}

\newcommand\xx{\raisebox{0.2ex}{\smaller ++}\xspace}
\newcommand\Cpp{\textsf{C\xx}\xspace}
\newcommand\Cppeleven{\textsf{C\raisebox{0.2ex}{\smaller ++}11}\xspace}
\newcommand\plainC{\textsf{C}\xspace}
\newcommand\Python{\textsf{Python}\xspace}
\newcommand\python{\Python}
\newcommand\Fortran{\textsf{Fortran}\xspace}
\newcommand\YAML{\textsf{YAML}\xspace}

\newcommand{\capgen}{\textsf{Capt'n General}\xspace}

\newcommand\beq{\begin{equation}}
\newcommand\eeq{\end{equation}}

\renewcommand{\url}[1]{\href{#1}{#1}}

\begin{document}

\begin{frontmatter}



\title{GAMBIT and its Application in the Search for Physics Beyond the Standard Model}


\author[imperial,oslo]{Anders Kvellestad}
\author[imperial,uq]{Pat Scott}
\author[adelaide]{Martin White}
\address[imperial]{\imperial}
\address[oslo]{\oslo}
\address[uq]{\uq}
\address[adelaide]{\adelaide}

\begin{abstract}
The Global and Modular Beyond-Standard Model Inference Tool (\gambit) is an open source software framework for performing global statistical fits of particle physics models, using a wide range of particle and astroparticle data. In this review, we describe the design principles of the package, the statistical and sampling frameworks, the experimental data included, and the first two years of physics results generated with it. This includes supersymmetric models, axion theories, Higgs portal dark matter scenarios and an extension of the Standard Model to include right-handed neutrinos. Owing to the broad spectrum of physics scenarios tackled by the \gambit community, this also serves as a convenient, self-contained review of the current experimental and theoretical status of the most popular models of dark matter.
\end{abstract}


\end{frontmatter}

\section{Introduction}

The core of the scientific method in the physical sciences is the identification of mathematical theories that describe some aspect of our Universe in terms of a number of free parameters. The use of global statistical fits, in either a Bayesian or frequentist framework, allows us to find the preferred parameter values of a candidate theory given experimental data, and to compare the abilities of different models to describe that data. Our current knowledge of particle physics is enshrined in the Standard Model (SM), and global fit techniques are routinely used to provide the most accurate estimates the parameters of the neutrino sector~\cite{Bari13,Tortola14,NuFit15}, the CKM matrix~\cite{CKMFitter}, and the electroweak sector~\cite{ZFitter,GFitter11}.

Despite its incredible successes in explaining experimental data, the SM still faces a number of experimental and theoretical challenges.  Many if not all of these can be explained by new physics Beyond the Standard Model (BSM).  Such physics could show up in a number of experiments, including direct searches for new particles at high energy particle colliders~\cite{ATLAS_diphoton,ATLAS15,CMS_SMS}, measurements of rare Standard Model processes~\cite{gm2exp,BelleII,CMSLHCb_Bs0mumu}, direct searches for dark matter~\cite{XENON2013,PICO60,LUX2016}, indirect astroparticle searches for distant annihilation or decay of dark matter~\cite{BringmannWeniger,LATdwarfP8,IC79_SUSY}, and cosmological observations~\cite{Planck15cosmo,Slatyer15a,keVsterile_whitepaper}. Unfortunately, despite the existence of many candidate theories beyond the SM, there is no unambiguous prediction of what we expect to observe, or in which experimental field we expect to observe it.  It is therefore highly likely that the next theory of particle physics will have to be pieced together by combining clues from a number of disparate fields and experiments. In the process, it is essential to also consistently combine \emph{null} results in experiments that had the potential to discover a given candidate theory, but failed to do so. Even in the complete absence of positive discoveries in the near future, it is essential to determine which candidate BSM theories are now comfortably excluded, and which regions of which candidate theories are now the most amenable to future discovery.

Global fits of BSM theories have thus been a very active area of research for well over a decade \cite{Baltz04,Allanach06,SFitter, Ruiz06}, with increases in computing power opening the option of exploring models with larger and larger parameter spaces. Nevertheless, it remains a considerable challenge to efficiently explore the high-dimensional parameter spaces of candidate theories whilst rigorously calculating likelihoods for a large range of experiments, each of which may require a costly simulation procedure. To further complicate matters, one must consistently handle systematic uncertainties that may be correlated across different datasets, resulting from either instrumental effects, or our imprecise knowledge of the nuclear, astro- or particle physics relevant to a given set of experiments. Prior to 2017, most global fits were focussed on supersymmetric theories, involving dedicated software that was built from the ground up with a knowledge of the supersymmetric parameters~\cite{Baltz04,Allanach06,SFitter, Ruiz06,Strege15,Fittinocoverage,Catalan:2015cna,MasterCodeMSSM10,2007NewAR..51..316T,2007JHEP...07..075R,Roszkowski09a,Martinez09,Roszkowski09b,Roszkowski10,Scott09c,BertoneLHCDD,SBCoverage,Nightmare,BertoneLHCID,IC22Methods,SuperbayesXENON100,SuperBayesGC, Buchmueller08,Buchmueller09,MasterCodemSUGRA,MasterCode11,MastercodeXENON100,MastercodeHiggs,Buchmueller:2014yva,Bagnaschi:2016afc,Bagnaschi:2016xfg,Allanach:2007qk,Abdussalam09a,Abdussalam09b,Allanach11b,Allanach11a,Farmer13,arXiv:1212.4821,Fowlie13,Henrot14,Kim:2013uxa,arXiv:1503.08219,arXiv:1604.02102,Han:2016gvr, Bechtle:2014yna, arXiv:1405.4289, arXiv:1402.5419, MastercodeCMSSM, arXiv:1312.5233, arXiv:1310.3045, arXiv:1309.6958, arXiv:1307.3383, arXiv:1304.5526, arXiv:1212.2886, Strege13, Gladyshev:2012xq, Kowalska:2012gs, Mastercode12b, arXiv:1207.1839, arXiv:1207.4846, Roszkowski12, SuperbayesHiggs, Fittino12, Mastercode12, arXiv:1111.6098, Fittino, Trotta08, Fittino06,
arXiv:1608.02489, arXiv:1507.07008, Mastercode15, arXiv:1506.02499, arXiv:1504.03260, Mastercode17}. These results typically covered low-dimensional subsets of the minimal supersymmetric SM (MSSM) or, in some cases the next-to-minimal variant, with relatively few global studies of other theories completed~\cite{Cheung:2012xb,Arhrib:2013ela,Sming14,Chowdhury15,Liem16,LikeDM,Banerjee:2016hsk,Matsumoto:2016hbs,Cuoco:2016jqt,Cacchio:2016qyh,BertoneUED,Chiang:2018cgb,hepfit,Matsumoto:2018acr}.

In 2017, the \gambit collaboration released the Global and Modular Beyond-Standard Model Inference Tool (\gambit) \cite{gambit}, an open-source package able to produce results in both the Bayesian and frequentist statistical frameworks, and easily extendible to new BSM models and new experimental datasets. A fully modular design enables much of the code to be reused when changing the theoretical model of interest. \gambit includes a wide variety of efficient sampling algorithms for posterior evaluation and optimisation, and ensures computational efficiency through massive, multi-level parallelisation, both of the sampling algorithms and individual likelihood calculations.

The purpose of this article is to give a brief introduction to the \gambit software and science programme, reviewing the most important results obtained with \gambit in the first few years since its initial release. These serve to illustrate the versatility of the code in attacking completely different BSM models, and the constraining power of the highly detailed and rigorous simulations of different particle and astroparticle datasets in \gambit. Given the centrality of dark matter (DM) in the current search for BSM physics, this review also serves as a convenient summary of the status of the most widely-studied DM candidates.

In Section~\ref{sec:gambit}, we describe the structure and design of the \gambit package, including the core framework, the means by which \gambit supports generic BSM models, the sampling and statistics module, and the various physics modules able to produce theoretical predictions and experimental likelihoods. In Section~\ref{sec:physics}, we summarise the results of recent \gambit global fits of various supersymmetric theories, Higgs portal and axion DM models, and a right-handed neutrino extension of the SM.  We then conclude in Section~\ref{sec:summary}.

\section{The \gambit software}
\label{sec:gambit}

The core \gambit software is written in \Cppeleven, but interfaces seamlessly with extensions and existing physics codes written in \python, \Mathematica, \Fortran and \plainC.  Since the release of \gambit \textsf{1.0.0} in 2017 \cite{gambit}, the most notable updates have been versions \textsf{1.1} (adding support for \Mathematica) \cite{gambit_addendum}, \textsf{1.2} (adding support for \python and higher-spin Higgs portal models) \cite{HP}, \textsf{1.3} (adding support for axion and axion-like particles) \cite{Axions} and \textsf{1.4} (adding support for right-handed neutrinos) \cite{RHN}.  The current public release is \textsf{v1.4.2}.  The source code is openly available from \href{http://gambit.hepforge.org}{http://gambit.hepforge.org} under the 3-clause BSD license.

\subsection{Core design}
\label{sec:core}

The core principles of \gambit's software design are modularity and flexibility.  All theoretical predictions and experimental likelihood evaluations are separated into a series of smaller, self-contained sub-calculations, with each sub-calculation represented by a single function. Each function is assigned a metadata string that identifies the physical quantity that the function is able to calculate.  Examples might be the mass of the lightest Higgs boson, or the likelihood for the latest run of the LUX direct detection experiment. Functions are further tagged with additional metadata strings indicating any other physical inputs required for them to run.  In the case of the Higgs mass, one might require e.g. the SM electroweak vacuum expectation value and the masses and couplings of various other particles.  In the case of the LUX likelihood, one might require the number of events observed by LUX, its detector efficiency as a function of nuclear recoil energy, and the theoretically predicted event rate.
At runtime, \gambit identifies which functions are actually required for the analysis of a given theory, and connects them dynamically in order to enable the calculation in the most efficient manner possible.

The individual functions are grouped together according to physics theme, into seven different \textbf{physics modules}.  We describe these specific modules in Section \ref{sec:modules} below.  Individual functions are thus referred to as \textbf{module functions}.  The module functions are the true building blocks of a \gambit analysis, allowing the code to automatically adapt itself to incorporate new observables, likelihoods, theories and experimental datasets.  The metadata string associated with the output of a module function is referred to as its \textbf{capability}, and the metadata associated with the required inputs are referred to as \textbf{dependencies}.  The process of dynamically connecting the outputs of module functions to the inputs of others at runtime thus consists of matching dependency strings to the capability strings of other functions (and ensuring that their \Cpp types also match).  This process is known as \textbf{dependency resolution}, and is performed by the \gambit dependency resolver.

\begin{figure}[t]
	\centering
	\includegraphics[width = 0.75\textwidth]{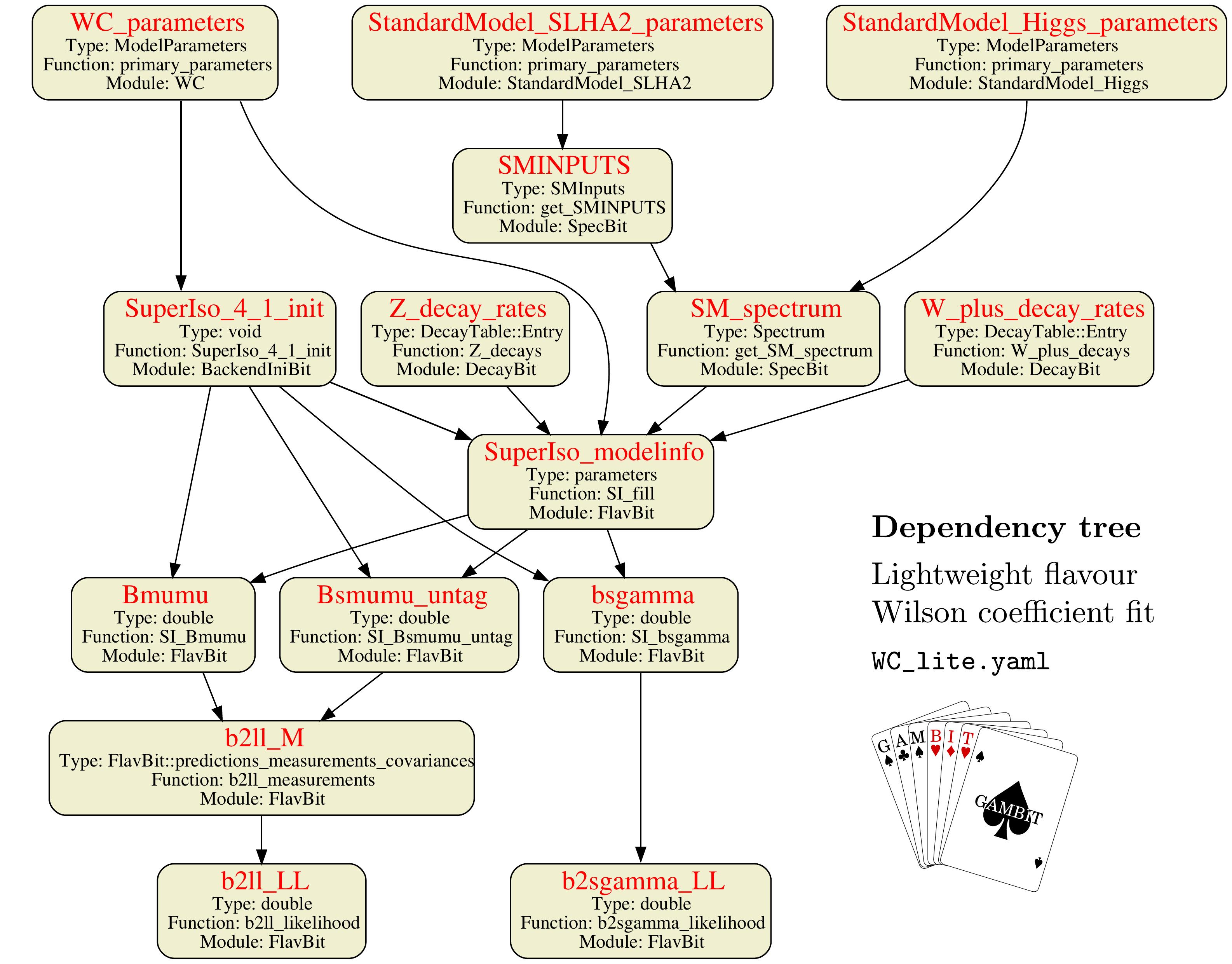}
	\caption{An example \gambit dependency tree for a simple fit of flavour Wilson coefficients to $b\to s\gamma$ and $B\to ll$ data. Boxes (graph nodes) correspond to single module functions.  Function capabilities are marked in red, and return types of the functions, their actual function names and enveloping modules are indicated in black.  Arrows (graph edges) indicate the direction of information flow, from the capability (output) of one function to the dependencies (inputs) of others.  The input file used to instigate this fit (\texttt{WC\_lite.yaml}) is one of the example files distributed with \gambit.  This particular fit makes use of the \gambit modules \flavbit, \specbit and \decaybit, as well as the backend (external package) \superiso \textsf{4.1} \cite{Mahmoudi:2007vz,Mahmoudi:2008tp,Mahmoudi:2009zz}.}
	\label{fig:dep_tree}
\end{figure}

The result is a (potentially extremely large) directed graph connecting the outputs and inputs of different module functions, known as a \textbf{dependency tree}.  The module functions themselves constitute graph nodes, and their resolved dependencies graph edges.  An example of such a graph is shown in Fig.\ \ref{fig:dep_tree}.  For such a graph to constitute a viable computational pathway to all theoretical predictions and experimental likelihoods of interest, two things are required.  The first is for the actual numerical values of some dependencies to be known in advance.  These are the parameters of the theory that the user wishes to analyse, and must be chosen `from on high' before the dependency tree can be evaluated.  These are selected by the user's chosen statistical parameter sampling algorithm, discussed below in Section \ref{sec:stats}.

From the values of a model's parameters, all other intermediate quantities can be obtained, as long as the second criterion is also met. This condition is that no closed loops exist in the graph, i.e.\ there are no dependencies of any module function upon things that can only be computed by knowing the result of the function. Many algorithms exist within graph theory for taking such a directed acyclic graph and obtaining a linear ordering of its nodes that respects the underlying structure of the graph.  \gambit uses the \textsf{Boost::Graph} library to obtain such an ordering, and then employs that ordering to evaluate the module functions in turn.  This ensures that all module functions run before any other functions that depend upon their results.  Within topologically equivalent subsets of the ordering, \gambit also further dynamically optimises the module function evaluation order for speed, according to previous function evaluation times and likelihoods.

Module functions may also make use of functions provided by external packages, or \textbf{backends}.  These are also connected dynamically at runtime to module functions by the dependency resolver, in much the same way as it ensures that dependencies upon the results of other module functions are fulfilled.  This layer of abstraction allows \gambit to provide its module functions with seamless and interchangeable access to functions from external codes written in \plainC, \Cpp, \python\textsf{2}, \python\textsf{3}, \Mathematica and all variants of \Fortran.  The \gambit build system allows users to select and automatically download, configure and build whatever combination of backends they prefer to use, and the dependency resolver automatically adapts to the presence or absence of different backends when selecting which functions to connect to others.  Backends presently supported in version \textsf{1.4.2} of \gambit are \capgen \cite{HP}, \darksusy \cite{darksusy4,darksusy}, \ddcalc \cite{DarkBit,HP}, \feynhiggs \cite{Heinemeyer:1998yj,Heinemeyer:1998np,Degrassi:2002fi,Frank:2006yh,Hahn:2013ria}, \gamlike \cite{DarkBit}, \gmtwocalc \cite{gm2calc}, \higgsbounds \cite{Bechtle:2008jh,Bechtle:2011sb,Bechtle:2013wla}, \higgssignals \cite{HiggsSignals}, \micromegas \cite{Belanger:2001fz,Belanger:2004yn,Belanger:2006is,Belanger:2008sj,Belanger:2010gh,Belanger:2013oya,micromegas}, \nulike \cite{IC22Methods,IC79_SUSY}, \pythia \cite{Sjostrand:2006za,Sjostrand:2014zea}, \spheno \cite{Porod:2003um,Porod:2011nf}, \superiso \cite{Mahmoudi:2007vz,Mahmoudi:2008tp,Mahmoudi:2009zz}, \susyhd \cite{Vega:2015fna} and \susyhit \cite{Djouadi:2006bz}.  Many more are also already supported in the current development version, which will be released in 2020.

\begin{figure}[tbp]
	\centering
	\includegraphics[width = \textwidth]{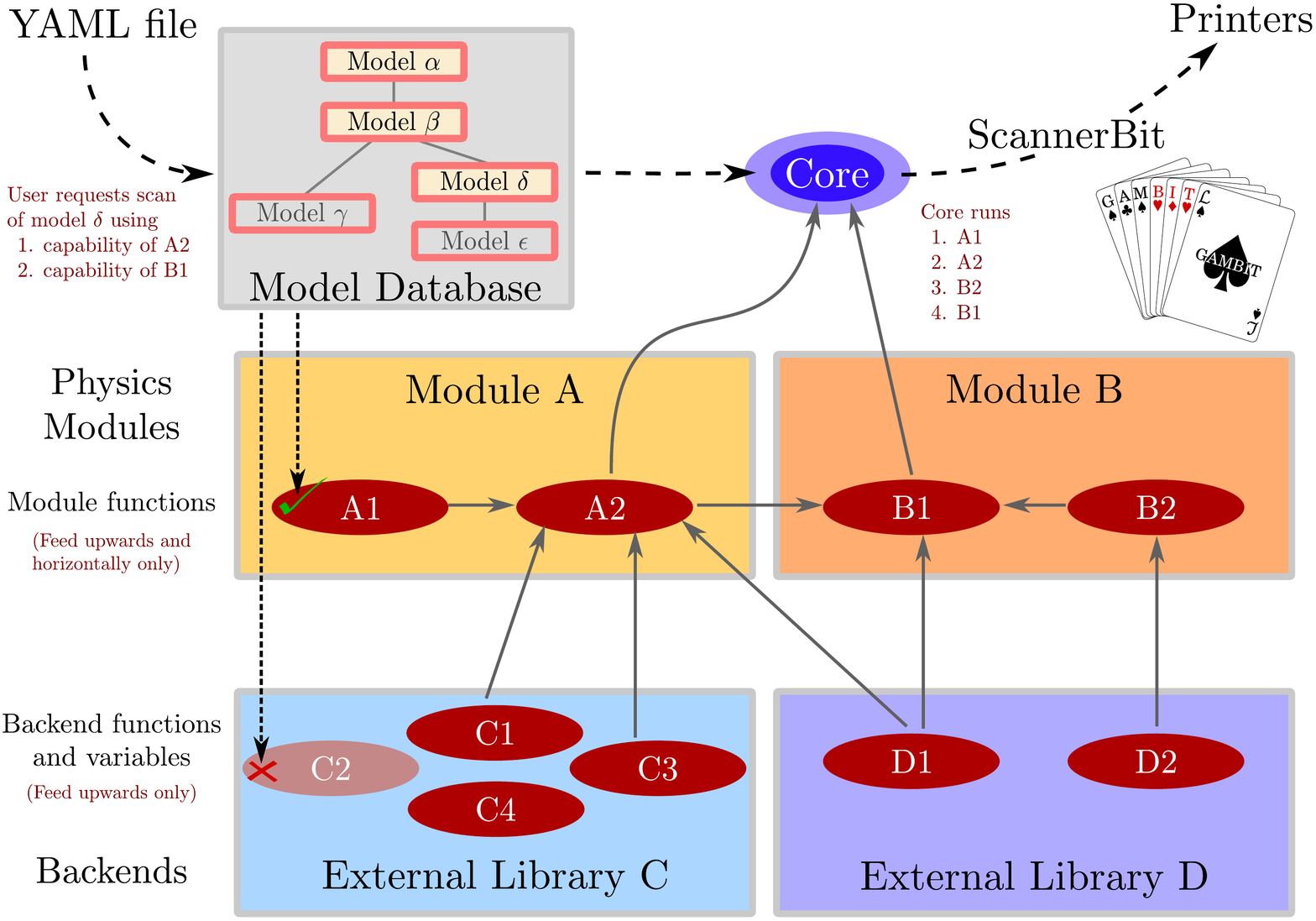}
	\caption{The overall structure of a \gambit run, illustrating the roles of the input \YAML file, modules, module functions, capabilities, dependencies, backends, backend functions, dependency resolver, hierarchical model database and the sampling machinery.  The user specifies one or more models to scan in the input \YAML file, and chooses likelihoods and observables to compute in the scan, making their choice by capability rather than by choosing specific functions.  The dependency resolver automatically identifies and connects appropriate module and backend functions in order to facilitate the computation of the requested likelihoods and observables, and the scanning machinery (\scannerbit) selects parameter combinations to pass through the resulting dependency tree.  From \cite{gambit}.}
	\label{fig:corechain}
\end{figure}

Actual runs of \gambit are driven by a single input file, in \YAML format.  In this file, the user selects the model(s) to analyse, gives details of which algorithms to use in order to sample the models' parameters, and provides a list of all likelihoods and physical observables that should be calculated in the scan.  The model parameter values constitute one boundary condition for dependency resolution (the dependency tree must begin from the parameters), and the target likelihoods and observables the other (the final outputs of the tree must be the required likelihoods and observables).  The dependency resolver is then responsible for identifying and filling in all the required steps in between.  To help direct this process and break degeneracies in the valid choices available to the dependency resolver at each step, the \YAML input file may also set rules that the dependency resolver must respect. These may be e.g.\ restrictions about which functions should be selected to fill which specific dependencies, or which version of a given backend should be used throughout the run.  These rules can be arbitrarily complicated, general or specific.  A rule can also contain explicit keyword options that will be passed to all module functions that fulfil the rule, allowing enormous control to be exercised over the details of the individual calculations from a single input file.

The core design elements of \gambit described so far in this section are module functions, backend abstraction, dependency resolution, and an input format that borders on its own programming language.  Together, these combine to provide an extremely flexible and extendible framework for performing global analyses of theories for BSM physics.  Fig.\ \ref{fig:corechain} illustrates how all of these features work together to enable a \gambit scan.  Further technical details can be found in Ref.\ \cite{gambit}.

\subsection{Model support}
\label{sec:models}

Another feature illustrated in Fig.\ \ref{fig:corechain} is the \gambit hierarchical model database.  Models are defined both in terms of their parameters, and in terms of their relationships to each other via parameter translation routines.  Models may descend from one another, meaning that a parameter combination in a child model can be translated `up' its family tree to a point in an appropriate subspace of its parent model, or in any other more distant ancestor model.  Cross-family `friend' translation pathways can also be defined.  These pathways allow module functions to be designed to work with one model, but to be used with another model without further alteration, so long as a translation pathway exists from one model to the other.

Module functions, backend functions and all rules set in \YAML files can be endowed with model-specific restrictions.  This allows the model-dependence of every sub-calculation to be tracked explicitly, and for the dependency resolver to explicitly ensure that the entire dependency tree of every scan is validated for use with the model under investigation.

\begin{sidewaysfigure}[tbp]
	\centering
	\includegraphics[width = \textheight]{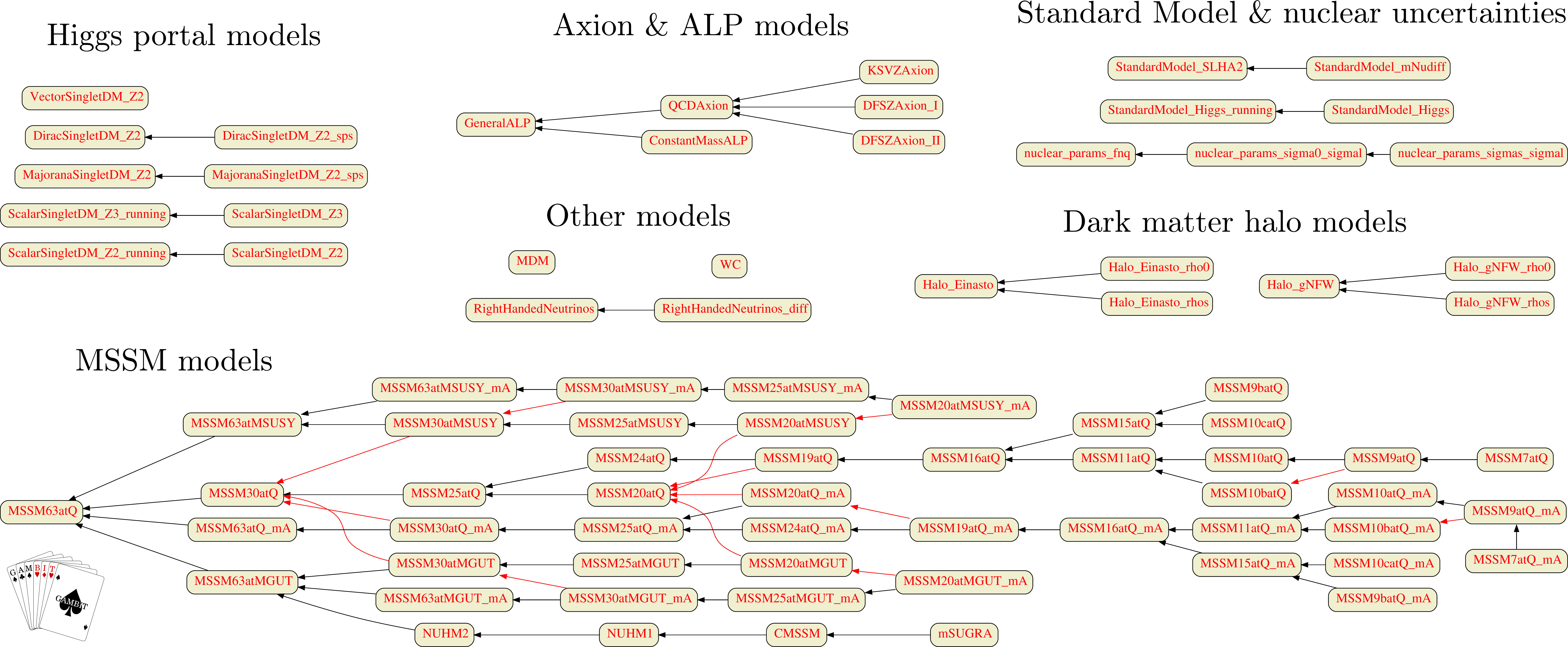}\vspace{2mm}
	\caption{Hierarchical model database of \gambit \textsf{1.4.2}. Models are shown as boxes, child-to-parent translations as black arrows, and friend translations as red arrows.}
	\label{fig:model_tree}
\end{sidewaysfigure}

The complete model database of \gambit \textsf{1.4.2} is shown in Fig.\ \ref{fig:model_tree}.

\subsection{Sampling and statistics}
\label{sec:stats}

In carrying out a global statistical analysis of a BSM theory, one may be interested in determining which parameter combinations are able to explain the totality of observed data within a given model, and to what extent -- or one may be more interested in using the experimental data to choose between entire theories.  The first of these tasks is parameter estimation, whereas the second is model comparison.  There are two philosophically distinct ways of posing both these questions:
\begin{enumerate}
\item How probable is it that we would have observed the data that we have, if a model and a specific combination of its parameters were true?
\item How probable is it that a model (or a specific combination of its parameters) is true, given the data that we have observed to date?
\end{enumerate}
Question 1 concerns frequentist statistics, whereas Question 2 is fundamentally Bayesian.

In the context of parameter estimation, the choice of question dictates whether the appropriate quantity to consider is the frequentist profile likelihood, or the Bayesian posterior.  The profile likelihood for some parameters of interest $\boldsymbol{\theta}$ is the maximum value of the likelihood at each parameter combination $\boldsymbol{\theta}$, regardless of the values of any other parameters $\boldsymbol{\alpha}$:
\begin{equation}
  \hat{\mathcal{L}}(\boldsymbol{\theta}) = \mathrm{max}_{\boldsymbol{\alpha}}\,\mathcal{L}(\boldsymbol{\theta},\boldsymbol{\alpha}),
\end{equation}
where the parameters $\boldsymbol{\alpha}$ are other `nuisance' parameters not of direct interest. Conversely, the Bayesian posterior probability distribution for the parameters $\boldsymbol{\theta}$ is given by Bayes' Theorem as the integral of the likelihood over $\boldsymbol\alpha$, weighted by ones prior belief $\pi(\boldsymbol{\theta},\boldsymbol{\alpha})$ as to the plausibility of different values of $\boldsymbol{\theta}$ and $\boldsymbol{\alpha}$:
\begin{equation}
\mathcal{P}(\boldsymbol\theta) = \int \mathcal{P}(\boldsymbol\theta,\boldsymbol{\alpha})\,\mathrm{d}\boldsymbol{\alpha} = \frac{1}{\mathbb{Z}}\int \mathcal{L}(\boldsymbol\theta,\boldsymbol{\alpha})\pi(\boldsymbol\theta,\boldsymbol{\alpha})\,\mathrm{d}\boldsymbol{\alpha}.
\end{equation}
Here $\mathbb{Z} \equiv \int \mathcal{L}(\boldsymbol\theta,\boldsymbol{\alpha})\pi(\boldsymbol\theta,\boldsymbol{\alpha})\,\mathrm{d}\boldsymbol{\alpha}\,\mathrm{d}\boldsymbol{\theta}$ is a normalisation factor referred to as the model evidence; taking ratios of evidences of different models is the most common method of Bayesian model comparison.  In contrast, frequentist model testing typically involves building up the distribution of the likelihood or other test statistic by simulation, in order to determine the precise probability of obtaining the observed (or worse-fitting) data if the model is assumed to be correct.

The choice of Bayesian posterior or profile likelihood has strong implications for the required design of the algorithm with which to sample the model parameters: efficiently obtaining converged estimates of profile likelihood and posterior distributions requires drastically different sampling distributions.  In neither case is random sampling at all sufficient nor correct, whether for accurate estimation of statistical properties nor for making statements about what is `typical' or `normal' within the parameter space of a given theory.  Efficient profile likelihood evaluation requires fast location of and convergence towards the maximum likelihood, whereas efficient posterior and evidence evaluation requires samples obtained with a density approximately proportional to the value of the posterior itself (as indeed is the case in most other numerical integration problems).

Sampling and statistical considerations in \gambit are handled mostly by the \scannerbit module \cite{ScannerBit}.  It contains all the tools necessary to transform the likelihood function provided by the dependency resolver into converged profile likelihoods and Bayesian posteriors.  It also facilitates Bayesian model comparison by calculating evidences (see e.g.\ Refs.\ \cite{HP,Axions} for recent examples), and frequentist model testing by providing information that can be used to perform statistical simulations (see Ref.\ \cite{EWMSSM} for a detailed example).

For Bayesian analyses, \scannerbit provides a series of different prior transformations, allowing the user to choose what assumptions to make about the probabilities of different model parameter at the beginning of a run, and to sample accordingly.

\scannerbit contains interfaces to various built-in and external implementations of a number of leading sampling algorithms.  These include algorithms optimised for profile likelihood evaluation, and algorithms optimised for posterior and evidence calculations.  Amongst these are \twalk \cite{ScannerBit}, a built-in ensemble Markov Chain Monte Carlo (MCMC) well suited to posterior evaluation, \great \cite{great}, a regular MCMC, \diver \cite{ScannerBit}, a differential evolution optimiser able to efficiently map profile likelihoods, and \multinest \cite{Feroz:2007kg,Feroz:2008xx} and \textsf{polychord} \cite{Handley:2015}, nested samplers well suited to evidence and posterior evaluation.  Detailed performance comparisons between the different samplers can be found in Ref.\ \cite{ScannerBit}.

For consistency and the convenience of module function writers, \gambit also provides a series of relatively simple pre-profiled and pre-marginalised likelihood functions \cite{gambit}.  These functions provide likelihoods where the influence of one or more Gaussian or log-normally distributed nuisance parameters is profiled or integrated out without the assistance of explicit sampling by \scannerbit.

\subsection{Physics modules}
\label{sec:modules}
The physics content of \gambit currently resides in seven modules, which contain the module and backend functions relevant for all necessary theoretical calculations, simulations of particle astrophysics experiments and likelihood calculations. Future \gambit updates will both refine the code in each module, and add new modules for new branches of physics (such as the forthcoming \textsf{CosmoBit} module).

\subsubsection{SpecBit}

BSM physics theories necessarily introduce new particles.  The first step in evaluating the likelihood of any parameter combination in a new theory is typically to calculate the masses and decay branching fractions of the new particles. These calculations get very complicated once one moves beyond tree level, as loop corrections can involve any number of new states in the theory, and loop corrections that shift the masses and decays of the existing SM particles must also be taken into account.

Particle mass and coupling calculations are handled in the \specbit module, which includes module functions for obtaining the pole masses and mixings of all new physical states in a model, scheme-dependent quantities such as those defined in the \DRbar and \MSbar schemes, and SM masses and couplings (e.g. couplings of the SM-like Higgs). Generally, this information is obtained by running an appropriate spectrum generator but, in the simple case that the pole masses of a model are specified as input parameters, \specbit simply formats the information to match that expected from a spectrum generator. In any case, it is important to realise that a spectrum cannot be stored simply as a set of numbers, since different experimental likelihoods may require predictions of running particle properties at different physical scales. Thus, \specbit facilitates the passing of a spectrum object to module functions that contains knowledge of the renormalisation group equations of a model, allowing module functions in other parts of the \gambit code to locally run the \DRbar or \MSbar parameters to the appropriate scale. Although \specbit can be extended to include any model, its development has tended to proceed through updates that add functionality for the specific models targeted in \gambit physics papers. To date, this includes functions for GUT-\cite{CMSSM} and weak-scale \cite{MSSM,EWMSSM} parameterisations of the MSSM, singlet DM models with either a scalar, fermion or vector DM candidate \cite{SSDM,SSDM2,HP}, minimal electroweak triplet and quintuplet DM \cite{Piteration,McKay2}, and a low-energy object that holds SM-like particle information. A range of backends is used to supply the \specbit calculations, including \SPheno \cite{Porod:2003um,Porod:2011nf} and \FlexibleSUSY \cite{Athron:2014yba} for BSM mass spectrum calculations. The latter is typically used for all spectrum generation requirements outside of the MSSM, including for the scalar singlet model examples described in this review. The Higgs and $W$ masses can also be calculated via the \FeynHiggs \cite{Heinemeyer:1998yj,Heinemeyer:1998np,Degrassi:2002fi,Frank:2006yh,Hahn:2013ria,Bahl:2016brp} backend.

\subsubsection{DecayBit}

Particle decay calculations are handled by the \decaybit module, after accepting the masses and couplings of particles from \specbit. These are used to calculate decay widths and branching fractions for each particle, which are stored in a single decay table entry for each particle. The collection of entries is then gathered into a full \GB decay table, which is passed on to other \GB modules.

\decaybit includes known SM particle decays, modifications of SM particle decays through new physics effects, and the decays of BSM particles. For the SM, \decaybit contains the Particle Data Group results for the total widths for the $W$, $Z$, $t$, $b$, $\tau$ and $\mu$ (plus antiparticles), and for the most common mesons $\pi^0$, $\pi^\pm$, $\eta$, $\rho^0, \rho^\pm$ and $\omega$~\cite{PDB}. In addition, partial widths to all distinct final states are provided for $W$, $Z$, $t$, $b$, $\tau$, $\mu$, $\pi^0$ and $\pi^\pm$. These ``pure SM'' decays are used in \gambit whenever an SM decay acquires no BSM contribution in a model, or when the only effect of the BSM physics is to introduce a new decay channel, in which case the pure SM decays can be appended to the new list of decay channels. For the pure SM Higgs boson, the user can decide whether to calculate the partial and total decay widths at the predicted value of the Higgs mass with \feynhiggs, or to use pre-computed tables provided in \decaybit, sourced from Ref.\ \cite{YellowBook13}.

BSM decays are handled on a model-by-model basis. For Higgs portal DM models, \decaybit contains analytic expressions for the partial width for a Higgs decay to two DM particles, and this decay is added to the list of SM Higgs partial widths, before rescaling the decay branching fractions and the total width. For MSSM variants, \decaybit calculates both the decays of all sparticles and additional Higgs bosons, and the SUSY corrections to the decays of the SM-like Higgs boson and the top quark. Higgs decay results may be sourced from either \HDECAY via \SUSYHIT, or \FeynHiggs, whilst top quark decays are only available via \FeynHiggs. Sparticle decays are obtained from \SDECAY via \SUSYHIT, but we note that a patch to the code is required to allow \GB to call functions from a shared library, and to solve problems with negative decay widths for some models due to large and negative 1-loop QCD corrections. Full details are given in~\cite{SDPBit}; the patch is applied automatically when \SUSYHIT is retrieved and built from within \gambit.

A recent update of \decaybit has seen the addition of routines for observables relating to right-handed neutrino studies. This includes the invisible decay width of the $Z$ boson $\Gamma_{\rm{inv}}$, and the leptonic $W$ boson decay widths $\Gamma_{W\to e \bar\nu_e}$, $\Gamma_{W\to \mu \bar\nu_\mu}$ and $\Gamma_{W\to \tau \bar\nu_\tau}$. Measurements and uncertainties are taken from Ref.~\cite{PDG17}, whilst theoretical results are taken from Refs.~\cite{Drewes:2015iva,Antusch:2014woa,Antusch:2015mia,Ferroglia:2012ir,Antusch:2015mia,Fernandez-Martinez:2015hxa,Abada:2013aba,Dubovyk:2018rlg,Antusch:2014woa}.

\subsubsection{PrecisionBit}

Some of the most severe constraints on BSM physics scenarios come from precision measurements of the electroweak sector, and other SM quantities. In \gambit, these are handled by the \precisionbit module, which provides nuisance likelihoods for SM quantities such as the top quark mass and strong coupling constant, which have been measured with high precision. A related function is the calculation of the BSM corrections to SM observables such as the mass of the $W$ boson and the weak mixing angle, and the provision of likelihood functions that compare these predictions with experimental data.

\begin{figure}[t]
\centering
\includegraphics[width=0.59\textwidth]{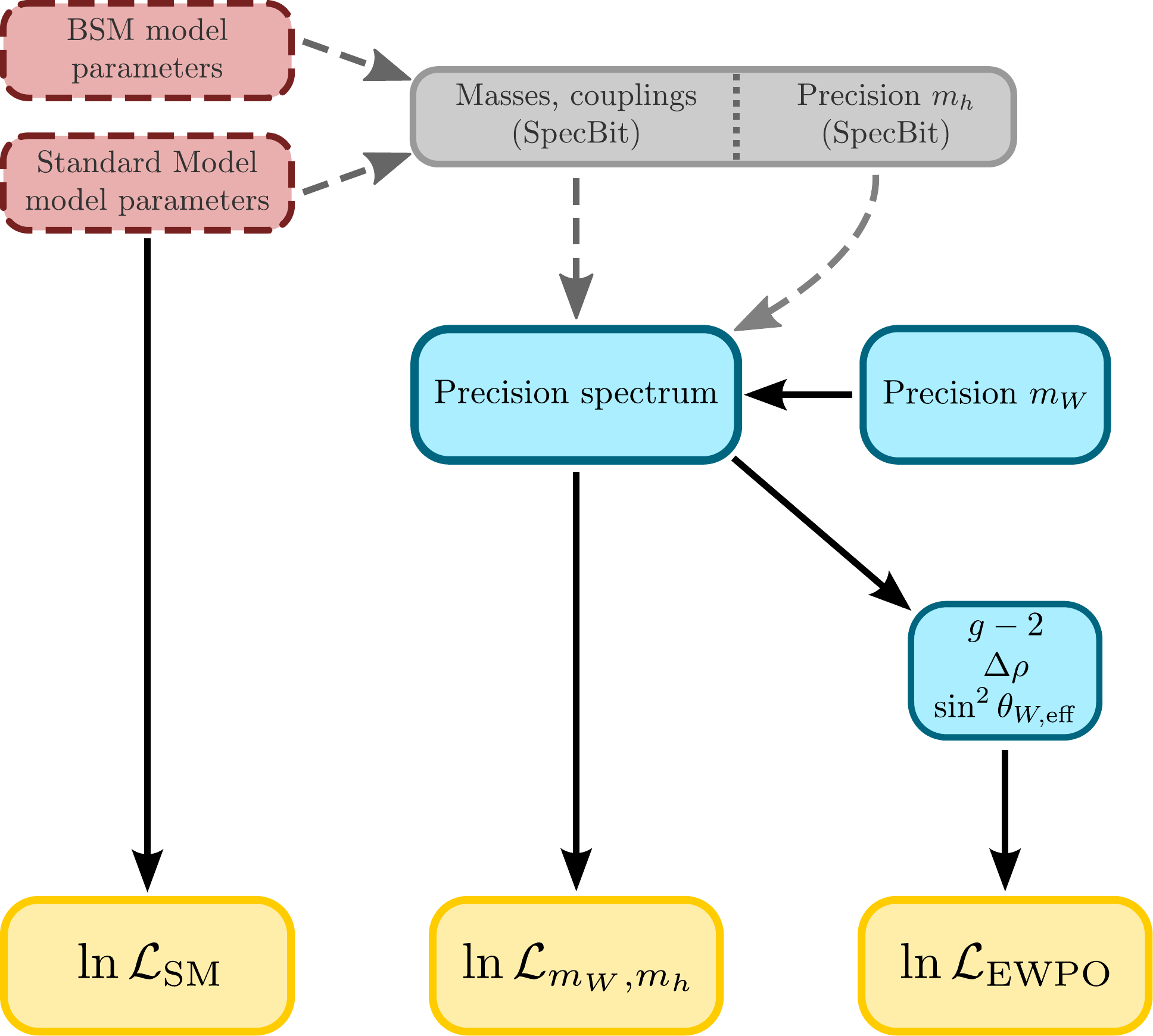}
\caption{Schematic representation of the structure of \precisionbit. From \cite{SDPBit}.}
\label{fig:precisionbit}
\end{figure}

A schematic representation of \precisionbit is shown in Fig.\ \ref{fig:precisionbit}, providing our first interesting example of the interaction between \gambit modules. Standard Model parameters that do not require BSM correction calculations are provided directly by the \gambit SM model, and are used in the calculation of SM nuisance likelihoods. The BSM parameters, meanwhile, are first used by \specbit in the calculation of particle masses, couplings and precision Higgs properties. \precisionbit then updates the results to form a precision-updated spectrum (including a dedicated calculation of the $W$ mass) which is used for calculating Higgs and $W$ mass likelihoods, in addition to a suite of electroweak precision observables (EWPO).

Likelihoods exist for the Fermi coupling constant ($G_F$), the fine-structure constant ($\alpha_{\mathrm{em}}$), the \MSbar light quark ($u$,$d$,$s$) masses at $\mu=2$~GeV, the charm ($m_c(m_c)$) and bottom ($m_b(m_b)$) masses, and the $W$, $Z$ and Higgs boson masses. There are also calculations and likelihoods for other precision observables such as the anomalous magnetic moment of the muon $a_\mu=\frac12(g-2)_\mu$, the effective leptonic weak mixing angle sin$^2\theta_{W,eff}$, and the departure from 1 of the ratio of the Fermi constants implied by the neutral and weak currents $\Delta\rho$. Note that, for the full suite of observables, calculations are currently only included for the MSSM; calculations for other models will be added as the corresponding models are implemented in \gambit.

Like \decaybit, \precisionbit has also recently been updated to include observables for right-handed neutrino studies. These consist of right-handed neutrino contributions to $m_W$ and sin$^2\theta_{W,eff}$.

\subsubsection{DarkBit}
BSM physics models that include particle DM candidates can potentially give rise to observable consequences in a wide range of astrophysical DM experiments.

In gamma-ray indirect detection, the \darkbit module contains a dedicated signal yield calculator, along with an interface to \gamlike, a likelihood calculator for current and future gamma-ray experiments. This combination can cope with signatures that result from an arbitrary mixture of final states, which significantly extends previous tools.

Further indirect detection constraints come from an interface to the \nulike neutrino telescope likelihood package~\cite{IC79_SUSY}.

Direct DM search experiments are handled by the dedicated \ddcalc package, which can be extended to include the effects of generic interactions between Weakly Interacting Massive Particles (WIMPs) and nucleons, as parameterised through effective operators. This includes both spin-dependent and spin-independent scattering.  The package models a wide range of direct search experiments including Xenon100, SuperCDMS, SIMPLE, LUX, PandaX, PICO-60 and PICO-2L.

Finally, the relic density of dark matter can be computed via interfaces to \ds and \micromegas~\cite{darksusy,micromegas_nu}, and used to constrain models by computing a likelihood based on the value observed by \textit{Planck} \cite{Planck18cosmo}.

\begin{figure*}[t]
  \centering
  \includegraphics[width=0.60\linewidth]{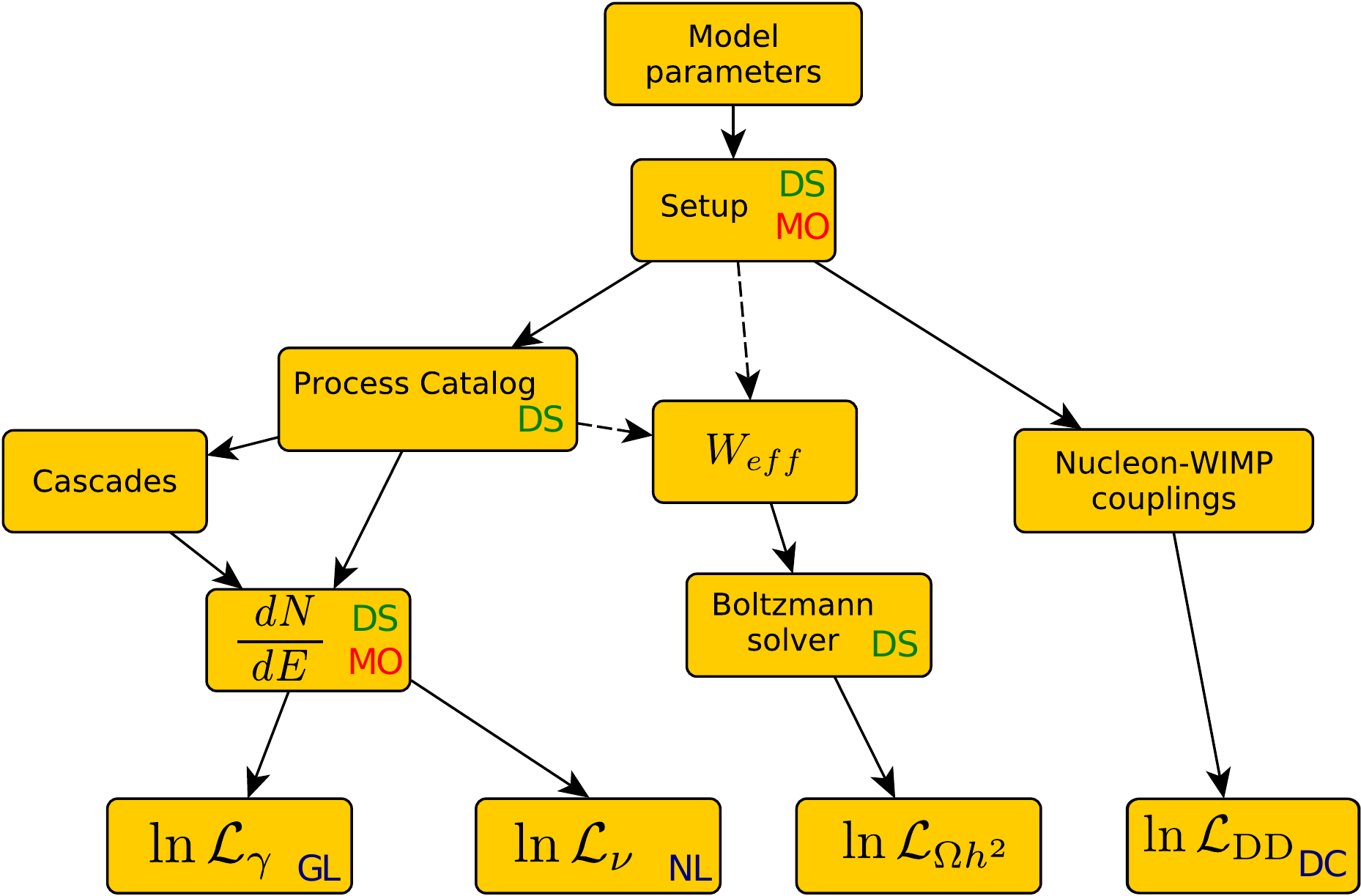}
  \caption{Schematic overview of the \darkbit module. The two-letter insets indicate what backend codes can
    be used:  \ds (DS), \micromegas (MO), \gamLike (GL), \nulike (NL) and
    \ddcalc (DC).  From \cite{DarkBit}.
    }
  \label{fig:flow}
\end{figure*}

The basic structure of \darkbit applicable to WIMP theories is sketched in Fig.\ \ref{fig:flow}, providing a good example of \gambit's modular design principle. None of the likelihoods requires knowledge of the BSM physics parameters, instead only requiring knowledge of derived quantities that can be shared between likelihood calculations. The first step in \darkbit is to create a Process Catalogue containing information on particle annihilation processes, using the particle masses and couplings provided by \specbit. For indirect detection calculations, this is used to create the gamma ray or neutrino spectrum of the annihilation products, via a weighted sum of indiviual contributions. For long decay chains, a native cascade decay Monte Carlo generator is used. This final annihilation spectrum is then passed to the likelihood calculators for gamma ray and neutrino telescope experiments. The Process Catalogue is also used to provide the effective annihilation rate for relic density calculations, which is then passed to a Boltzmann solver, followed by the relic density likelihood calculator. For direct dectection signatures, the model parameters are used to set the WIMP-nucleon couplings, which are then used in the calculation of the direct detection likelihood via the \ddcalc package.

A recent update of \darkbit has added various module functions required for the calculation of axion observables and likelihoods. The included observables are detailed in Section~\ref{sec:axionL}.

An important additional function of \darkbit is to constrain nuisance parameters for various astrophysical unknowns that strongly affect direct and indirect searches for DM. \darkbit contains likelihoods for the parameters of the local DM spatial and velocity distributions, plus the nuclear matrix elements that enter direct search WIMP-nucleon scattering calculations.

\subsubsection{FlavBit}

A very powerful indirect probe of BSM physics comes from the measurement of flavour physics processes, as theoretical predictions for these observables would be shifted by loop corrections from new particles. The excellent precision of flavour phyics measurements allows them to be sensitive to much higher energy scales than direct searches for new particles. Indeed, recent measurements from the LHCb experiment~\cite{Aaij:2016flj,Aaij:2015esa,Aaij:2015yra,Aaij:2014ora,Aaij:2013qta,Aaij:2019wad,Aaij:2014pli} and from $B$ factories~\cite{Lees:2012xj,Lees:2013uzd,Huschle:2015rga,Abdesselam:2016cgx,Abdesselam:2016llu,Aubert:2006vb,Lees:2015ymt,Wei:2009zv,Wehle:2016yoi} show tensions with the SM that are generating a considerable amount of theoretical interest.

\flavbit implements flavour physics constraints from rare decay observables using the effective Hamiltonian approach, in which the cross-sections for transitions from initial states $i$ to final states $f$ are proportional to the squared matrix elements $|\langle f |{\cal H}_{\rm eff}|i\rangle|^2$.  For example, an effective Hamiltonian for $b \rightarrow  s$ transitions given by
\begin{equation}
\mathcal{H}_{\rm eff}  =  -\frac{4G_{F}}{\sqrt{2}} V_{tb} V_{ts}^{*} \sum_{i=1}^{10} \Bigl(C_{i}(\mu) \mathcal{O}_i(\mu)+C'_{i}(\mu) \mathcal{O}'_i(\mu)\Bigr)\;.
\end{equation}
The local operators $\mathcal{O}_i$ represent long-distance interactions.  The Wilson coefficients $C_i$ can be calculated using perturbative methods, by requiring matching between the high-scale theory and the low-energy effective theory, at some scale $\mu_W$ which is of the order of $m_W$. The Wilson coefficients can then be evolved to the characteristic scale for $B$ physics calculations ($\mu_b$, of the order of the $b$ quark mass) using the renormalisation group equations of the \emph{effective} field theory. A similar approach can be taken to $b\rightarrow d$ transitions, using a different basis of low-energy operators. The original list of observables in \flavbit was divided into four categories:

\begin{itemize}
\item {\bf Tree-level leptonic and semi-leptonic decays: }includes decays of $B$ and $D$ mesons to leptons, including $B^\pm \to \tau \nu_\tau$, $B \to D^{(*)} \tau \nu_\tau$ and $B \to D^{(*)} \ell \nu_\ell$.
\item {\bf Electroweak penguin transitions: }includes measurements of rare decays of the form $B \to M \ell^+\ell^-$ (where $M$ is a meson with a smaller mass than the parent meson), such as angular observables of the decay $B^0 \to K^{*0} \mu^+\mu^-$.
\item {\bf Rare purely leptonic decays: }includes $B$ decays with only leptons in the final state, such as $B^0_{(s)} \to \mu^+ \mu^-$.
\item {\bf Other observables: }includes $b\to s$ transitions in the radiative decays $B \to X_s \gamma$, the mass difference ($\Delta M_s$ between the heavy $B_H$ and light $B_L$ eigenstates of the $B^0_s$ system, and kaon and pion decays (e.g. the leptonic decay ratio ${\cal B}(K^\pm\to \mu \nu_\mu)/{\cal B}(\pi^\pm\to \mu \nu_\mu)$).
\end{itemize}

Theoretical calculations for these processes are handled via an interface to \superiso~\cite{Mahmoudi:2007vz,Mahmoudi:2008tp,Mahmoudi:2009zz}. Experimental results used in the calculation of likelihoods come from a variety of sources, including the PDG, the BaBar and Belle experiments, the HFAG collaboration and the LHCb experiment. Full details are given in~\cite{FlavBit}.

More recently, \flavbit has been updated with observables relevant to right-handed neutrinos. These include:
\begin{itemize}
\item {\bf Lepton-flavour violating (LFV) muon and tau decay searches} performed by the MEG, BaBar, Belle, ATLAS, LHCb and SINDRUM collaborations~\cite{TheMEG:2016wtm,Aubert:2009ag,Hayasaka:2007vc,Lees:2010ez,Hayasaka:2010np,Aad:2016wce,Aaij:2014azz,Bellgardt:1987du}. LFV processes can also result in a neutrinoless $\mu - e$ conversion inside a nucleus, and these are included in the form of three results using Ti, Pb and Au nuclei obtained by the SINDRUM II experiment~\cite{Kaulard:1998rb,Honecker:1996zf,Bertl:2006up}.
\item {\bf Tests of lepton universality violation} in the semileptonic decays of $B$ mesons $B^{0/\pm} \to X^{0/\pm} l^+ l^-$, as performed by LHCb \cite{Aaij:2014ora,Aaij:2017vbb}.
\end{itemize}

A forthcoming major update to the \flavbit module will add an interface for \superisofour, with added support for theory uncertainty covariance matrices. The experimental likelihoods will also receive an update, via a new interface to the \heplike package \cite{HEPLike}.

\begin{figure}[tp]
\centering
\includegraphics[width=0.60\linewidth]{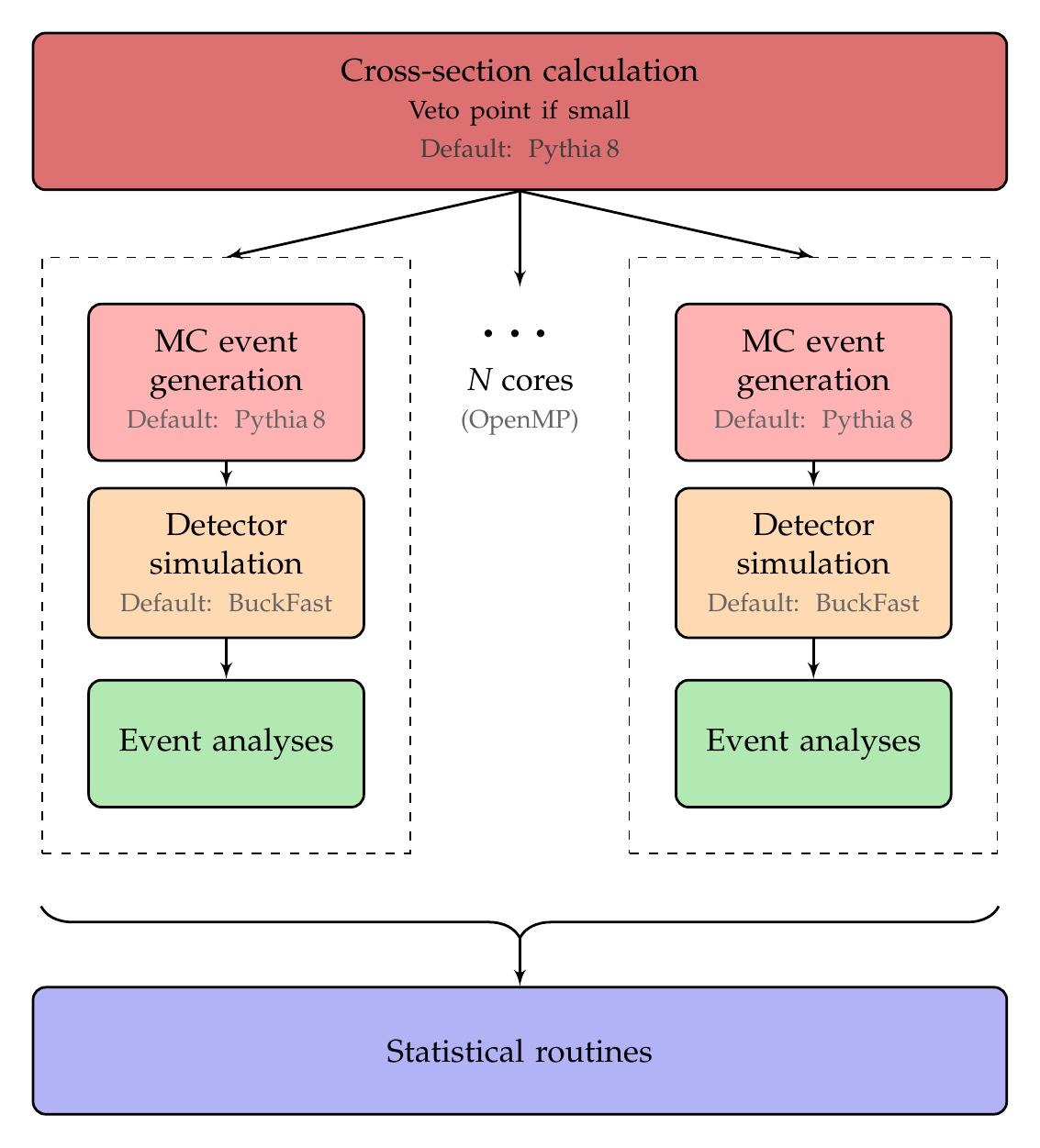}
\caption{Schematic diagram of the \colliderbit processing chain for LHC likelihoods. From \cite{ColliderBit}.}
\label{fig:lhcchain}
\end{figure}

\subsubsection{ColliderBit}
\label{sec:colliderbit}
A leading source of constraints on BSM physics models comes from high-energy collider searches for new particles, plus the relatively recent measurements of the Higgs boson mass and decay branching fractions. The \colliderbit module includes the most comprehensive list of recent LHC particle searches of any public package, alongside a new interpolation of LEP results for supersymmetric particle searches. Higgs signal strength and mass measurements (including limits on possible signatures arising from new Higgs bosons) are handled via an interface to the \higgssignals~\cite{HiggsSignals} and \higgsbounds~\cite{Bechtle:2008jh,Bechtle:2011sb} packages, which includes data from LEP, the Tevatron and the LHC.

LHC constraints are particularly difficult to model rigorously for general models. Searches for new particles are often optimised on, and interpreted in terms of, so-called ``simplified models'', which feature only a few options from the much broader phenomenology of the parent model. For example, searches for supersymmetric particles might assume that only a particular pair of sparticles is ever produced, with decays fixed to a particular final state. The resulting exclusion limit will never apply directly to a more general model, although one can obtain approximate limits by scaling individual simplified model limits by the known cross-sections and branching ratios for each parameter point~~\cite{Kraml:2013mwa,Papucci:2014rja}.

In \colliderbit, we provide more rigorous limits by performing an actual reproduction of the ATLAS and CMS limit-setting procedures, as shown in Fig.\ \ref{fig:lhcchain}. This includes a cross-section calculation for new particle production processes, followed by Monte Carlo simulation of LHC events for each parameter point using a custom parallelised version of the \pythiaeight generator~\cite{Sjostrand:2006za,Sjostrand:2014zea}. The results can either be fed at the truth level into code that reproduces the kinematic selections of a wide range of LHC analyses, or passed through a custom detector simulation based on four-vector smearing before analysis. Cross-sections are currently taken at leading order (plus leading log) from \pythiaeight, but a forthcoming update will allow user-specified cross-sections. The final step of the process is to calculate a combined likelihood by either taking the signal region in a given final state for each experiment that is expected to have the highest sensitivity to the model in question, or by using a covariance matrix for analyses in cases where this is published by the relevant experimental collaboration. The list of \colliderbit analyses is continually updated, and currently includes a broad selection of searches for supersymmetric particles, plus monojet searches for DM particles.

\subsubsection{NeutrinoBit}
\label{sec:neutrinobit}
The \neutrinobit module contains a variety of module functions for calculating observables and likelihoods in the neutrino sector, both for SM(-like) neutrinos and for right-handed neutrinos (RHNs). RHNs could cause observable consequences in a number of experiments, although it should be noted that the recent \gambit study focussed on models that are capable of explaining the light neutrino oscillation data, which excludes most sterile neutrino dark matter models. This is because long-lived RHNs would require very small couplings with SM matter, in which case their contribution to light neutrino mass generation is negligible.

\neutrinobit currently contains likelihoods dealing with the following classes of experimental data:

\begin{itemize}
\item \textbf{Active neutrino mixing: }\neutrinobit includes likelihoods for the 3-flavour SM-like active neutrino mixing observables $\theta_{12}$, $\theta_{13}$, $\theta_{23}$ (mixing angles), $\delta_{\mathrm{CP}}$ (CP-phase) and $\Delta m^2_{21}$ and $\Delta m^2_{3\ell}$ (mass splittings) with $\ell = 1$ for normal mass ordering and $\ell = 2$ for inverted mass ordering. The likelihoods use the one-dimensional $\Delta\chi^2$ tables provided by the NuFIT collaboration~\cite{Esteban:2016qun,NuFit}.  These in turn include results from the solar neutrino experiments Homestake (chlorine)~\cite{Cleveland:1998nv}, Gallex/GNO~\cite{Kaether:2010ag}, SAGE~\cite{Abdurashitov:2009tn}, SNO~\cite{Aharmim:2011vm}, the four phases of Super-Kamiokande~\cite{Hosaka:2005um,Cravens:2008aa,Abe:2010hy} and two phases of Borexino~\cite{Bellini:2011rx,Bellini:2008mr,Bellini:2014uqa}.  They also include results from the atmospheric experiments IceCube/DeepCore~\cite{Aartsen:2014yll}, the reactor experiments KamLAND~\cite{Gando:2013nba}, Double-Chooz~\cite{An:2016srz}, Daya-Bay~\cite{An:2016ses} and Reno~\cite{reno}, the accelerator experiments MINOS~\cite{Adamson:2013whj,Adamson:2013ue}, T2K~\cite{t2k} and NO$\nu$A~\cite{nova}, and the cosmic microwave background results from Planck~\cite{Ade:2015xua}.
\item \textbf{Lepton universality: }\neutrinobit contains likelihoods for lepton universality violation in fully leptonic decays of charged mesons, $X^+ \to l^+ \nu$.
\item \textbf{CKM unitarity: }The determination of the CKM matrix elements usually relies on the assumption that the active-sterile neutrino mixing matrix is zero. The presence of non-trivial mixing thus modifies the CKM matrix elements, and the experimentally-observed values can be used to simultaneously constrain the true CKM element values, and the active-sterile mixing matrix $\Theta$. \neutrinobit constructs a likelihood based on the deviations of the true values of $(V_{CKM})_{us}$ and $(V_{CKM})_{ud}$ from their experimentally-measured values.
\item \textbf{Neutrinoless double-beta decay: }In a double-beta decay process, two neutrons decay into two protons, with the emission of two electrons and two anti-neutrinos. Majorana neutrinos would give rise to lepton number violation, resulting in neutrinoless double-beta decay ($0\nu\beta\beta$). In addition, the exchange of RHNs can modify the effective neutrino mass $m_{\beta\beta}$, which is constrained by half-life measurements of $0\nu\beta\beta$ decay. The best upper limits currently come from the GERDA experiment (Germanium)~\cite{Agostini:2017iyd} with $m_{\beta\beta}<0.15-0.33\;\text{eV}$ (90\% CL), and KamLAND-Zen (Xenon)~\cite{KamLAND-Zen:2016pfg}, $m_{\beta\beta}<0.061-0.165\;\text{eV}$ (90\% CL). \neutrinobit uses these values to define one-sided Gaussian likelihoods, with theoretical calculations for RHN models taken from Refs.~\cite{Drewes:2016lqo,Faessler:2014kka}
\item \textbf{Big Bang Nucleosynthesis: }RHNs can affect the abundances of the primordial elements if they decay shortly before BBN, as the typical energy of the decay products is significantly higher than the plasma energy at that time. This can lead to the dissociation of formed nuclei, or the creation of deviations from thermal equilibrium.  The requirement that RHNs decay before BBN implies an upper limit on their lifetime which, in turn, results in a constraint on the total mixing with the active neutrinos. \neutrinobit currently includes a basic BBN likelihood that uses decay expressions from Refs.~\cite{Gorbunov:2007ak,Canetti:2012kh}, and requires the lifetime of each RHN to be less than 0.1s~\cite{Ruchayskiy:2012si}. A more comprehensive update will be released in future, associated with the new \textsf{CosmoBit} module.
\item \textbf{Direct RHN searches: }Direct searches for RHNs can be performed by looking for peaks in the lepton energy spectrum of a meson decay, looking for evidence of production in beam dump experiments, and by studying the decay of vector bosons or mesons in $e^+e^-$ or $pp$ colliders. \neutrinobit contains likelihoods for RHN searches at the PIENU~\cite{PIENU:2011aa}, PS-191~\cite{Bernardi:1987ek}, CHARM~\cite{Bergsma:1985is}, E949~\cite{Shaykhiev:2011zz,Artamonov:2014urb}, NuTeV~\cite{Vaitaitis:1999wq}, DELPHI~\cite{Abreu:1997uq}, ATLAS~\cite{Aad:2015xaa} and CMS~\cite{Sirunyan:2018mtv} experiments.
\end{itemize}

\section{Applications to new physics}
\label{sec:physics}
\subsection{Supersymmetry}
\label{sec:SUSY}

Supersymmetry (SUSY) has long been one of the leading candidates for BSM physics, owing to its potential for simultaneously answering several of the questions left open by the SM. In particular, the hierarchy problem and the dark matter ``WIMP miracle'' suggest the possible existence of SUSY states around the weak scale.

Most phenomenological explorations of SUSY take the MSSM as their starting point. On top of its minimal supersymmetrisation of the SM, the MSSM effectively parameterises our ignorance about the high-scale mechanism responsible for breaking SUSY. This is done by including in the Lagrangian all gauge-invariant and renormalisable terms that break SUSY ``softly'', that is, without re-introducing the quadratic Higgs mass divergences that gave rise to the hierarchy problem. In this way the MSSM provides a unified framework for exploring a wide range of possible manifestations of SUSY, but at the price of a vast parameter space: if no further assumptions are made the soft SUSY-breaking terms introduce more than one hundred free parameters.

Many different assumptions have been employed in the literature to reduce this parametric freedom and improve predictability.  The resulting models broadly fit in two categories.

The first category consists of high-scale models that take inspiration from the fact that SUSY can provide gauge coupling unification at some high Grand Unified Theory (GUT) scale, typically around $10^{16}$\,GeV. In these models a small number of unified mass and coupling parameters are defined at the GUT scale and then run down to the electroweak scale where phenomenological predictions are calculated. Thus, the assumption of high-scale unification constrains the model to a low-dimensional subspace of the full MSSM space, effectively imposing a set of characteristic correlations among the many MSSM parameters at the weak scale.

Probably the most studied SUSY model in this category is the Constrained MSSM (\CMSSM)~\cite{Nilles:1983ge}. Here the parameter space is reduced to only four continuous parameters and a sign choice: the unified soft-breaking scalar mass, $m_0$; the unified soft-breaking gaugino mass, $m_{1/2}$; the unified trilinear coupling, $A_0$; the ratio of the vacuum expectation values for the two Higgs doublets, $\tan\beta\equiv v_\mathrm{u}/v_\mathrm{d}$; and the sign of the supersymmetric Higgsino mass parameter $\mu$. The \CMSSM has been studied in global fits for over a decade~\cite{Han:2016gvr, Bechtle:2014yna, arXiv:1405.4289, arXiv:1402.5419, MastercodeCMSSM, arXiv:1312.5233, arXiv:1310.3045, arXiv:1309.6958, arXiv:1307.3383, arXiv:1304.5526, arXiv:1212.2886, Strege13, Gladyshev:2012xq, Kowalska:2012gs, Mastercode12b, arXiv:1207.1839, arXiv:1207.4846, Roszkowski12, SuperbayesHiggs, Fittino12, Mastercode12, arXiv:1111.6098, Fittino, Trotta08, Ruiz06, Allanach06, Fittino06, Baltz04, SFitter}, most recently in the \gambit analysis in~\cite{CMSSM}.

Two much-studied generalisations of the \CMSSM are the Non-Universal Higgs Mass models 1 and 2 (\NUHMone and \NUHMtwo)~\cite{Matalliotakis:1994ft,Olechowski:1994gm,Berezinsky:1995cj,Drees:1996pk,Nath:1997qm}. These models loosen the tight link in the \CMSSM between the Higgs sector and the sfermions by separating out the soft-breaking mass parameters of the Higgs sector from the common scalar mass parameter $m_0$. This is achieved by introducing either one (\NUHMone) or two (\NUHMtwo) additional parameters at the GUT scale. In recent years the \NUHMone and \NUHMtwo have been studied in several global fit analyses~\cite{MastercodeCMSSM, arXiv:1312.5233, Strege13, Fittino12, Mastercode15, Buchmueller:2014yva, arXiv:1405.4289}. \gambit global fits of the \NUHMone and \NUHMtwo were performed along with the fit of the \CMSSM in~\cite{CMSSM}. In Section~\ref{sec:GUT-SUSY} we summarise the \gambit results for these GUT-scale SUSY models.

The second category of MSSM sub-models are the weak-scale models. Here the focus is on exploring a broad range of weak-scale phenomenological scenarios in an economical manner, by varying only the MSSM parameters that most directly impact the observables under study. With all MSSM parameters defined near the weak scale, these models are mostly agnostic to questions concerning physics at very high scales, such as grand unification. The models are often labeled as \textsf{MSSM}$n$ (or as \textsf{pMSSM}$n$ for the \textit{phenomenological} \textsf{MSSM}$n$), with $n$ specifying the number of weak-scale MSSM parameters that are treated as free parameters.

Various such weak-scale models have been subjected to global fit analyses in the past few years~\cite{arXiv:1608.02489, arXiv:1507.07008, Mastercode15, arXiv:1506.02499, arXiv:1504.03260, Mastercode17}. The \gambit analyses in this category are~\cite{MSSM}, which looks at a seven-dimensional MSSM parameterisation (\MSSMseven), and~\cite{EWMSSM}, in which the fast LHC simulation capabilities of \colliderbit are used for a collider-focused fit of the four-dimensional MSSM ``electroweakino'' (chargino and neutralino) sector (\EWMSSM). We summarise the \gambit results for the \MSSMseven in Section~\ref{sec:MSSM7} and for the \EWMSSM in Section~\ref{sec:EWMSSM}.

The phenomenological richness of the MSSM means that a wide range experimental results are relevant for constraining the parameter space. The mass and signal strength measurements for the $125$\,GeV Higgs boson and the measurement of the relic density of dark matter are of particular importance. We note that the impact of the relic density measurement depends strongly on whether the SUSY model is assumed to account for the full relic density, or, more conservatively, some arbitrary fraction of it. The \gambit studies reviewed here all take the latter approach.

Measurements of electroweak precision observables such as $m_W$ and the muon $g-2$, and of flavour observables like $BR(B \rightarrow X_s \gamma)$ and $BR(B_{(s)} \rightarrow \mu^+ \mu^-)$, introduce further important requirements on the SUSY parameter space. Finally, the null results from direct and indirect dark matter searches, and from collider searches for sparticles and additional Higgs bosons, essentially rule out some parts of SUSY parameter space. Though, to determine the exact implications of such null-result collider searches is far from trivial, as will be illustrated by the \EWMSSM results discussed in Section~\ref{sec:EWMSSM}.

\subsubsection{Results for the \CMSSM, \NUHMone and \NUHMtwo}
\label{sec:GUT-SUSY}

The \gambit global fits of the \CMSSM, \NUHMone and \NUHMtwo in~\cite{CMSSM} are interpreted in terms of frequentist profile likelihood maps, identifying the best-fit point and the $1\sigma$ and $2\sigma$ preferred regions relative to this point. As the results for \NUHMtwo are qualitatively similar to those for \NUHMone, we here focus on the \CMSSM and \NUHMone results.

\begin{figure}
  \centering
  \includegraphics[width = 0.49\textwidth]{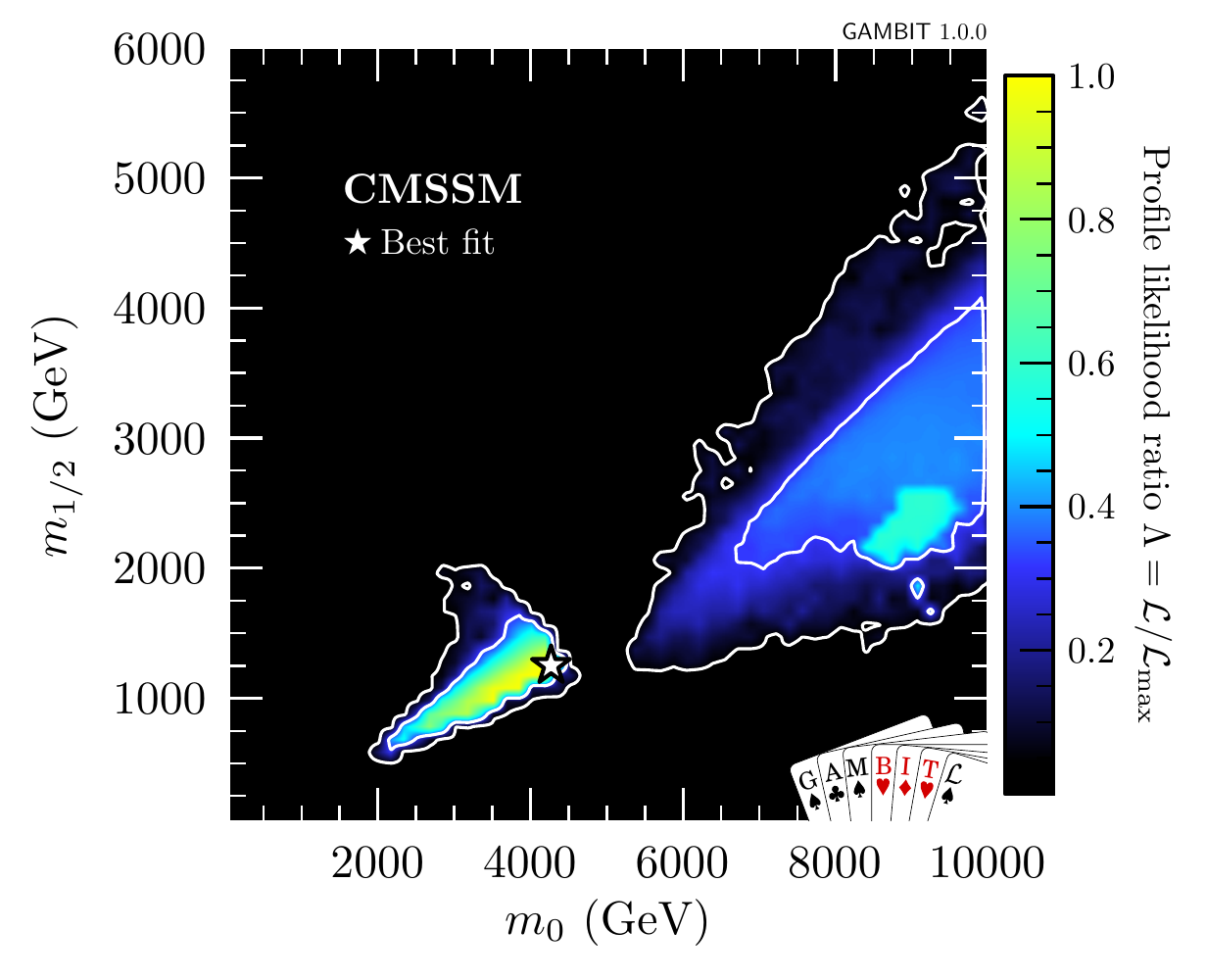}
  \includegraphics[width = 0.49\textwidth]{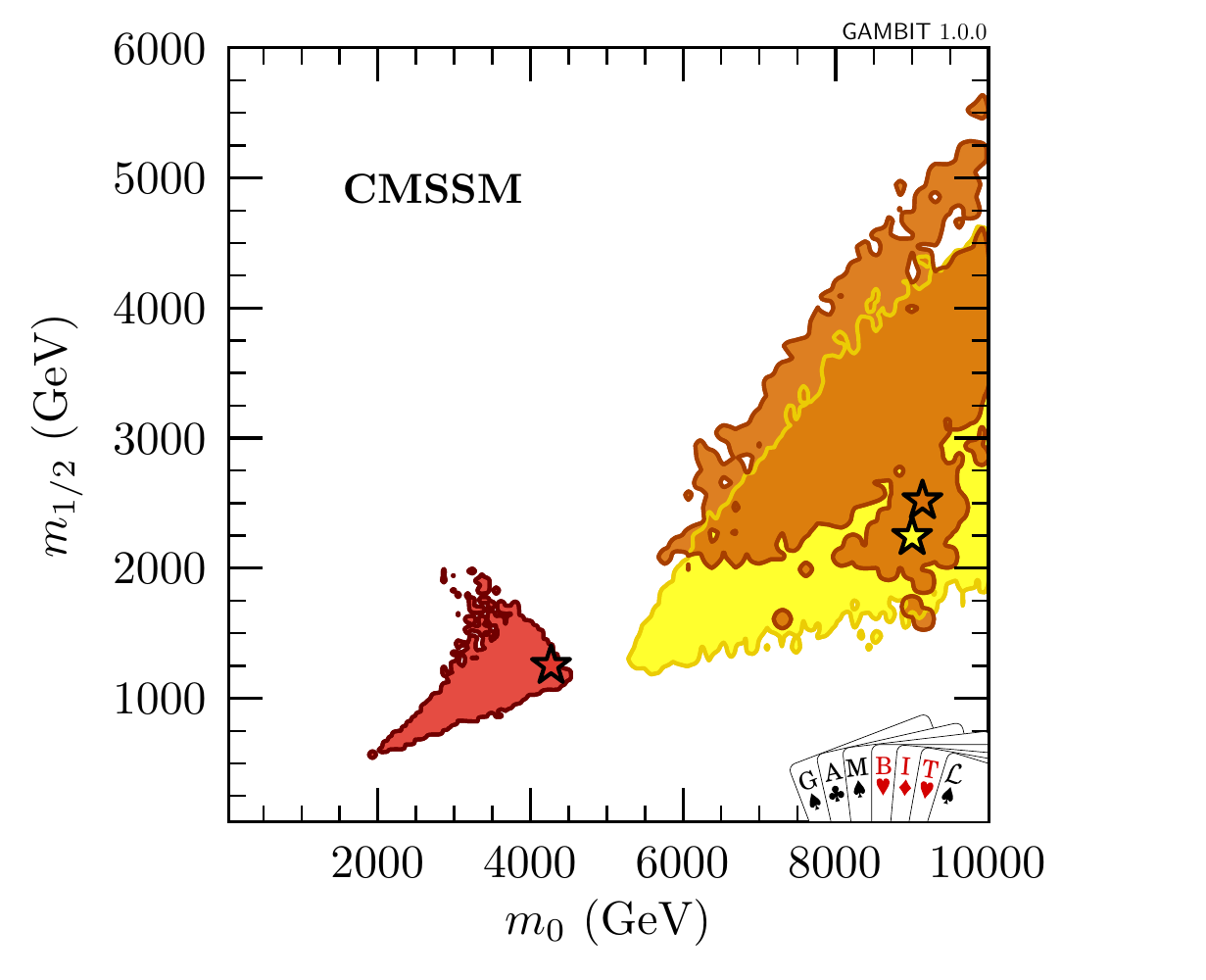}\\
  \includegraphics[width = 0.49\textwidth]{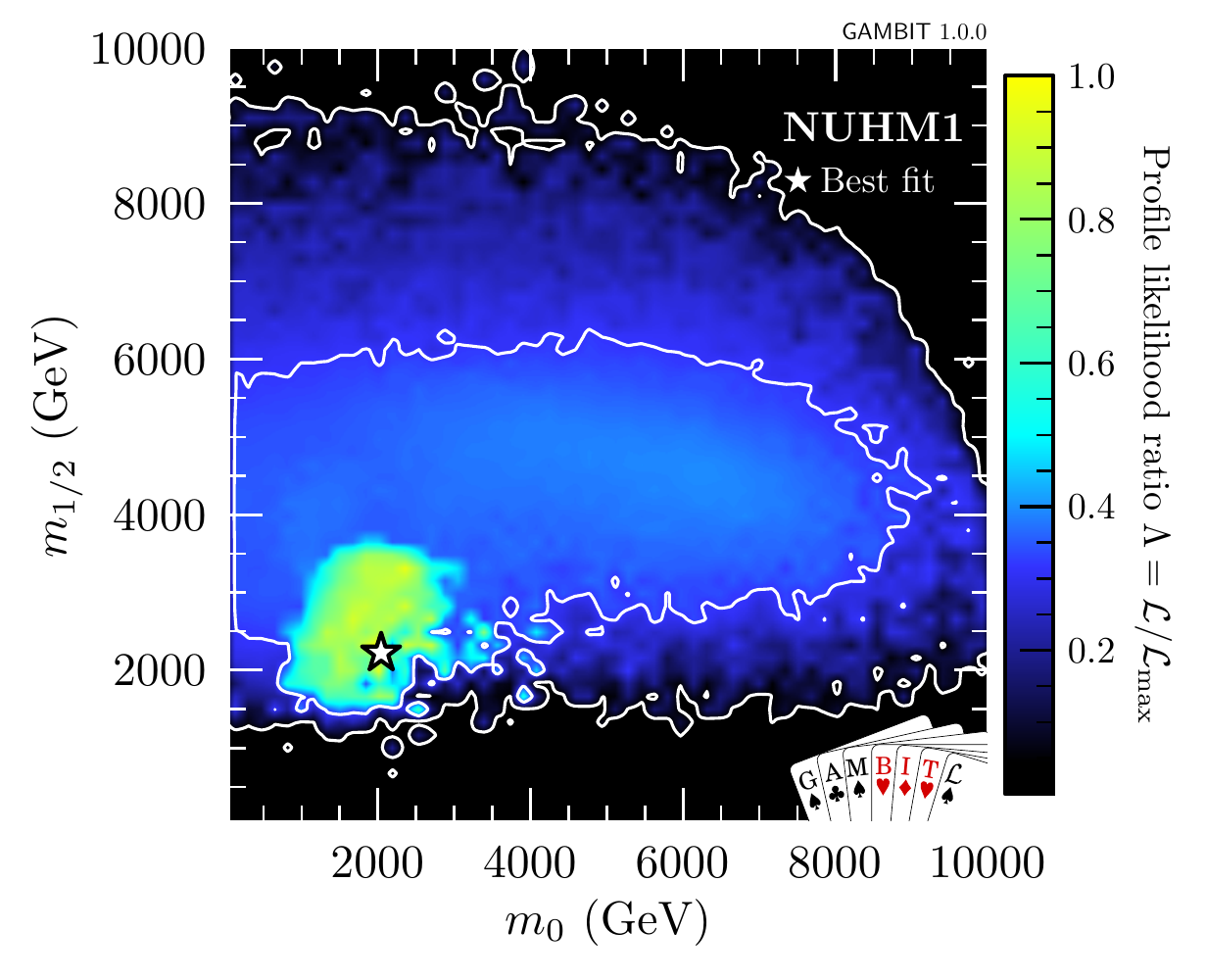}
  \includegraphics[width = 0.49\textwidth]{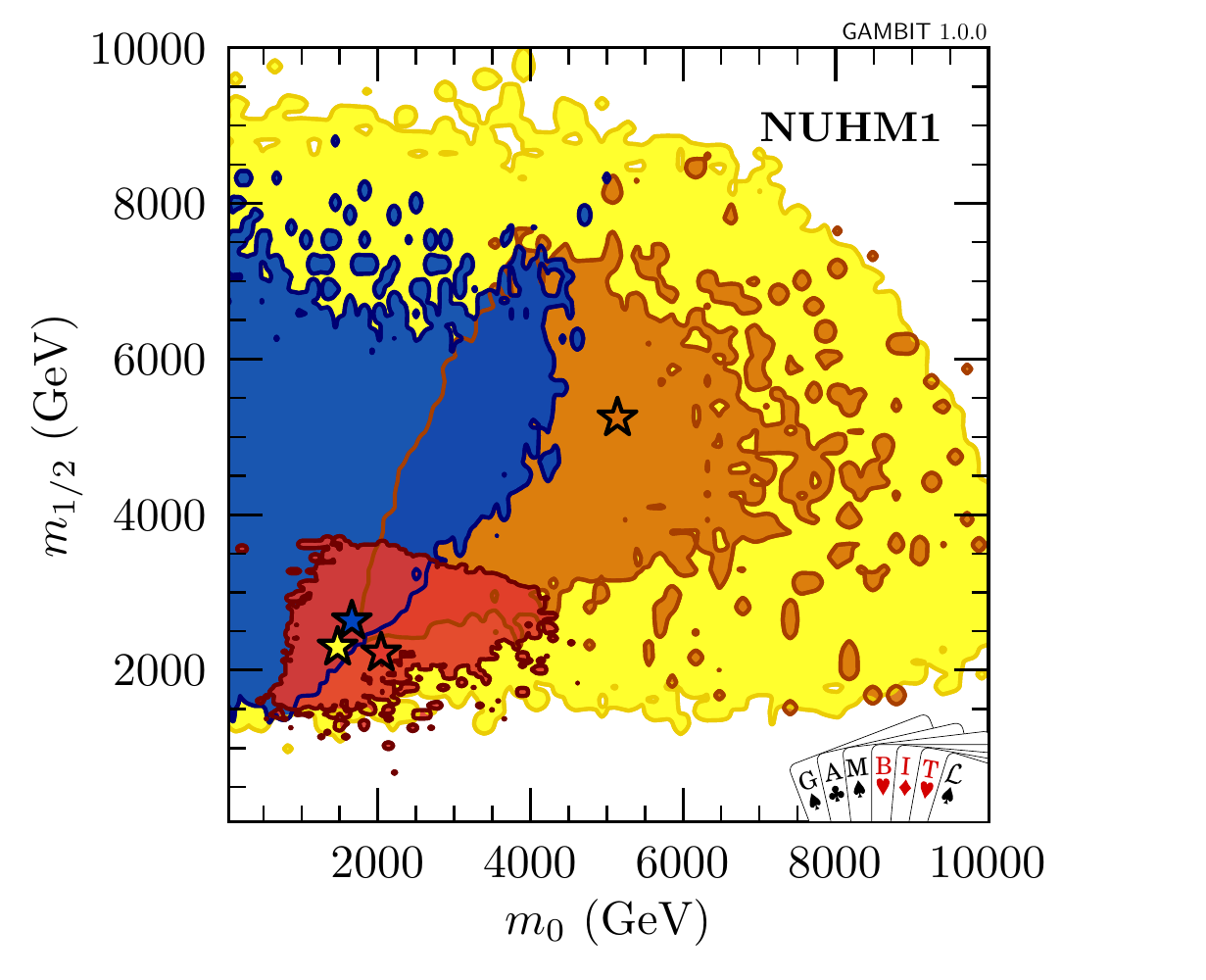}\\
  \includegraphics[height=3.1mm]{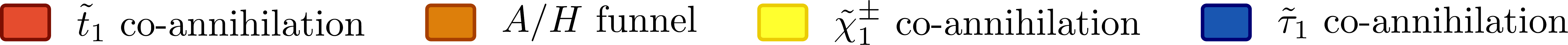}
  \caption{
  Profile likelihood in the $(m_0,m_{1/2})$ plane in the \CMSSM (\textit{top left}) and the \NUHMone (\textit{bottom left}). The right-hand panels show the mechanisms that contribute to bringing the predicted DM relic density close to or below the observed value. The white contours show the $1\sigma$ and $2\sigma$ preferred regions relative to the best-fit point (white star). From~\cite{CMSSM}.
  }
  \label{fig:CMSSM_NUHM1_m0_m12}
\end{figure}

The profile likelihood maps for the $(m_0,m_{1/2})$ planes of the \CMSSM and \NUHMone are shown in the left panels of Fig.\ \ref{fig:CMSSM_NUHM1_m0_m12}. The \NUHMone plane is clearly less constrained compared to the \CMSSM. The underlying reason is the additional parametric freedom in the Higgs sector of the \NUHMone, where the MSSM Higgs parameters $m_{H_u}^2$ and $m_{H_d}^2$ are not unified with $m_0^2$ at the GUT scale, but are rather set by an independent parameter $m_H$ through the GUT-scale requirement $m_{H_u} = m_{H_d} \equiv m_H$. (In the \NUHMtwo, $m_{H_u}$ and $m_{H_d}$ are taken as independent parameters at the GUT scale.) We note that as $m_H$ is taken to be a real parameter, we have $m_{H_u}^2 = m_{H_d}^2 > 0$ at the GUT scale. The correct shape of the Higgs potential at the weak scale must therefore be generated through radiative corrections, as is the case for the \CMSSM.

The right-hand panels in Fig.\ \ref{fig:CMSSM_NUHM1_m0_m12} help us understand the preferred parameter space in more detail. In these panels different sub-regions of the $2\sigma$ region are coloured according to which mechanism(s) contribute to keeping the DM relic density close to or below the observed value. The following criteria are used to define the DM mechanism regions in~\cite{CMSSM} and in the \MSSMseven study in~\cite{MSSM}:
\begin{itemize}
\item stop co-annihilation: $m_{\tilde{t}_1} \leq 1.2\,m_{\NeutOne}$,
\item sbottom co-annihilation: $m_{\tilde{b}_1} \leq 1.2\,m_{\NeutOne}$,
\item stau co-annihilation: $m_{\tilde{\tau}_1} \leq 1.2\,m_{\NeutOne}$,
\item chargino co-annihilation: $\NeutOne$ $\ge50\%$ Higgsino,\footnote{For brevity we refer to this mechanism simply as ``chargino co-annihilation'', though it also includes co-annihilations with the next-to-lightest neutralino. Further, for many points in this region the most important effect is simply enhanced $\NeutOne$--$\NeutOne$ annihilations, owing to the dominantly Higgsino $\NeutOne$ composition.}
\item $A/H$ funnel: $1.6\,m_{\NeutOne} \leq \textrm{($m_A$ or $m_H$)} \leq 2.4\,m_{\NeutOne}$,
\item $h/Z$ funnel: $1.6\,m_{\NeutOne} \leq \textrm{($m_Z$ or $m_h$)} \leq 2.4\,m_{\NeutOne}$.
\end{itemize}
The coloured regions overlap for parameter points where more than one mechanism contributes.

In the \CMSSM the overall highest-likelihood point is found in the stop co-annihilation region, at $m_0 \lesssim 4.5$\,TeV. This region is associated with large, negative values for the trilinear coupling, $A_0 \lesssim -5$\,TeV, and $\tan\beta \lesssim 16$. Only two other DM mechanisms are active within the best-fit parameter space, namely the $A/H$ funnel and chargino co-annihilation. Thus, in contrast with earlier \CMSSM fits, these results show that the stau co-annihilation region has fallen out of the preferred parameter space. This is mainly driven by the likelihood contribution from the LHC Higgs measurements, which penalise the lower-$m_0$ region where the lightest stau get sufficiently close in mass to the lightest neutralino.

As discussed above, the link between $m_0$ and the Higgs sector is relaxed in the \NUHMone. This opens up the parameter space at lower $m_0$, allowing the stau co-annihilation region back within the $2\sigma$ preferred region, as seen in the lower right panel of Fig.\ \ref{fig:CMSSM_NUHM1_m0_m12}. A second consequence of $m_0$ being decoupled from the Higgs sector in the \NUHMone is that the allowed chargino co-annihilation region is extended to arbitrarily small $m_0$ values, compared to in the \CMSSM. We can understand this by investigating the \CMSSM case: the chargino co-annihilation DM mechanism is important when the MSSM Higgsino mass parameter $\mu$ is smaller than the bino mass parameter $M_1$ at the weak scale, as in that case the $\NeutOne$ will be the lightest state in a triplet of near mass-degenerate Higgsinos (two neutralinos and one chargino).\footnote{In the general MSSM it is also possible to have chargino co-annihilation between a pair of wino-dominated $\NeutOne$ and $\CharOne$, when $|M_2| < |M_1|, |\mu|$. However, this mechanism is not available in the models discussed here, as the GUT-scale relation $M_1 = M_2 = M_3 \equiv m_{1/2}$ leads to $M_2 \sim 2 M_1$ at the weak scale.}
In the \CMSSM, the MSSM Higgsino mass parameter $\mu$ is strongly linked to $m_0$ via the conditions for EWSB; reducing $m_0$ effectively increases $\mu$. The bino mass parameter $M_1$ is on the other hand controlled by $m_{1/2}$ via the GUT-scale relation $M_1 = M_2 = M_3 \equiv m_{1/2}$. For a fixed value of $m_{1/2}$, lowering $m_0$ therefore eventually leads to $M_1 \ll |\mu|$, resulting in a bino-dominated $\NeutOne$ significantly lower in mass than the Higgsino-dominated neutralinos/chargino. In the \NUHMone, on the other hand, the $\mu$ parameter is mostly controlled by $m_H$. This allows for $|\mu| < M_1$, and thus chargino co-annihilation, also in the low-$m_0$ region.

\begin{figure}[t]
  \centering
  \includegraphics[width = 0.49\textwidth]{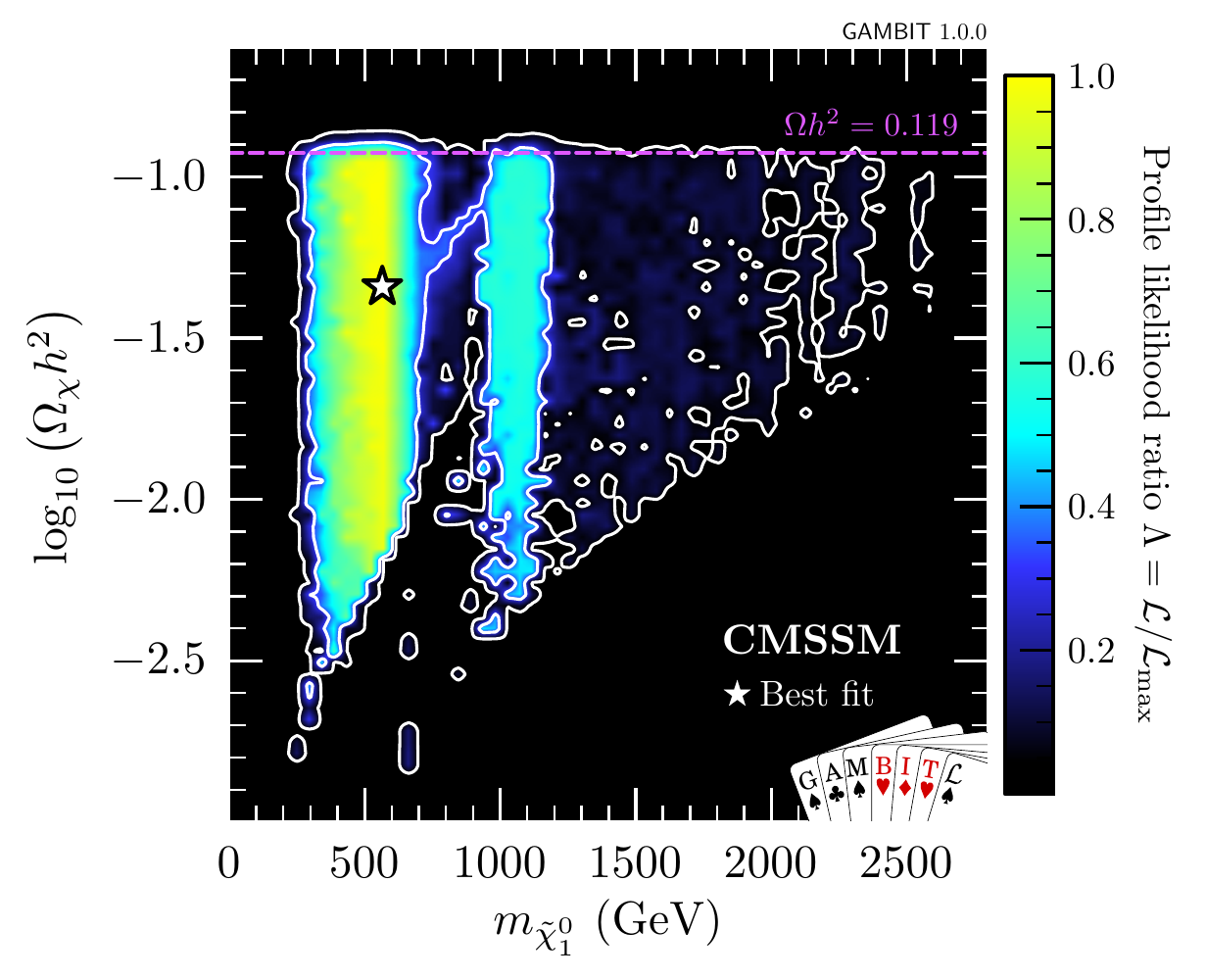}
  \includegraphics[width = 0.49\textwidth]{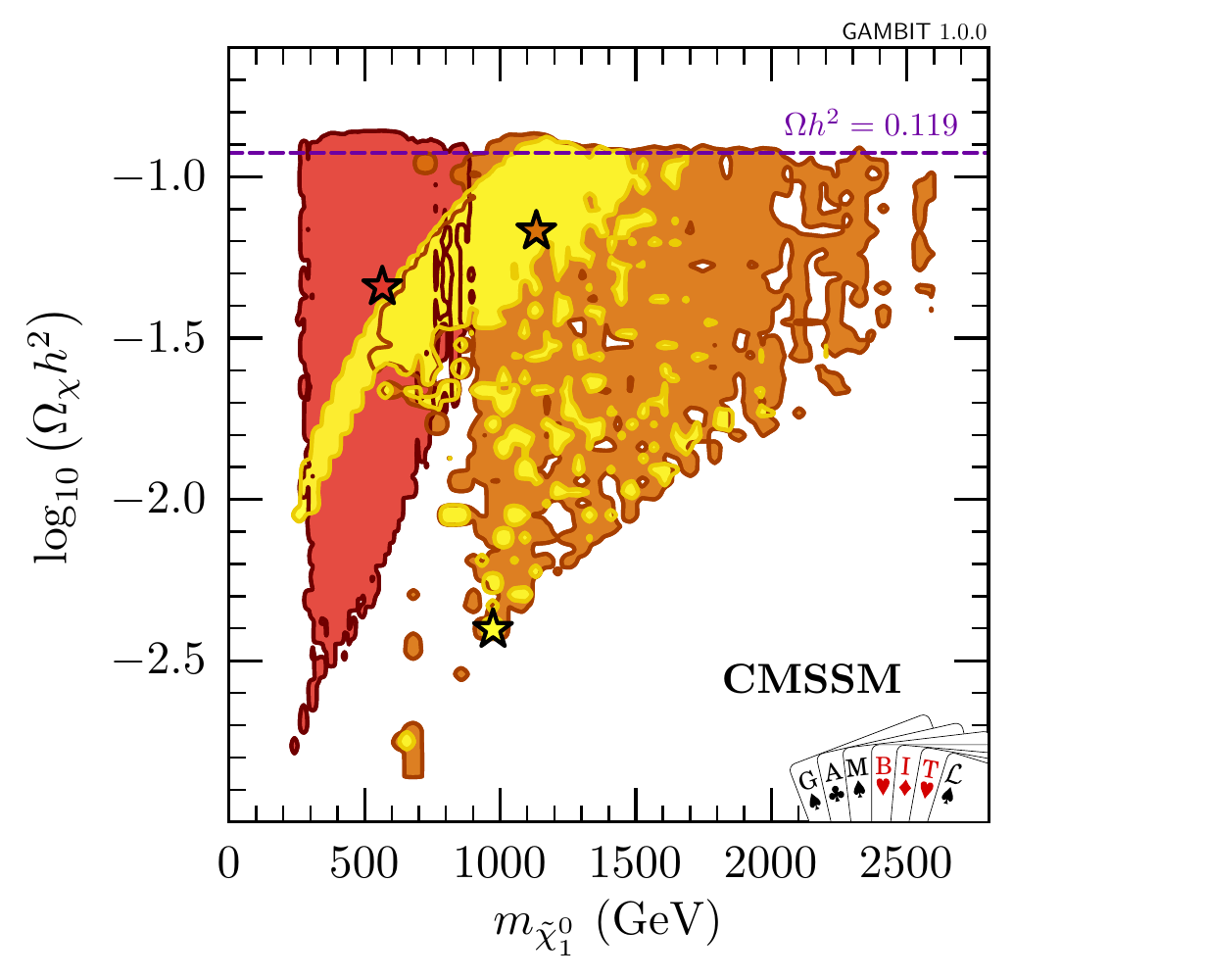}\\
  \includegraphics[height=3.1mm]{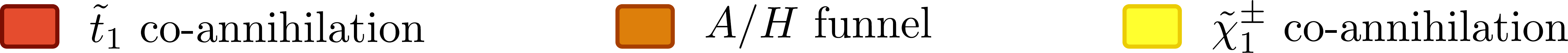}
  \caption{
  Profile likelihood in the $(m_{\NeutOne},\Omega_\chi h^2)$ plane of the \CMSSM (\textit{left}), and the mechanisms that bring the predicted relic density close to or below the measured value (\textit{right}). The stars show the best-fit points, while the white contours outline the $1\sigma$ and $2\sigma$ regions. From~\cite{CMSSM}.
  }
  \label{fig:CMSSM_mN1_oh2}
\end{figure}

As mentioned above, in these fits the observed DM relic density is only imposed as an upper bound, to leave open the possibility for non-MSSM contributions in the observed DM density. While this choice broadens the allowed parameter space, it is worth noting that the parameter regions that fully explain the relic density can have equally high likelihoods as those with a lower predicted relic density. This can be seen in the left panel of Fig.~\ref{fig:CMSSM_mN1_oh2}, which shows the profile likelihood in the \CMSSM plane of the neutralino mass $m_{\NeutOne}$ and the predicted relic density $\Omega_{\chi} h^2$. For most $m_{\NeutOne}$ values there is little variation in the profile likelihood when moving up to a point where the prediction saturates the observed value (dashed purple line).

The right-hand panel in Fig.~\ref{fig:CMSSM_mN1_oh2} shows that, in the \CMSSM, the lowest predicted neutralino masses are found within the stop and chargino co-annihilation regions, extending down to $m_{\NeutOne} \sim 250$\,GeV. In the \NUHMone and \NUHMtwo, the chargino co-annihilation and stau co-annihilation regions extend further down, to $m_{\NeutOne} \sim 150$\,GeV. The chargino co-coannihilation region in Fig.~\ref{fig:CMSSM_mN1_oh2} also illustrates the well-known result that a dominantly Higgsino $\NeutOne$ produces the entire observed relic density when $m_{\NeutOne} \sim 1$\,TeV. Moving along the observed relic density towards higher neutralino masses, additional contributions from resonant $A/H$-funnel annihilations become more and more important.

\begin{figure}[t]
  \centering
  \includegraphics[width=0.49\textwidth]{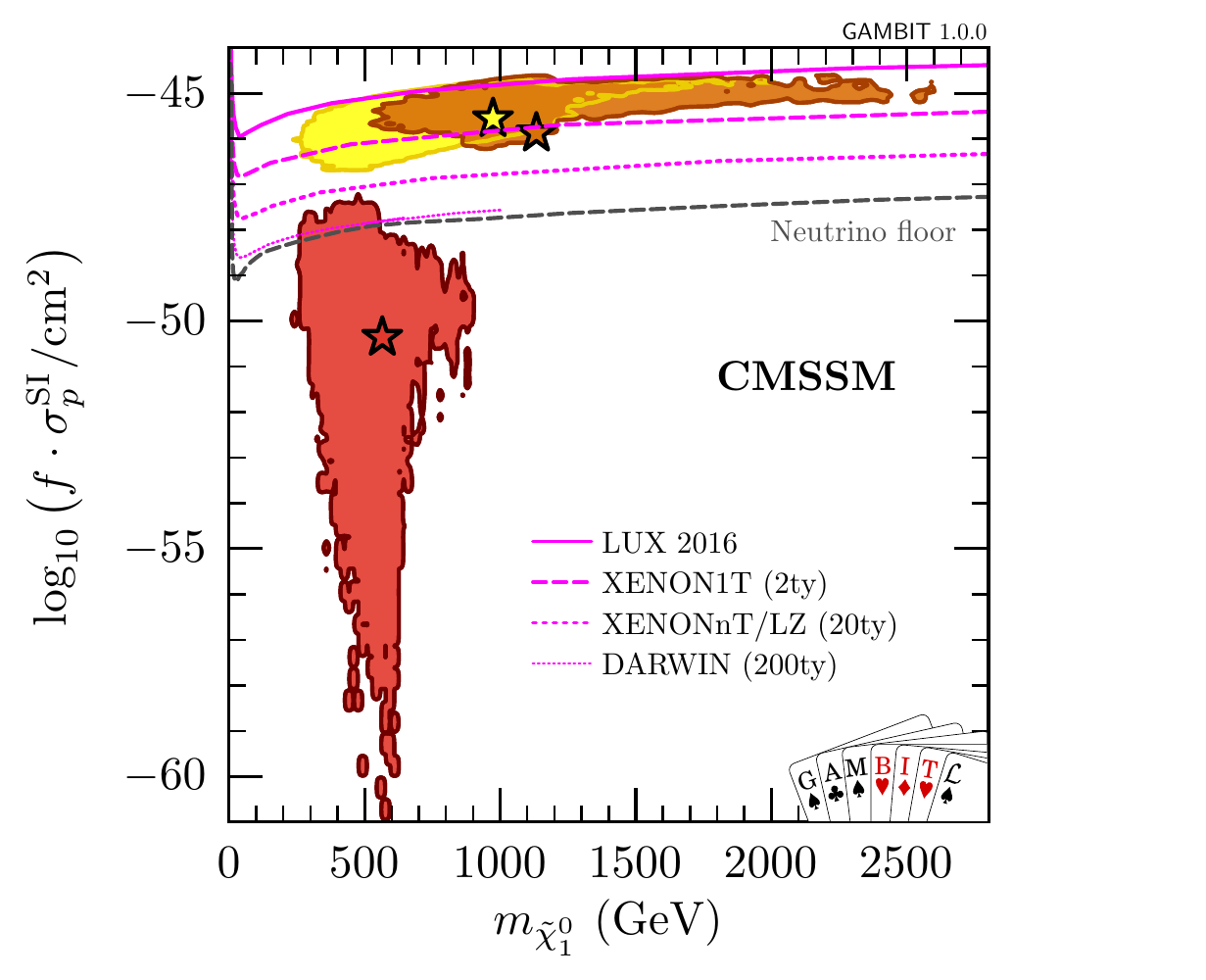}
  \includegraphics[width=0.49\textwidth]{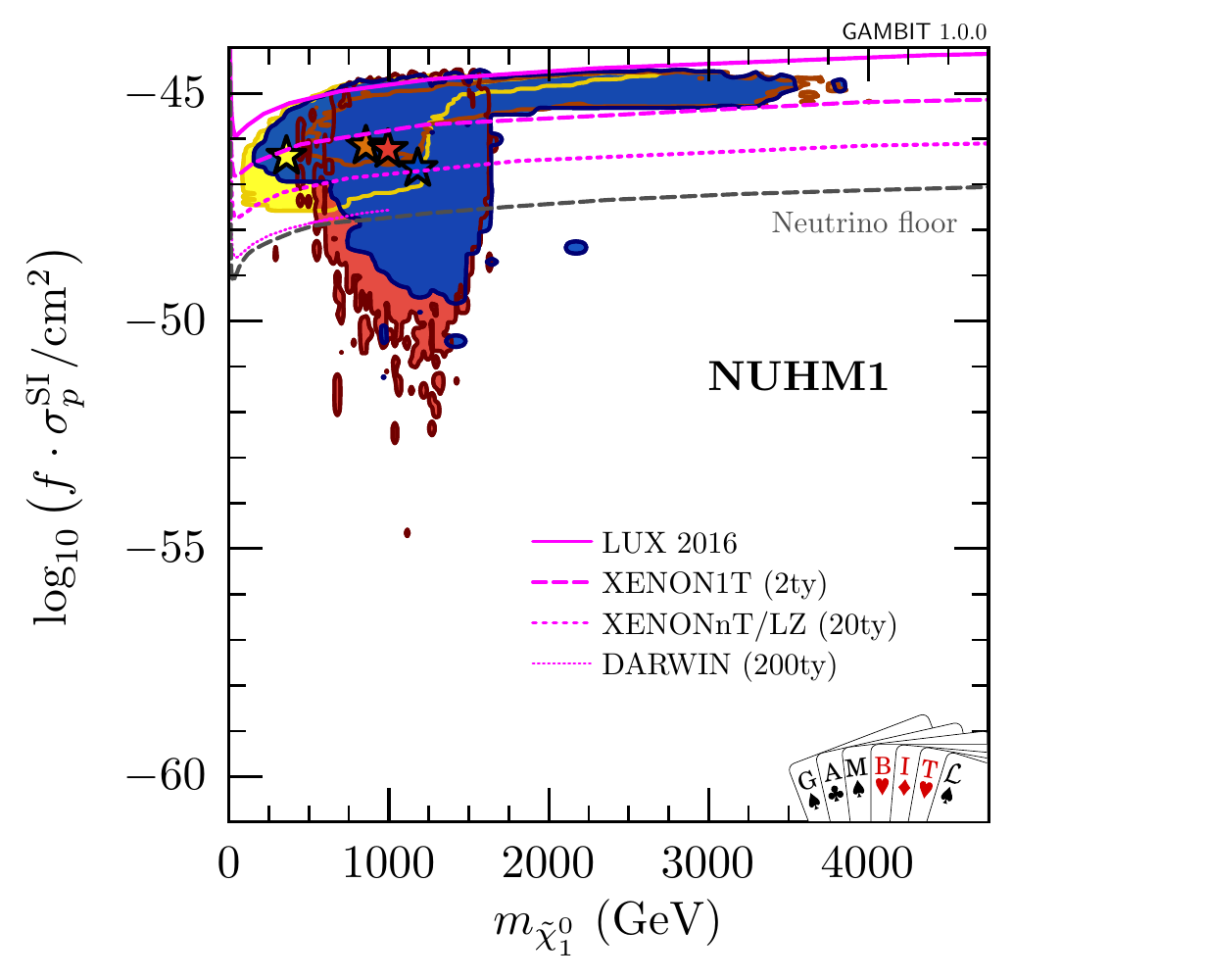}\\
  \includegraphics[height=3.1mm]{figures/susy/rdcolours4.pdf}
  \caption{
  The $2\sigma$ preferred regions in the plane of the spin-independent neutralino-proton cross-section versus the neutralino mass for the \CMSSM (\textit{left}) and the \NUHMone (\textit{right}), coloured according to the mechanism(s) that limit the predicted DM relic density. The pink lines show the observed 90\% CL exclusion limit from LUX~\cite{LUXrun2} and projected limits for XENON1T (two tonne-years of exposure), XENONnT/LZ (20 tonne-years of exposure)~\cite{XENONnTLZ} and DARWIN (200 tonne-years of exposure)~\cite{DARWIN}. The $1\sigma$ and $2\sigma$ regions are shown as white contours; best-fit points are marked by stars. From~\cite{CMSSM}.
  }
  \label{fig:CMSSM_NUHM1_direct_detection}
\end{figure}

Direct detection DM searches seem the most promising experimental probe for the SUSY scenarios preferred in these fits. In Fig.\ \ref{fig:CMSSM_NUHM1_direct_detection} the preferred \CMSSM (left) and \NUHMone regions are shown in the plane of the lightest neutralino mass versus the spin-independent neutralino-proton cross-section. The predicted cross-section is scaled by the fraction $f$ of the full DM relic density that the given parameter point attributes to neutralinos. The solid pink line shows the $90\%$\,CL exclusion limit from the LUX 2016 result~\cite{LUXrun2}, which was included as a likelihood component in these fits. The dashed and dotted lines show projected $90\%$\,CL limits for the XENON and DARWIN experiments~\cite{XENONnTLZ, DARWIN}. While the stop co-annihilation regions will largely remain out of reach, as will much of the stau co-annihilation region in the \NUHMone, both the chargino co-annihilation and the $A/H$ funnel regions can be fully probed in future direct detection searches.

Finally, we note that the \CMSSM, \NUHMone and \NUHMtwo fit results in~\cite{CMSSM} indicate that these models no longer hold much promise for resolving the observed discrepancy in the muon anomalous magnetic moment. The strong constraints on the low-mass parameter space -- in particular from LHC sparticle searches, DM direct detection and the LHC Higgs measurements -- push the fits towards heavier sfermion and electroweakino spectra, thus diminishing the possible SUSY contribution to the muon $(g-2)$.

\subsubsection{Results for the \MSSMseven}
\label{sec:MSSM7}

We now move on to the weak-scale parameterisations of the MSSM, starting with the \gambit analysis of the \MSSMseven in~\cite{MSSM}. Here the free parameters are the wino mass parameter, $M_2$; the $(3,3)$ elements of the $\mathbf{A}_u$ and $\mathbf{A}_d$ MSSM trilinear coupling matrices, $(\mathbf{A}_u)_{33} \equiv A_{u_3}$ and $(\mathbf{A}_d)_{33} \equiv A_{d_3}$ (the other trilinear couplings are set to 0); the soft-breaking Higgs mass parameters, $m_{H_u}^2$ and $m_{H_u}^2$; a common parameter $m_{\tilde{f}}^2$ for the sfermion soft-breaking mass parameters; and the ratio of the Higgs vacuum expectation values, $v_u/v_d \equiv \tan\beta$. All the parameters are defined at the scale $Q = 1$\,TeV, except $\tan\beta$, which is defined at $Q = m_Z$.

While this model is a weak-scale MSSM parameterisation, the GUT-inspired relation
\begin{align}
\frac{3}{5}\cos^2\theta_\mathrm{W}M_1 = \sin^2\theta_\mathrm{W}M_2 = \frac{\alpha}{\alpha_\mathrm{s}}M_3,
\label{eq:GUT_relation}
\end{align}
is imposed to limit the dimensionality of the parameter space. Equation~\ref{eq:GUT_relation} represents an expected weak-scale relation between $M_1$, $M_2$ and $M_3$ if they originate from a common GUT-scale parameter, like $m_{1/2}$ in the \CMSSM.

As in the GUT-scale models, the Higgsino mass parameter $\mu$ is determined from the input parameters -- most importantly $m_{H_u}^2$ and $m_{H_u}^2$ -- and the requirements for EWSB. Since Eq.\ \ref{eq:GUT_relation} implies that $|M_1| < |M_2|$, we again have three \textit{a priori} possibilities for the composition of the neutralino state: dominantly bino ($|M_1| < |\mu|$), dominantly Higgsino ($|\mu| < |M_1|$), or a bino-Higgsino mixture ($|M_1| \sim |\mu|$).

\begin{figure}
  \centering
  \includegraphics[width=0.49\textwidth]{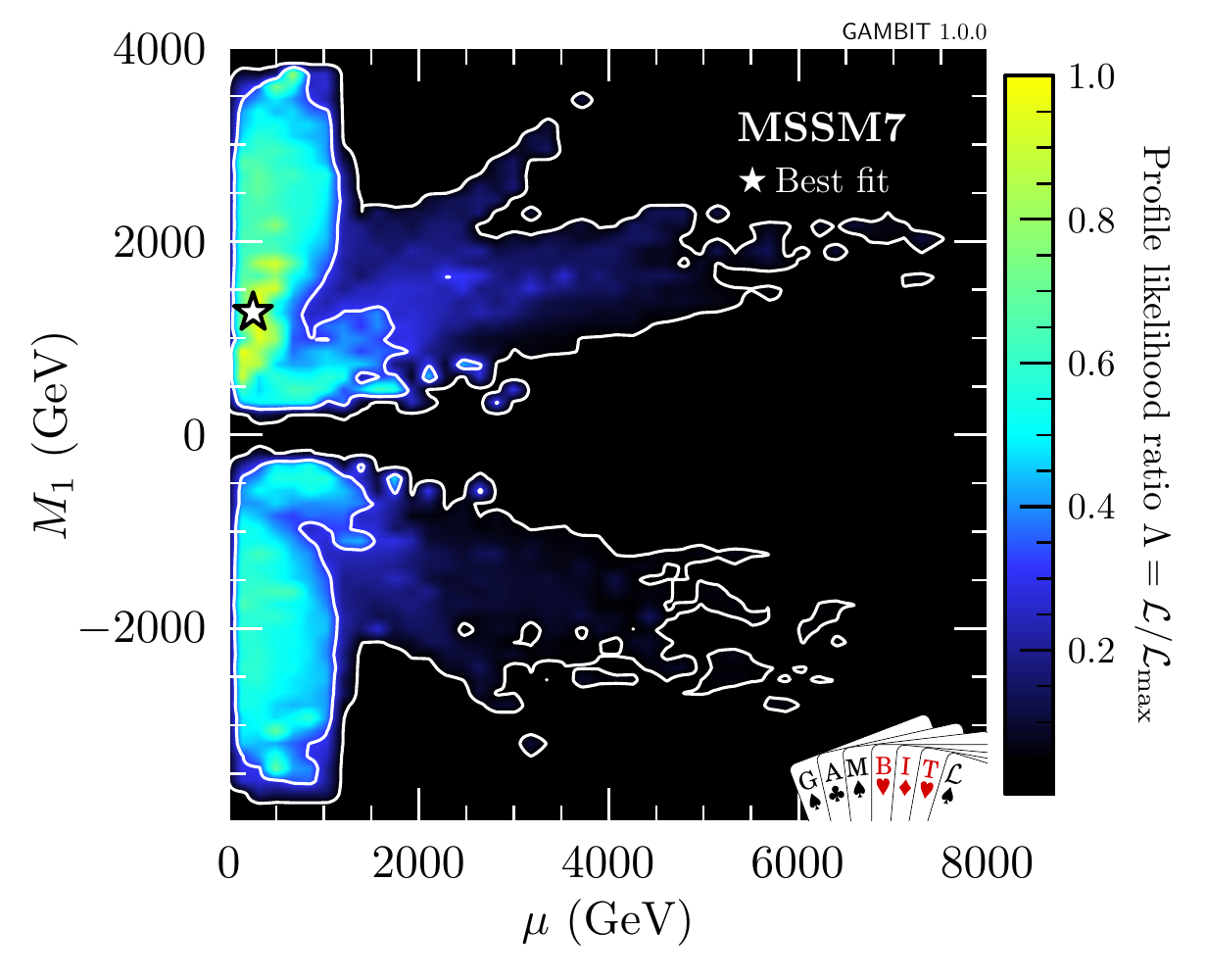}
  \includegraphics[width=0.49\textwidth]{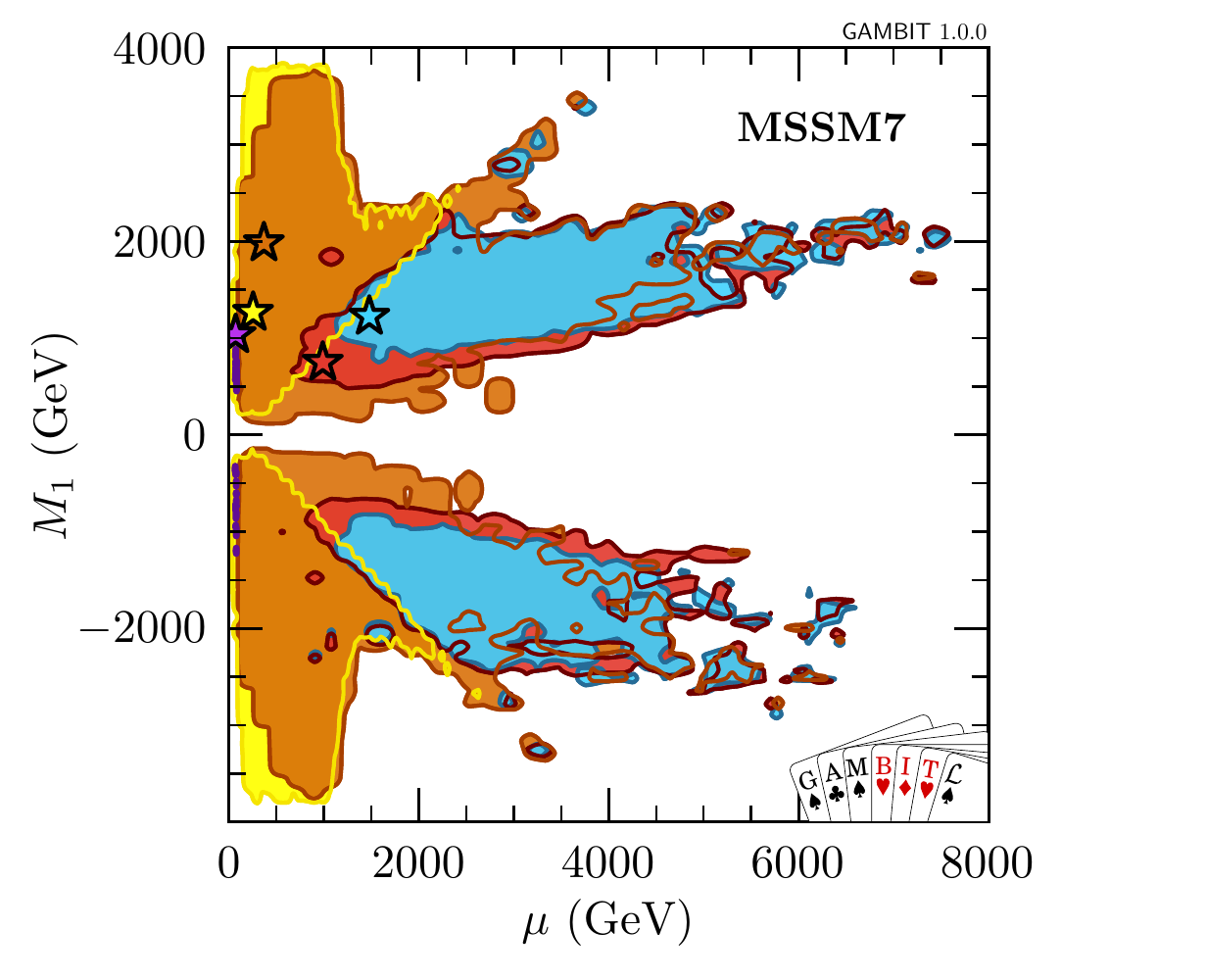}\\
  \includegraphics[width=0.49\textwidth]{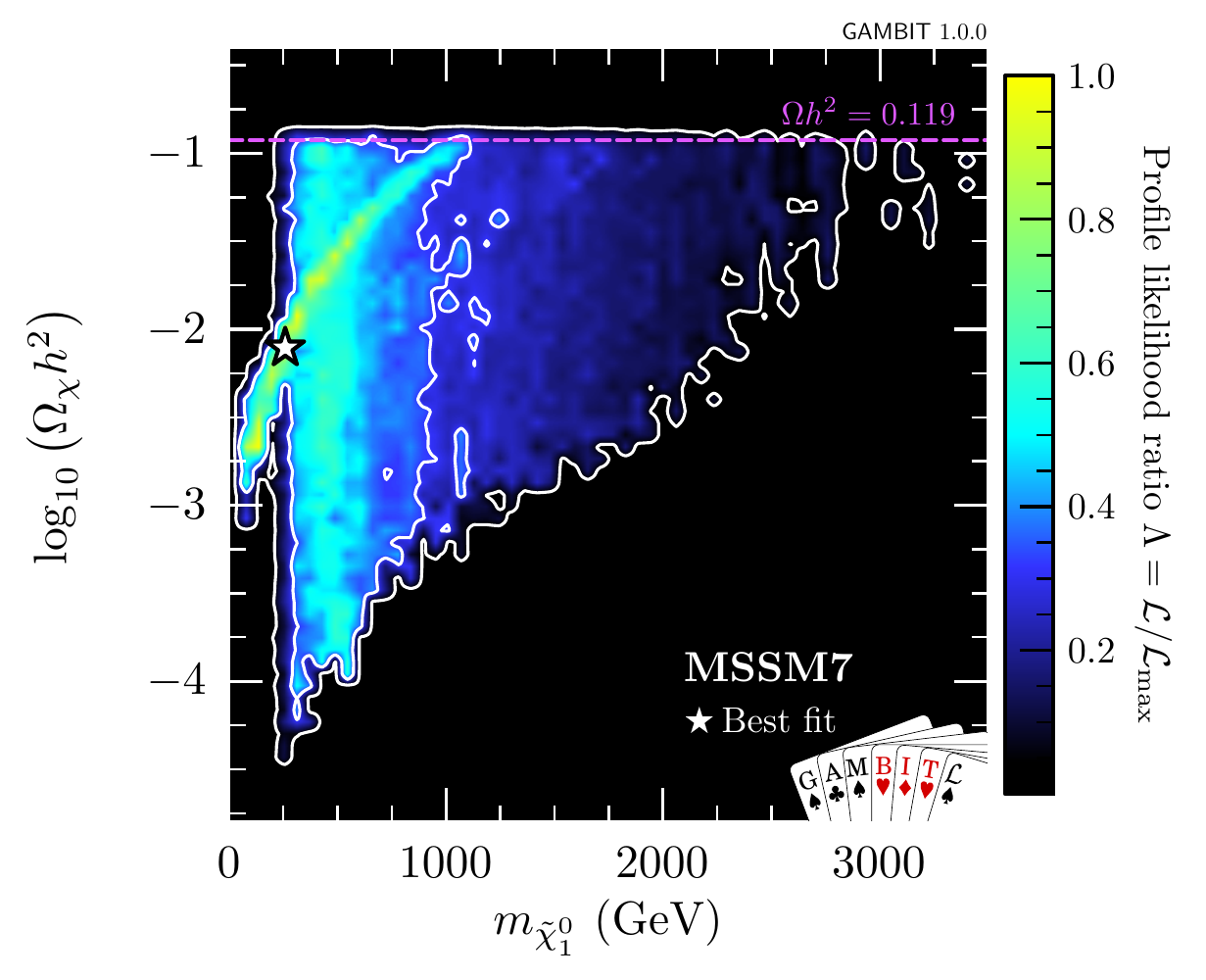}
  \includegraphics[width=0.49\textwidth]{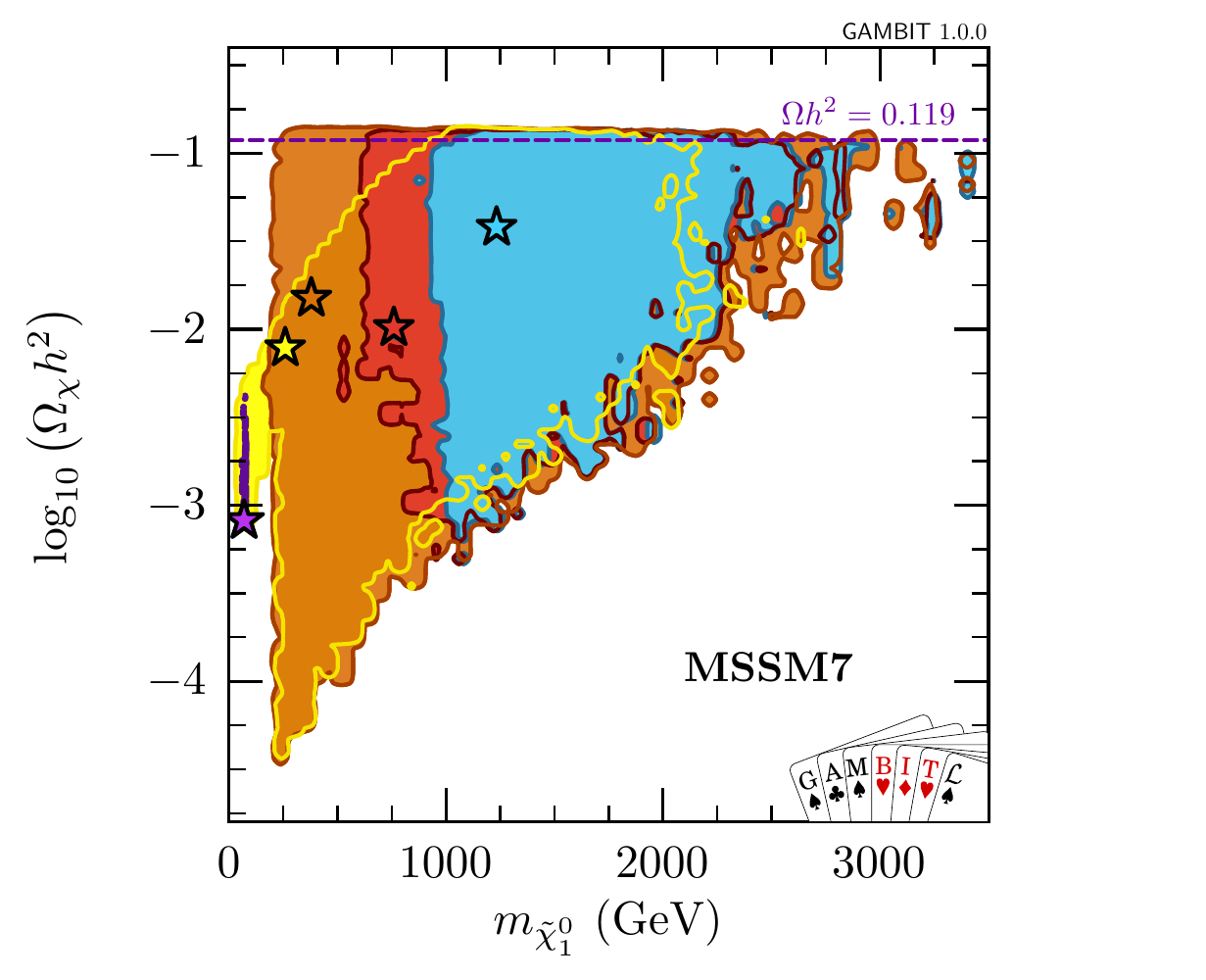}\\
  \includegraphics[height=3.1mm]{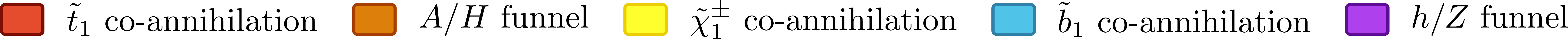}
  \caption{
  Profile likelihoods in the $(\mu,M_1)$ plane (\textit{top left}) and the $(m_{\NeutOne},\Omega_\chi h^2)$ plane of the \MSSMseven. The right-hand panels show the $2\sigma$ preferred parameter regions coloured according to which mechanism(s) contribute to limit the relic density. The stars mark the best-fit points, while the white contours show the $1\sigma$ and $2\sigma$ preferred regions. From~\cite{MSSM}.
  }
  \label{fig:MSSM7}
\end{figure}

The global fit analysis in~\cite{MSSM} finds that all these three neutralino scenarios are allowed within the $2\sigma$ preferred parameter space of the \MSSMseven. This can be seen in the top panels of Fig.\ \ref{fig:MSSM7}, showing the profile likelihood in the $(\mu, M_1)$ plane (left) and the active mechanisms that bring the relic density close to or below the observed value (right). In the $\mu < |M_1|$ regions of the plane, corresponding to a mostly Higgsino $\NeutOne$, the chargino co-annihilation and $A/H$ funnel mechanisms dominate. Moving towards larger $\mu$ we enter the bino-Higgsino mixture scenario at $\mu \sim |M_1|$, before reaching the bino-$\NeutOne$ scenario at $\mu > |M_1|$. Here the chargino co-annihilation mechanism is no longer relevant, so an acceptable relic density must be achieved either through efficient $A/H$ funnel annihilations, co-annihilations with the lightest stop or sbottom, or a combination of these mechanisms.\footnote{The lack of a stau co-annihilation region in the \MSSMseven is related to the assumption of a common sfermion mass parameter defined at the low scale of $Q=1$\,TeV. The differences in sfermion masses are then mostly determined by the amount of L/R mixing in the sfermion mass matrices, rather than RGE running of mass parameters. Since the L/R mixing terms for both up-type and down-type sfermions are proportional to the corresponding Yukawa couplings, the light stop ends up being the lightest sfermion across much of parameter space, and the light sbottom is always lighter than then light stau.}

The overall best-fit point in the \MSSMseven is found in the chargino co-annihilation region, with $m_{\NeutTwo} \approx m_{\CharOne} \approx m_{\NeutOne} \approx 260$\,GeV. As can be seen in the lower panels of Fig.\ \ref{fig:MSSM7}, the predicted neutralino relic density at this point can only explain around 10\% of the observed DM relic density. However, with only slightly heavier neutralino masses there are \MSSMseven scenarios that achieve close to the same likelihood values -- well within the $1\sigma$ region -- and account for the full relic density. These are scenarios with a mostly bino $\NeutOne$ and efficient $\NeutOne$--$\NeutOne$ annihilations through the $A/H$ funnel.

The cutoff of this $A/H$ funnel region at $m_{\NeutOne} \sim 250$\,GeV, corresponding to $m_{A/H} \sim 500$\,GeV, is due to several independent likelihood contributions that penalize the lower-mass scenarios. In particular, the constraint on BSM contributions to $BR(B \rightarrow X_s \gamma)$ plays an important role here, as the $A^0$ mass is closely related to the $H^\pm$ mass, and a light charged Higgs will induce sizable SUSY contributions to this decay. Further important constraints on this region come from the LHC Higgs measurements, and also from LHC gluino searches, as the gluino mass parameter $M_3$ is connected to $M_1$ via Eq.\ \ref{eq:GUT_relation}, giving $M_3 \sim 5 M_1$.

\begin{figure}[t]
  \centering
  \includegraphics[width=0.49\textwidth]{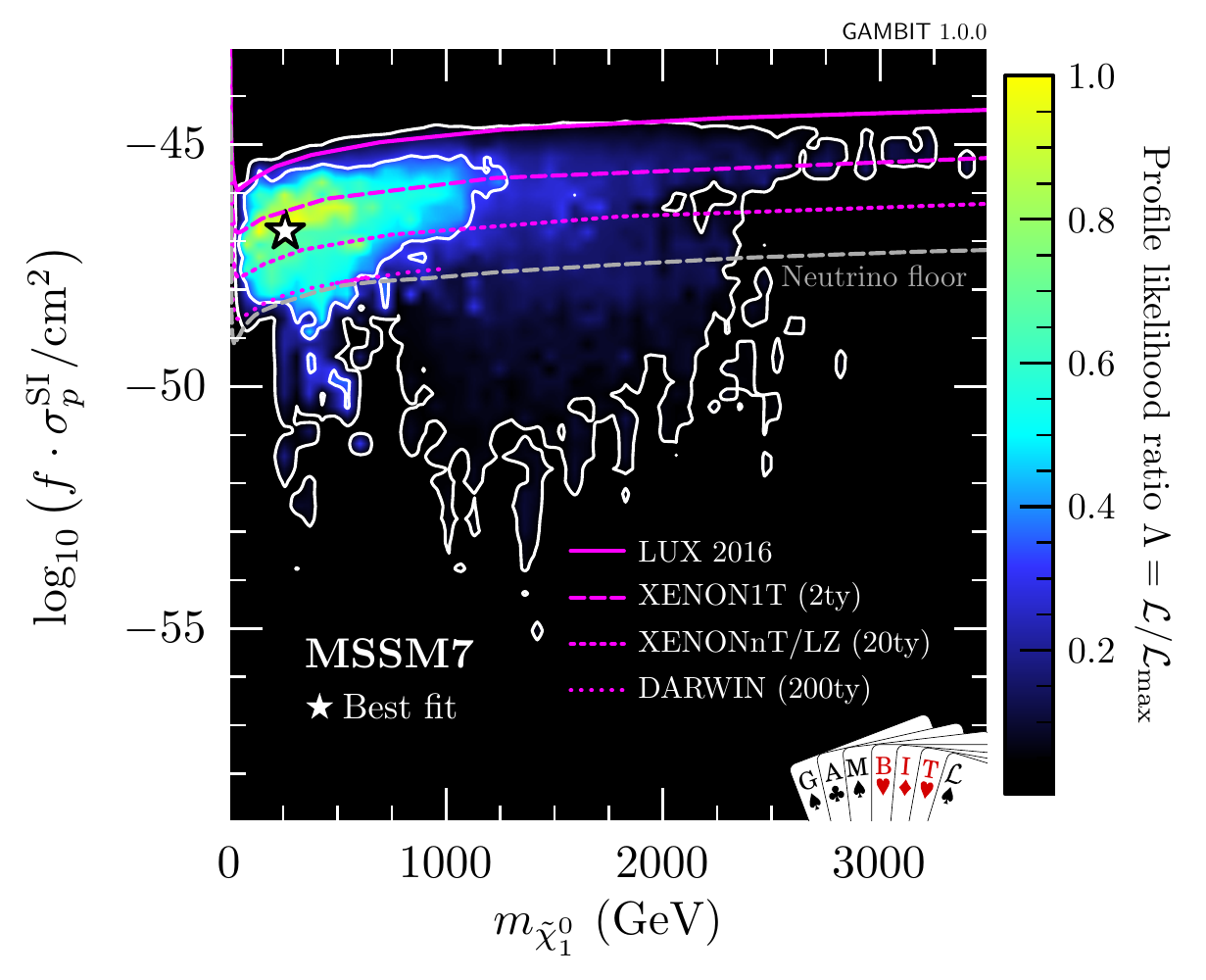}%
  \includegraphics[width=0.49\textwidth]{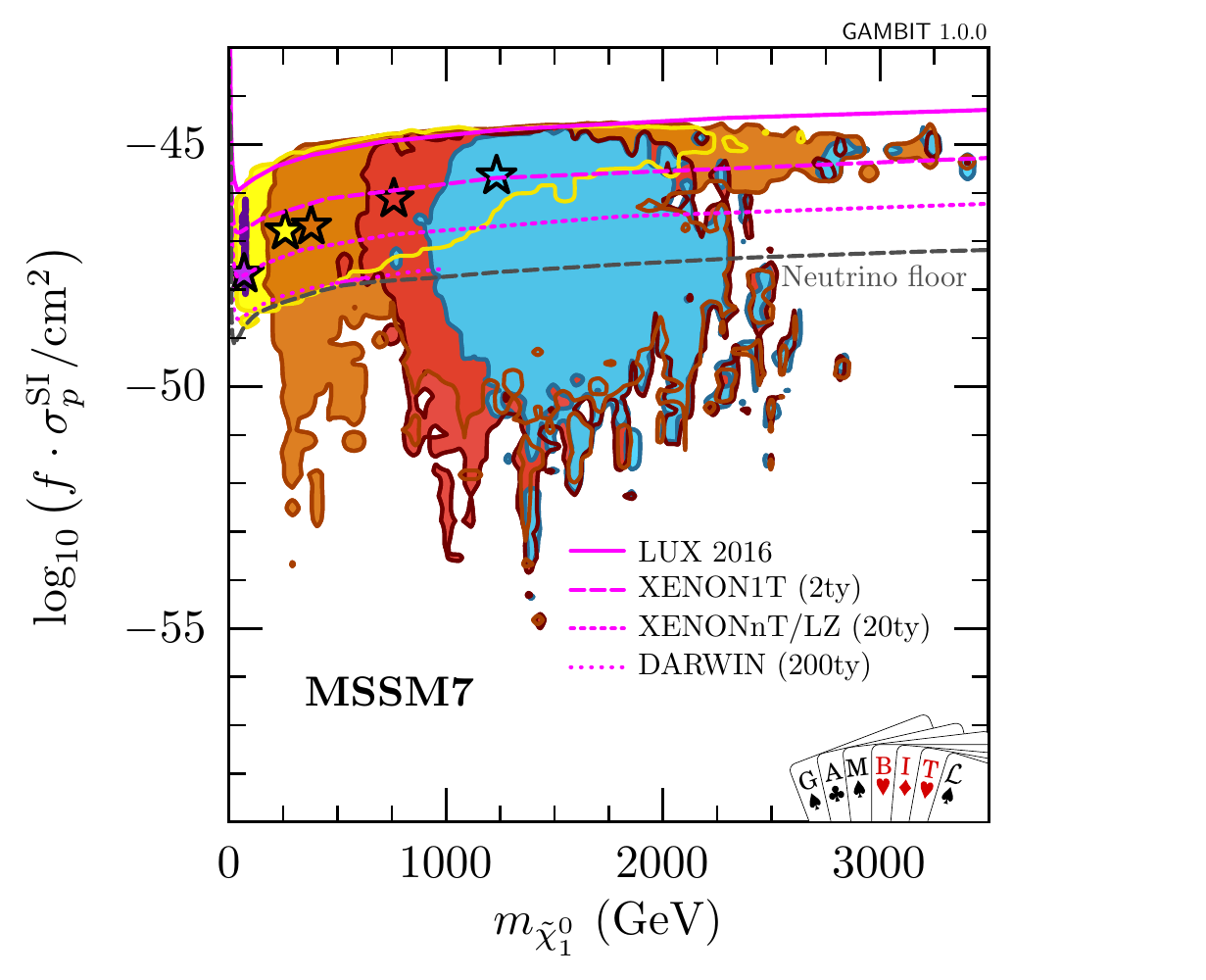}\\
  \includegraphics[height=3.1mm]{figures/susy/MSSM7_rdcolours5.pdf}
  \caption{
	Profile likelihood in the plane of the neutralino mass versus the spin-independent neutralino-proton cross-section in the \MSSMseven (left), and the relic density mechanisms that are active in different parts of the $2\sigma$ region (right). The predicted neutralino-proton cross-section is rescaled at each point by the fraction $f$ of the observed DM relic density that the neutralino relic prediction accounts for. 90\% CL exclusion limits are shown for the full LUX exposure~\cite{LUXrun2} and the projected reach for for XENON1T (two tonne-years of exposure), XENONnT/LZ (20 tonne-years of exposure)~\cite{XENONnTLZ} and DARWIN (200 tonne-years of exposure)~\cite{DARWIN}. The $1\sigma$ and $2\sigma$ regions are outlined by white contours. The stars mark the best-fit points. From~\cite{MSSM}.
   }
  \label{fig:MSSM7_direct_detection}
\end{figure}

We note that even the $h/Z$ funnel mechanisms are present within the $2\sigma$ parameter regions, for $m_{\NeutOne} \approx 45$\,GeV and $m_{\NeutOne} \approx 62$\,GeV. However, the allowed scenarios in this low-$m_{\NeutOne}$ region have an almost pure Higgsino $\NeutOne$ anyway, so this alone ensures a predicted relic density far below the observed value, also explaining how the otherwise strong constraints from DM direct detection are avoided.

As for the GUT-scale models discussed in the previous section, direct DM searches seem the most promising probe of the \MSSMseven scenarios preferred by this fit. Figure \ref{fig:MSSM7_direct_detection} shows the profile likelihood (left) and the active DM mechanisms (right) across the plane of the neutralino mass and the spin-independent neutralino-proton cross-section. We see that future direct detection experiments will explore not only the full chargino co-annihilation region, but almost the entire $1\sigma$ region preferred in the \gambit fit.

Concerning the muon $(g-2)$ discrepancy, the fit in~\cite{MSSM} shows that there is little hope that the \MSSMseven can provide an explanation. This is not particularly surprising: because the model dimensionality is kept low, relating all sfermion mass parameters to the common $m_{\tilde{f}}^2$ parameter at the weak scale, it is impossible to get sufficiently light smuons and muon sneutrinos without simultaneously causing significant tension with other observables such as LHC squark searches.

\subsubsection{Results for the \EWMSSM}
\label{sec:EWMSSM}

Current SUSY searches by the ATLAS and CMS experiments at the LHC are usually optimised and interpreted assuming a simplified model. These models typically include only two or three different sparticles and assume 100\% of decays occur to the signal processes. Such theory simplifications are a necessary compromise given the level of detail and complexity in experimental searches. Nevertheless it leaves open an important question: what impact do the results from ATLAS and CMS SUSY searches have on the parameter space of more realistic models like the MSSM?

The \gambit analysis in~\cite{EWMSSM} takes on this question in the context of LHC searches for neutralinos and charginos. The canonical simplified model for these searches is one that assumes production of a purely wino $\NeutTwo \CharOne$ pair, with subsequent decays to a purely bino $\NeutOne$ via $\NeutTwo \rightarrow Z \NeutOne$ and $\CharOne \rightarrow W^\pm \NeutOne$. This gives motivation for a search for events with leptons, jets and missing energy (see e.g.\ \cite{Aaboud:2018jiw, Sirunyan:2017lae}). The \gambit study assumes a phenomenologically far richer model, referred to as the \EWMSSM.  This is the effective theory obtained when assuming that all sparticles except the MSSM electroweakinos are too heavy to affect current collider searches. The \EWMSSM is thus a model with six sparticles -- four neutralinos and two charginos -- controlled by only four free MSSM parameters: $M_1$, $M_2$, $\mu$ and $\tan\beta$. Loosely speaking, the bino soft-mass $M_1$ controls the mass of one neutralino, the wino soft-mass $M_2$ controls the masses of one neutralino and one chargino, and the Higgsino mass parameter $\mu$ sets the masses of two neutralinos and one chargino.

In contrast to the global fits discussed in the previous two sections, the fit in \cite{EWMSSM} focuses exclusively on collider constraints. This choice allows the fit to explore the full range of possible collider scenarios in the \EWMSSM without further enlarging the model parameter space. Keeping the dimensionality of the parameter space fairly low is of critical importance, due the large computational expense of this fit: for each sampled \EWMSSM parameter point, \colliderbit is used to run full Monte Carlo simulations of the relevant ATLAS and CMS searches. While running full simulations at each point in a global fit is always computationally challenging, it is particularly so when simulating electroweakino searches due to the low signal acceptance rates in these searches.\footnote{For most of the included LHC searches there is no public information on how background estimates are correlated across signal regions. In these cases the single signal region with the best expected sensitivity must be identified at each \EWMSSM parameter point in the fit. This adds to the already substantial computational cost, as distinguishing between ``competing'' signal regions often requires higher Monte Carlo statistics than what is needed to get reasonable signal estimates for each individual signal region alone.}

The analysis in \cite{EWMSSM} includes \colliderbit simulations of most of the $13$\,TeV electroweakino searches that were available at the time of the study~\cite{Aaboud:2018jiw,Aaboud:2018sua,Aaboud:2018htj,Aaboud:2018zeb,CMS:2017fth,Sirunyan:2018iwl,Sirunyan:2017qaj,CMS-PAS-SUS-16-039}. The combined likelihood obtained from these simulations is the main component in the fit likelihood function. The other collider observables going into the total likelihood are a collection of SUSY cross-section limits from LEP and the invisible decay widths of the $Z$ and the $125$\,GeV Higgs.

\begin{figure}[t]
  \centering
  \includegraphics[width=0.49\textwidth]{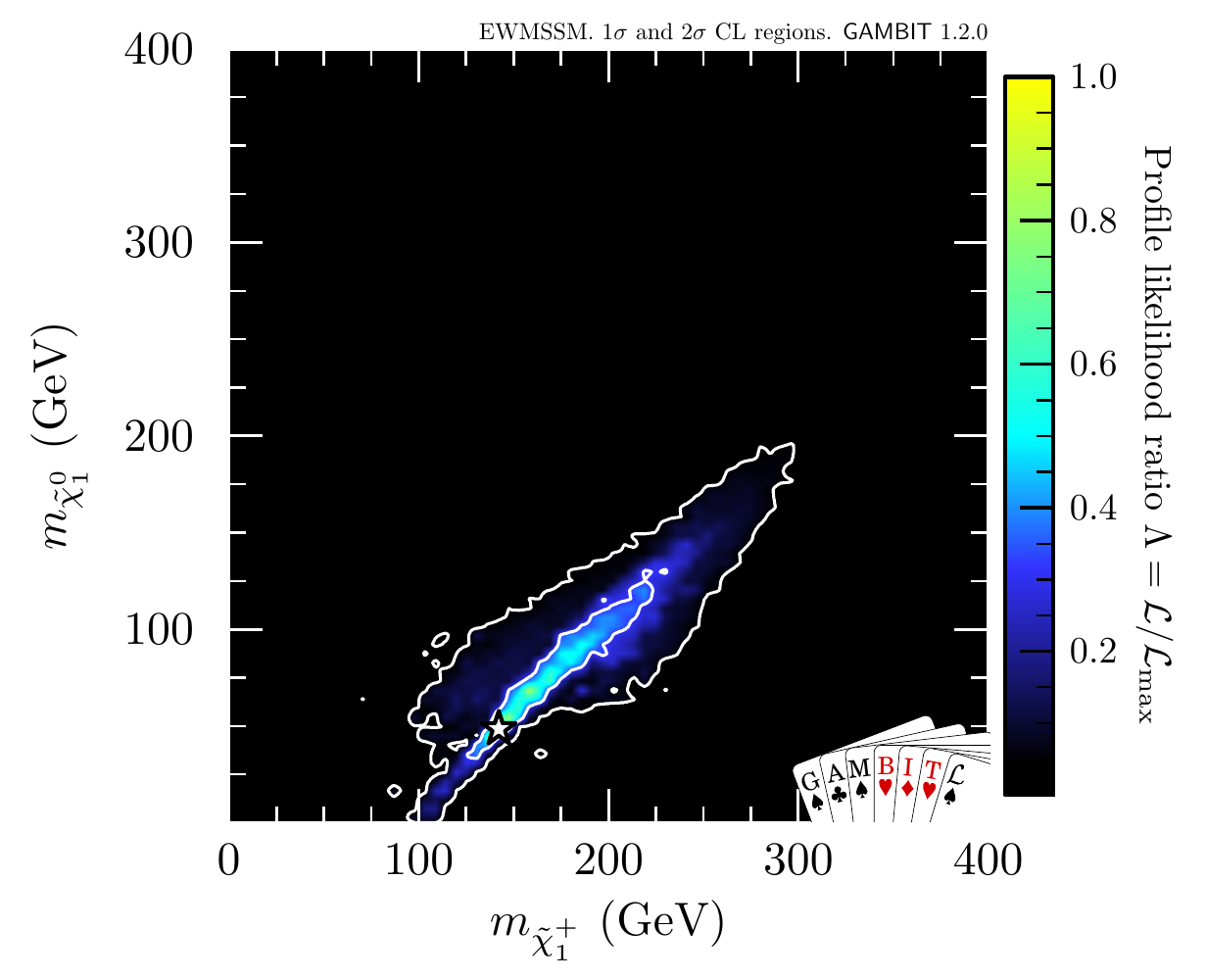}
  \includegraphics[width=0.49\textwidth]{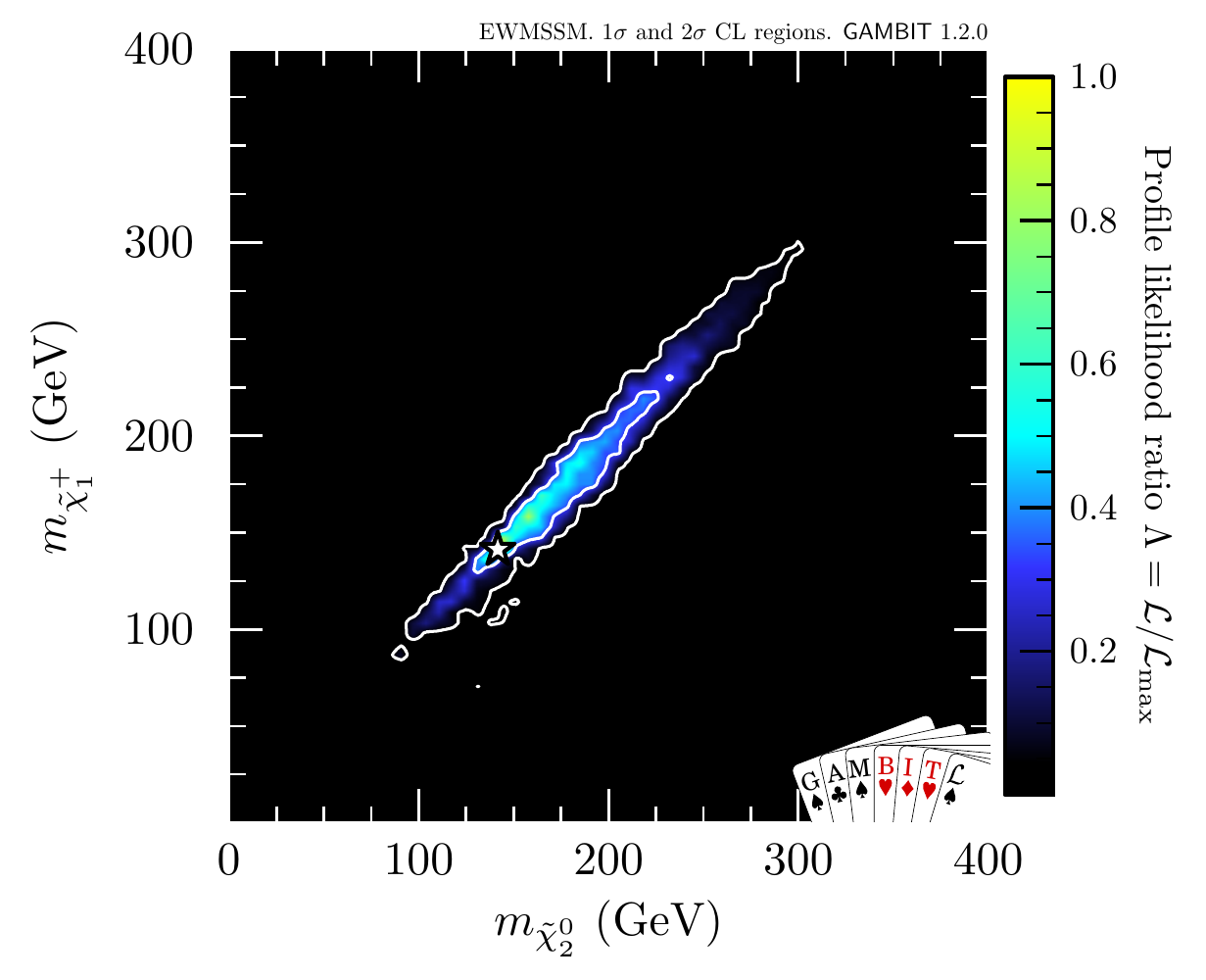}\\
  \includegraphics[width=0.49\textwidth]{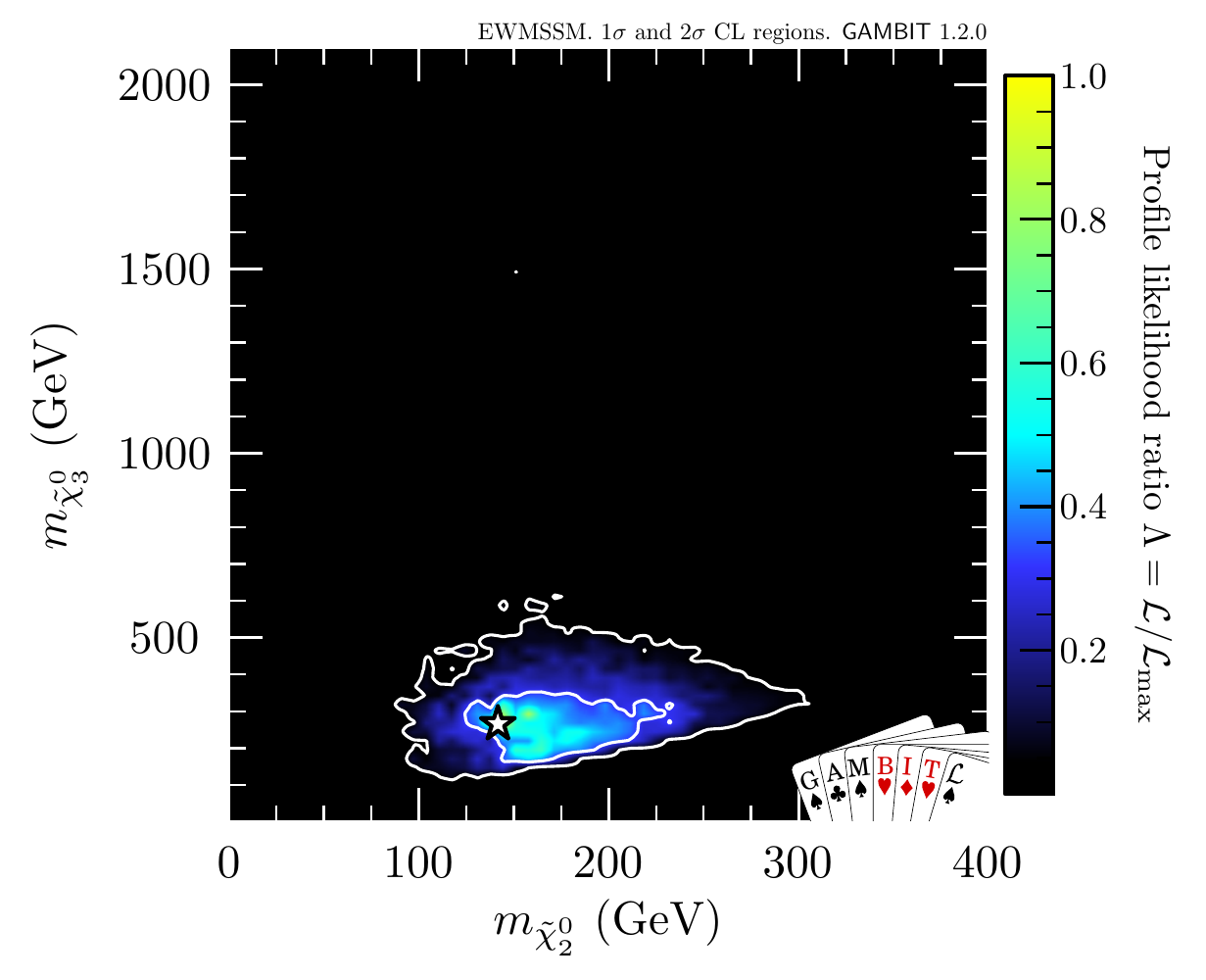}
  \includegraphics[width=0.49\textwidth]{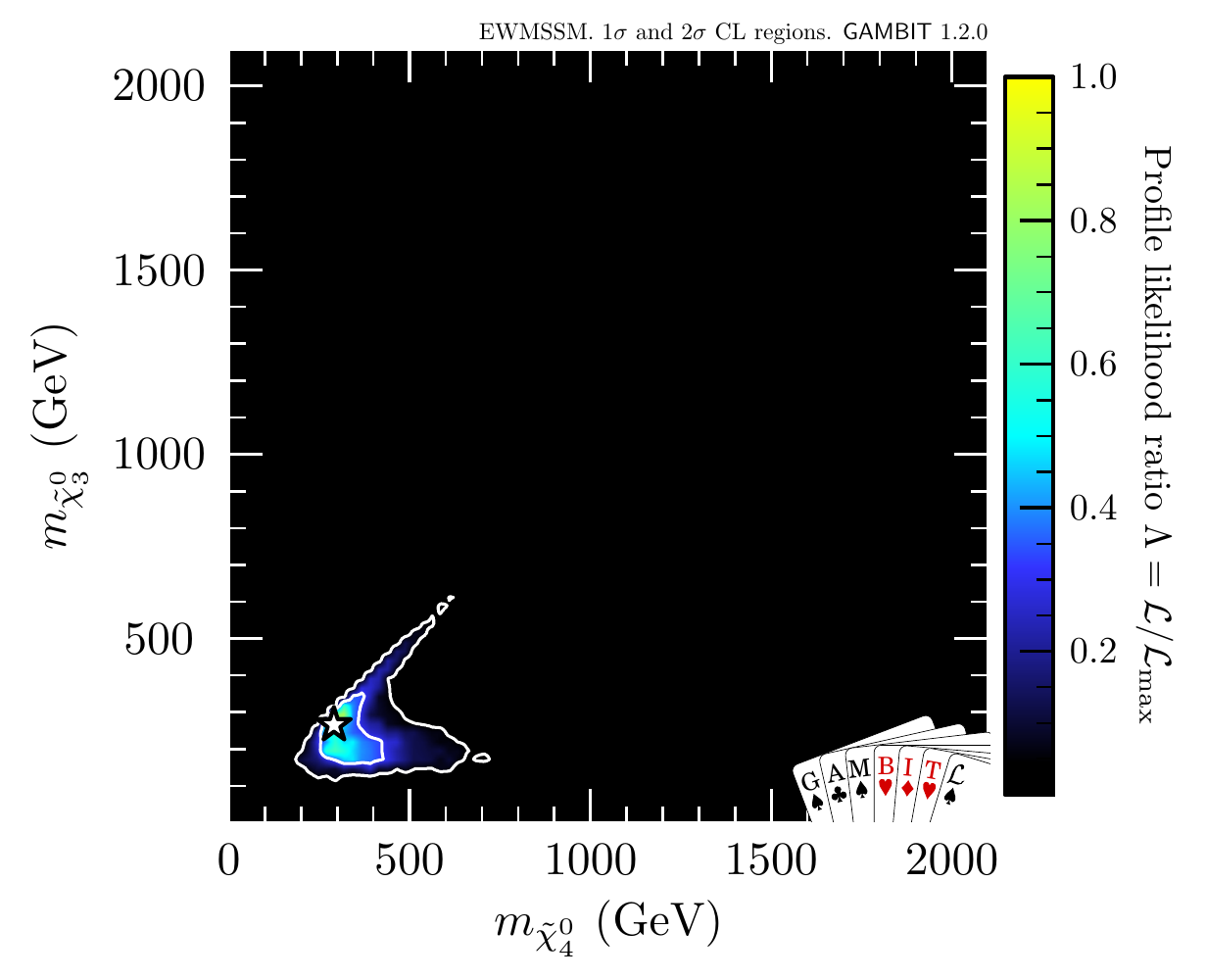}
  \caption{
  Profile likelihood in four different \EWMSSM mass planes: the $(m_{\CharOne},m_{\NeutOne})$ plane (top left), the $(m_{\NeutTwo},m_{\CharOne})$ plane (top right), the $(m_{\NeutTwo},m_{\NeutThree})$ plane (bottom left), and the $(m_{\NeutFour},m_{\NeutThree})$ plane (bottom right). The white contours show the $1\sigma$ and $2\sigma$ preferred regions. The star marks the best-fit point. From~\cite{EWMSSM}.
  }
  \label{fig:EWMSSM_mass_planes}
\end{figure}

The main result from \cite{EWMSSM} is that, when combined, the ATLAS and CMS electroweakino results prefer \textsf{EWMSSM} scenarios with a distinct pattern of relatively light neutralino and chargino masses (Fig.\ \ref{fig:EWMSSM_mass_planes}). The preferred $2\sigma$ parameter region has all six neutralinos and charginos below $\sim$$700$\,GeV, with the lightest neutralino below $\sim$$200$\,GeV. The lightest neutralino is always dominantly bino, but it also has a non-negligible wino or Higgsino component. Further, the best-fit parameter region predicts two characteristic $\gtrsim m_Z$ gaps in the mass spectrum: the first between the mostly bino $\NeutOne$ and the mostly wino (Higgsino) $\NeutTwo$/$\CharOne$, and the second between $\NeutTwo$/$\CharOne$ and the mostly Higgsino (wino) $\NeutFour$/$\CharTwo$.\footnote{In the preferred scenarios, $\NeutThree$ is always mostly Higgsino and thus fairly close in mass to the other Higgsino-dominated states, i.e.\ either $\NeutTwo$/$\CharOne$ or $\NeutFour$/$\CharTwo$.}

At first sight this result may seem surprising. None of the included ATLAS and CMS searches have seen a convincing SUSY signal, yet when combined they prefer the low-mass region over the decoupling region, where all \EWMSSM collider predictions would align with SM expectations. The reason is that the \EWMSSM is able to simultaneously fit a pattern of small excesses across several of the simulated LHC searches, while at the same time avoiding generating too much tension with the other searches. The excesses that mostly drive this result come from searches for 2-, 3-, and 4-lepton final states in ATLAS \cite{Aaboud:2018jiw,Aaboud:2018sua,Aaboud:2018zeb}, specifically in signal regions that target leptons from on-shell $Z$ and $W$ decays. This explains the preference in the fit for electroweakino mass spectra with two $\gtrsim m_Z$ mass gaps.

\begin{figure}
  \centering
  \includegraphics[width=0.32\textwidth]{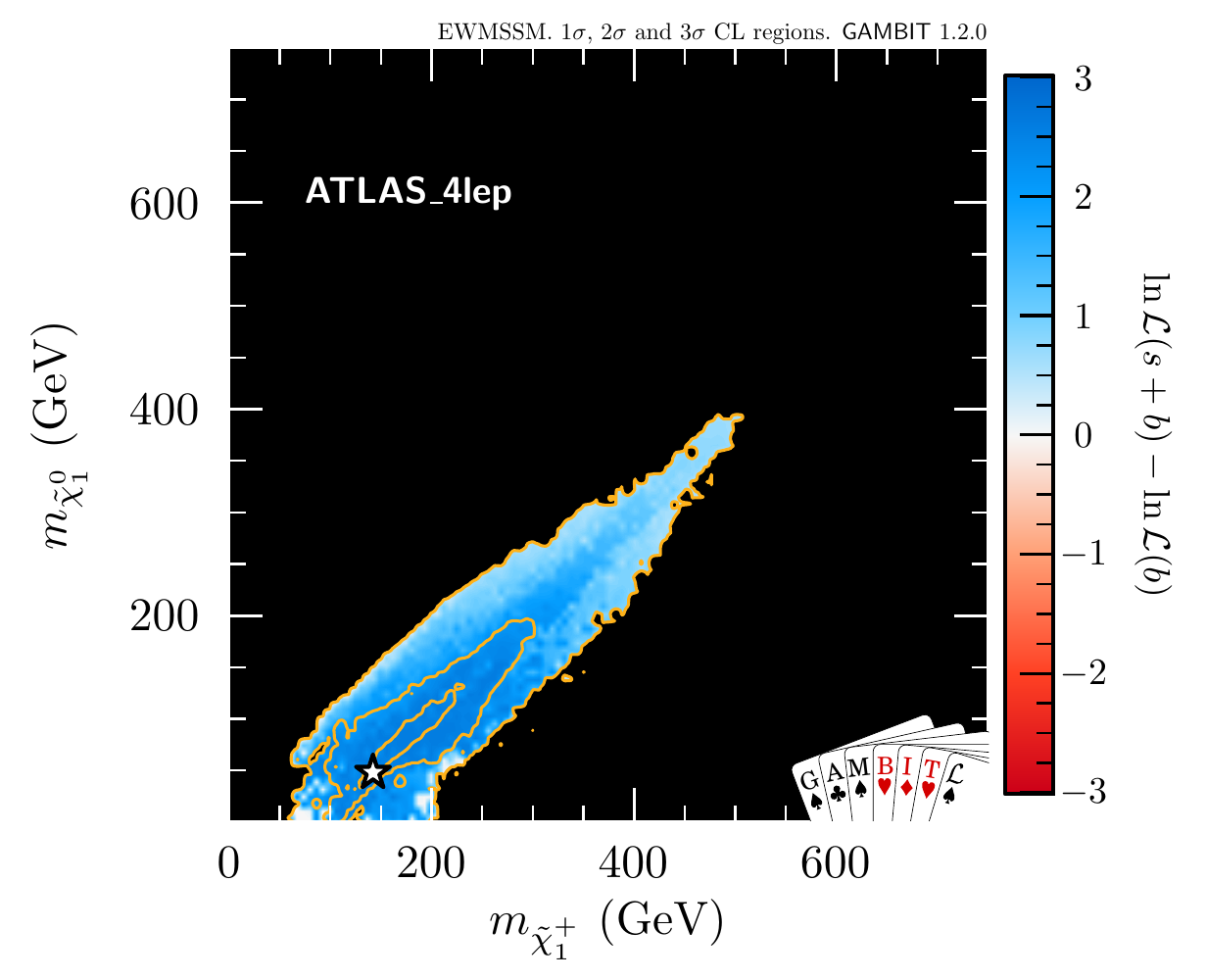}
  \includegraphics[width=0.32\textwidth]{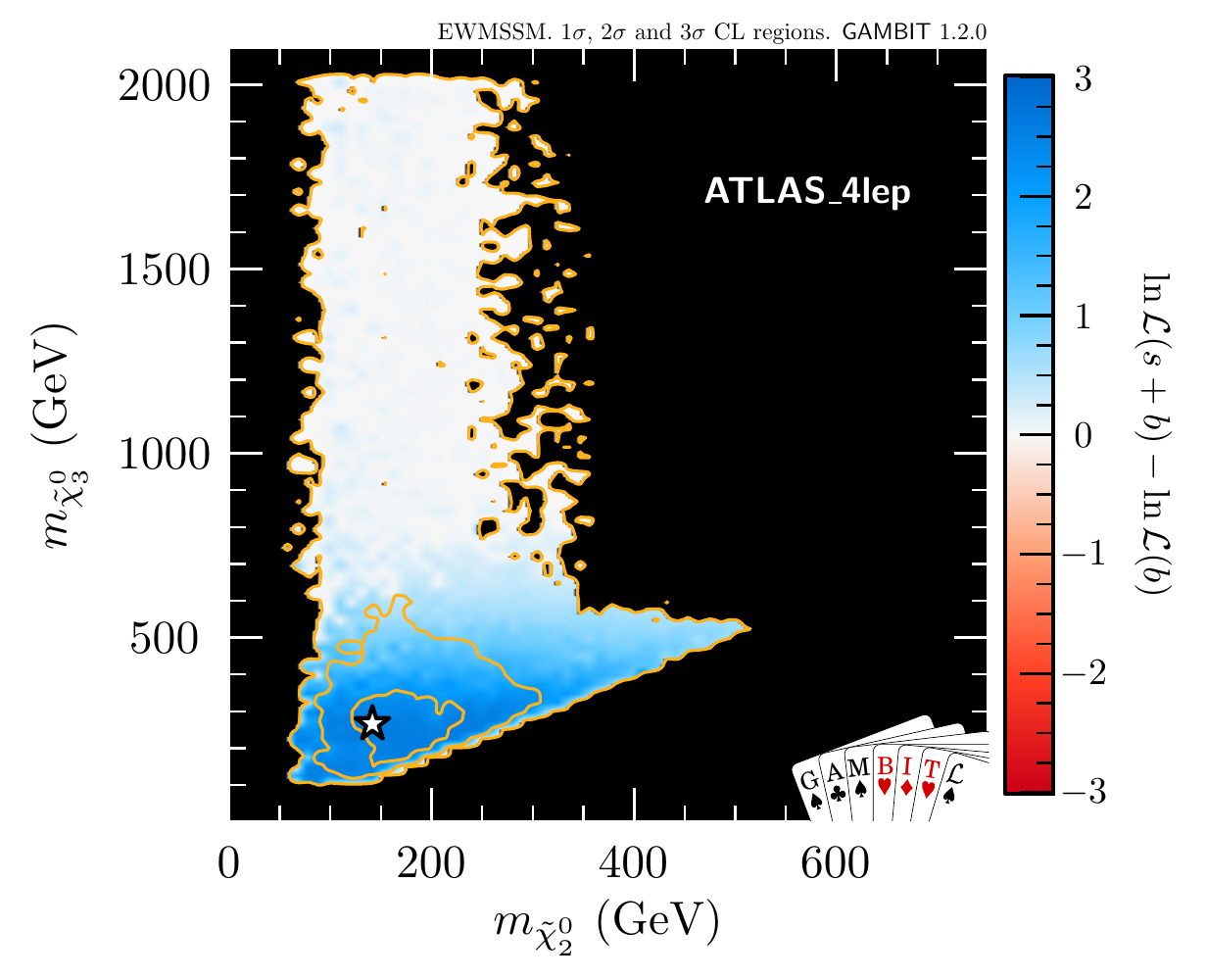}
  \includegraphics[width=0.32\textwidth]{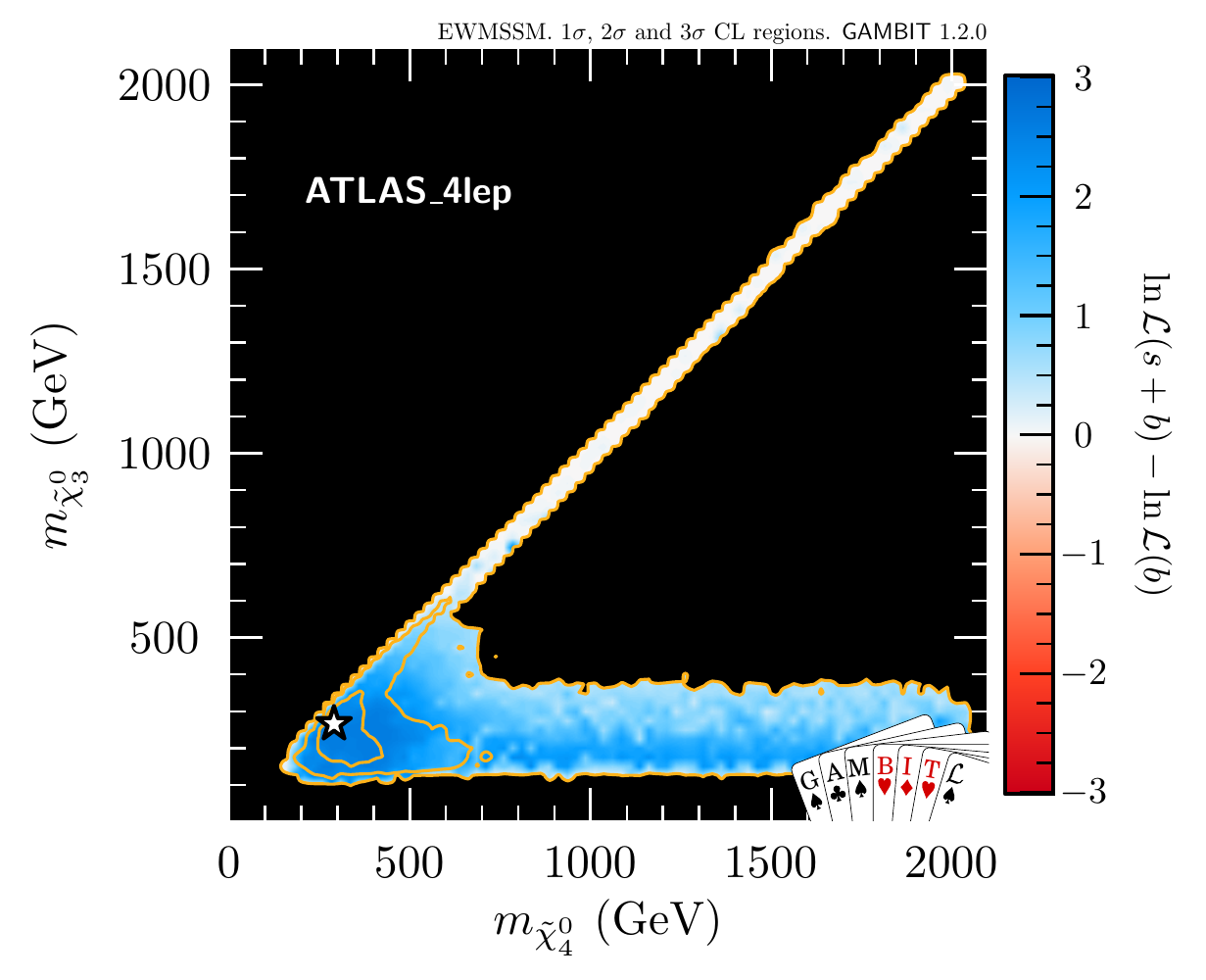}
  \includegraphics[width=0.32\textwidth]{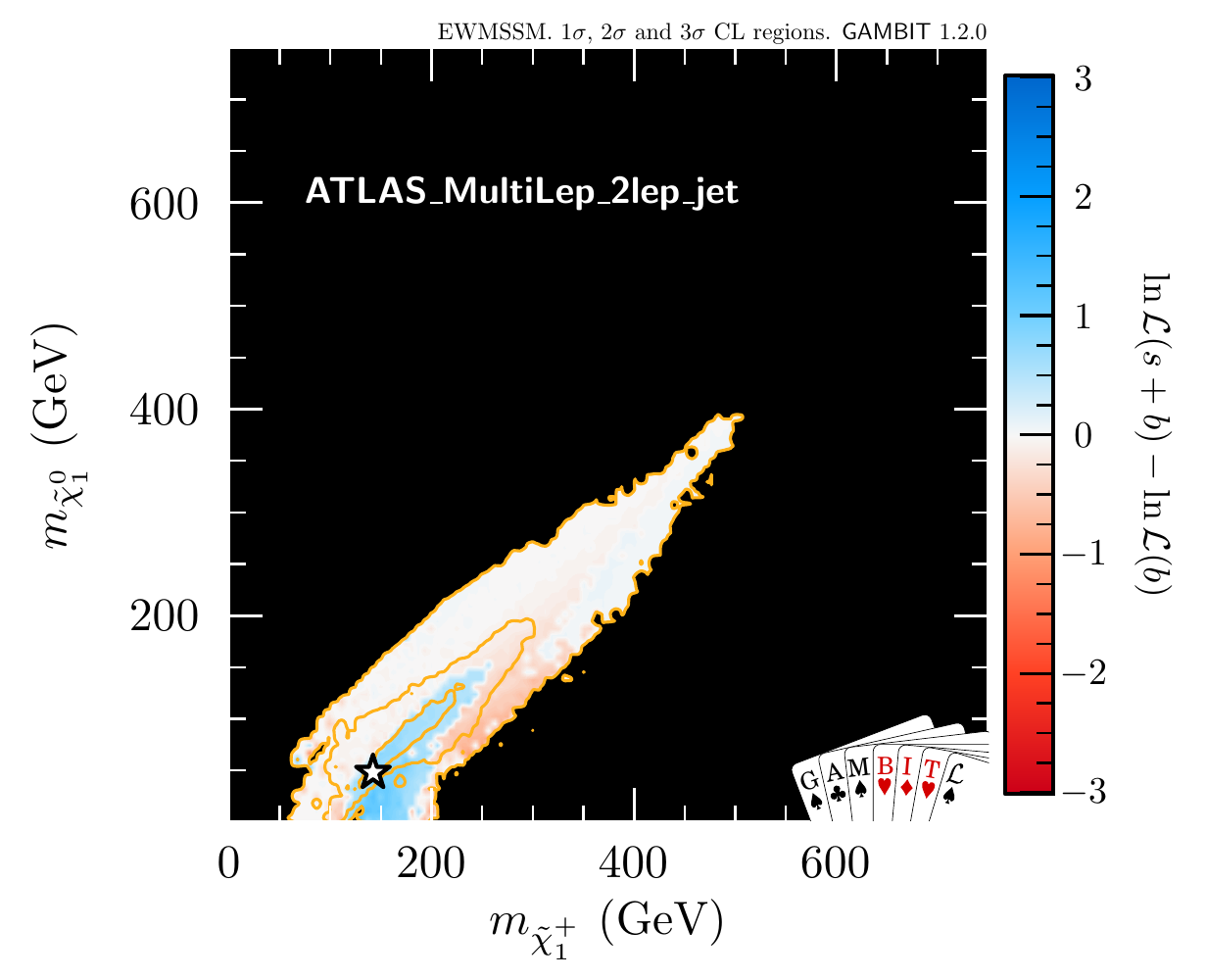}
  \includegraphics[width=0.32\textwidth]{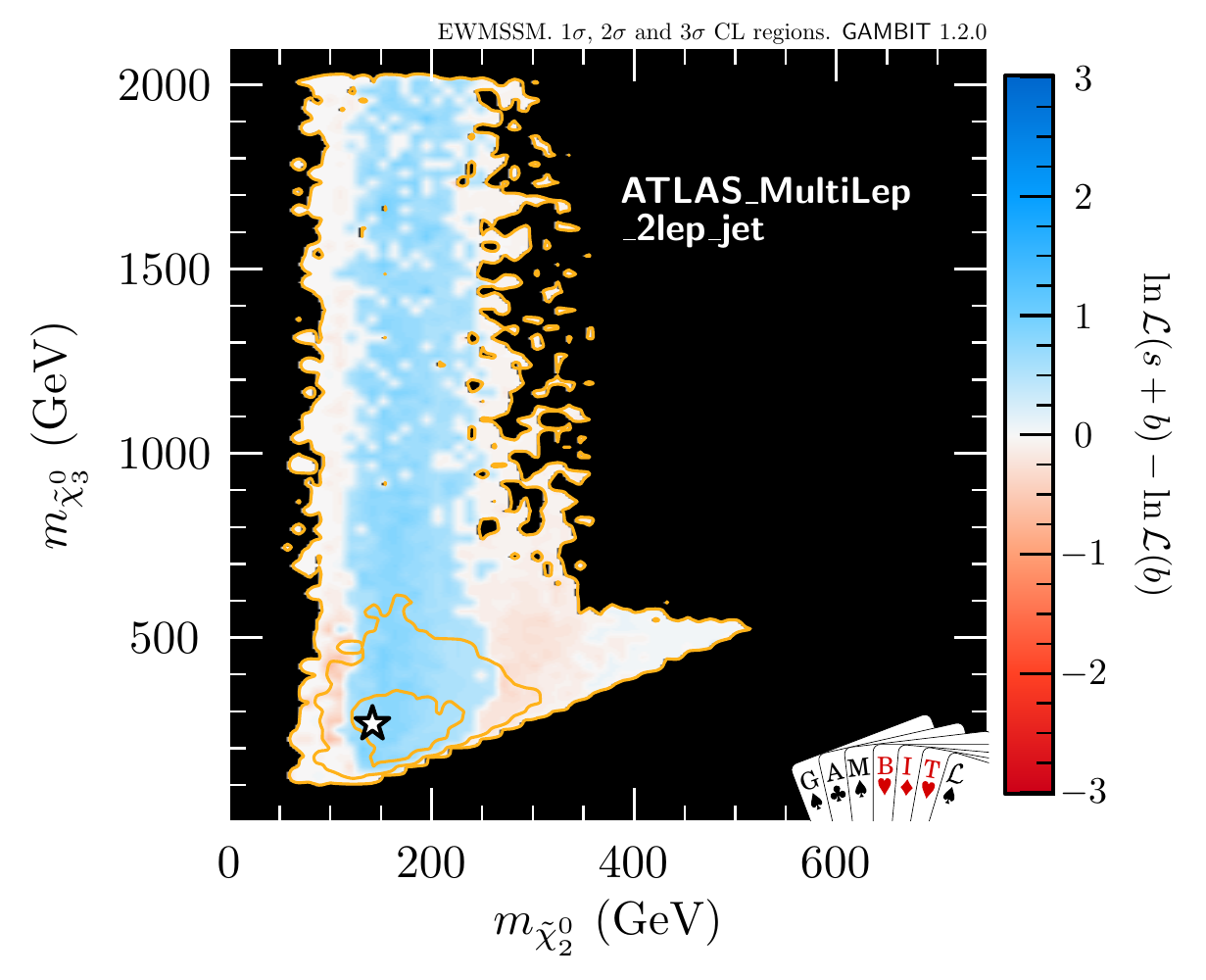}
  \includegraphics[width=0.32\textwidth]{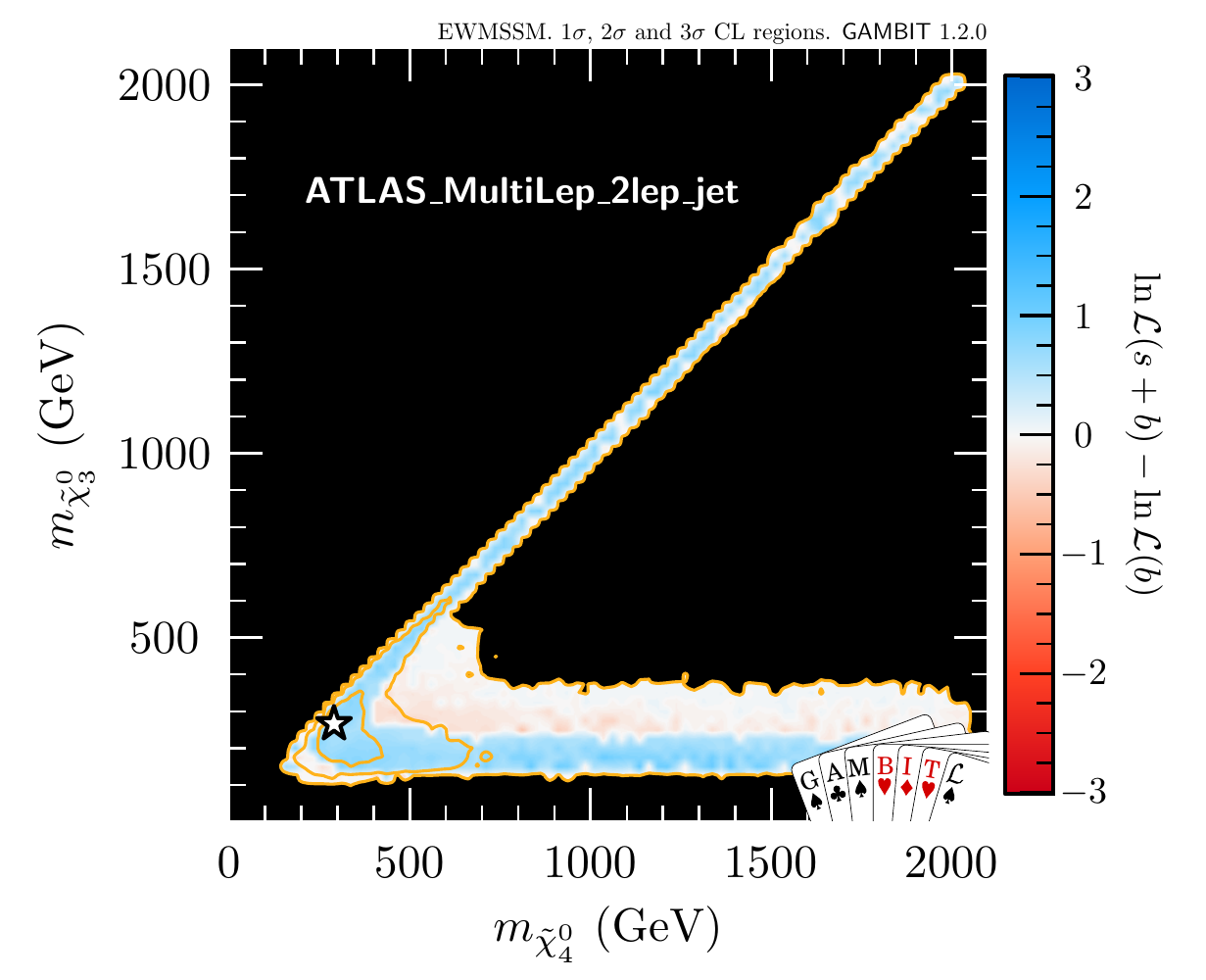}
  \includegraphics[width=0.32\textwidth]{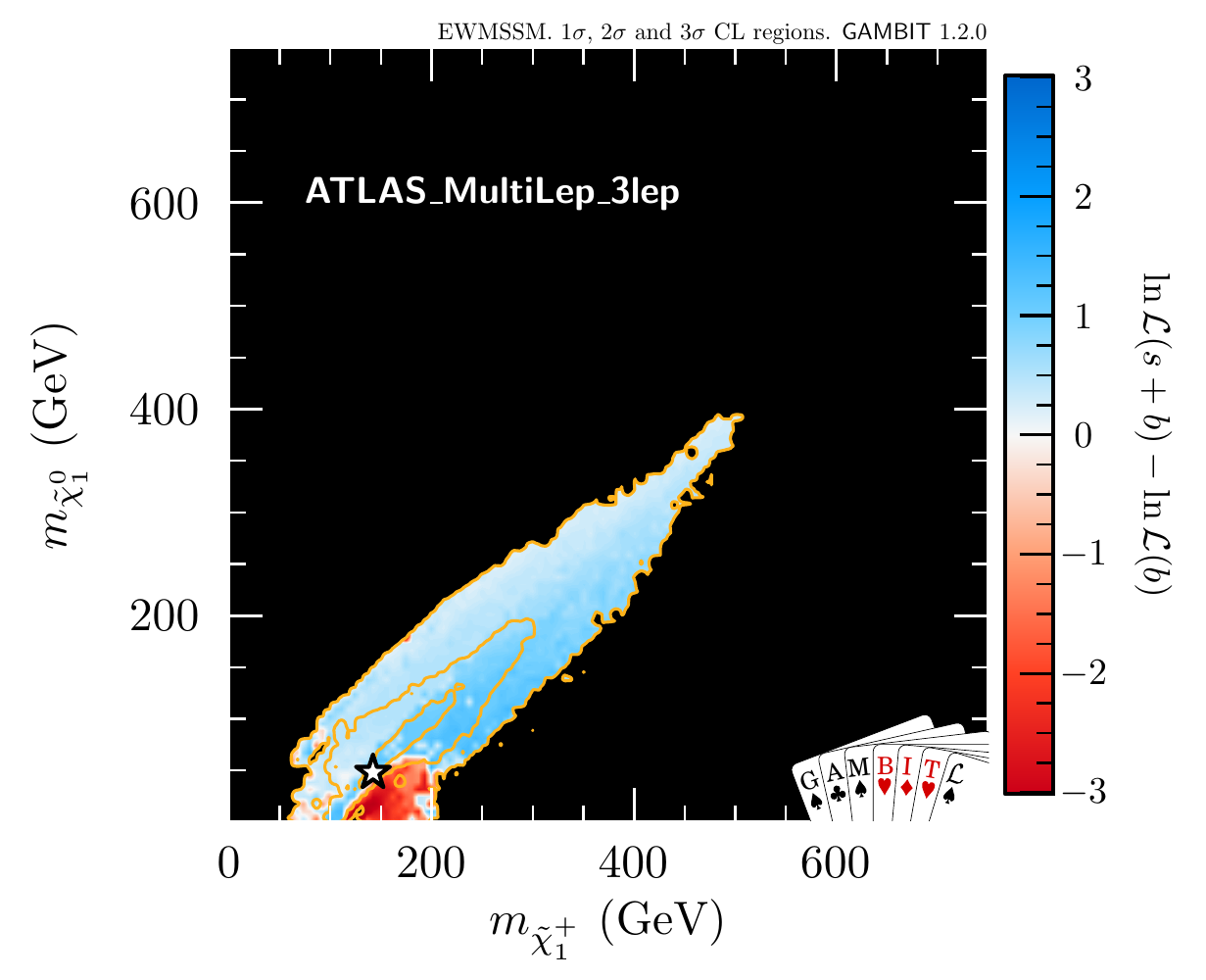}
  \includegraphics[width=0.32\textwidth]{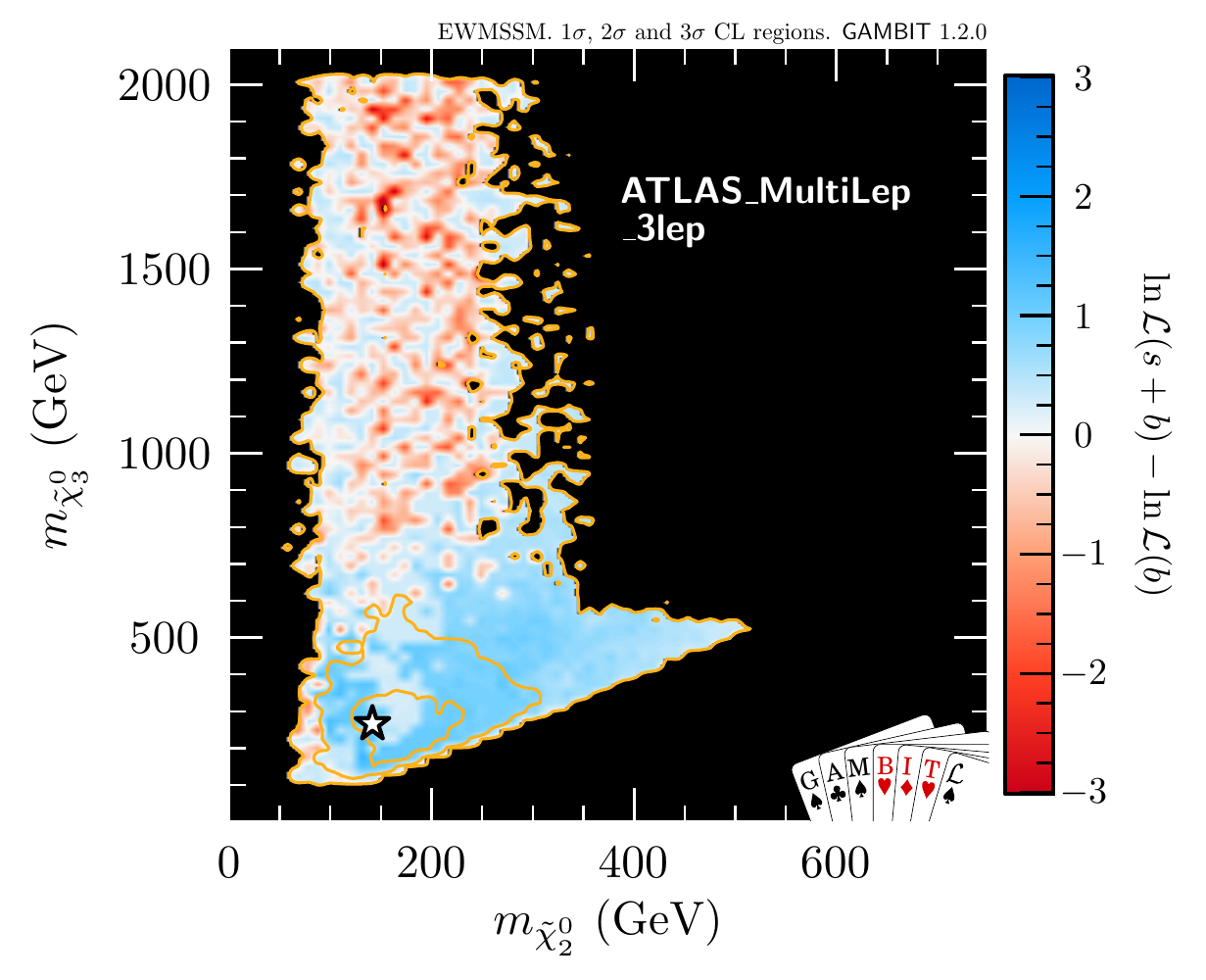}
  \includegraphics[width=0.32\textwidth]{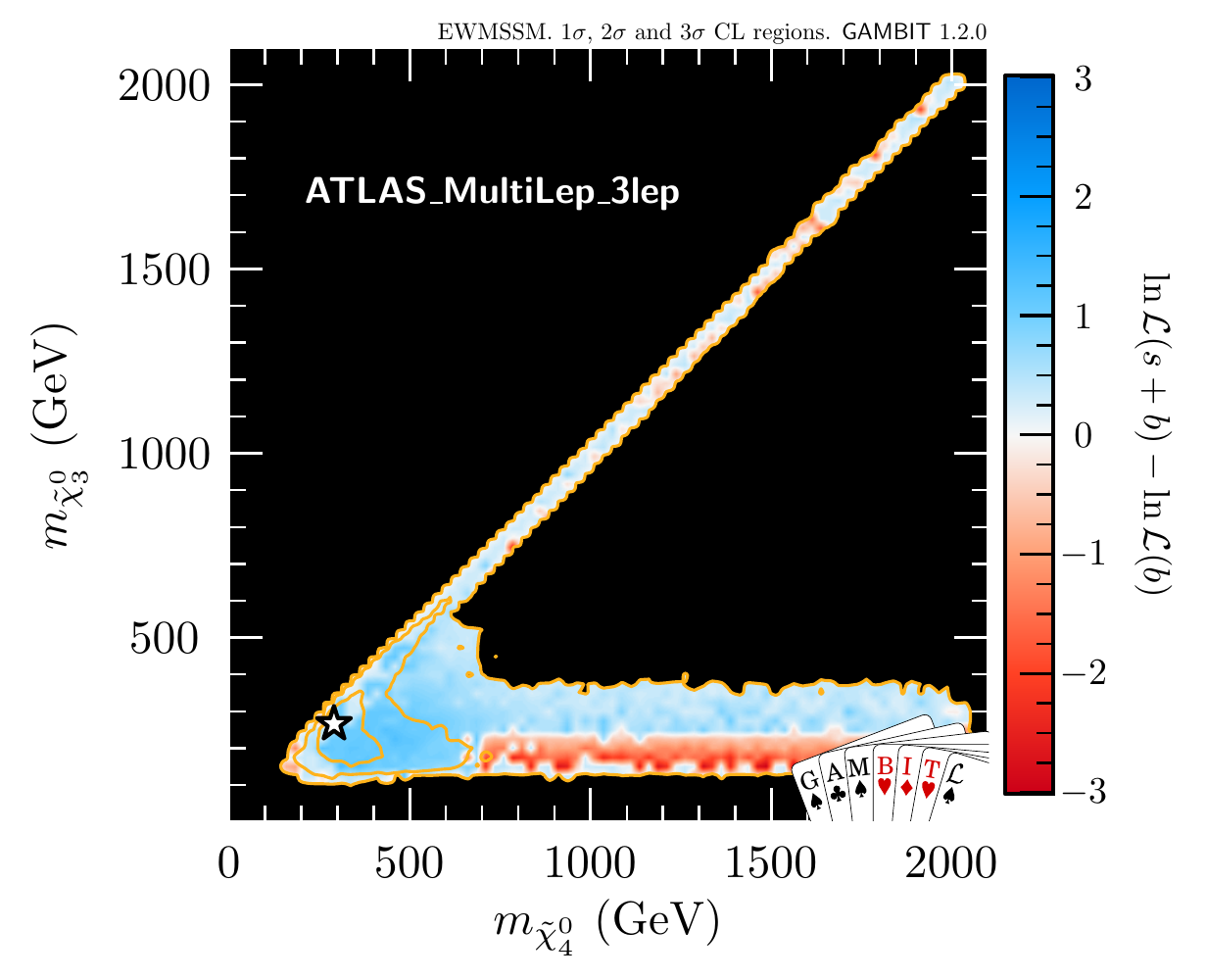}
  \includegraphics[width=0.32\textwidth]{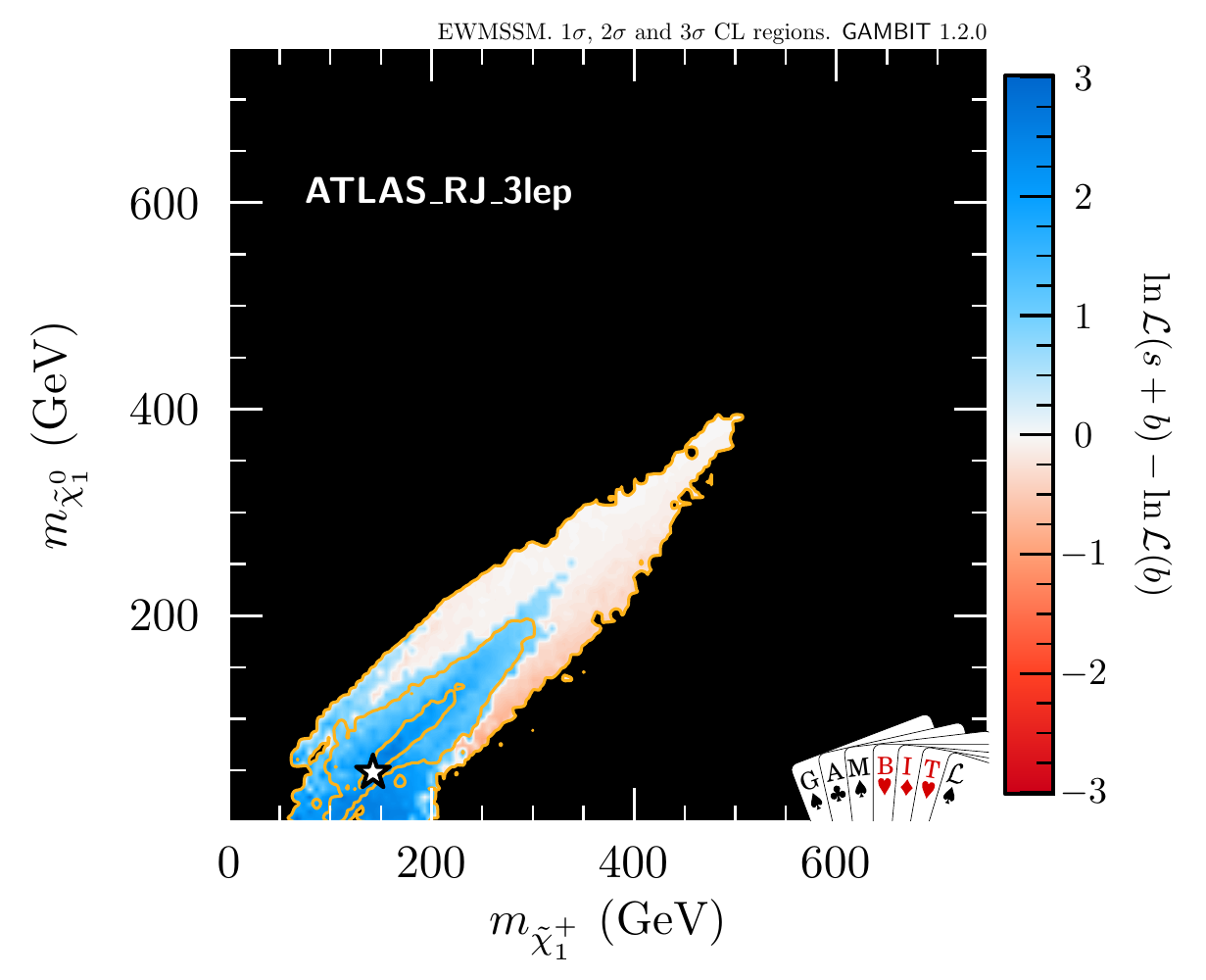}
  \includegraphics[width=0.32\textwidth]{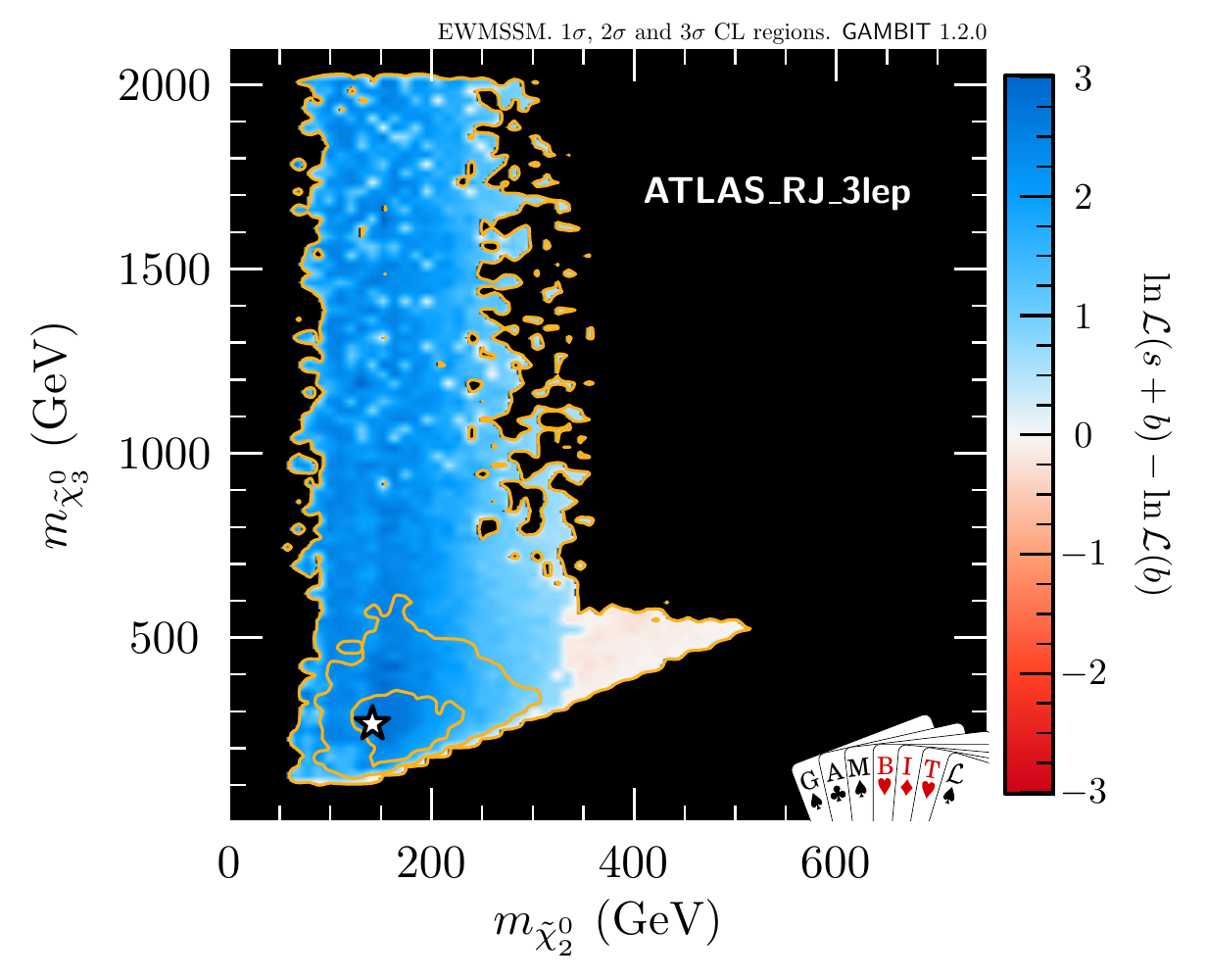}
  \includegraphics[width=0.32\textwidth]{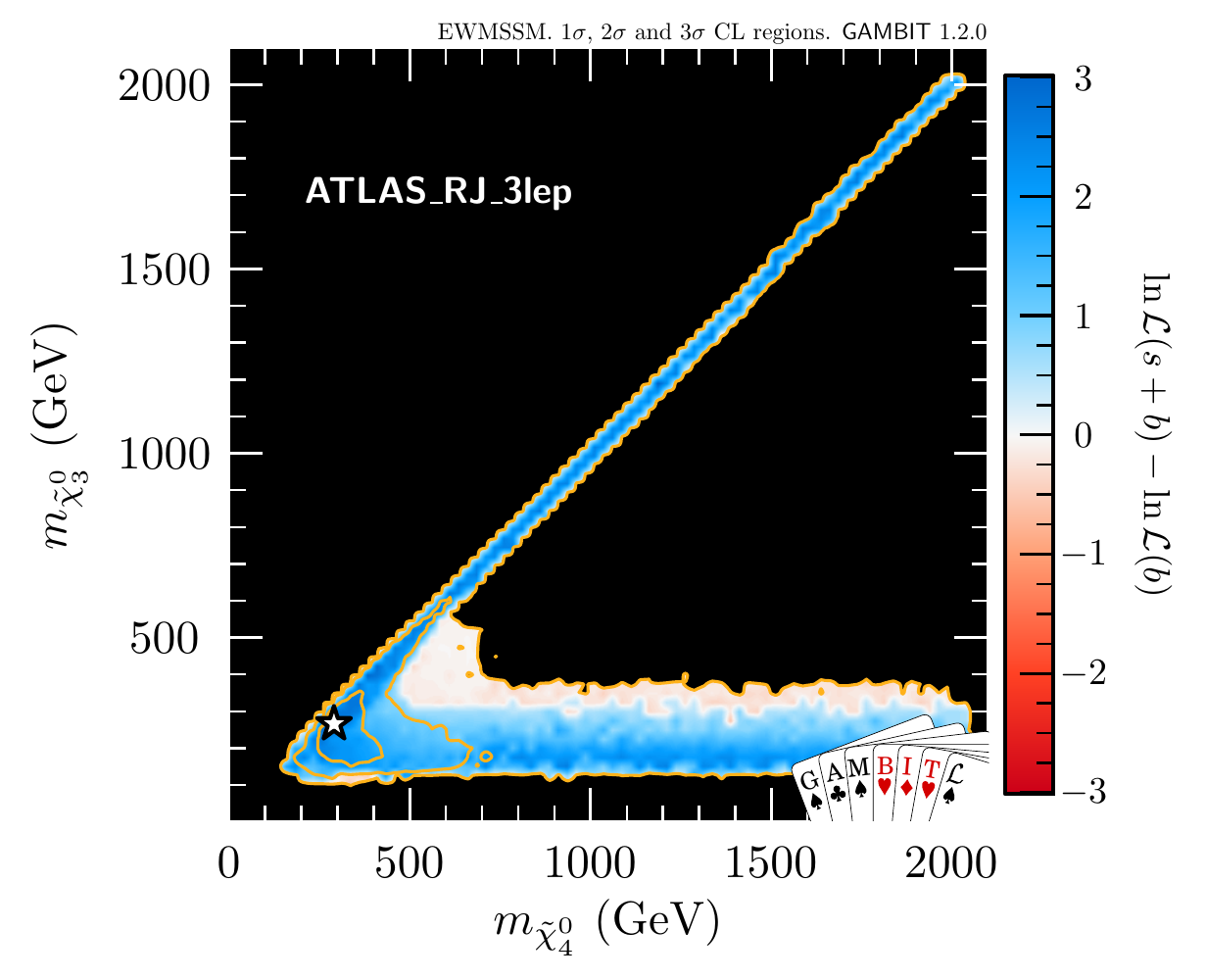}
  \caption{
  Contributions to the total fit likelihood from the ATLAS searches in Ref.\ \cite{Aaboud:2018zeb} (top), Ref.\ \cite{Aaboud:2018jiw} (second and third rows), and Ref.\ \cite{Aaboud:2018sua} (bottom), shown across the full $3\sigma$ regions in the $(m_{\CharOne}, m_{\NeutOne})$ plane (left), the $(m_{\NeutTwo}, m_{\NeutThree})$ plane (middle), and the $(m_{\NeutFour}, m_{\NeutThree})$ plane (right). In the blue regions a non-zero signal prediction in the given search improves the overall fit, while in red regions the signal prediction worsens the fit. In the white regions the given search is not sensitive. The orange contours outline the $1\sigma$, $2\sigma$ and $3\sigma$ regions preferred in the fit. The white star marks the best-fit point. From~\cite{EWMSSM}.
  }
  \label{fig:EWMSSM_mass_planes_per_analysis}
\end{figure}

To understand the interplay between the analyses contributing to the excess, we can look at their individual likelihood contributions across the combined best-fit surface. This is done in Fig.\ \ref{fig:EWMSSM_mass_planes_per_analysis}, where the contributions from four ATLAS results are displayed across the preferred $3\sigma$ regions in three different mass planes. When reading these plots it is important to keep in mind that the plotted points are those parameter samples picked out by profiling the \textit{total} likelihood. The sharp changes in analysis likelihood seen in some plots are due to abrupt changes in what scenarios are picked out by this profiling, which again changes which signal region is selected to set the analysis likelihood value.

One example of the interplay between analyses is seen by comparing the middle panels on the first and third rows. The first of these show the likelihood contribution from an ATLAS search for 4-lepton final states, with the leptons coming from two $Z$ bosons~\cite{Aaboud:2018zeb}. We see that fitting a 4-lepton excess in the \EWMSSM relies on having non-negligible production of $\NeutThree$, as this allows for signal leptons from the decays $\NeutThree \rightarrow Z \Neut_{1,2}$. The second of these panels is for an ATLAS search for 3-lepton final states~\cite{Aaboud:2018jiw}, designed to target $\NeutTwo \CharOne$ production. For a given $m_{\NeutTwo}$, reducing $m_{\NeutThree}$ to $\lesssim 600$\,GeV (as preferred by the 4-lepton search) also improves the fit to this 3-lepton search, which for high $m_{\NeutThree}$ sees some tension with the data. At lower $m_{\NeutThree}$ production processes with $\NeutThree$ come into play, involving more complicated event topologies. At the same time the production cross-section for the $\NeutTwo \CharOne$ pair is reduced somewhat, due to a higher Higgsino component. The combined effect is a change in which 3-lepton signal region is identified as having the best expected sensitivity.

The combined excess in the $13$\,TeV searches is estimated in~\cite{EWMSSM} to have a local significance of $3.3\sigma$. The impact of $8$\,TeV LHC results on the preferred low-mass scenarios is investigated by post-processing all parameter samples in the $1\sigma$ region with simulations of relevant ATLAS and CMS electroweakino searches at $8$\,TeV \cite{Aad:2015jqa,ATLAS:2LEPEW_20invfb,ATLAS:3LEPEW_20invfb,CMS:3LEPEW_20invfb}. The result is an upwards shift in the best-fit mass spectrum, by $\sim20$\,GeV in all masses, and a small reduction of the estimated significance of the excess, to $2.9\sigma$.

We also note that even though the \EWMSSM fit did not include DM constraints, parts of the preferred parameter space do give acceptable relic density predictions while avoiding exclusion from current direct and indirect DM searches. This is possible for scenarios with $m_{\NeutOne}$ close to $m_Z/2$ or $m_h/2$, where resonant annihilations via the $Z/h$ funnel can bring the predicted relic density close to or below the observed value.

While the small excess seen in the \EWMSSM fit is quite possibly due to background fluctuations, the fit demonstrates two important points. First, that LHC constraints on light SUSY can be significantly weaker in realistic SUSY such as the MSSM than in simplified models.\footnote{While not discussed here, the analysis in~\cite{EWMSSM} shows that for every mass hypothesis in the $(m_{\NeutTwo}, m_{\NeutOne})$ plane -- not just for points in the best-fit region -- there is a point in the \EWMSSM parameter space that fits the \textit{combined} collider results at least as well as the SM expectation.} Second, that proper statistical combinations of collider searches can be a powerful tool to uncover suggestive patterns in BSM parameter spaces.

\subsection{Higgs Portal models for dark matter}

No definitive evidence has yet been uncovered for non-gravitational interactions of DM with the SM.  At some level however, such interactions must be inevitably generated by effective operators connecting Lorentz-invariant, gauge singlet combinations of SM particles to equivalently symmetric combinations of DM fields.  The lowest-dimension such operator in the SM is the Higgs bilinear $H^\dagger H$.  Depending on the spin and gauge representation of a DM candidate $X$, the lowest-order Lorentz- and gauge-invariant DM operator may be either the bilinear $X^\dagger X$, or a lone DM field. Operators linear in $X$ are only consistent if $X$ is itself a Lorentz invariant (i.e.\ a scalar), and a gauge singlet.  If it is to be a viable DM candidate however, $X$ must be stable on cosmological timescales. The most straightforward way to achieve this is for $X$ to hold a different charge to SM particles under some new unbroken (typically discrete) symmetry.  This has the effect of forbidding terms linear in $X$, preventing the field from decaying.

The lowest-order operator connecting $X$ to the SM guaranteed to exist at some level is therefore the so-called `Higgs portal' operator $X^\dagger X H^\dagger H$.  Following electroweak symmetry breaking, this operator gives rise to a mass term for $X$ proportional to $v_0^2$ (with $v_0$ the vacuum expectation value of the Higgs field), a Higgs-DM-DM vertex proportional to $v_0$, and a direct four-particle vertex between two Higgses and two DM particles.  The new 3-particle and 4-particle interactions of $X$ with the Higgs boson lead to DM annihilation (enabling thermal production and possible indirect detection), spin-independent DM-nucleon scattering (leading to possible direct detection), DM production at colliders (with the possibility for signals in e.g.\ monojet searches), and invisible decays of the Higgs to two DM particles when $m_X < m_h/2$.

Depending on the Lorentz representation of DM, $X^2 H^2$ may be a fully renormalisable dimension 4 operator (if $X$ is a scalar), an effective dimension 4 operator (if $X$ is a vector), or an effective dimension 5 operator (if $X$ is a fermion).  All three of these cases have been considered in detail in the literature, with a particular focus on models where $X$ is itself a gauge singlet and the $X^2H^2$ term is therefore the sole link between DM and the SM.  The most commonly studied cases have been the $\mathbb{Z}_2$-symmetric scalar [\citenum{SilveiraZee,McDonald94,Burgess01,Davoudiasl:2004be,Goudelis09,Yaguna09,Profumo2010a,Andreas:2010dz,Arina11,Mambrini11, Raidal:2011xk,Mambrini:2011ik,He:2011de,Drozd:2011aa,Okada:2012cc,Cheung:2012xb,Okada:2013bna,Cline:2013gha,Chacko:2013lna, Endo:2014cca,Craig:2014lda, Feng15,Duerr15,arXiv:1510.06165,Duerr16,He:2016mls,Han:2016gyy,Dupuis:2016fda,Cuoco:2016jqt,Binder:2017rgn,Ghorbani:2018yfr,Chiang:2018gsn,Stocker:2018avm,Hardy:2018bph,Bernal:2018kcw,Glioti:2018roy,Urbano:2014hda,Escudero:2016gzx,Kanemura:2011nm,Djouadi:2011aa,Djouadi:2012zc,Bishara:2015cha,Ko:2016xwd,Beniwal:2015sdl,Kamon:2017yfx,Dutta:2017sod,Dick:2018lqx}; \gambit analyses \citenum{SSDM,SSDM2}], vector [\citenum{Djouadi:2011aa,Kanemura:2011nm,Djouadi:2012zc,Bishara:2015cha,Chen:2015dea,DiFranzo:2015nli, Beniwal:2015sdl,Ko:2016xwd,Kamon:2017yfx,Dutta:2017sod,arXiv:1704.05359,Dick:2018lqx,Baek:2014jga}; \gambit analysis \citenum{HP}] and fermionic [\citenum{Djouadi:2011aa,Kanemura:2011nm,LopezHonorez:2012kv,Djouadi:2012zc,Urbano:2014hda, Baek:2014jga,Bishara:2015cha,Beniwal:2015sdl,Ko:2016xwd,Fedderke:2014wda,Matsumoto:2014rxa,arXiv:1506.04149, arXiv:1506.08805,arXiv:1506.06556, Escudero:2016gzx,Kamon:2017yfx,Dutta:2017sod,Dick:2018lqx,Matsumoto:2018acr}; \gambit analysis \citenum{HP}] variants, along with the $\mathbb{Z}_3$-symmetric scalar [\citenum{Belanger2013a,Kang:2017mkl,2017JHEP...10..088B,Hektor:2019ote,Kannike:2019mzk}; \gambit analysis \citenum{SSDM2}].

\subsubsection{$\mathbb{Z}_2$-symmetric scalar singlet}

The simplest Higgs portal model for DM, and indeed probably the most minimal of all models for particle DM, is a single, real, gauge-singlet scalar field $S$, protected from decay by a $\mathbb{Z}_2$ symmetry.  The only new renormalisable Lagrangian terms allowed by gauge, Lozentz and $\mathbb{Z}_2$ symmetry are
\begin{equation}
\mathcal{L}_{\mathbb{Z}_2} = \frac12 \mu_{\sss S}^2 S^2 + \frac14\ls S^4 + \frac12\lhs S^2|H|^2.
\label{L_S}
\end{equation}
The model is fully specified by the $S$ bare mass $\mu_{\sss S}$, the dimensionless $S$ quartic self-coupling $\ls$, and the dimensionless Higgs portal coupling $\lhs$.  For the most part, the $S$ quartic coupling has little impact on the phenomenology of the model, as it leads only to DM self-interactions, which are not sufficiently constrained by existing data to place strong limits on $\ls$.  A key exception, however, is the impact of $\ls$ on the running of gauge couplings under renormalisation group flow, which can have important implications for stability of the electroweak vacuum.

Denoting the physical SM Higgs field by $h$, following electroweak symmetry breaking $H \rightarrow \left[0, (v_0+h)/\sqrt{2}\right]^\text{T}$.  This generates new vertices of the form $v_0hS^2$ and $h^2S^2$, and induces a shift to the $S$ bare mass, such that at tree level
\begin{equation}
\ms = \sqrt{\mu_{\sss S}^2 + \frac12{\lhs v_0^2}}.
\label{ms}
\end{equation}

\begin{figure}[t]
\centering
\includegraphics[width=0.6\columnwidth]{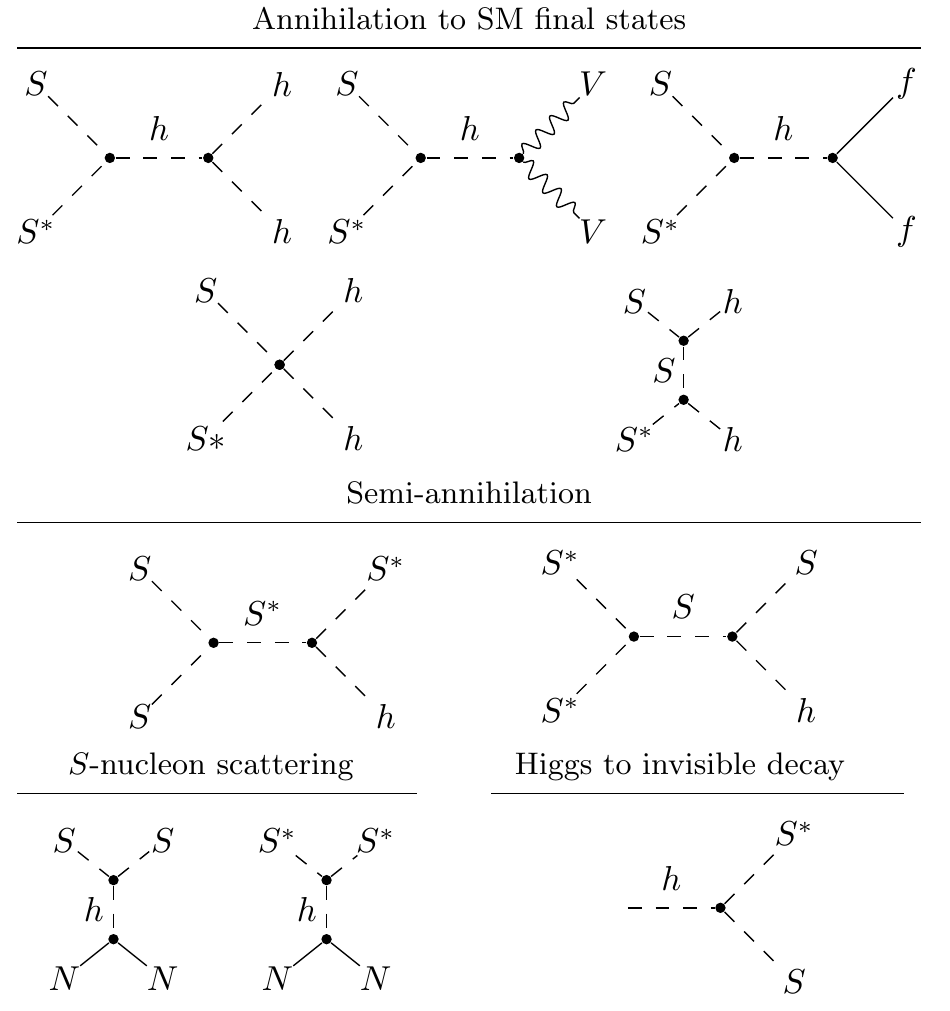}
\caption{Feynman diagrams for annihilation, semi-annihilation, nuclear scattering and Higgs decays in scalar singlet Higgs portal models.  $N, f$ and $V$ refer to nucleons, fermions and SM electroweak vector bosons ($Z$ and $W$), respectively. Diagrams are shown for the $\mathbb{Z}_3$-symmetric case, where DM exists in $S$ and anti-$S$ (i.e. $S^*$) states, but the same diagrams apply in the $\mathbb{Z}_2$-symmetric case with $S=S^*$, except for semi-annihilation (which is absent in the $\mathbb{Z}_2$ model).  The same diagrams also apply to $\mathbb{Z}_2$-symmetric vector and fermionic Higgs portal models (with $S$ replaced by the relevant DM particle and semi-annihilation also forbidden by the $\mathbb{Z}_2$ symmetry). From \cite{SSDM2}.}
\label{fig:diagrams}
\end{figure}

\begin{figure}[t]
	\centering
  \includegraphics[width=0.495\columnwidth]{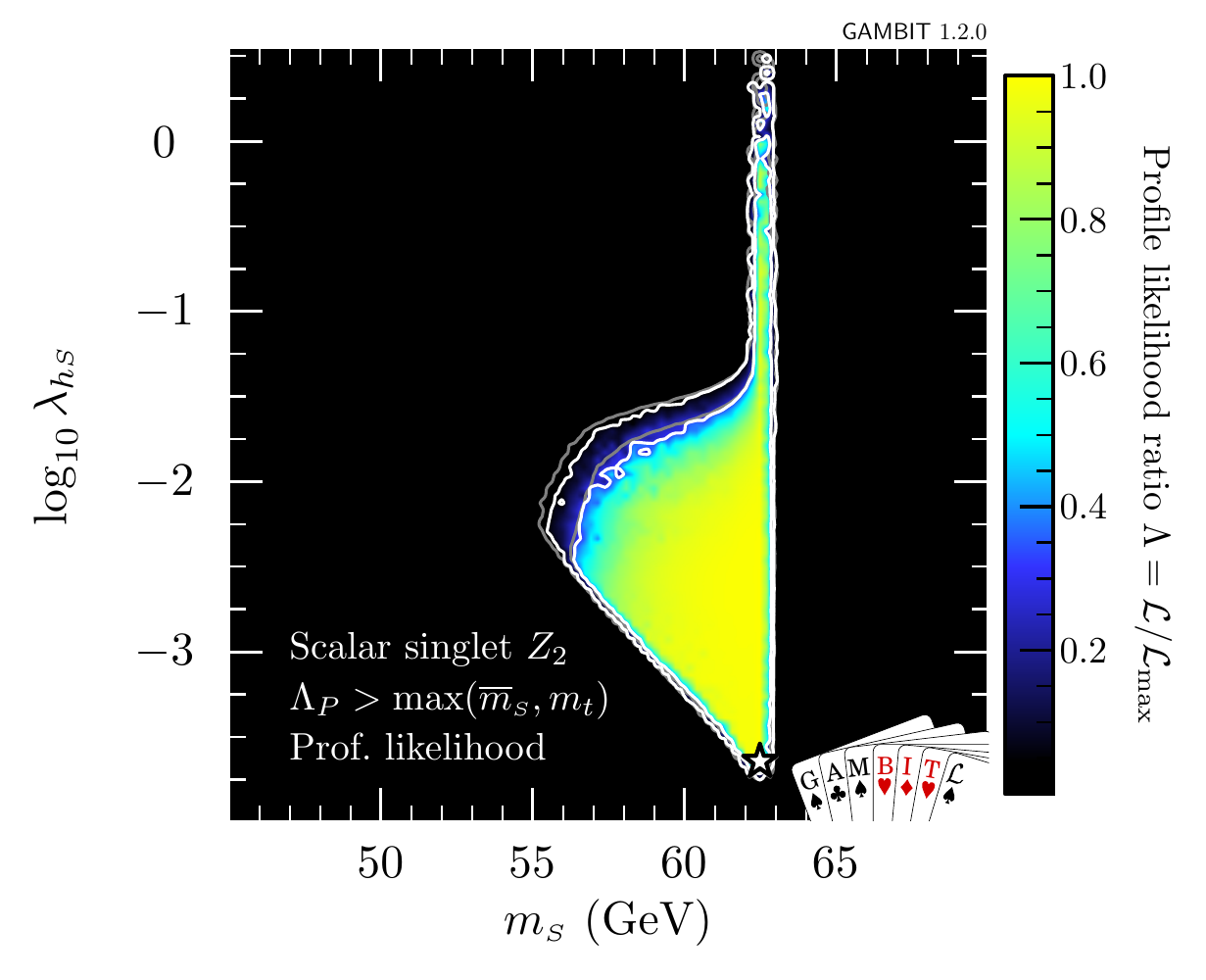}
  \includegraphics[width=0.495\columnwidth]{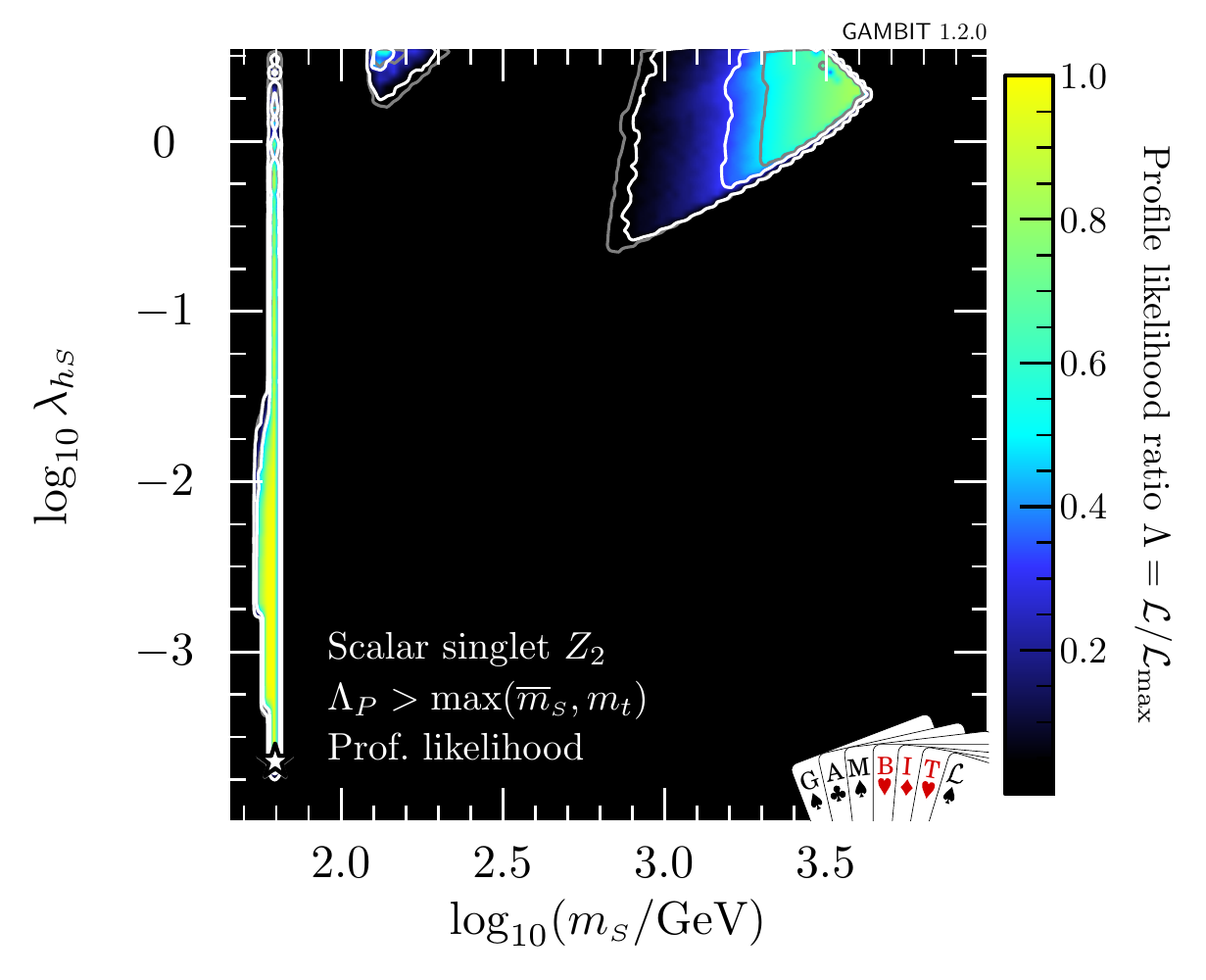}
  \caption{Profile likelihoods of parameters in the $\mathbb{Z}_2$-symmetric scalar singlet Higgs portal dark matter model, including constraints from direct and indirect detection, the relic density of dark matter and LHC searches for invisible decays of the Higgs boson, along with various Standard Model, dark matter halo and nuclear uncertainties.  \textit{Left}: the low-mass resonance region.  \textit{Right}: the full mass range.  Contours show 1 and 2$\sigma$ confidence regions, with white corresponding to the main scan (including the 2018 XENON1T direct search \cite{Aprile:2018dbl}) and grey to a secondary scan using the 2017 XENON1T result \cite{Aprile:2017iyp}.  White stars indicate the location of the best-fit point.  From \protect\cite{SSDM2}.}
	\label{fig:z2scalar1}
\end{figure}

\begin{figure}[t]
	\centering
  \includegraphics[width=0.495\columnwidth]{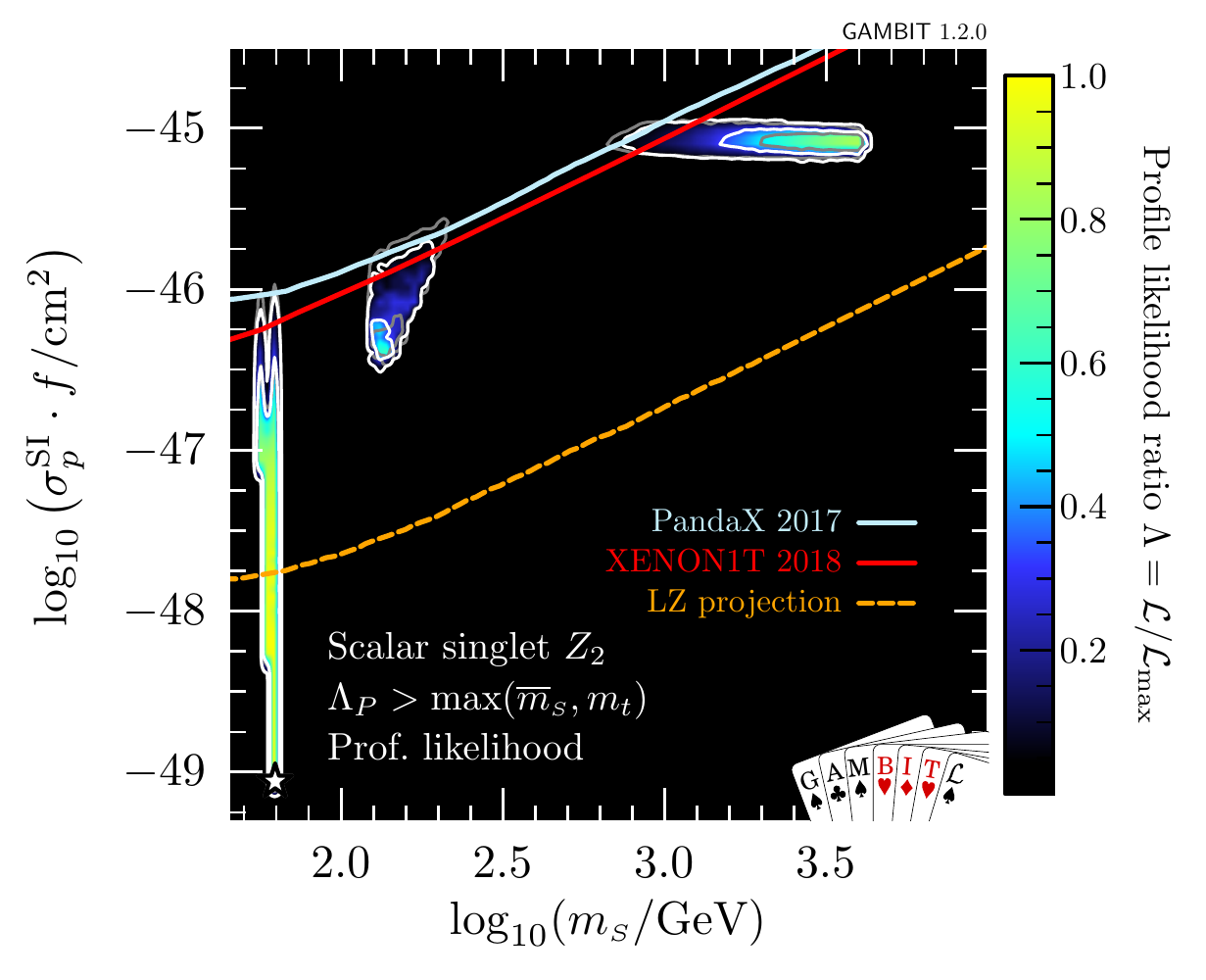}
  \caption{Results from the same analysis of the $\mathbb{Z}_2$-symmetric scalar singlet Higgs portal dark matter model as shown in Fig.\ \ref{fig:z2scalar1}, but plotted in the plane of the effective spin-independent nuclear scattering cross-section and the scalar mass, in order to compare directly to the sensitivity of direct detection experiments.  All models have their effective cross-section defined as $f\sigma_\mathrm{SI}$, where $f\equiv \Omega_S / \Omega_\mathrm{DM}$ is the fraction of the relic density constituted by the scalar singlet.  Experiments assume $f=1$ when publishing their results. Contours show 1 and 2$\sigma$ confidence regions, and stars best fits. From \protect\cite{SSDM2}.}
	\label{fig:z2scalar2}
\end{figure}

\begin{figure}[t]
	\centering
  \includegraphics[width=0.495\columnwidth]{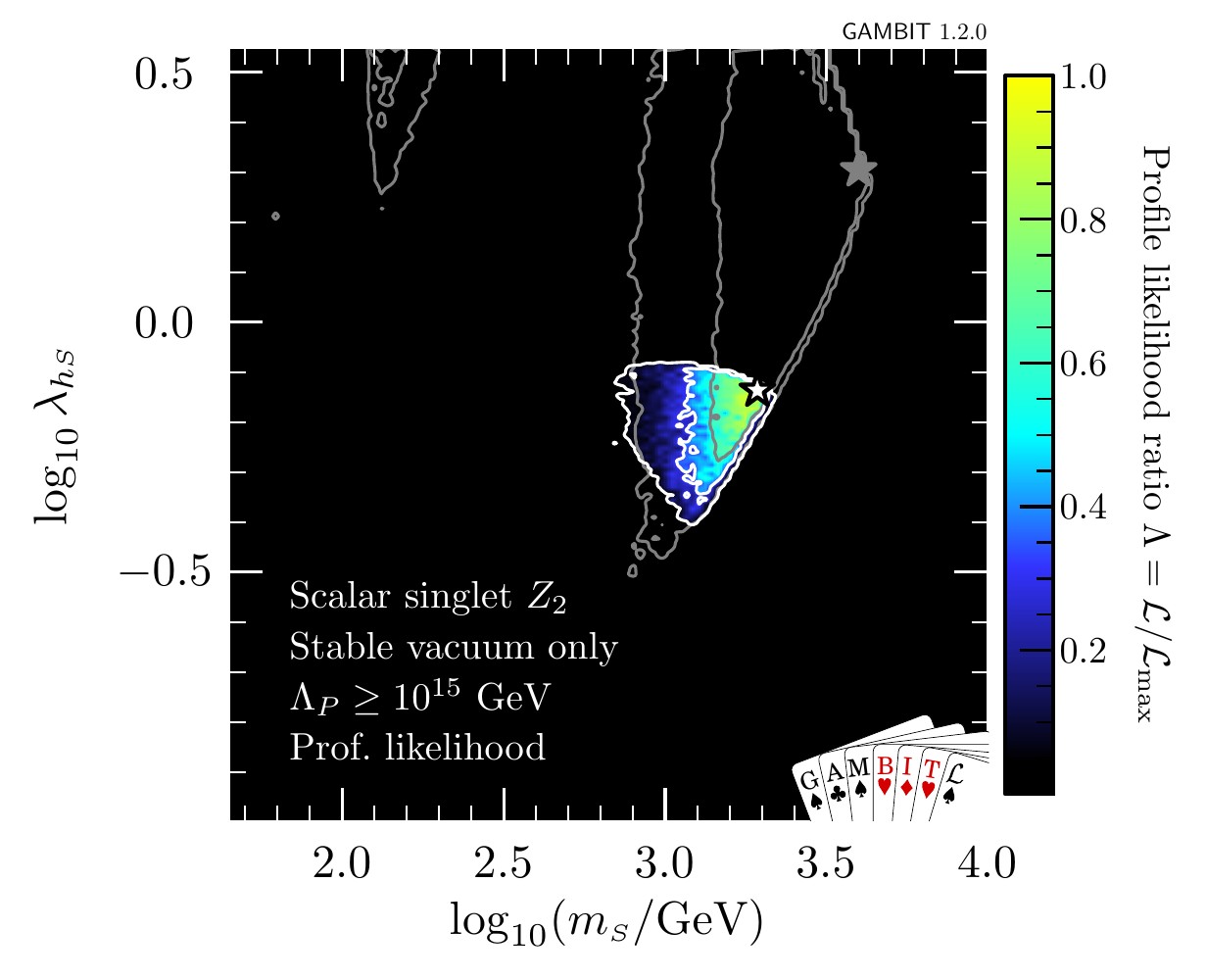}
  \includegraphics[width=0.495\columnwidth]{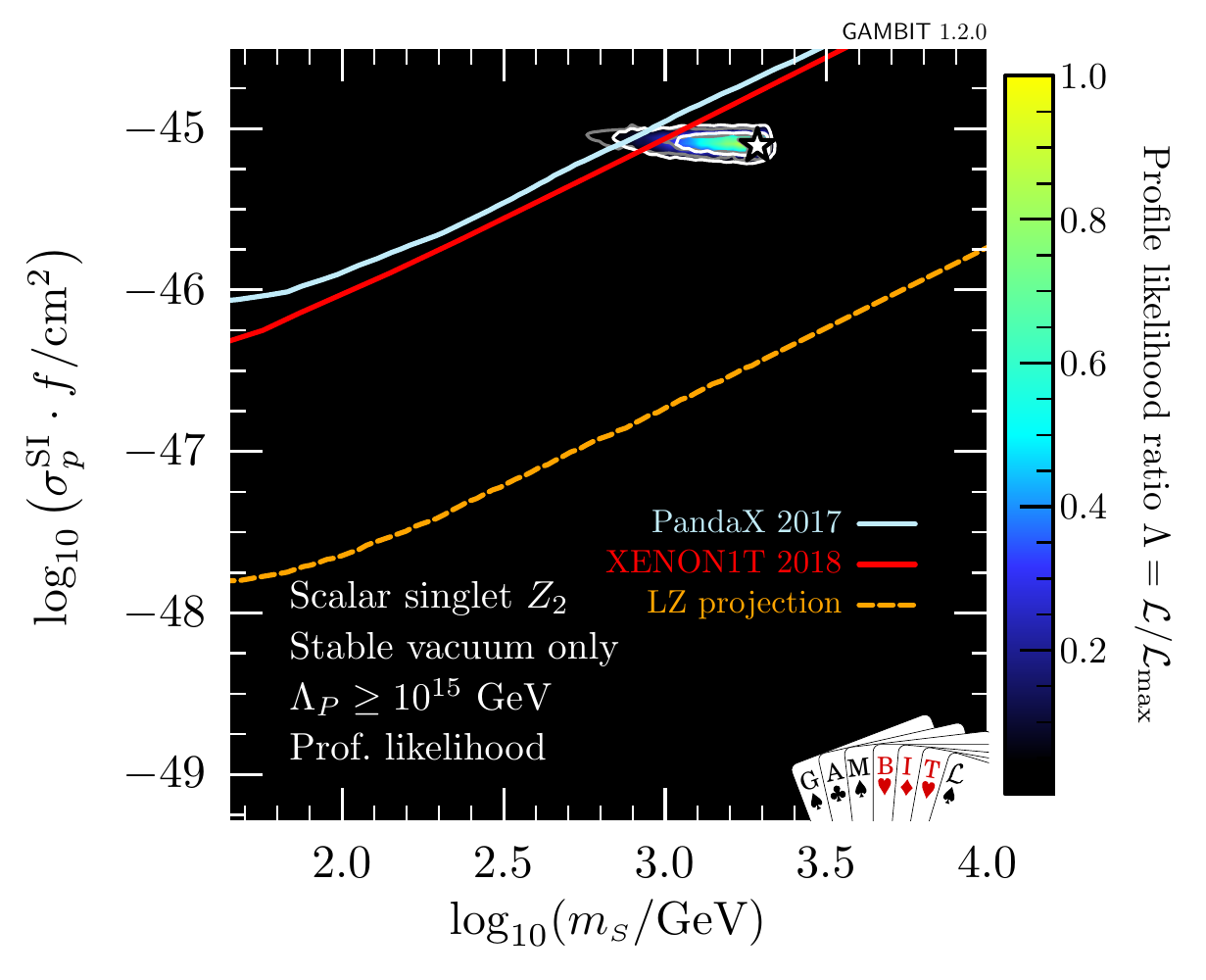}
  \caption{Regions in the $\mathbb{Z}_2$-symmetric scalar singlet model that satisfy all experimental constraints, stabilise the electroweak vacuum and remain perturbative up to scales of $10^{15}$\,GeV. Contours show 1 and 2$\sigma$ confidence regions, and stars best fits. Grey contours show the allowed regions without the requirements of vacuum stability and perturbativity. From \protect\cite{SSDM2}.}
	\label{fig:z2scalar3}
\end{figure}

The interaction with the physical Higgs endows $S$ with essentially all of the classic phenomenology of WIMP DM, via the diagrams shown in Fig.\ \ref{fig:diagrams} -- along with the added possibility of Higgs decays $h\to SS$ where $\ms \le \mh/2$.  The leading constraints on the model come from searches for gamma rays from dark matter annihilation in dwarf spheroidal galaxies \cite{LATdwarfP8}, the observed relic density of dark matter \cite{Planck18cosmo}, direct searches performed by the XENON1T \cite{Aprile:2018dbl} and PandaX \cite{Cui:2017nnn} experiments, and searches for invisible Higgs decays at the LHC \cite{Belanger:2013xza,CMS-PAS-HIG-17-023}.

The resulting preferred regions of parameter space are shown in Fig.\ \ref{fig:z2scalar1}.  These results explicitly allow models where $S$ is only a fraction of the observed DM, and include a fully self-consistent rescaling of the predicted signals at direct and indirect searches according to the fraction $f \le 1$ of DM constituted by $S$ at each point in the parameter space.  The allowed parameter space splits into three regions: one at high masses where direct detection loses sensitivity, a second at intermediate mass where the 4-boson vertex boosts the annihilation cross-section and depletes the relic density, and another at and immediately below $\ms = \mh/2$, where $S$ annihilates highly efficiently via an $s$-channel resonance mediated by the Higgs, depleting the relic density to below the observed value even for very small values of $\lhs$.

The Higgs invisible width constraint rules out large couplings $\lhs$ at singlet masses below the resonance.  The thermal relic density of $S$ provides the lower limit of the low-mass and high-mass allowed regions.  Indirect detection plays the leading role only on the high-mass edge of the resonance, where thermal effects in the early Universe push annihilation slightly off resonance but late-time annihilation remains strongly boosted.  Direct detection plays a significant role throughout the parameter space, as can be seen in Fig.\ \ref{fig:z2scalar2}.  Except for the very bottom of the resonance region, the entirety of the model will soon be probed by direct detection.

Gamma-ray lines do not provide any meaningful constraint, as the partial annihilation cross-section for $SS\to\gamma\gamma$ is only appreciable in parts of the parameter space where the relic density is significantly suppressed.  Likewise, monojet searches only constrain very large values of $\lhs$ already excluded by other constraints or expected to lead to new strong dynamics.  Indeed, both these points also apply to all other Higgs portal models that we discuss in this review.

Given that the Higgs portal operator is not just an effective interaction, but a fully renormalisable operator in this model, it is also important to consider the UV behaviour of the theory.  Due to the observed values of the top and Higgs masses, the SM posesses a second minimum in its scalar potential at $\gtrsim \mathcal{O}(10^{15})$\,GeV, causing the low-scale vacuum in which we reside to be metastable.  Adding an additional scalar to the SM impacts the running of the Higgs quartic coupling, raising its value at high scales.  This can prevent the quartic coupling from running negative, and make the low-scale minimum a global rather than a local one.  The catch is that $\ls$ must be relatively large in order to achieve this effect.  Fig.\ \ref{fig:z2scalar3} shows the parts of the parameter space, consistent with all experimenal constraints, where $\ls$ can be pushed high enough to stabilise the SM vacuum, but without pushing any of the couplings non-perturbative below a scale of $10^{15}$\,GeV. Clearly, the $\mathbb{Z}_2$-symmetric scalar singlet can solve the vacuum stability problem without introducing new strong dynamics, and satistfy all experimental constraints, but only in a region around \mbox{$\ms = 1$--2\,TeV} and $\sigma_\mathrm{SI} \sim 10^{-45}$\,cm$^2$.  Curiously, this is also in the region consistent with the (admittedly very small) excess seen in the most recent XENON1T results \cite{Aprile:2018dbl}.  In any case, this hypothessis will clearly be tested very quickly in the upcoming runs of the LZ and XENONnT \cite{Akerib:2015cja,XENONnTLZ} experiments.

The results in Figs.\ \ref{fig:z2scalar1}--\ref{fig:z2scalar3} are based on profile likleihood analyses, and illustrate what is possible in each parameter plane, were one able to freely vary the other parameters of the theory (including nuisance parameters) in order to achieve the best possible fit to all available data.  If one instead carries out a Bayesian analysis, looking instead at the posterior probability density for these parameters, a different picture emerges.  In this case, parameter combinations become more likely if they can provide a good fit for a broader range of values of the other parameters of the theory, i.e.\ if they can fit the data with less fine tuning.  In this case, the low-mass resonance region is strongly disfavoured, as `hitting' the resonance and avoiding the relic density constraint for a given value of $\ms$ requires some fine-tuning of various SM nuisance parameters such as $\mh$; the same is true to a lesser extent for the intermediate-mass region as well.  We therefore see that from a Bayesian perspective, the region where the singlet model stabilises the SM vacuum is in fact favoured over the other regions of the theory, even before considering the implications for vacuum stability.

\begin{figure}[t]
	\centering
  \includegraphics[width=0.495\columnwidth]{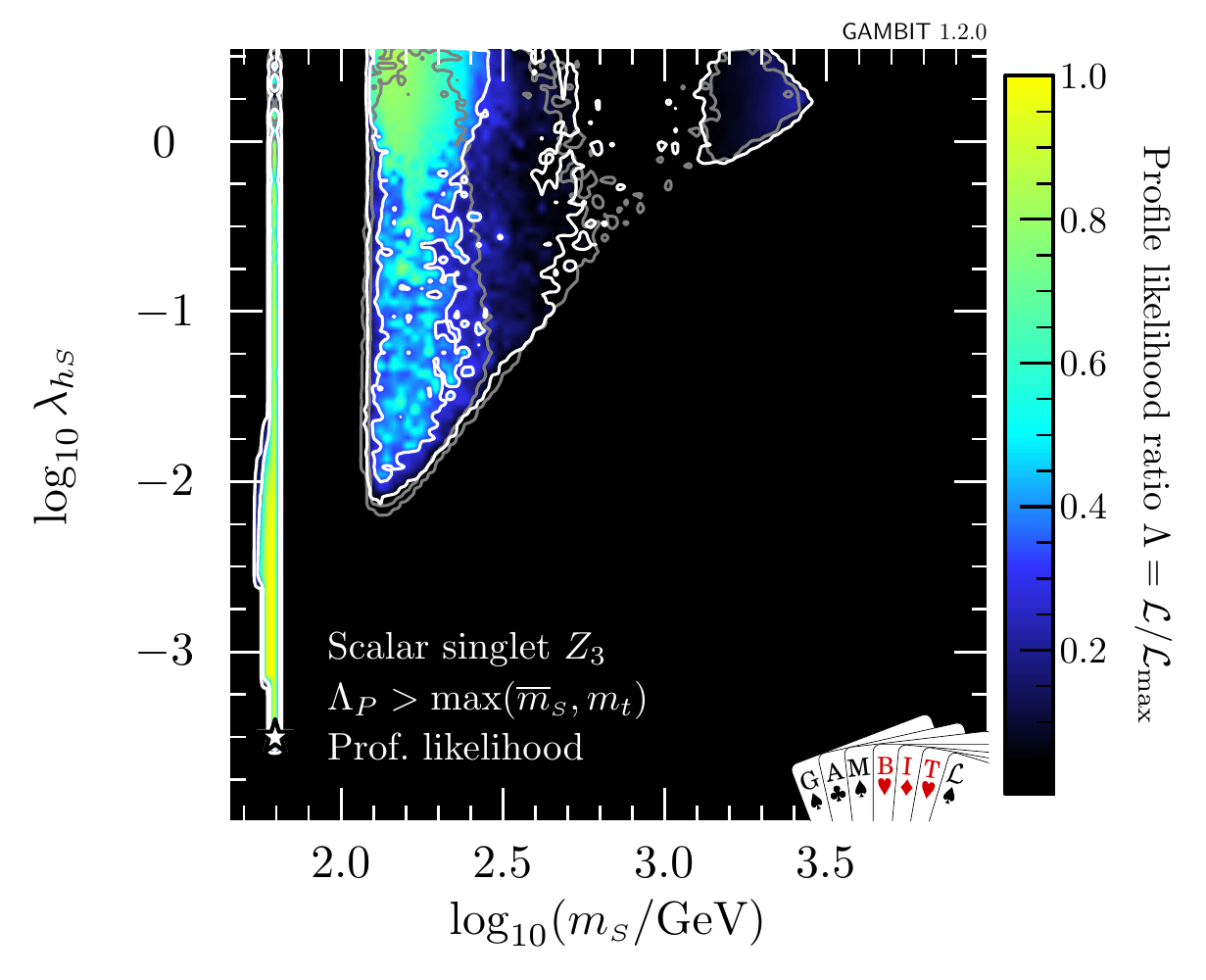}
  \includegraphics[width=0.495\columnwidth]{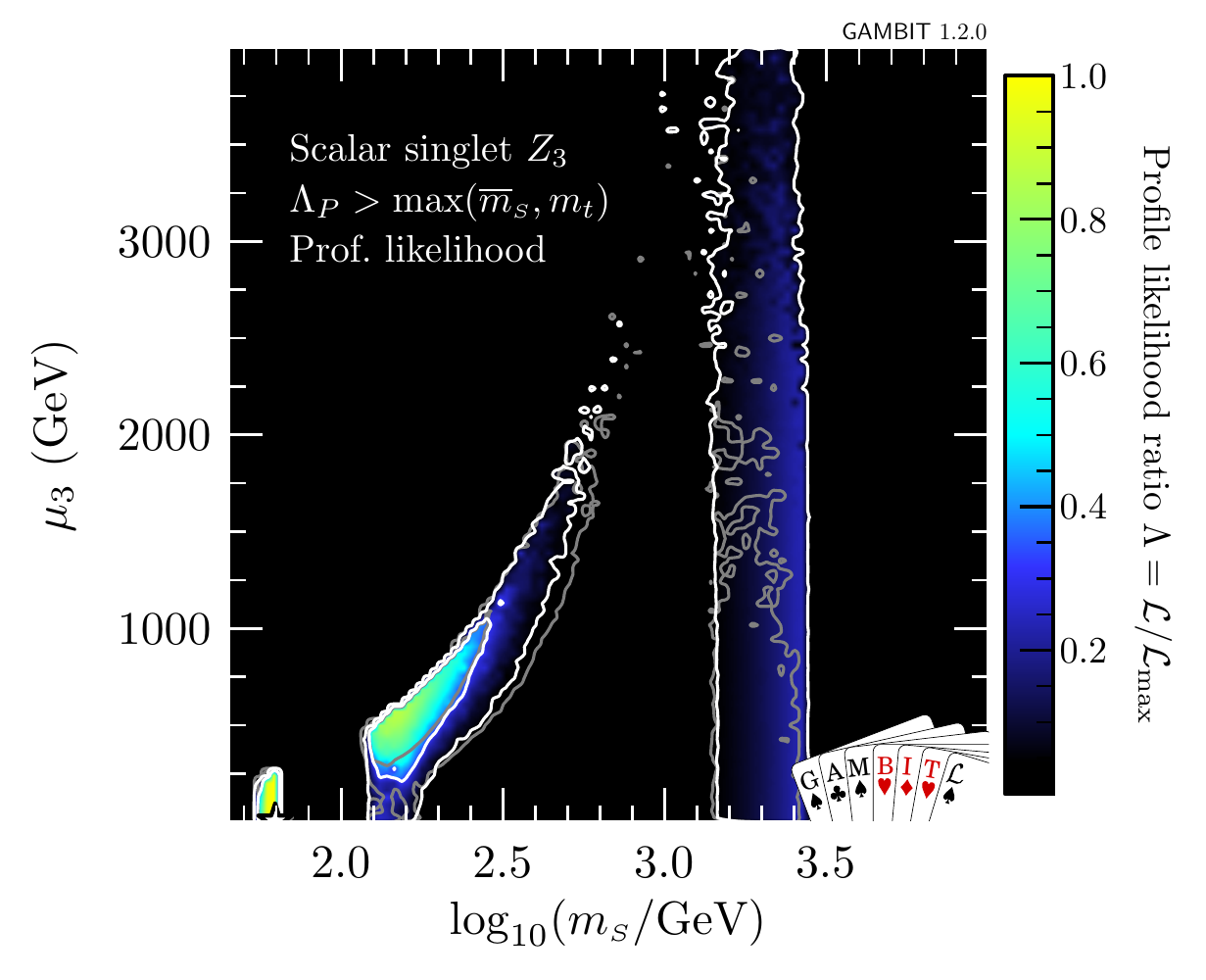}
  \caption{Profile likelihoods of parameters in the $\mathbb{Z}_3$-symmetric scalar singlet Higgs portal dark matter model, including constraints from direct and indirect detection, the relic density of dark matter and LHC searches for invisible decays of the Higgs boson, along with various Standard Model, dark matter halo and nuclear uncertainties.  Contours show 1 and 2$\sigma$ confidence regions, with white corresponding to the main scan (including the 2018 XENON1T direct search \cite{Aprile:2018dbl}) and grey to a secondary scan using the 2017 XENON1T result \cite{Aprile:2017iyp}.  White stars indicate the location of the best-fit point.  From \protect\cite{SSDM2}.}
	\label{fig:z3scalar1}
\end{figure}

\begin{figure}[t]
	\centering
  \includegraphics[width=0.495\columnwidth]{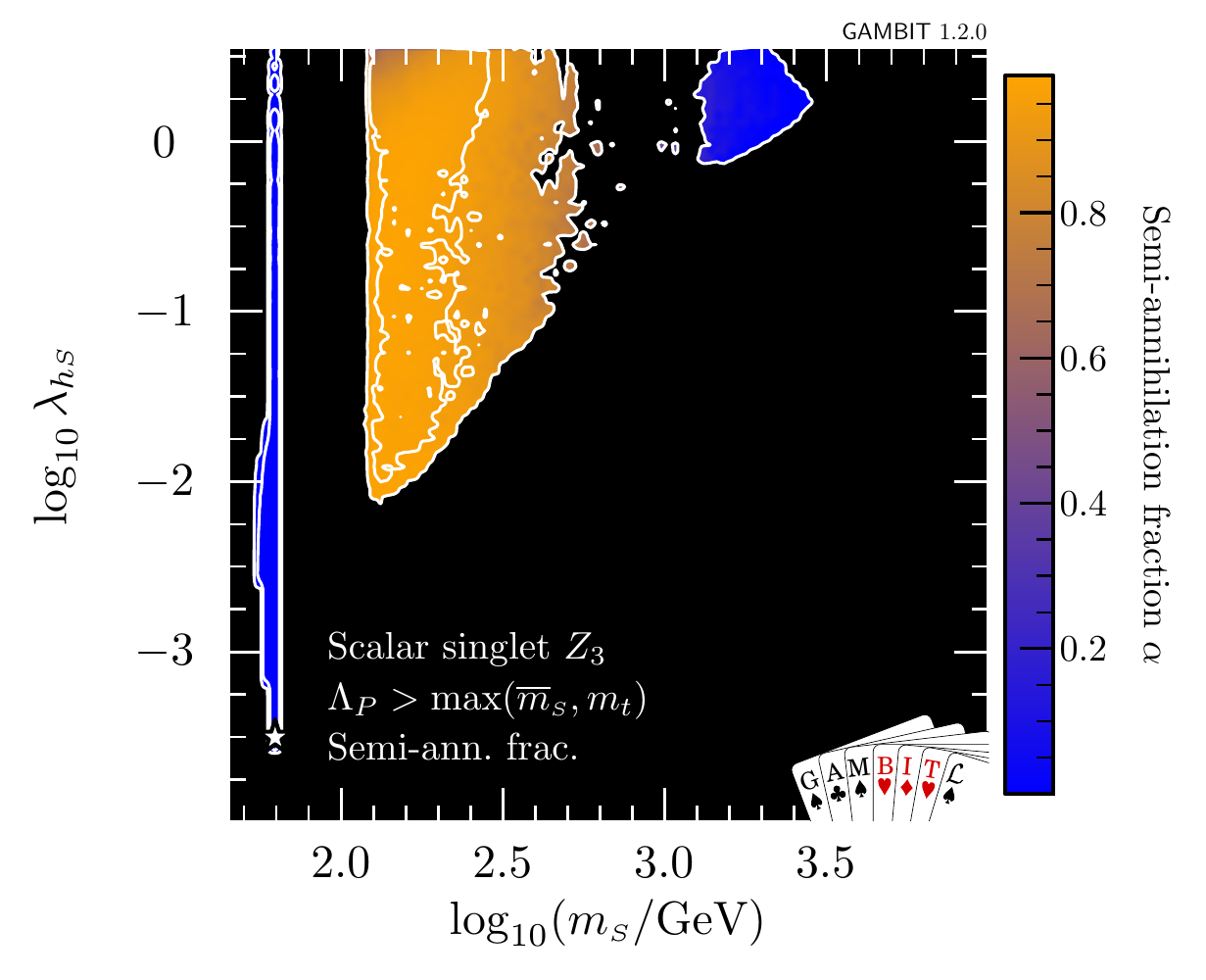}
  \includegraphics[width=0.495\columnwidth]{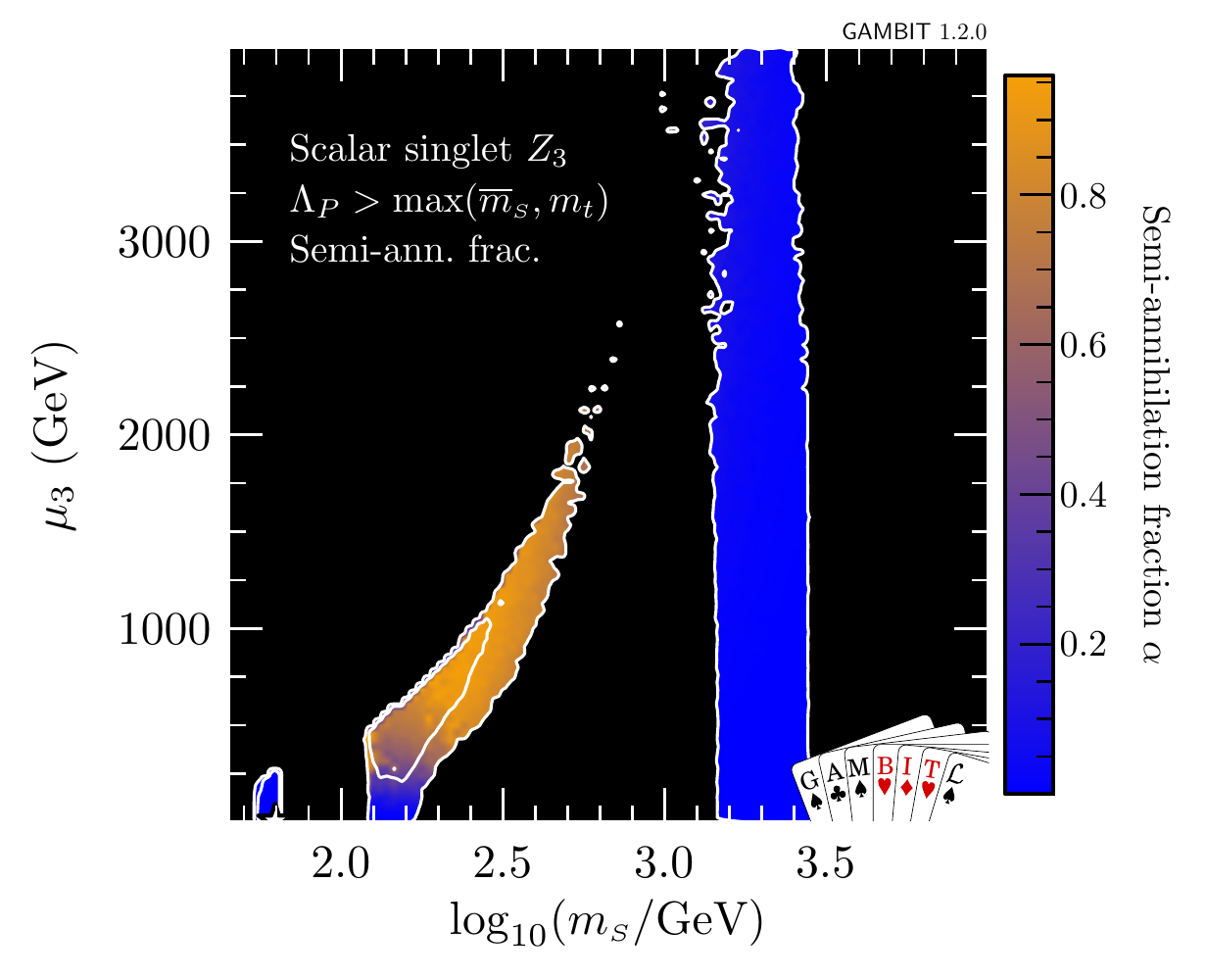}
  \caption{Results from the same analysis of the $\mathbb{Z}_3$-symmetric scalar singlet Higgs portal dark matter model as shown in Fig.\ \ref{fig:z3scalar1}, but shaded according to the semi-annihilation fraction $\alpha$ (Eq.\ \protect\ref{eqn:sa_fraction}). From \protect\cite{SSDM2}.}
	\label{fig:z3scalar2}
\end{figure}

\subsubsection{$\mathbb{Z}_3$-symmetric scalar singlet}

In contrast to the self-adjoint $\mathbb{Z}_2$-symmetric scalar singlet, a $\mathbb{Z}_3$ symmetry leads to a complex scaler DM candidate, with both DM ($S$) and anti-DM ($S^*$) states contributing to the relic density.  This symmetry also allows an additional cubic term in the Lagrangian,
\begin{equation}
\mathcal{L}_{\mathbb{Z}_3} = \mu_{\sss S}^2 S^\dagger S + \ls (S^\dagger S)^2 + \frac{\mu_3}{2}(S^{\dagger 3}+S^3) + \lhs S^\dagger S|H|^2,
\end{equation}
where we have introduced the new dimension-1 $S$ cubic coupling $\mu_3$.  This new coupling allows for so-called semi-annihilation processes $SS\to S^*h$ and $S^*S^*\to Sh$, shown in Fig.\ \ref{fig:diagrams}.

Compared to the $\mathbb{Z}_2$-symmetric model, semi-annihilation is able to deplete the relic density of DM at intermediate masses and open up an entirely new region of viable parameter space.  This is shown in terms of the profile likelihood in Fig.\ \ref{fig:z3scalar1}, and highlighted in terms of the semi-annihilation fraction
\begin{equation}
\alpha=\frac{1}{2}\frac{\langle\sigma v_\mathrm{rel}\rangle_{SS\rightarrow hS}}{\langle\sigma v_\mathrm{rel}\rangle+\frac{1}{2}\langle\sigma v_\mathrm{rel}\rangle_{SS\rightarrow hS}},\label{eqn:sa_fraction}
\end{equation}
in Fig.\ \ref{fig:z3scalar2}.  Here $\langle\sigma v_\mathrm{rel}\rangle$ is the thermally averaged (semi-)annihilation cross-section weighted by the relative velocity between annihilating particles.

The vacuum structure of the theory is also more complicated than that of the $\mathbb{Z}_2$-symmetric model, as regions where $\mu_3 \geq 2\sqrt{\ls}\ms$ or $\mu_{\sss S}^2<0$ and $\lhs$ is large can possess a second, $\mathbb{Z}_3$-breaking minimum.  The results shown in Figs.\ \ref{fig:z3scalar1} and \ref{fig:z3scalar2} avoid these regions, demanding that $S$ does not itself obtain a VEV, and that the potential remains bounded from below.

Like the $\mathbb{Z}_2$-symmetric variant, the $\mathbb{Z}_3$-symmetric model can in principle completely stabilise the SM vacuum.  However, because of the various factors of 2 introduced relative to the $\mathbb{Z}_2$ case, by virtue of DM not being self-adjoint, the region where this is possible is in fact in strong tension with the results from both XENON1T \cite{Aprile:2018dbl} and PandaX \cite{Cui:2017nnn}.  $\mathbb{Z}_3$-symmetric models that stabilise the SM vacuum and produce the entire observed DM relic density are ruled out at 99\% confidence; those constituting only a fraction of DM are ruled out at 98\% confidence.  The same is expected of other $\mathbb{Z}_N$-symmetric models with $N>3$, which also feature non-self-adjoint DM.

As in the $\mathbb{Z}_2$-symmetric case, a Bayesian analysis prefers the higher-mass part of the parameter space, due to the fine-tuning needed to achieve agreement with all experimental data in both the resonance and semi-annihilation (intermediate mass) regions.  In this case, the additional tuning in $\mu_3$ required to satisfy the condition $\mu_3 \leq 2\sqrt{\ls}\ms$ -- and to achieve sufficient semi-annihilation in the intermediate-mass region -- further penalises these regions.

\begin{figure}[t]
	\centering
  \includegraphics[width=0.495\columnwidth]{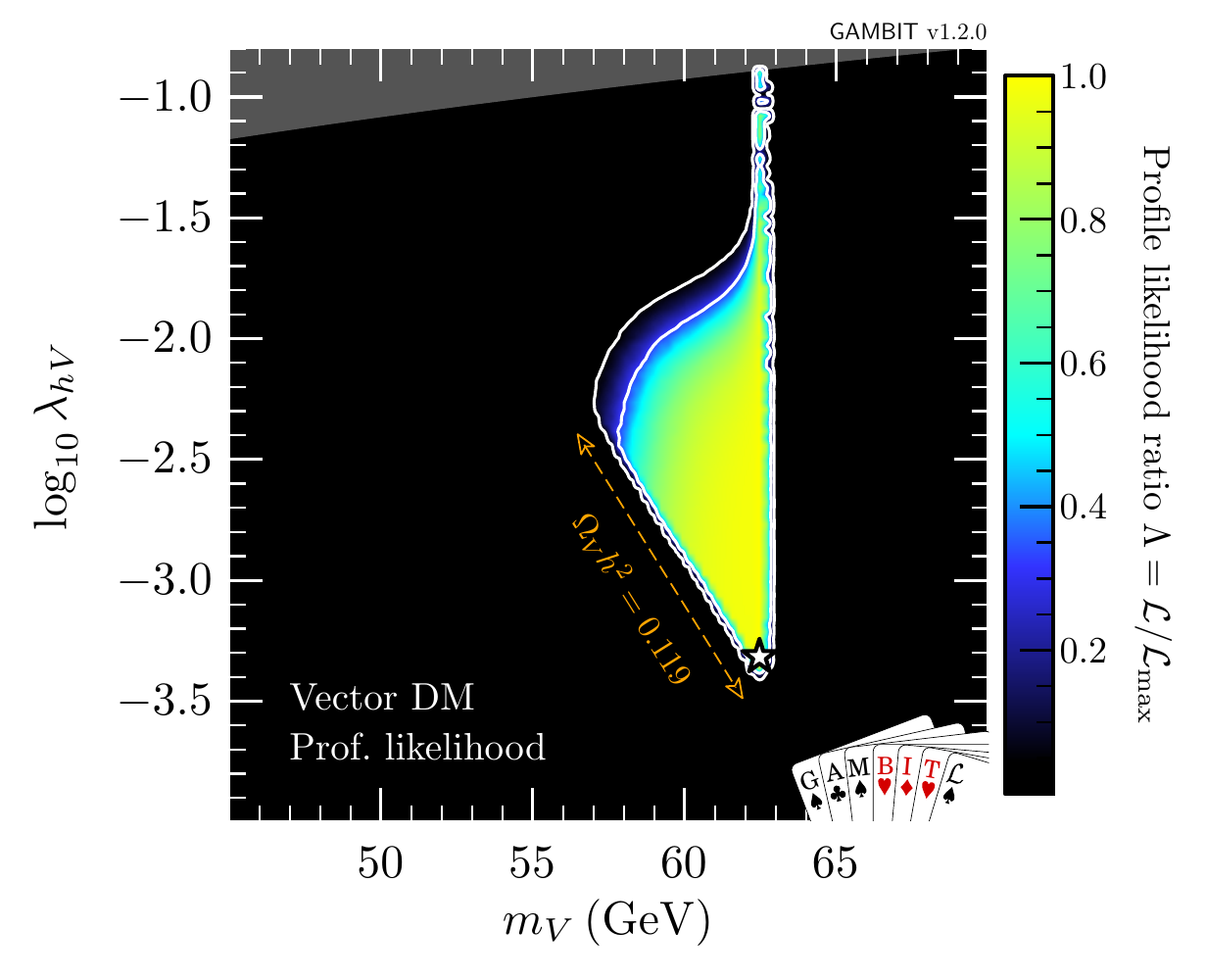}
  \includegraphics[width=0.495\columnwidth]{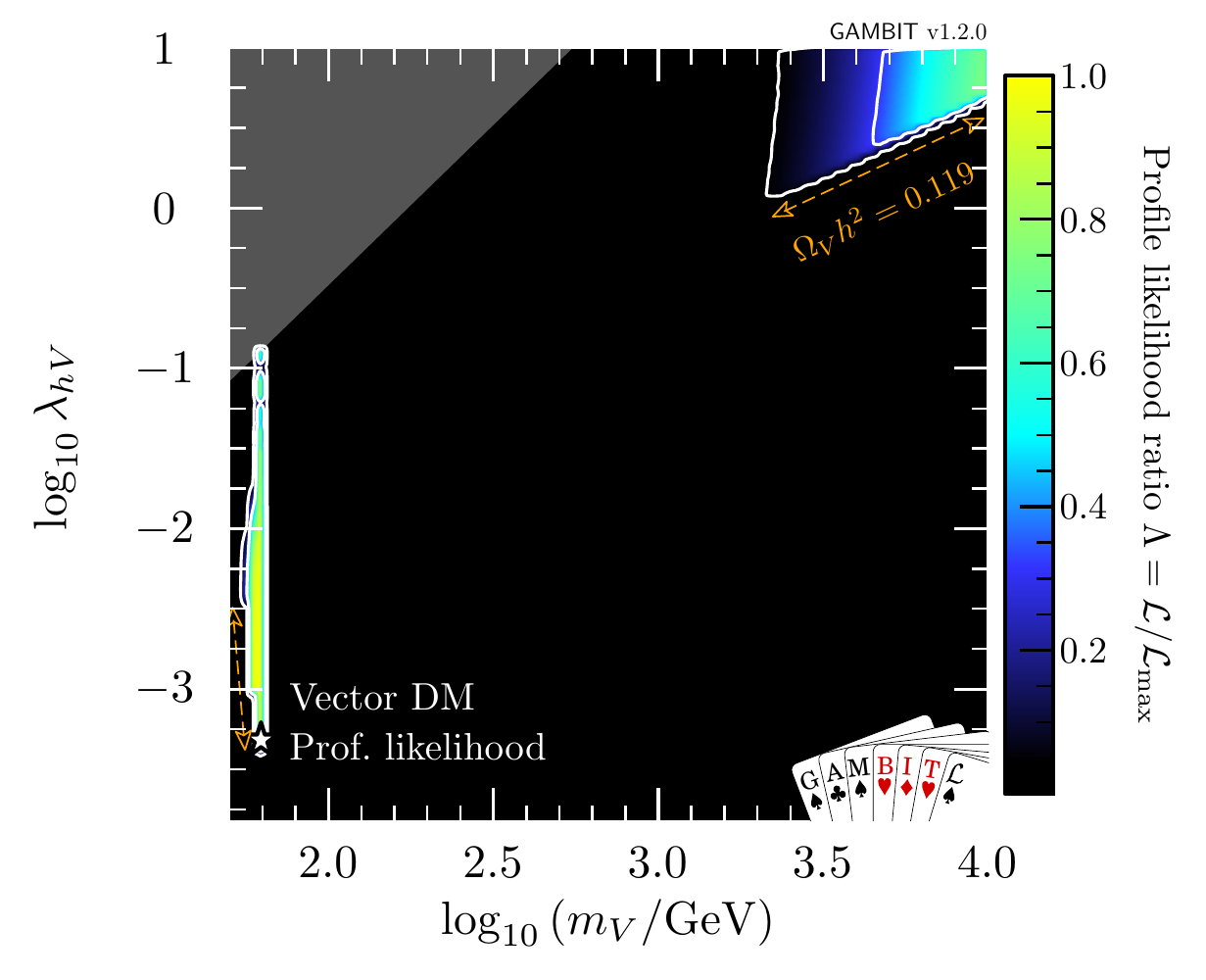}
  \caption{Profile likelihoods of parameters in the $\mathbb{Z}_2$-symmetric vector singlet Higgs portal dark matter model, including constraints from direct and indirect detection, the relic density of dark matter and LHC searches for invisible decays of the Higgs boson, along with various Standard Model, dark matter halo and nuclear uncertainties.  \textit{Left}: the low-mass resonance region.  \textit{Right}: the full mass range.  Grey shading indicates the area that fails the unitarity cut (Eq.\ \ref{eq:vec_unitarity}). Orange annotations indicate the edge of the allowed parameter space along which the model reproduces the entire cosmological abundance of dark matter. Contours show 1 and 2$\sigma$ confidence regions. White stars indicate the location of the best-fit point. From \protect\cite{HP}.}
	\label{fig:vector}
\end{figure}

\subsubsection{$\mathbb{Z}_2$-symmetric vector singlet}

If DM is a $\mathbb{Z}_2$-symmetric vector singlet $V_\mu$ interacting with the SM via the Higgs portal, its effective Lagrangian takes the form
\begin{equation}
\mathcal{L_V} = -\frac{1}{4} W_{\mu\nu} W^{\mu\nu} + \frac{1}{2} \mu_V^2 V_\mu V^\mu - \frac{1}{4!} \lambda_{V} (V_\mu V^\mu)^2 + \frac{1}{2} \lambda_{hV} V_\mu V^\mu H^\dagger H.
\label{eq:Lag_V}
\end{equation}
Here $W_{\mu\nu} \equiv \partial_\mu V_\nu - \partial_\nu V_\mu$ is the field strength tensor for the new vector.  The tree-level DM mass has exactly the same form as Eq.\ \ref{ms}.  Although all terms here are dimension 4, the theory is not renormalisable, as it possesses an explicit mass term for $V_\mu$.  Perturbative unitarity is violated at energies above this mass.  In the \gambit analysis \cite{HP}, this issue was avoided by excluding the region of parameter space
\begin{equation}
0 \le \lambda_{hV} \le 2m_V^2/v_0^2
\label{eq:vec_unitarity}
\end{equation}
from the analysis.

The phenomenology of the vector model is very similar to that of the $\mathbb{Z}_2$-symmetric scalar variant, with the only major difference being the absence of the intermediate-mass solution due to the unitarity requirement (Fig.\ \ref{fig:vector}; the region excluded from the analysis due to the unitarity condition is shown in grey).  The Bayesian analysis once again prefers the high-mass region due to the fine-tuning of nuisance parameters required in the resonance region.

\subsubsection{$\mathbb{Z}_2$-symmetric Dirac \& Majorana fermionic singlets}

The Lagrangians of the fermionic singlet Higgs portal models are
\begin{align}
  \mathcal{L}_{\chi} &= \frac{1}{2} \overline{\chi} (i\slashed{\partial} - \mu_\chi) \chi - \frac{1}{2}\frac{\lambda_{h\chi}}{\Lambda_\chi} \Big(\cos\theta \, \overline{\chi}\chi + \sin\theta \, \overline{\chi}i\gamma_5 \chi \Big) H^\dagger H,
  \label{eq:Lag_chi}\\
  \mathcal{L}_{\psi} &= \overline{\psi} (i \slashed{\partial} - \mu_\psi) \psi \nonumber - \frac{\lambda_{h\psi}}{\Lambda_\psi} \Big(\cos\theta \, \overline{\psi}\psi + \sin\theta \, \overline{\psi}i\gamma_5 \psi \Big) H^\dagger H,
  \label{eq:Lag_psi}
\end{align}
with the Majorana variant denoted $\chi$ and the Dirac variant $\psi$.  These noticeably possess dimension-5 effective portal operators suppressed by the scale of new physics $\Lambda$, with both scalar ($CP$-even) and pseudoscalar ($CP$-odd) couplings.  The degree to which the portal interaction violates $CP$ is dictated by the mixing angle $\theta$, where $\theta = 0$ corresponds to pure $CP$ conservation and $\theta = \frac\pi2$ to maximal $CP$ violation.

As in the scalar and vector models, the portal interaction produces terms quadratic in the DM field following electroweak symmetry breaking.  The pseudoscalar coupling leads to an imaginary mass term, which must be rotated away with the field transformation $X \rightarrow e^{i\gamma_5 \alpha/2} X$ for $X \in \{\chi, \psi\}$, in order to arrive at the physical (real) mass.  This introduces a new parameter $\alpha$.  The physical masses are then
\begin{equation}
m_{X}^2 = \left(\mu_{X} + \frac{1}{2}\frac{\lambda_{hX}}{\Lambda_{X}} v_0^2 \cos\theta \right)^2 + \left(\frac{1}{2}\frac{\lambda_{X}}{\Lambda_{X}}v_0^2 \sin\theta \right)^2.
\end{equation}
The rotation parameter $\alpha$ is fixed by the requirement that the mass be real, so all phenomenology can be described by three parameters: $m_X$, $\lambda_X/\Lambda_X$ and $\xi \equiv \theta + \alpha$.  Notably, the pure $CP$-conserving theory ($\theta = 0$) remains $CP$-conserving after electroweak symmetry breaking ($\xi = 0$), but maximal $CP$ violation before electroweak symmetry breaking does not correspond to maximal violation after the symmetry is broken (i.e. $\theta = \frac\pi2 \notimplies \xi = \frac\pi2$).

Whilst the $CP$-even Higgs portal coupling leads to the familiar velocity and momentum-independent nuclear scattering cross-section, the $CP$-odd coupling gives rise to an interaction suppressed by $q^2$, the square of the momentum exchanged in the scattering event.  This leads to an overall suppression of direct detection signals and corresponding constraints for $\xi \rightarrow \frac\pi2$.  Conversely, the $CP$-odd coupling produces a velocity and momentum-independent annihilation cross-section, whereas the $CP$-even coupling gives rise to a velocity-suppressed annihilation cross-section.

\begin{figure}[tbp]
	\centering
  \includegraphics[width=0.495\columnwidth]{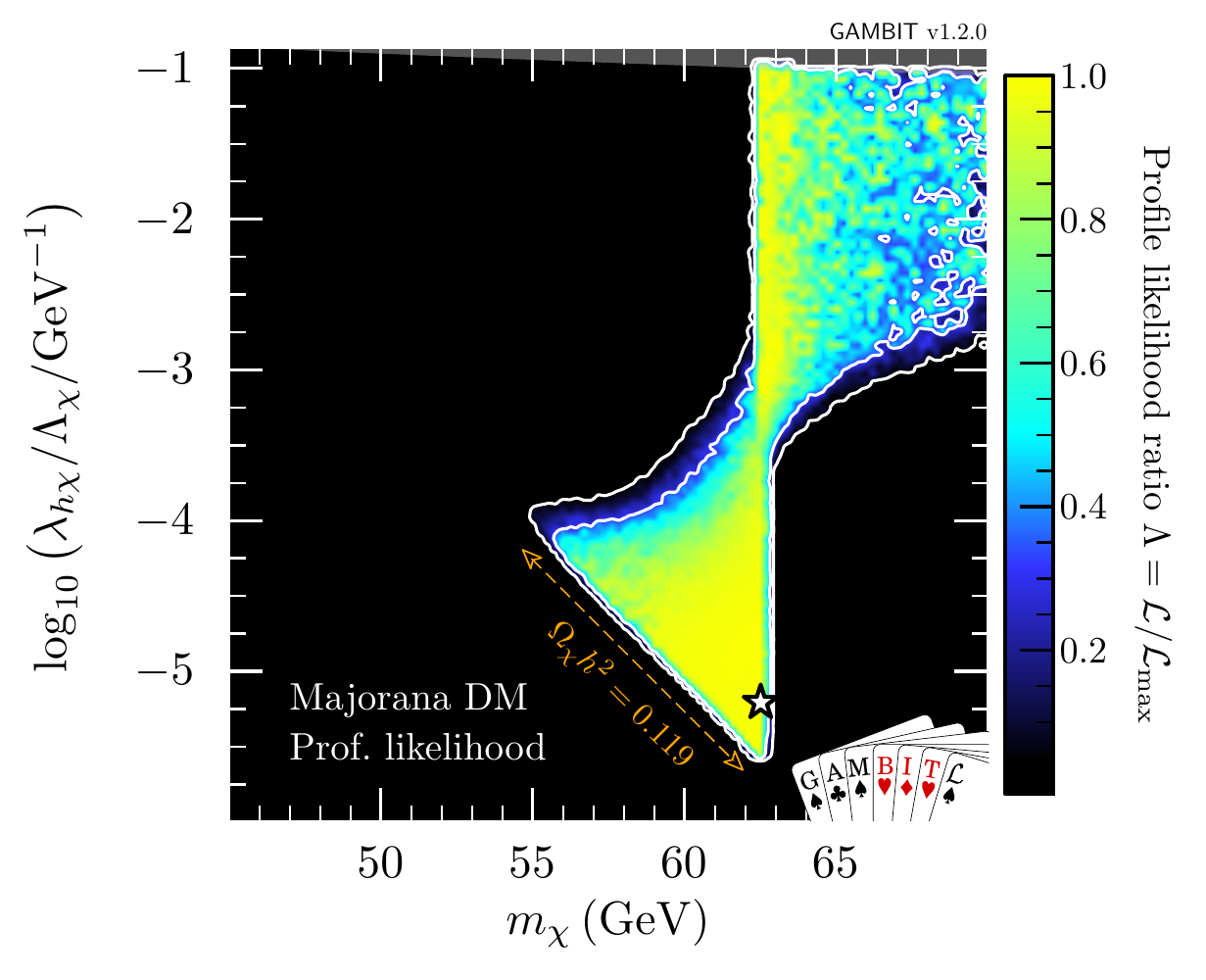}
  \includegraphics[width=0.495\columnwidth]{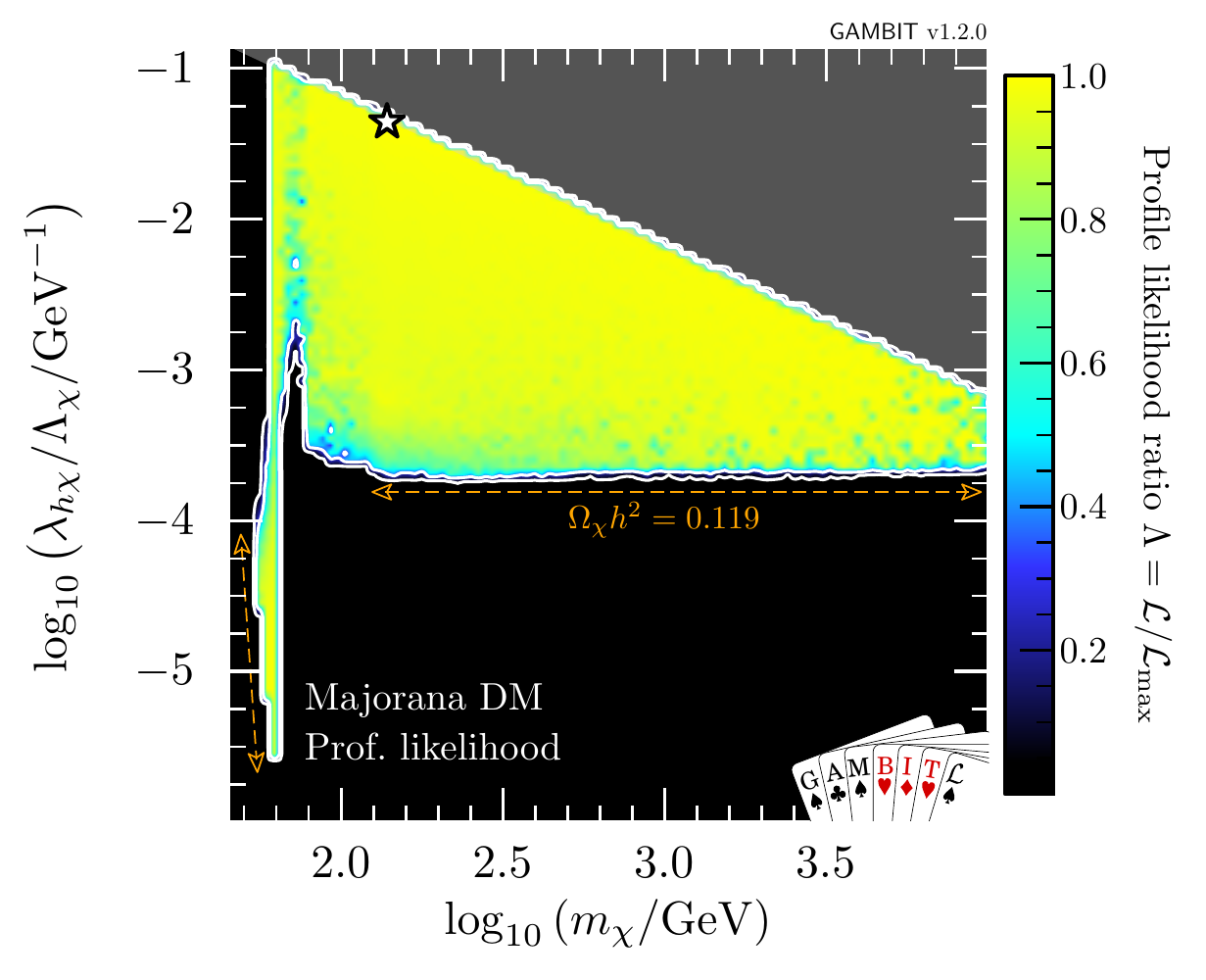}\\
  \includegraphics[width=0.495\columnwidth]{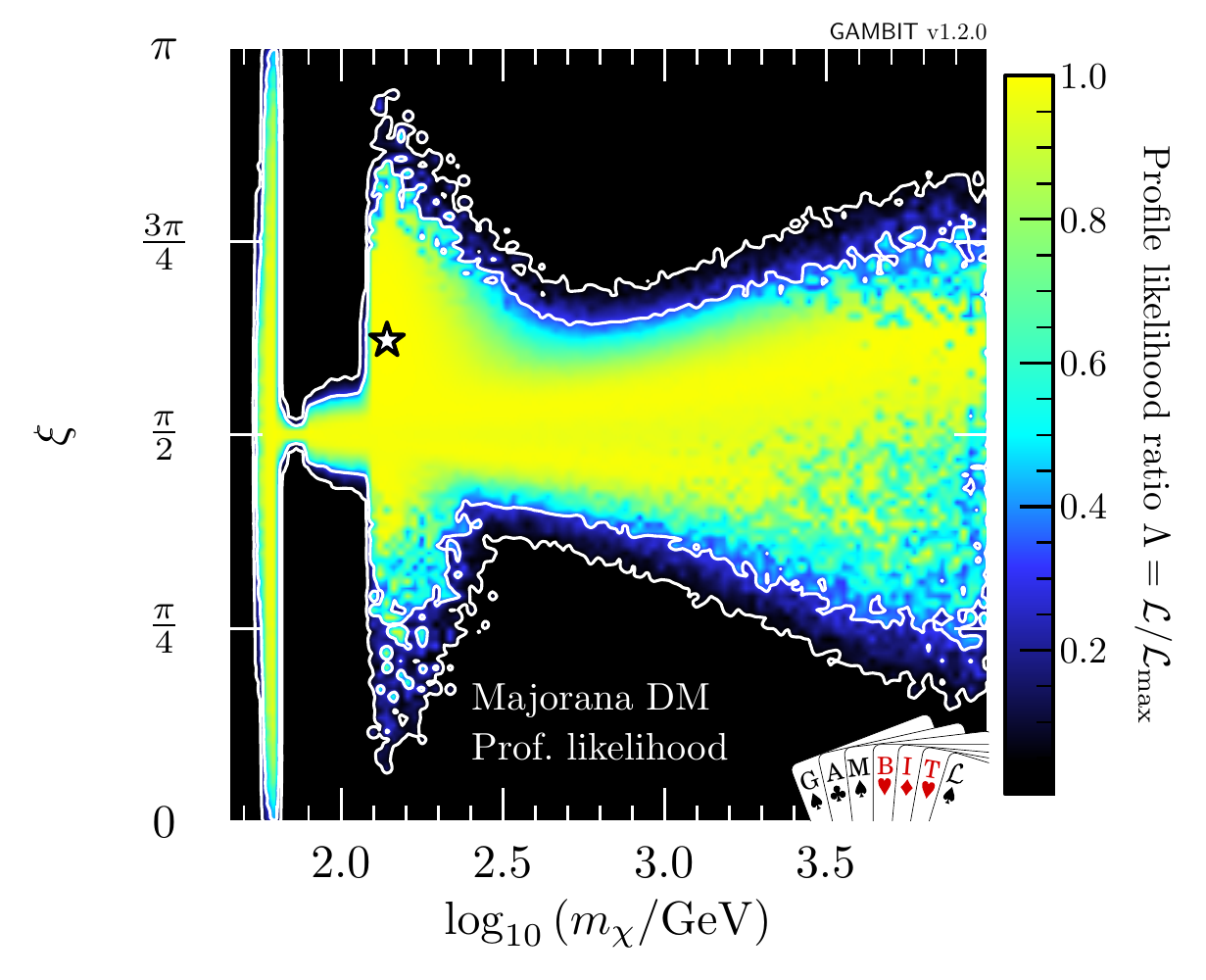}
  \includegraphics[width=0.495\columnwidth]{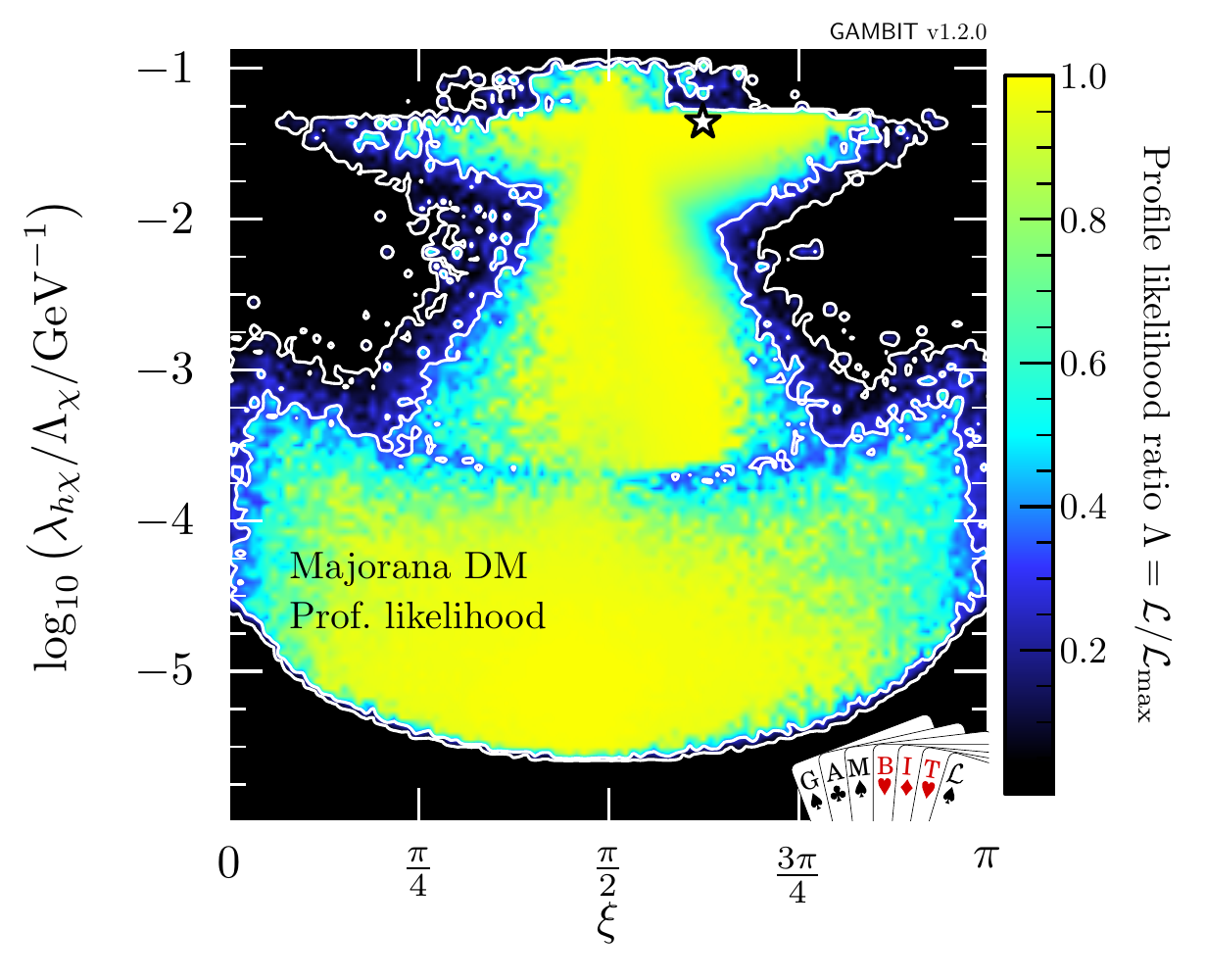}
  \caption{Profile likelihoods of parameters in the $\mathbb{Z}_2$-symmetric Majorana fermion singlet Higgs portal dark matter model, including constraints from direct and indirect detection, the relic density of dark matter and LHC searches for invisible decays of the Higgs boson, along with various Standard Model, dark matter halo and nuclear uncertainties.  The upper-left panel shows a zoomed-in view of the low-mass resonance region.  Grey shading indicates the area that fails the unitarity cut (Eq.\ \ref{eq:fermion_unitarity}). Orange annotations indicate the edge of the allowed parameter space along which the model reproduces the entire cosmological abundance of dark matter. Contours show 1 and 2$\sigma$ confidence regions.  White stars indicate the location of the best-fit point. From \protect\cite{HP}.}
	\label{fig:fermion1}
\end{figure}

\begin{figure}[tbp]
	\centering
  \includegraphics[width=0.495\columnwidth]{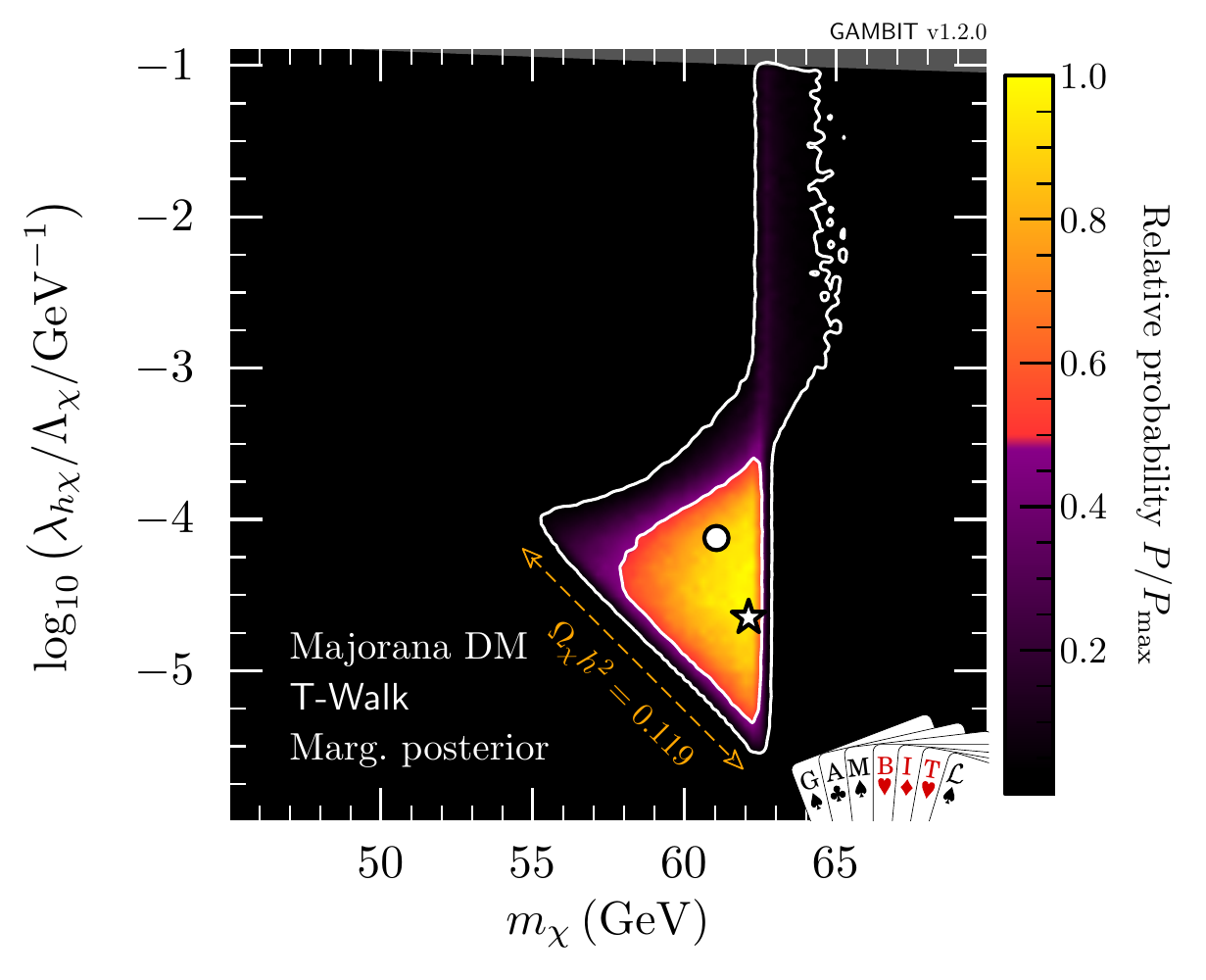}
  \includegraphics[width=0.495\columnwidth]{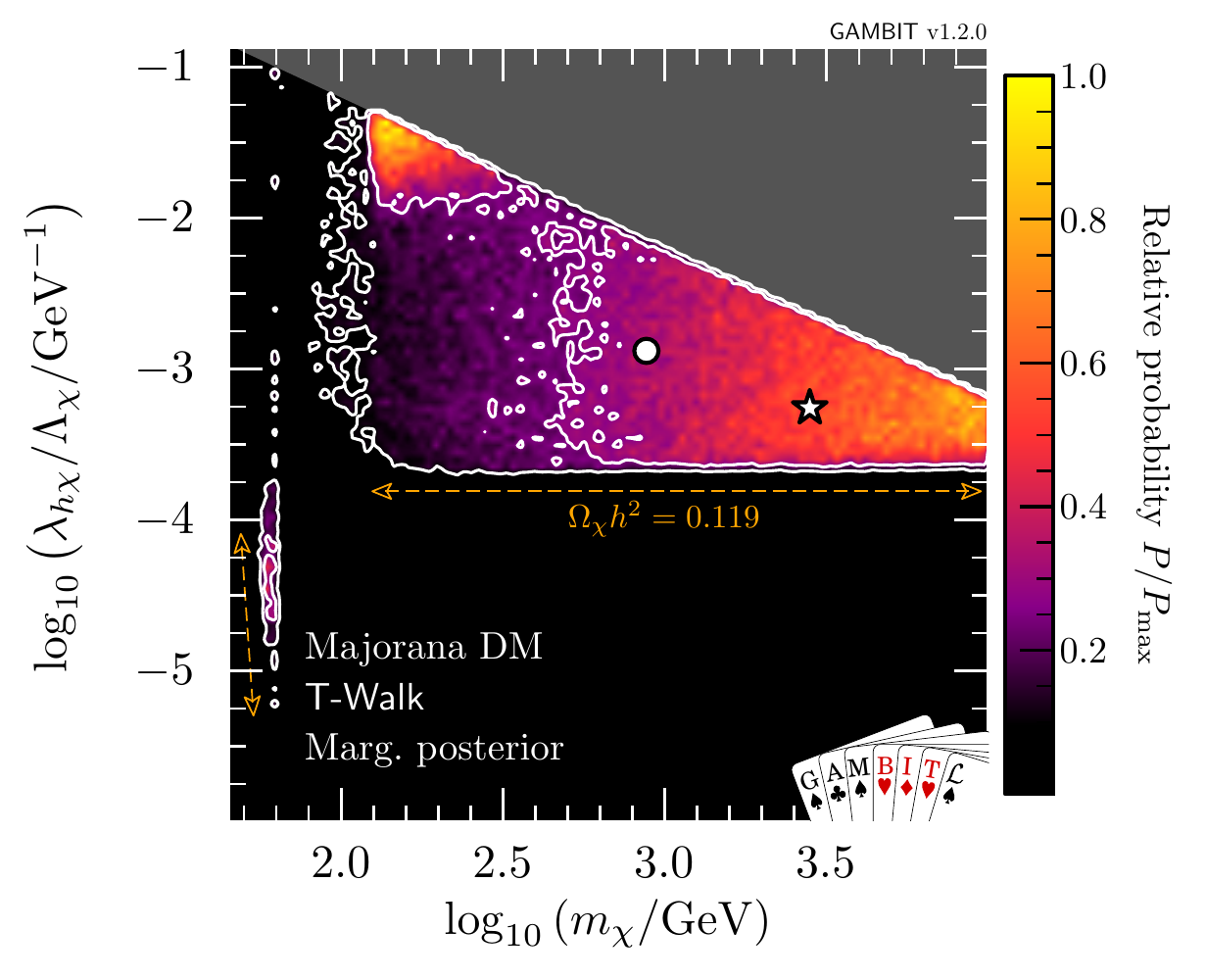}
  \includegraphics[width=0.495\columnwidth]{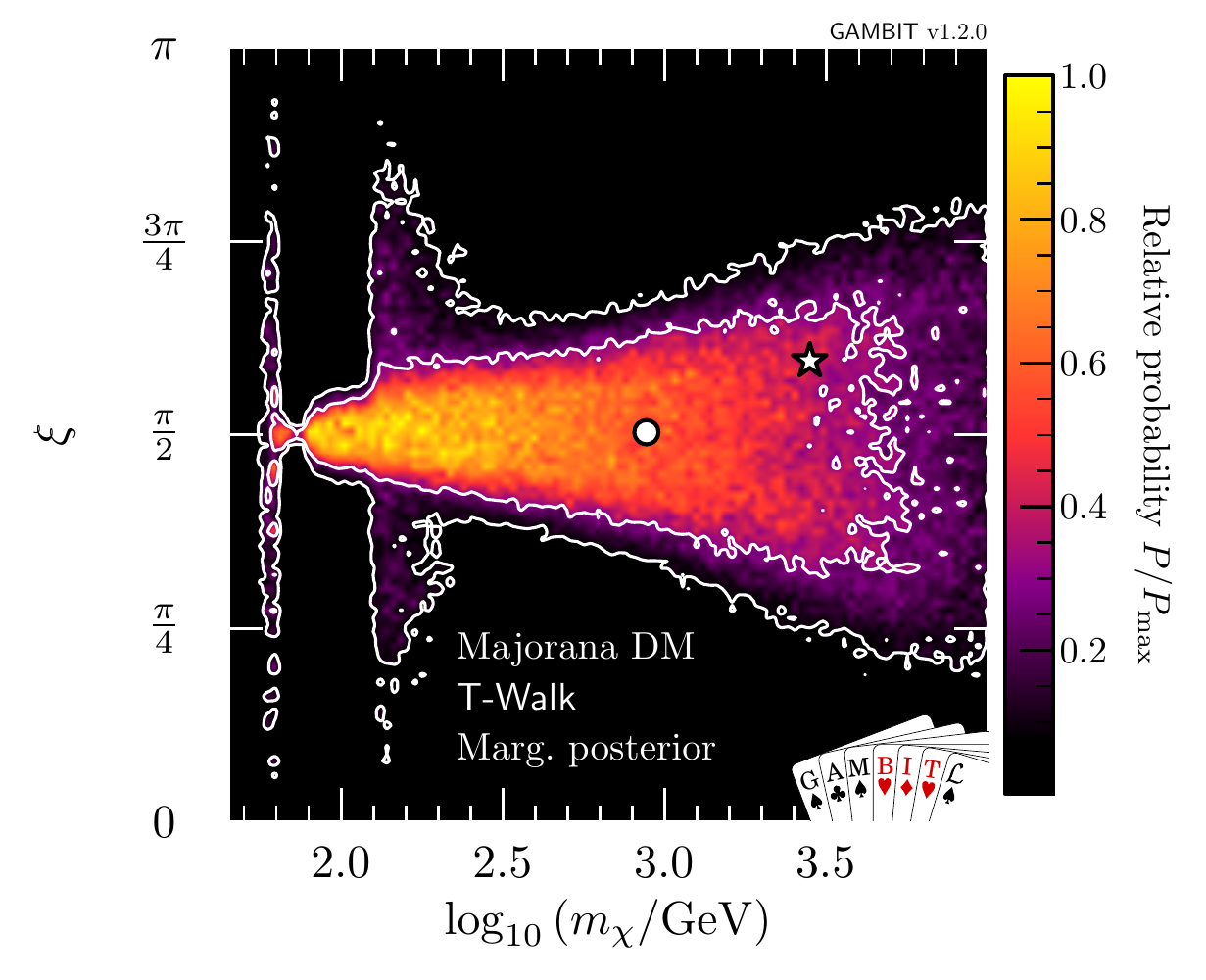}
  \includegraphics[width=0.495\columnwidth]{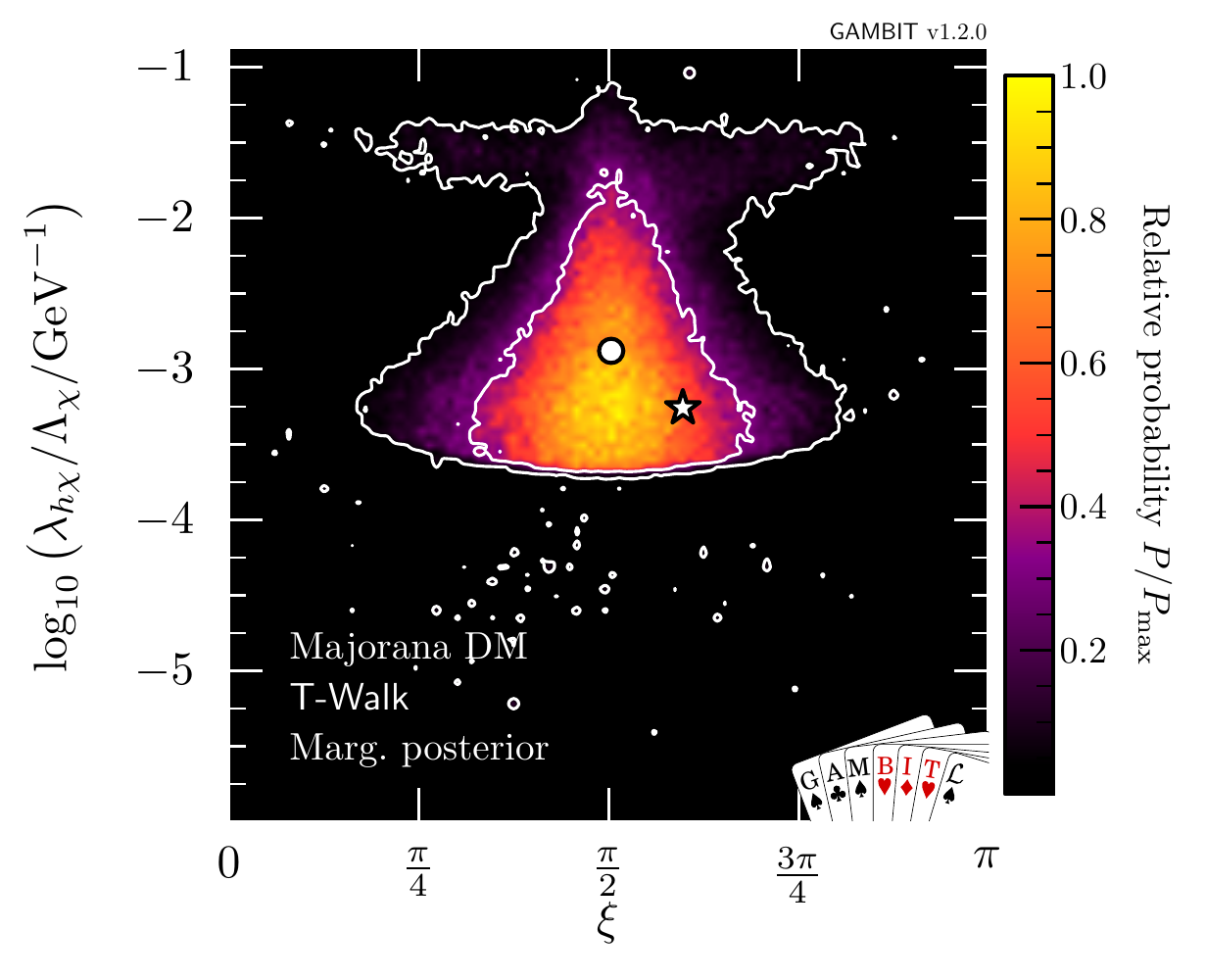}
  \caption{Posterior probability densities from a Bayesian analysis of the $\mathbb{Z}_2$-symmetric Majorana fermion singlet Higgs portal dark matter model, using the same likelihood functions as Fig.\ \protect\ref{fig:fermion1}.  White bullets indicate posterior means; other annotations are as in Fig.\ \protect\ref{fig:fermion1}. From \protect\cite{HP}.}
	\label{fig:fermion2}
\end{figure}

Profile likelihoods from the global fit to the Majorana fermion model are shown in Fig.\ \ref{fig:fermion1}.  Results for Dirac fermion dark matter are broadly very similar, and differ from the Majorana case only in the exact location of the border of the allowed parameter space, reflecting the essentially inconsequential nature of the relative factors of 2 between the two Lagrangians.  Grey regions correspond to the regime
\begin{equation}
\lambda_{hX}/\Lambda_{X} \geq 2\pi/m_X,
\label{eq:fermion_unitarity}
\end{equation}
where the validity of the EFT becomes questionable.  Further discussion on this issue can be found in Ref.\ \cite{HP}; it would also be possible to unitarise the theory, and draw further constraints in this region, using the $K$-matrix formalism \cite{Bell:2016obu,Balaji:2018qyo}.

The preferred regions in the mass-coupling plane (upper panels of Fig.\ \ref{fig:fermion1}) include the now-familiar resonance and high-mass regions.  However, unlike the vector and scalar models, these are fully connected by valid models at all masses, with the preferred region bounded from below mostly by the relic density constraint, supported by indirect detection.  This is because profiling over $\xi$ allows for the selection of $CP$-violating couplings in order to avoid constraints from direct detection.  The degree of tuning in $\xi$ required to achieve this is apparent in the lower panels of Fig.\ \ref{fig:fermion1}, where it is clear that good fits can be found for any value of $\xi$ in the resonance region, but that higher masses require some degree of $CP$ violation in order to avoid direct detection.  This becomes even clearer in the equivalent Bayesian results shown in Fig.\ \ref{fig:fermion2}, where intermediate masses and couplings are disfavoured relative to other regions, due to the need to make $CP$ violation nearly maximal in order to avoid direct detection.

\begin{figure}[t]
	\centering
  \includegraphics[width=0.495\columnwidth]{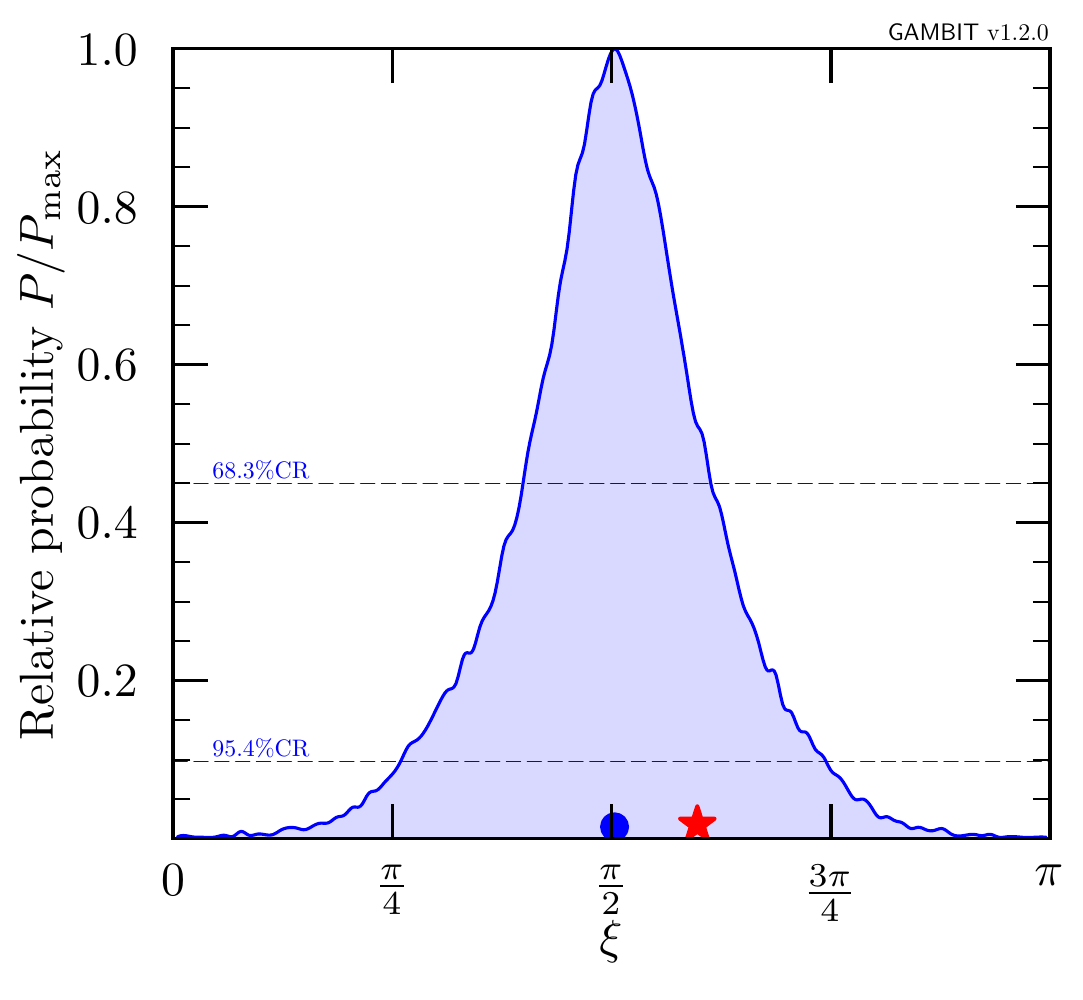}
  \caption{Marginalised one-dimensional posterior probability density for the $CP$-mixing parameter $\xi$ in the $\mathbb{Z}_2$-symmetric Majorana fermion singlet Higgs portal dark matter model.  This result has been extracted from the same analysis as that shown in Fig.\ \protect\ref{fig:fermion2}.  The value $\xi = 0 = \pi$ corresponds to $CP$ conservation; a clear preference for violation of $CP$ symmetry is evident. The blue bullet indicates the posterior mean value of $\xi$, and the red star the value of $\xi$ at the best-fit sample.  From \cite{HP}.}
	\label{fig:fermion3}
\end{figure}

Integrating the posterior over all parameters other than $\xi$ (Fig.\ \ref{fig:fermion3}), there is a clear preference for $CP$ violation.  This reflects the fact that the more $CP$ violation permitted, the broader the range of other parameters able to give good fits to the combined data of all experiments.  Performing Bayesian model comparison between the full model and its pure $CP$-conserving subspace (i.e.\ $\xi = 0$) results in Bayes factors of between 70:1 and 140:1, depending on the adopted priors.  This indicates a strong preference for $CP$ violation in fermionic Higgs portal models.  Bayesian model comparison between the scalar, vector and fermionic variants of the Higgs portal DM model reveals essentially equal odds for each of the scalar and fermionic models, but a 6:1 preference for all of these models over the vector variant.

\subsection{Axions}

\subsubsection{Axion models and their implementation in \gambit}

Axions are an intriguing theoretical possibility due to their ability to solve the strong-$CP$ problem of the SM whilst providing a credible DM candidate~\cite{Preskill:1982cy,Abbott:1982af,Dine:1982ah,1986_turner_axiondensity}. One can also use axion-like particles to reconcile various tensions between astrophysical observations and theory, including the cooling of white dwarfs~\cite{Isern:1992gia,1205.6180,1211.3389,1512.08108,1605.06458,1605.07668,1708.02111}, and the transparency of the Universe to gamma rays~\cite{0707.4312,0712.2825,1001.0972,1106.1132,1201.4711,1302.1208}.

The strong-$CP$ problem is ultimately a fine-tuning problem, arising from the fact that the SM symmetries permit a $CP$-odd term in the SM Lagrangian density of the form:

\begin{equation}
\mrm{\lagrangian}{QCD} \supset - \frac{\mrm{\alpha}{S}}{8\otherpi} \mrm{\theta}{QCD} G_{\mu \nu}^a \widetilde{G}^{\mu \nu, a} \, , \label{eq:QCDLagrangian}
\end{equation}
where $G_{\mu \nu}^a$ is the gluon field strength tensor, $\widetilde{G}^{\mu \nu, a}$ is its dual (both of which have the $SU(3)$ gauge index $a$ explicitly shown), and $\mrm{\alpha}{S}$ is the strong coupling constant. The angle $\mrm{\theta}{QCD} \in [-\otherpi, \otherpi]$ is a free parameter. In the SM, the term also receives a contribution from the chiral anomaly which, for down- and up-type Yukawa matrices $Y_d$ and $Y_u$, replaces $\mrm{\theta}{QCD}$ by the effective angle
\begin{equation}
\mrm{\theta}{eff} \equiv \mrm{\theta}{QCD} - \arg \left [ \det (Y_dY_u) \right ] \, . \label{eq:thetaeff}
\end{equation}
A non-zero $\mrm{\theta}{eff}$ would result in $CP$-violating effects in strong interactions, which are severely constrained by observed upper limits on the electric dipole moment of the neutron, demanding $|\mrm{\theta}{eff}|\lsim \num{e-10}$ \cite{1509.04411}. Naively, this can only be avoided in the SM by fine-tuning the value of $\mrm{\theta}{QCD}$ to cancel the contribution from the chiral anomaly.

An alternative solution, first proposed by Peccei and Quinn~\cite{1977_pq_axion1,1977_pq_axion2}, is to add a new global, axial $U(1)$ symmetry spontaneously broken by the vacuum expectation value $v$ of a complex scalar field. This breaking has an associated pseudoscalar Nambu-Goldstone boson, $a(x)$, which supplements $\mrm{\theta}{eff}$ by a new term $Na(x)/v$, where the non-zero integer $N$ is the colour anomaly of the added symmetry. The Vafa-Witten theorem~\cite{1984_vafa_vafawitten1,1984_vafa_vafawitten2} can then be used to show that  $\mrm{\theta}{eff}+Na(x)/v$ is dynamically driven to zero, solving the strong $CP$ problem.

In the resulting theory of the QCD axion, the axion is practically massless until the time of the QCD phase transition, due to a shift symmetry of the $U(1)$ phase, which prevents a mass term in the Lagrangian. After this, however, it picks up a small, temperature-dependent mass due to breaking of the continuous shift symmetry by fluctuations of the gluon fields. This gives rise to an effective axion potential
\begin{equation}
	V(a) = \fa^2 \, \ma^2 \, \left[1 - \cos (a/\fa) \right] \, , \label{eq:axion_eff_pot}
\end{equation}
where $\ma$ is the temperature-dependent axion mass and $\fa \equiv v/N$. The zero-temperature axion mass, $\mazero$, can be calculated using next-to-leading order chiral perturbation theory, and it turns out to be inversely proportional to $f_a$ for the QCD axion. At higher temperatures, numerical estimates of the mass are available from lattice QCD results, which can be described to a good approximation by
\begin{equation}
\ma(T) =  \mazero
\begin{cases}
\hfil 1 \hfil & \mathrm{if \; } T \leq \Tcrit \\
\left ( \frac{\Tcrit}{T} \right )^{\beta/2} & \mathrm{otherwise}
\end{cases} \, . \label{eq:axionmass}
\end{equation}
$\Tcrit$ and $\beta$ are in principle calculable, but can be left as nuisance parameters in order to account for systematic uncertainties in the calculations.

In fact, QCD axions are only one instance of a general class of \emph{axion-like particles} (ALPs), which could generally result from the breaking of a $U(1)$ symmetry at some scale $f_a$, with mass generation occurring via the explicit breaking of the residual symmetry at some lower scale $\Lambda$~\cite{1987_kim_lightpseudoscalars,1002.0329,1801.08127}. It can be shown that in a Friedmann-Robertson-Walker-{Lema\^itre} universe, a QCD axion or ALP field $\theta(t)=a(t)/\fa$ satisfies the equation of motion
\begin{equation}
  \ddot{\theta} + 3H(t) \, \dot{\theta} + \ma^2(t) \, \sin (\theta) = 0,
\label{eq:AxionFieldEq}
\end{equation}
where we have assumed the canonical axion potential of
\begin{equation}
  V(\theta) = \fa^2\ma^2\left[1-\cos (\theta)\right]. \label{eq:potential}
\end{equation}
This is subject to the boundary condition $\theta(\mrm{t}{i}) = \thetai$ and $\dot{\theta}(\mrm{t}{i}) = 0$, where $\thetai$ is called the \textit{initial misalignment angle}.

The \gambit collaboration completed a comprehensive study of axion and broader ALP theories in 2018~\cite{Axions}, using an extensive list of experimental constraints. These rely on the interactions of ALPs with SM matter, which can be studied in an effective field theory framework~\cite{Kaplan:1985dv,1985_srednicki_axioneft,1986_georgi_axioneft}.

\begin{figure}[bt]
	\centering
	\begin{tikzpicture}[
	sibling distance=3.5cm,
	level distance=2.6cm,
	treenode/.style = {draw, rectangle, font=\footnotesize, rounded corners=2pt, align=center, inner sep=3pt, execute at begin node=\setlength{\baselineskip}{1.25em}},
	lvl1/.style={treenode},
	lvl2/.style={treenode, yshift=-0.9cm}]

\node [treenode] (a0) {\highl{\textsf{GeneralALP (7)}}\\ 7~parameter model\\ $\fa$, $\mazero$, $\gagg$, $\gaee$, $\beta$, $\Tcrit$, $\thetai$}
	child { node [lvl1, xshift=-0.2cm] (a1) {\highl{\textsf{QCDAxion (4+4)}}\\ Free parameters:\\ $\fa$, $E/N$, $C_{aee}$, $\thetai$\\ Nuisance parameters:\\ $\LambdaQCD$, $\Tcrit$, $\beta$, $\widetilde{C}_{a\gamma\gamma}$}
		child { node [lvl2, xshift=-0.2cm] (a11) {\highl{\textsf{DFSZAxion-I (3+4)}}\\ Free parameters:\\ $\fa$, $\tan (\beta^\prime)$, $\thetai$\\ Nuisance parameters:\\ $\LambdaQCD$, $\Tcrit$, $\beta$, $\widetilde{C}_{a\gamma\gamma}$\\ Fixed parameters:\\ $E/N=8/3$} }
		child { node [lvl2] (a12) {\highl{\textsf{DFSZAxion-II (3+4)}}\\ Free parameters:\\ $\fa$, $\tan (\beta^\prime)$, $\thetai$\\ Nuisance parameters:\\ $\LambdaQCD$, $\Tcrit$, $\beta$, $\widetilde{C}_{a\gamma\gamma}$\\ Fixed parameters:\\ $E/N=2/3$} }
		child { node [lvl2, xshift=0.2cm] (a13) {\highl{\textsf{KSVZAxion (3+4)}}\\ Free parameters:\\ $\fa$, $E/N$, $\thetai$\\ Nuisance parameters:\\ $\LambdaQCD$, $\Tcrit$, $\beta$, $\widetilde{C}_{a\gamma\gamma}$\\ Fixed parameters:\\ $C_{aee}$} }
	}
	child { node [lvl1, xshift=0.2cm] (a2) {\highl{\textsf{ConstantMassALP (5)}}\\  Free parameters:\\ $\fa$, $\Lambda$, $C_{a\gamma\gamma}$, $C_{aee}$, $\thetai$\\ Fixed parameters:\\ $\Tcrit$ irrelevant, $\beta \equiv 0$}
	}
;

\end{tikzpicture}
	\caption{Family tree of axion models in \gambit. The numbers in brackets refer to the number of model parameters; $(n+m)$ indicates $n$ (largely unconstrained) fundamental parameters of the model and $m$ (typically well-constrained) nuisance parameters. From \cite{Axions}.}
	\label{fig:AxionModelTree}
\end{figure}
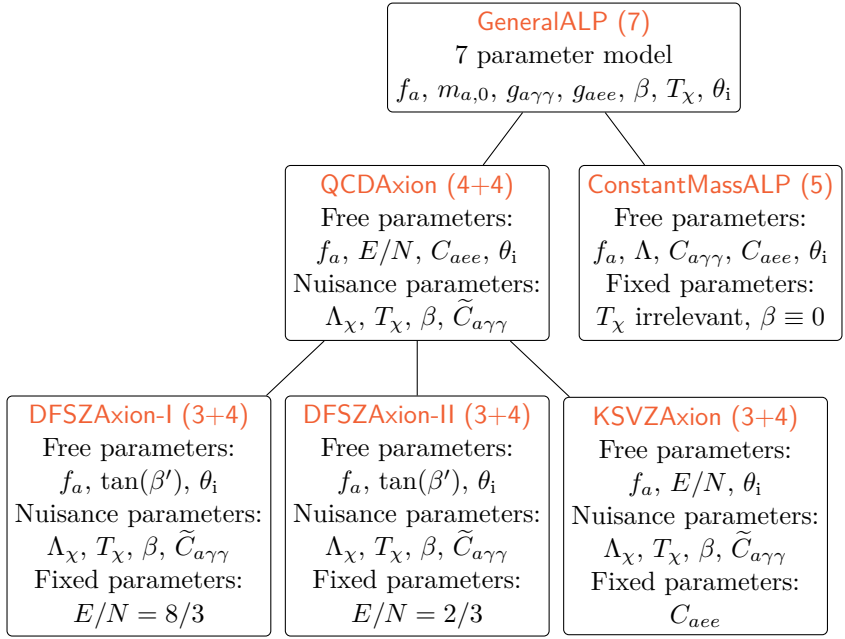

The most general axion/ALP model in \gambit assumes the effective Lagrangian density to take the form
\begin{equation}
	\lagrangian_a^\mathrm{int} = -\frac{\fa\gagg}{4} \theta F_{\mu\nu}\widetilde{F}^{\mu\nu} - \frac{\fa\gaee}{2m_e} \bar{e}\gamma^\mu\gamma_5e\del_\mu \theta \, .\label{eq:ax:lagrange}
\end{equation}
Note that this provides for possible axion-photon and axion-electron interactions, whilst ignoring terms for other interactions that do not currently give rise to interesting experimental observables. The complete family tree of \gambit axion/ALP models is shown in Fig.\ \ref{fig:AxionModelTree}, headed by the \genalp model, whose parameters have all now been defined. This provides a phenomenological description of axion physics that is not constrained to give physical solutions, as the couplings are not inversely proportional to $\fa$.

The \qcdaxion model appears as a child model, and differs from the more general case by having tight constraints on some parameters, arising from the known relationships with the QCD scale. The axion-electron coupling is traded for the model-dependent form factor~$\caee$
\begin{equation}
	\gaee = \frac{m_e}{\fa} \; \caee\, , \label{eq:qcdaxioncouplings1}
\end{equation}
whilst the axion-photon coupling is replaced by the model-dependent ratio of the electromagnetic and colour anomalies~$E/N$
\begin{equation}
	\gagg = \frac{\mrm{\alpha}{EM}}{2\otherpi \fa}\left(\frac{E}{N} - \caggtilde\right) \, .\label{eq:qcdaxioncouplings2}
\end{equation}
$\caggtilde$ is a model-independent contribution from axion-pion mixing, which is taken from Ref.~\cite{1511.02867}, and assigned a nuisance likelihood with a relevant uncertainty. Note that the ratio $E/N$ should in principle take discrete values, but it is sampled as a continuous parameter for convenience, seeing as the possible rational values that it can take are close together. The final nuisance parameter of the \qcdaxion model is $\LambdaQCD$, which results from replacing the parameter  $\mazero$ of the \genalp model by an energy scale such that
\begin{equation}
	\mazero \equiv \frac{\LambdaQCD^2}{\fa} \, .
\end{equation}
The value of $\LambdaQCD$ is taken from first-principle calculations of the zero-temperature axion mass provided in Ref.~\cite{1511.02867},\footnote{This value was later updated in \cite{Gorghetto:2018ocs}, after the appearance of Ref.\ \cite{Axions}.} and it is subject to a Gaussian nuisance likelihood.

The other models of interest for this review are the \ksvz and \dfsz model variants, which involve further field content being added to the SM. In \ksvz models~\cite{1979_kim_ksvz,1980_shifman_ksvz}, the SM is supplemented by one or more electrically neutral, heavy quarks, and there are no tree-level interactions between the axion and SM fermions. There is, however, still an axion-photon interaction, which generates an axion-electron interaction at one loop. The \gambit study investigated four different \ksvz models, distinguished only by the choice of $E/N$ from the set 0, 2/3, 5/3 and 8/3.

\dfsz models supplement the SM by an additional Higgs doublet~\cite{1980_zhitnitsky_dfsz,1981_dine_dfsz}, which results in direct axion-electron interactions. Defining the ratio of the two Higgs vacuum expectation values to be  $\tan (\beta^\prime)$, one can write two variants of the \dfsz scenario as
\begin{align}
	\begin{array}{lll}
		\caee = \sin^2 (\beta^\prime)\left/3\right., \quad \phantom{.} & E/N = 8/3  \quad \phantom{.}& (\dfszI) \, \\
		\caee = \left[1-\sin^2 (\beta^\prime) \right]\left/3\right., \quad \phantom{.} & E/N = 2/3 \quad \phantom{.} & (\dfszII) \,
	\end{array} \label{eq:dfsz:caee}.
\end{align}
It is thus convenient to replace the parameter~$\caee$ in the \qcdaxion model by $\tan (\beta^\prime)$.

\subsubsection{Experimental constraints on axions}
\label{sec:axionL}
Many experiments are sensitive to the axion theories described here, and current null results place tight constraints on axions for specific combinations of masses and coupling strengths. Here we provide a brief review of those constraints, referring the reader to Ref.~\cite{Axions} for a detailed description of the experimental likelihoods.

\begin{itemize}
\item \textbf{Light-shining-through-wall (LSW) experiments: }Photon-axion interactions would allow photons to pass through a wall by becoming an axion, only to convert back to a photon on the other side. LSW experiments attempt to observe this by shining laser light onto an opaque material in the presence of a strong magnetic field. The \gambit LSW likelihood uses the results from the ALPS-I experiment, using data for both evacuated and gas-filled magnets~\cite{1004.1313}.
\item \textbf{Helioscopes: }Axion production in the Sun can be probed by observing the solar disc with a long magnet contained in an opaque casing. Any axions produced in the Sun that made it to Earth would pass through the exterior, and potentially convert to photons within the magnetic field in the interior. The details of solar axion production depend on the solar model, in addition to the axion-photon and axion-electron couplings. The \gambit axion studies utilise the AGSS09met solar model~\cite{Serenelli09,AGSS} and its more recent iteration~\cite{1611.09867}, and utilise two separate likelihoods for the 2007 and 2017 results of the CAST experiment~\cite{hep-ex/0702006,1705.02290}.
\item \textbf{Haloscopes (cavity experiments): }Axion haloscopes aim to detect resonant axion-photon conversion inside a tunable cavity~\cite{1983_Sikivie, 1985_Sikivie}, with microwave cavities providing the greatest current sensitivity to axions. Unfortunately, the resonant nature of the experiment means that one obtains highly sensitive constraints only within a very narrow mass range. The ability of haloscope experiments to detect axions depends on their cosmological abundance, as well as the galactic DM velocity distribution~\cite{2011_Hoskins}.  The \gambit study combines separate likelihood terms for the Rochester-Brookhaven-Fermi (RBF)~\cite{DePanfilis:1987dk,Wuensch:1989sa}, University of Florida (UF)~\cite{Hagmann:1990tj}, ADMX 1998-2009~\cite{astro-ph/9801286,Asztalos:2001tf,astro-ph/0310042,astro-ph/0603108,0910.5914} and ADMX 2018~\cite{1804.05750} datasets.
\item \textbf{Dark matter relic density: }Although axions are not a thermal relic such as those encountered in WIMP models, the relic abundance of axion DM is calculable numerically via the details of the realignment mechanism that follow from the equation of motion given in Eq~\ref{eq:AxionFieldEq}. This can be compared with the observed value from the most recent \emph{Planck} analysis~\cite{Planck15cosmo}. The \gambit axion study applied this as both an upper limit (in which case axions are allowed to provide only a component of DM) and, in separate analyses of each model, a measurement. In the former case, anticipated yields in experiments that rely on the local DM density were scaled accordingly.
\item \textbf{Distortions of gamma-ray spectra: }Axion-photon conversions could occur in strong galactic or inter-galactic magnetic fields, resulting in a distortion of the spectra of distant sources~\cite{Raffelt:1987im,hep-ph/0111311,hep-ph/0204216,0704.3044}. There is a critical energy scale $E_{crit}$ at which photons will efficiently convert into axions, and it can be shown that spectral distortions only occur in real measurements when the critical energy lies within the spectral window of the instrument~\cite{1205.6428,1305.2114}. This has the effect of localising constraints from spectral distortion measurements to specific ranges of the axion mass. The \gambit axion study utilises a likelihood based on H.E.S.S studies of the active galactic nucleus PKS 2155-304~\cite{1311.3148}.
\item \textbf{Supernova 1987A: }If axions had been produced in the SN1987A supernova explosion, they could have been converted to photons in the Galactic magnetic field, and detected as a coincident gamma ray burst by the Solar Maximum Mission~\cite{Chupp:1989kx}. The absence of this observation has been used to constrain axion properties. The \gambit study uses a likelihood based on Ref.~\cite{1410.3747}.
\item \textbf{Horizontal Branch stars and the R parameter: }The existence of axions would provide an extra mechanism of energy loss for stars, causing them to cool faster~\cite{Sato:1975vy,Raffelt:1990yz,book_raffelt_laboratories}. This would affect the relative time that stars spend on the Horizontal Branch (HB) and upper Red Giant Branch (RGB), which in turn sets the observed ratio of the numbers of stars on these branches ($R = \mrm{N}{HB}/\mrm{N}{RGB}$). Theory suggests that axions would have the most significant impact on the lifetimes of HB stars, leading to a reduction in $R$. The \gambit $R$ parameter likelihood is based on the comparison of a calculation of the $R$ parameter for axion theories~\cite{1512.08108,1983A&A...128...94B,Raffelt:1989xu,1311.1669,1406.6053} with the observed value of $\mrm{R}{obs}=1.39 \pm 0.03$~\cite{1406.6053}, which is based on a weighted average of cluster count obervations \cite{astro-ph/0403600}.
\item \textbf{White Dwarf cooling hints: }White dwarfs (WDs) are intriguing axion laboratories for several reasons. The first is that energy loss via axion production in WDs can be probed experimentally by using measurements of the oscillations of their radii and luminosities. These can be related to their internal structure via astroseismology, and measurements of the decrease in the oscillation periods can be related to energy loss. The second reason is that WDs have electron-degenerate cores, allowing us to probe the axion-electron coupling rather than the electron-photon coupling. A number of previous studies have calculated the expected period decrease in the presence of axions.  The \gambit WD cooling likelihood is based on interpolation of the results and uncertainties found in Refs.~\cite{1205.6180,1211.3389,1605.06458,1605.07668}. Current evidence suggests that WDs actually require an additional cooling mechanism relative to standard models, but this remains controversial due to a number of experimental and theoretical issues. The \gambit axion paper thus contains studies generated both with and without WD cooling hints added to the combined likelihood.
\end{itemize}

\subsubsection{\gambit results for the \qcdaxion model}

Although the \gambit axion paper contained results for all of the models described above, we will here concentrate on the \qcdaxion results in the interests of brevity. The various parameters (including nuisance parameters) are shown in Table~\ref{tab:priors:QCDAxion}, along with the chosen priors and prior ranges. For each of the nuisance parameters, the prior range is chosen to cover a range of approximately $-5\sigma$ to $+5\sigma$ around the known central value, where $\sigma$ is the known uncertainty. The range of $E/N$ values is selected to encompass those encountered in a broad range of previous axion model studies, whilst the range on $\fa$ is driven by the requirement that the range of possible axion masses reaches from very small masses to the the largest mass allowed by bounds on hot DM. Our choice of a log prior for $\fa$ is motivated by the fact that the scale is unknown. $\caee$ is explored in a generous range around 1, whilst the causal structure of the early Universe motivates our use of a flat prior on the initial misalignment angle $\thetai$. The local DM density $\rho_0$ is given the same treatment as in previous \gambit studies.

\begin{table}[tbp]
	\caption{Prior choices for \qcdaxion models in \cite{Axions}.\label{tab:priors:QCDAxion}}
	\footnotesize
	\centering
	\begin{tabular}{@{}lccc}
		\toprule
		\textbf{Model} & \multicolumn{2}{l}{\textbf{Parameter range/value}} & \textbf{Prior type} \\
		\midrule
		\qcdaxion & \iuo{\fa}{\GeV} & \prrange{e6}{e16} & log \\
		& \iuo{\LambdaQCD}{\MeV}& \prrange{73}{78} & flat \\
		& $\caggtilde$ & \prrange{1.72}{2.12} & flat \\
		& $E/N$ & \prrange{-1.33333}{174.667} & flat \\
		& $\caee$ & \prrange{e-4}{e4} & log \\
		& $\thetai$ & \prrange{-3.14159}{3.14159} & flat \\
		& $\beta$ & \prrange{7.7}{8.2} & flat \\
		& \iuo{\Tcrit}{\MeV}& \prrange{143}{151} & flat \\
		\midrule
		Local DM density & \iuo{\rho_0}{\GeV\per\centi\metre^3} & \prrange{0.2}{0.8} & flat \\
		\bottomrule
	\end{tabular}
\end{table}

\begin{figure}[tbp]
	\centering
	{
		\includegraphics[width=0.49\linewidth]{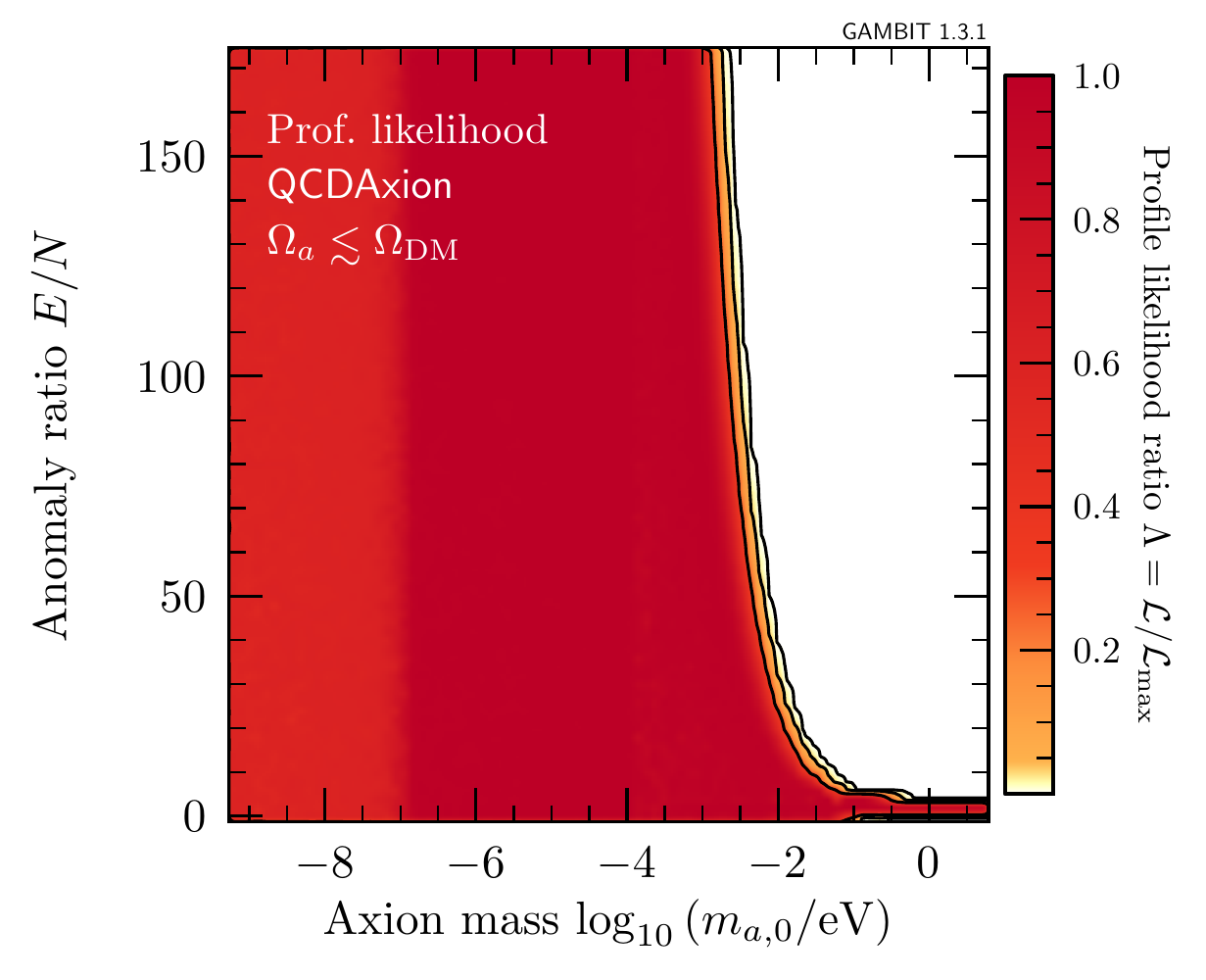}
		\hfill
		\includegraphics[width=0.49\linewidth]{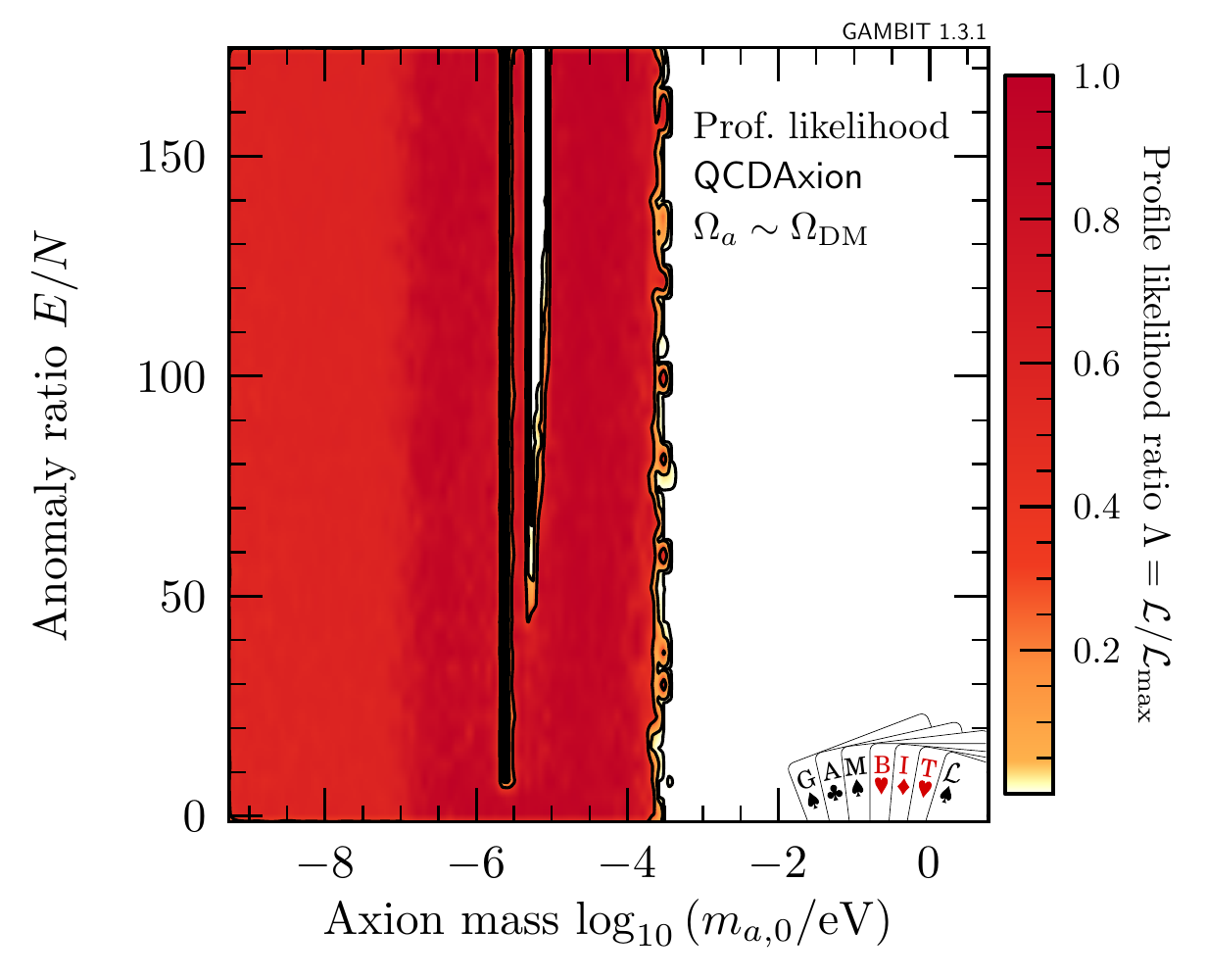}
	}
	{
		\includegraphics[width=0.49\linewidth]{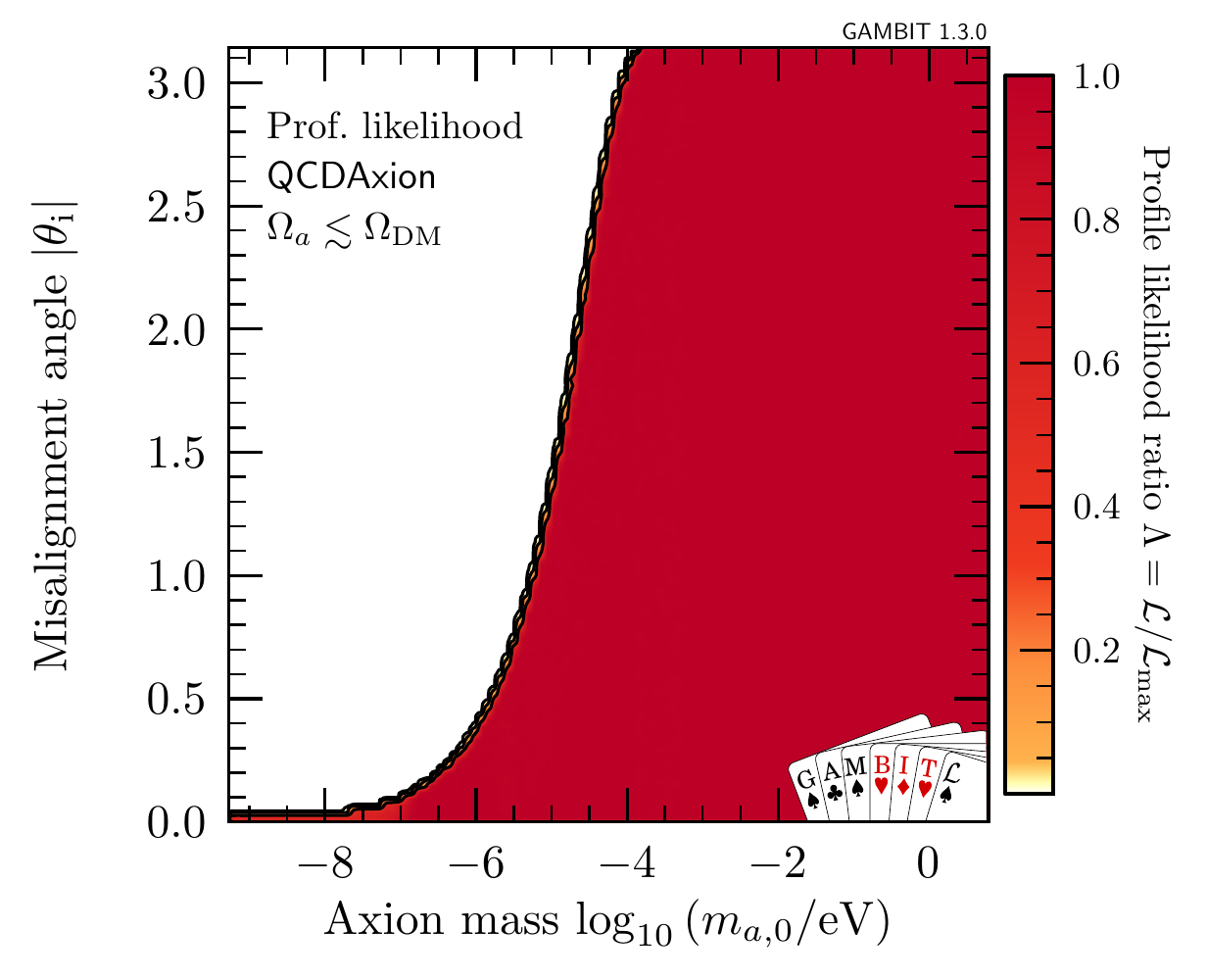}
		\hfill
		\includegraphics[width=0.49\linewidth]{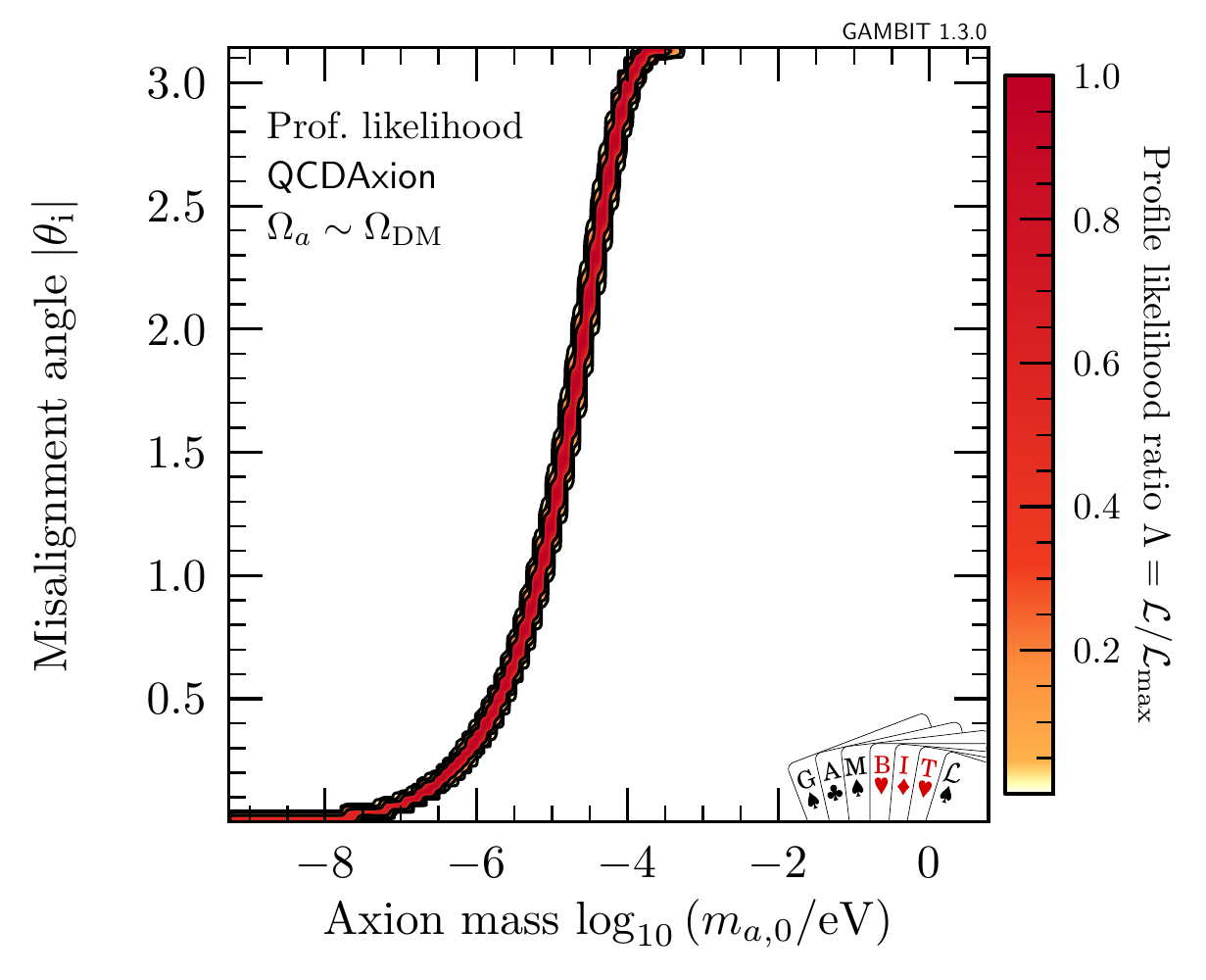}
	}
	\caption{Profile likelihoods~(from \diver) for \qcdaxion models with upper limits~(\textit{left}) and matching condition~(\textit{right}) for the observed DM relic density. The upper and lower panels show the constraints on the anomaly ratio, $E/N$, and the absolute value of the initial misalignment angle, $|\thetai|$, respectively. From \cite{Axions}. \label{fig:QCDAxion:frequentist}}
\end{figure}

Fig.\ \ref{fig:QCDAxion:frequentist} shows profile likelihood distributions in various planes of the \qcdaxion parameters, obtained without the presence of WD cooling hints in the combined likelihood. The left panels show the result of imposing the relic density constraint as an upper limit. The exclusion of the low-$f_a$ (high mass) region, except at very low values of the axion-photon coupling (which is related to $E/N$), arises from the $R$ parameter and CAST results. The slight reduction in the profile likelihood for masses lower than approximately \SI{0.1}{\micro\eV} also comes from the $R$ parameter likelihood; for such masses, the maximum allowed value for the axion-electron coupling ($\caee \leq \num{e4}$) is not large enough to perfectly satisfy the $R$-parameter constraint. If axions are assumed to saturate the relic abundance of DM (right top panel), the high-mass region is excluded entirely due to the fact that the realignment mechanism cannot produce enough DM. The bottom row of Fig.\ \ref{fig:QCDAxion:frequentist} shows the allowed values for the initial misalignment angle. In the case that axions supply all of DM, we recover the familiar result that $\left|\thetai\right| \ll 1$, for QCD axion masses of $\mazero \lsim \SI{0.1}{\micro\eV}$, a fine-tuning that we will discuss further in the context of a Bayesian analysis.

\begin{figure}
        \centering
        {
                \includegraphics[width=0.49\linewidth]{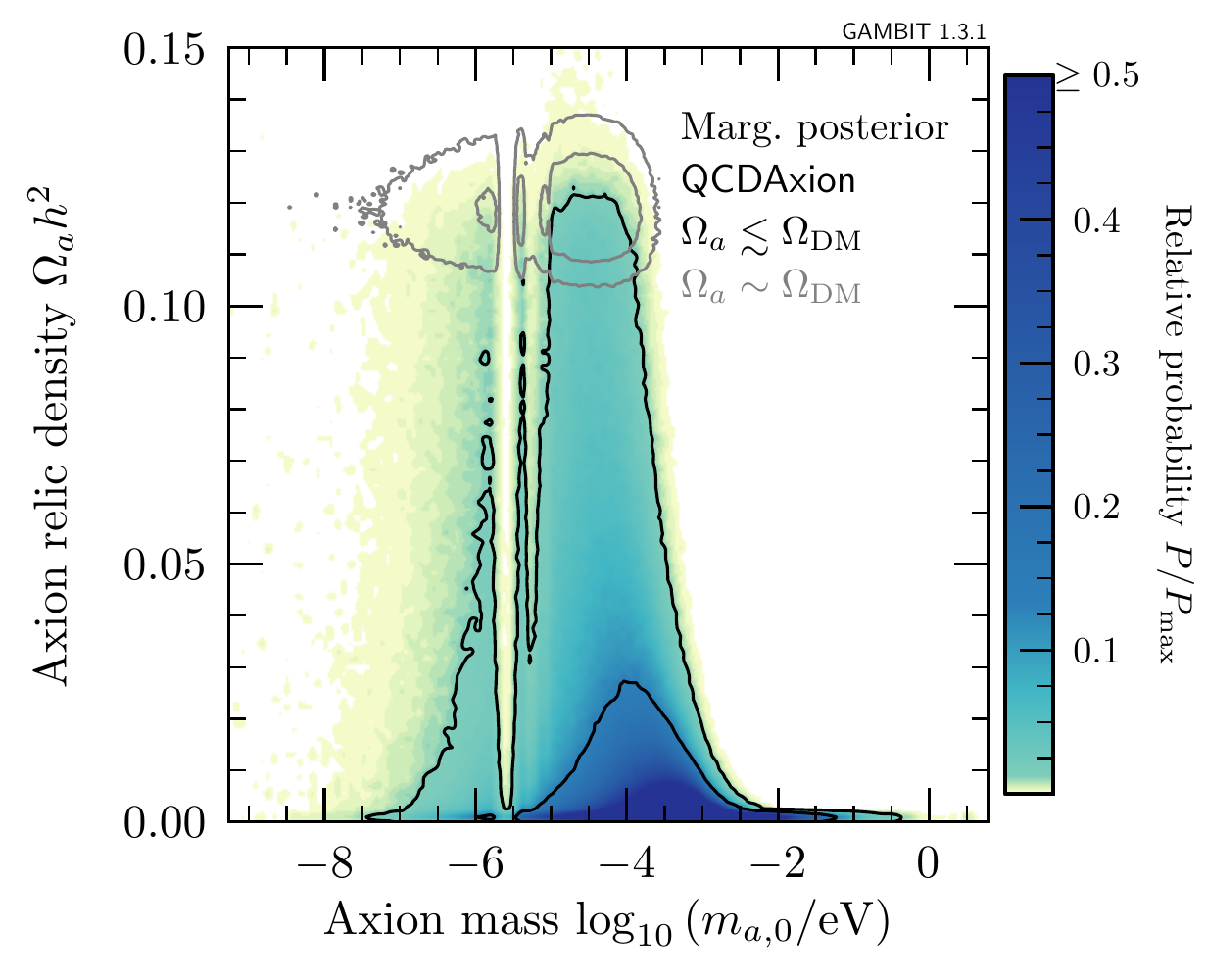}
                \includegraphics[width=0.49\linewidth]{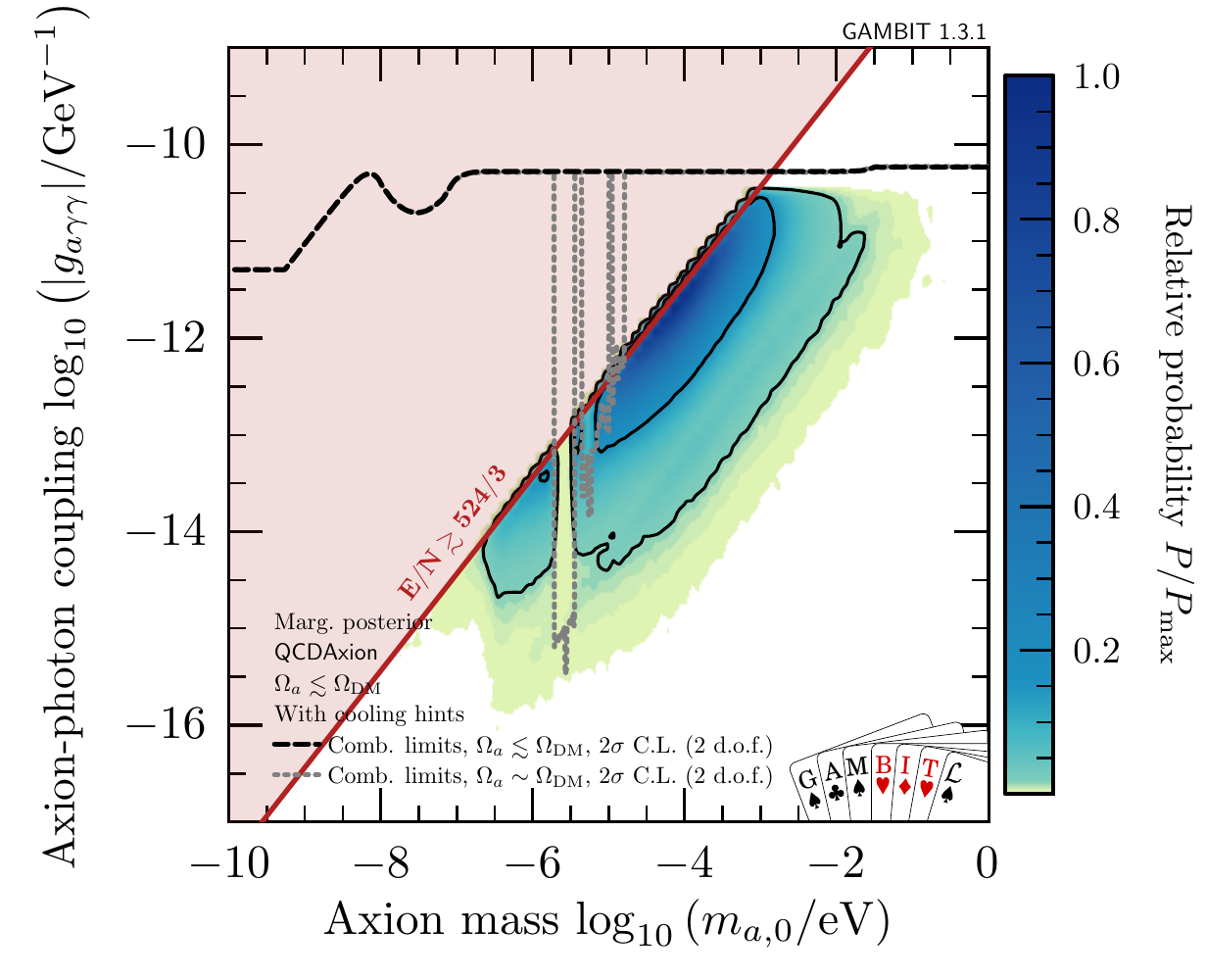}
        }
        \caption{Marginalised posterior for \qcdaxion models with the DM relic density constraint treated as an upper limit. \textit{Left}: constraints on the energy density in axions today, $\Omega_a h^2$, without the inclusion of WD cooling hints.  \textit{Right}: constraints on the absolute value of the axion-photon coupling, $|\gagg|$.  This panel also includes WD cooling hints, but they have little impact on the result.  For comparison, the right panel also shows the region for which QCD axions are not theoretically possible (red line and shading), as well as the frequentist $2\sigma$ C.L. constraints on more general ALP models (dashed lines). From \cite{Axions}. \label{fig:cooling:QCDAxion:RelicDens:twoD}}
\end{figure}

In the left panel of Fig.\ \ref{fig:cooling:QCDAxion:RelicDens:twoD}, we show the marginalised Bayesian posterior in the $\Omega_a h^2-m_{a,0}$ plane without WD cooling hints, demonstrating that the scan can find viable parts of the parameter space where axions consistent with all current experimental observations can account for a sizable fraction of dark matter. The situation is similar even when WD cooling hints are included (not shown). One can also observe an interesting bound on the axion mass. If the DM relic density constraint is applied as an upper limit, we find $\SI{0.73}{\micro\eV} \le \mazero \le \SI{6.1}{\milli\eV}$ at 95\% credibility (equal-tailed interval). Meanwhile, if axions must provide all of the observed dark matter, this changes to $\SI{0.53}{\micro\eV} \le \mazero \le \SI{0.13}{\milli\eV}$.

In the right panel of Fig.\ \ref{fig:cooling:QCDAxion:RelicDens:twoD}, we show the marginalised posterior in the $|\gagg|-m_{a,0}$ plane with the DM relic density constraint applied as an upper limit, and with WD cooling hints included. Also shown are the {na\"ive} bounds on the parameter space that result from phenomenological constraints on \genalp models and the maximum value of $E/N$. The shape of the preferred region is partly formed by the effect of fine-tuning. At low axion masses, this is required to avoid dark matter overproduction, whilst at large axion masses it is required to achieve low values of $|\gagg|$ through cancellations between $E/N$ and $\tilde{C}_{a\gamma\gamma}$. The preferred parameter region is localised within a few orders of magnitude in mass around $\mazero \sim \SI{100}{\micro\eV}$ and $\gagg \sim \SI{e-12}{\GeV^{-1}}$.

\begin{table}[tbp]
	\caption{Prior choices for \dfszI, \dfszII and \ksvz models in \cite{Axions}. Note that the priors listed in the first section of the table apply to all three models.\label{tab:priors:DFSZvsKSVZ}}
	\footnotesize
	\centering
	\begin{tabular}{lcccl}
		\toprule
		\textbf{Model} & \multicolumn{2}{l}{\textbf{Parameter range/value}} & \textbf{Prior type} & \textbf{Comments}\\
		\midrule
		& \iuo{\fa}{\GeV} & \prrange{e6}{e16} & log & Applies to all\\
		& \iuo{\LambdaQCD}{\MeV}& \prrange{73}{78} & flat & Applies to all\\
		& $\caggtilde$ & \prrange{1.72}{2.12} & flat & Applies to all\\
		& $\thetai$ & \prrange{-3.14159}{3.14159} & flat & Applies to all\\
		& $\beta$ & \prrange{7.7}{8.2} & flat & Applies to all\\
		& \iuo{\Tcrit}{\MeV}& \prrange{143}{151} & flat & Applies to all\\
		\midrule
		\dfszI & $E/N$ & $8/3$ & delta & \\
		& $\tan(\beta')$ & \prrange{0.28}{140.0} & log & \\
		\dfszII & $E/N$ & $2/3$ & delta & \\
		& $\tan(\beta^\prime)$ & \prrange{0.28}{140.0} & log & \\
		\ksvz & $E/N$ & $0$, $2/3$, $5/3$, $8/3$ & delta & Discrete\\
		\midrule
		Local DM density & \iuo{\rho_0}{\GeV\per\centi\metre^3} & \prrange{0.2}{0.8} & flat \\
		\bottomrule
	\end{tabular}
\end{table}

The \gambit Bayesian analysis of axion models also includes a model comparison of the \qcdaxion, \dfszI, \dfszII and \ksvz, based on scans of the latter models that use the priors defined in Table~\ref{tab:priors:DFSZvsKSVZ}. Bayesian evidence values $\mathbb{Z}(\mathcal{M})$ for each model $\mathcal{M}$ were calculated using the \MultiNest nested sampling package, before constructing the Bayes factor \cite{Jeffreys:1939xee,10.2307/2291091,10.2307/4356165}
\begin{equation}
	\mathcal{B} \equiv \frac{\mathbb{Z}(\mathcal{M}_1)}{\mathbb{Z}(\mathcal{M}_2)} \equiv \frac{\int \!  \mathcal{L}\left(\text{data}\left| \right. \boldsymbol{\theta_1} \right) \pi_1(\boldsymbol{\theta_1}) \, \dd \boldsymbol{\theta_1}}{\int \!  \mathcal{L}\left(\text{data}\left| \right. \boldsymbol{\theta_2} \right) \pi_2(\boldsymbol{\theta_1}) \, \dd \boldsymbol{\theta_2}} \, ,
\end{equation}
which relates two models $\mathcal{M}_1$ and $\mathcal{M}_2$ with parameters $\boldsymbol{\theta_1}$ and~$\boldsymbol{\theta_2}$. $\pi_1$ and $\pi_2$ are the priors on the parameters of the two models, and $\mathcal{L}$ is the likelihood. The Bayes factor is connected to the ratio of posterior probabilities of the models being correct
\begin{equation}
	\frac{\mathcal{P}\left(\mathcal{M}_1 \left| \text{data} \right.\right)}{\mathcal{P}\left(\mathcal{M}_2 \left| \text{data} \right.\right)} = \mathcal{B} \; \frac{\pi(\mathcal{M}_1)}{\pi(\mathcal{M}_2)},
\end{equation}
where the prior probabilities of the models themselves being correct are given by $\pi(\mathcal{M}_1)$ and $\pi(\mathcal{M}_2)$.  In the following, it is assumed that $\pi(\mathcal{M}_1)=\pi(\mathcal{M}_2)$, causing the the posterior odds ratio to be equal to the Bayes factor.

Without cooling hints, the odds ratios for pairs of models provide insufficient evidence to favour any particular scenario. However, if it is demanded that axions solve the DM and WD cooling problems simultaneously, there is a positive preference for the \qcdaxion model over the DFSZ- and KSVZ-type models, at a level of about 5:1. This results from the larger $\caee$ values allowed in the \qcdaxion model, which peaks at $\caee\approx 100$ in the one-dimensional marginalised posterior. Such a large coupling may cause a problem for model building. A frequentist analysis of the same scenario allows both the \dfsz and \ksvz models to be rejected with respect to the \qcdaxion model with better than 99\% confidence; if DM is instead allowed to consist only partially of axions, only \ksvz models can be rejected in this way.

\subsection{Right-handed neutrinos}
\subsubsection{Model definition}
The addition of right-handed ``sterile'' neutrinos to the SM has been proposed to explain the existence of neutrino flavour oscillations, which imply a non-zero neutrino mass.  They also serve an aesthetic theoretical purpose, as neutrinos are the only elementary fermions in the SM to not have both left- and right-handed incarnations. Moreover, sterile neutrinos are permitted to have both a Dirac mass term $\bar{\nu_L}M_D\nu_R$ and a Majorana mass term $\bar{\nu_R}M_M\nu_R^c$, and specific choices of the latter allow sterile neutrinos to solve cosmological problems such as the baryon asymmetry of the Universe~\cite{Fukugita:1986hr,Akhmedov:1998qx,Asaka:2005pn}, and the DM problem~\cite{Dodelson:1993je, Shi:1998km}.

A convenient parameterisation of a right-handed neutrino sector is the Casas-Ibarra parametrisation, amended to include 1-loop corrections to the left-handed neutrino mass matrix~\cite{Casas:2001sr,Lopez-Pavon:2015cga}. This involves writing a matrix that encodes the mixing among left-handed neutrinos (LHNs) and right-handed neutrinos (RHNs) as
\begin{align}
\Theta = iU_{\nu}\sqrt{m_{\nu}^\text{diag}}\mathcal{R}\sqrt{\tilde{M}^\text{diag}}^{-1},
\label{CItheta}
\end{align}
where $U_{\nu}$ is the PMNS matrix, $m_{\nu}^\text{diag}$ is a diagonalised, one-loop-corrected LHN mass matrix and $\tilde{M}^\text{diag}$ is the analogous RHN mass matrix. $\mathcal{R}$ is a complex, orthogonal matrix written as the product
\begin{align}
\mathcal{R} = \mathcal{R}^{23}\mathcal{R}^{13}\mathcal{R}^{12}\;,
\label{Rorder}
\end{align}
where the $\mathcal{R}^{ij}$ can, in turn, be parameterised by complex angles $\omega_{ij}$ with
\begin{align}
\mathcal{R}^{ij}_{ii} &= \mathcal{R}^{ij}_{jj} = \cos\omega_{ij}, \\
\mathcal{R}^{ij}_{ij} &= -\mathcal{R}^{ij}_{ji} = \sin\omega_{ij}, \\
\mathcal{R}^{ij}_{kk} &= 1; k \neq i,j\;.
\end{align}

Working in the flavour basis in which the Yukawa couplings of the charged leptons are diagonal by construction, the PMNS matrix $U_\nu$ can be written as
\begin{align}\label{UnuParameterisation}
  U_{\nu} = V^{23}U_{\delta}V^{13}U_{-\delta}V^{12}\mathrm{diag}(e^{i\alpha_1/2},e^{i\alpha_2/2},1)\;,
\end{align}
where $U_{\pm\delta} = \mathrm{diag}(e^{{\mp}i\delta/2},1,e^{{\pm}i\delta/2})$
and $V^{ij}$ is parameterised by the LHN mixing angles $\theta_{ij}$. The non-zero elements of $V^{ij}$ are analogous to those of $\mathcal{R}$.  $\alpha_1$, $\alpha_2$ and $\delta$ are $CP$-violating phases that are not excluded \emph{a priori}.

\begin{table}[t]
  \centering
  \begin{tabular}{l}
    \toprule
    \textbf{Parameter} \\
    \midrule
    Active neutrino parameters\\
    \quad$\theta_{12}$ [rad] \\
    \quad$\theta_{23}$ [rad] \\
    \quad$\theta_{13}$ [rad] \\
    \quad$m_{\nu_0}$ [eV] \\
    \quad$\Delta m^2_{21}$ $[10^{-5}\,\text{eV}^2]$ \\
    \quad$\Delta m^2_{3l}$ $[10^{-3}\,\text{eV}^2]$ \\
    \quad$\alpha_1$, $\alpha_2$ [rad] \\
    Sterile neutrino parameters\\
    \quad$\delta$ [rad] \\
    \quad Re $\omega_{ij}$ [rad] \\
    \quad Im $\omega_{ij}$ \\
    \quad$M_I$ [GeV] \\
    \quad$R_{\rm{order}}$ \\
    Nuisance parameters \\
    \quad$m_h$ [GeV]\\
    \bottomrule
  \end{tabular}
  \caption{The full list of scanned parameters for the \gambit right-handed neutrino study \cite{RHN}.}
  \label{tab:scanpars}
\end{table}

\begin{figure}[tbp]
  \centering
  \includegraphics[width=0.45\linewidth]{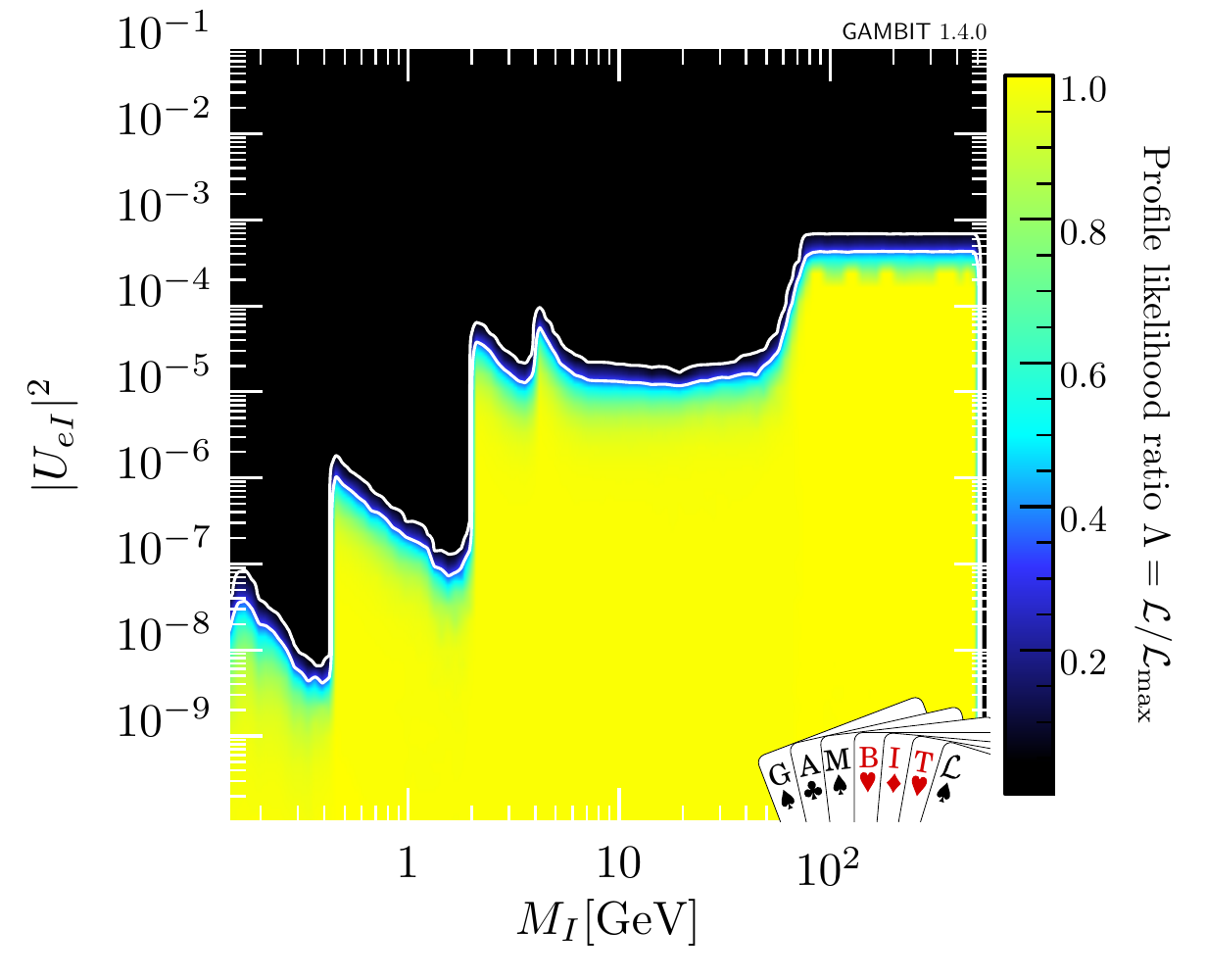}
  \includegraphics[width=0.45\linewidth]{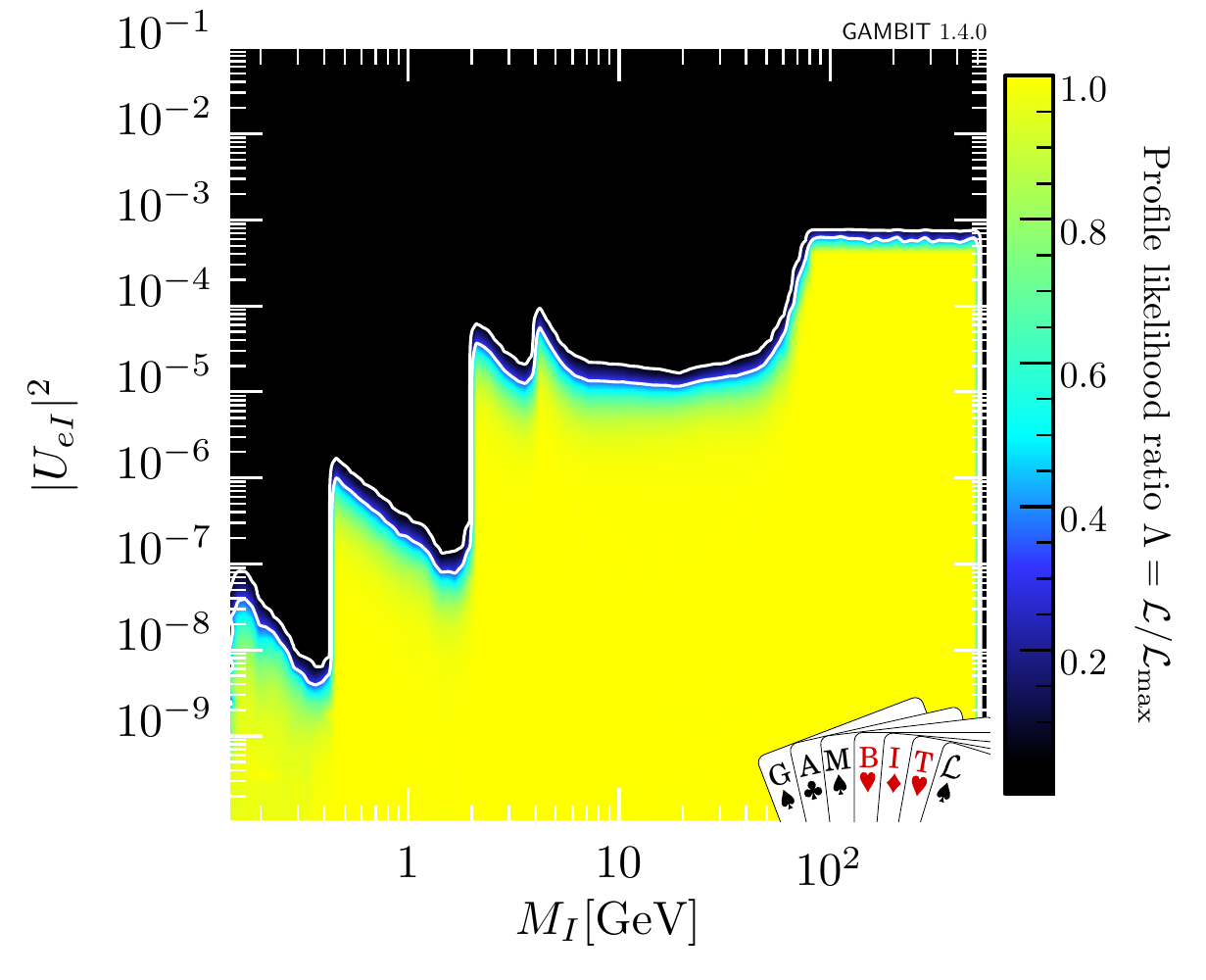}\\
  \includegraphics[width=0.45\linewidth]{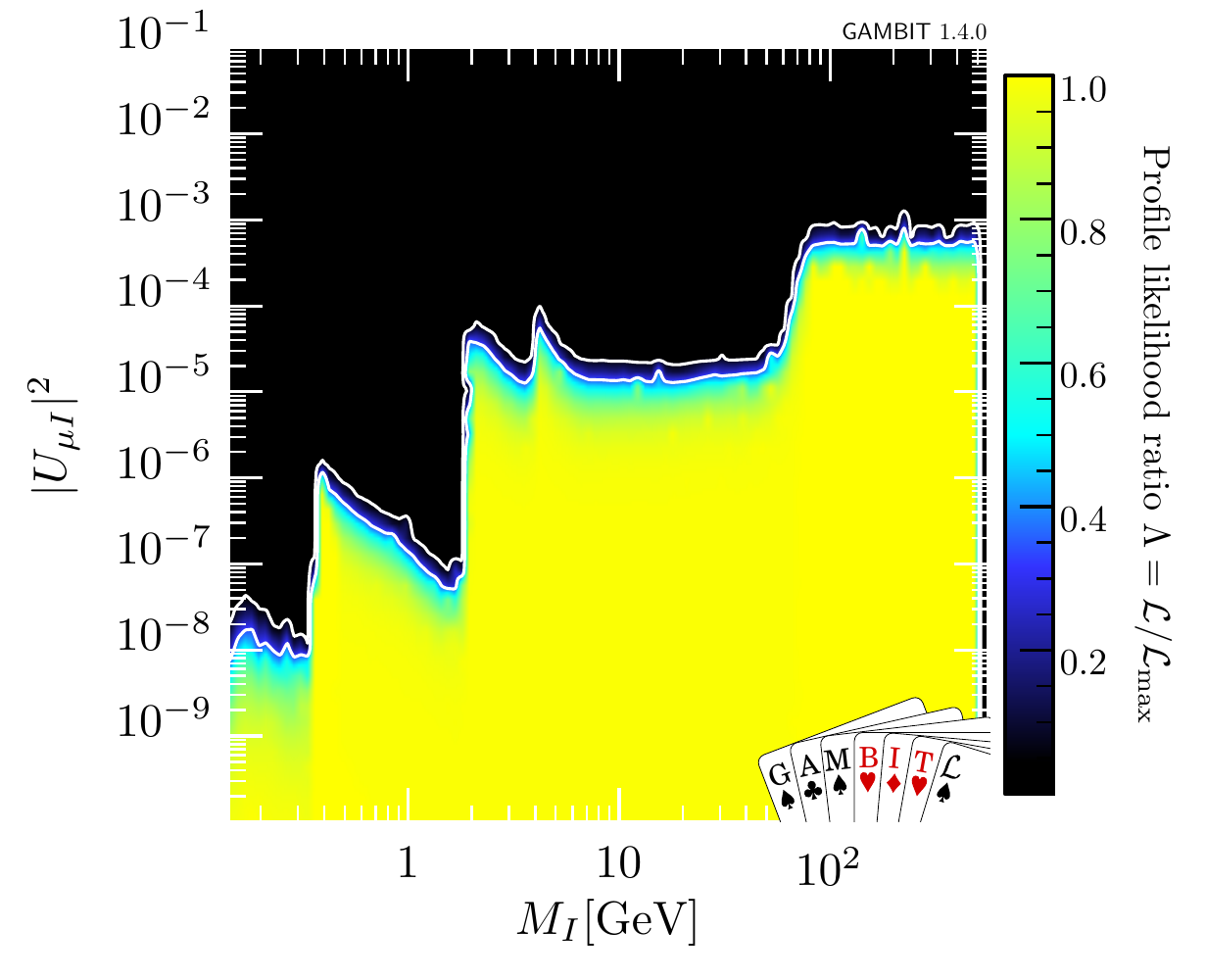}
  \includegraphics[width=0.45\linewidth]{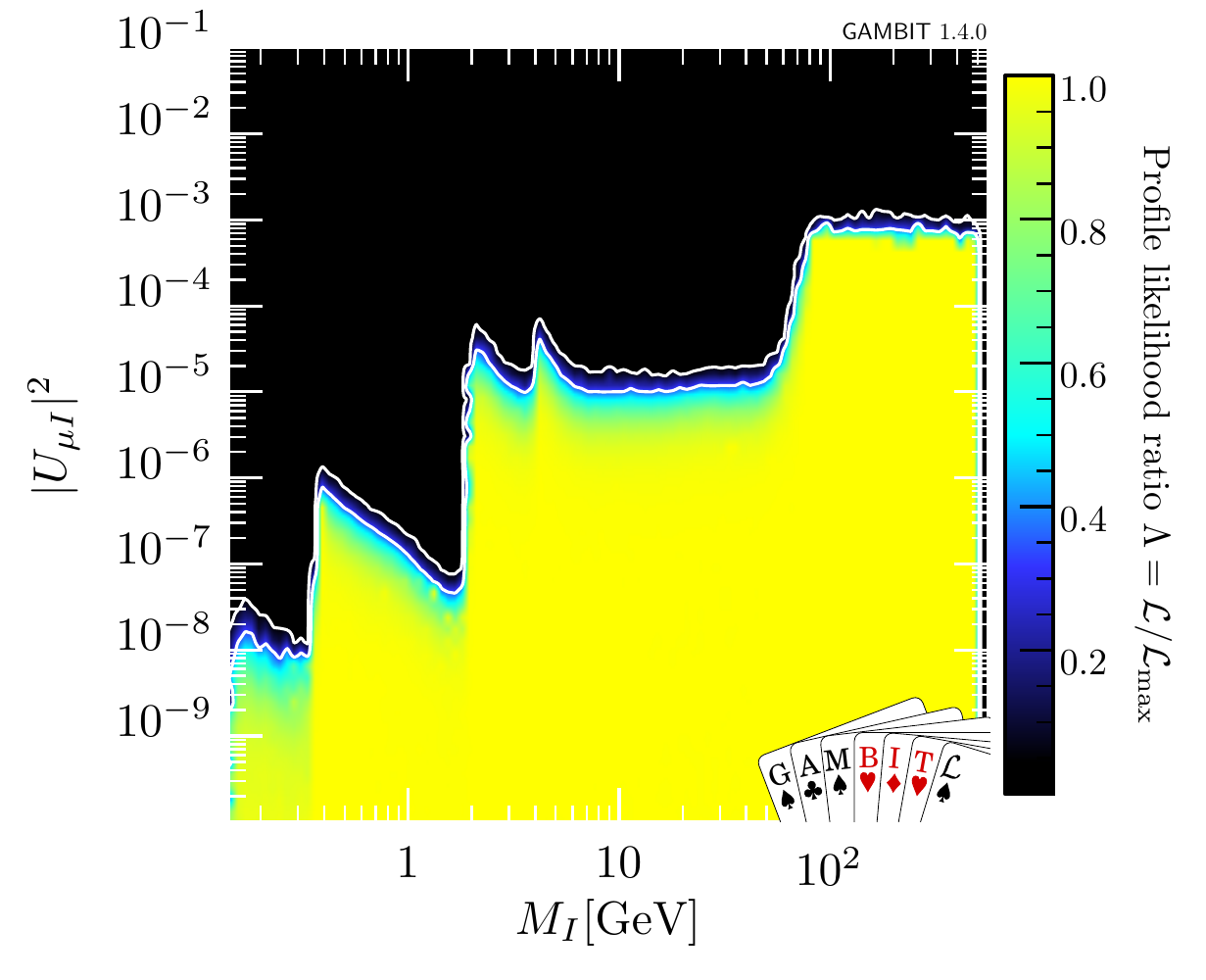}\\
  \includegraphics[width=0.45\linewidth]{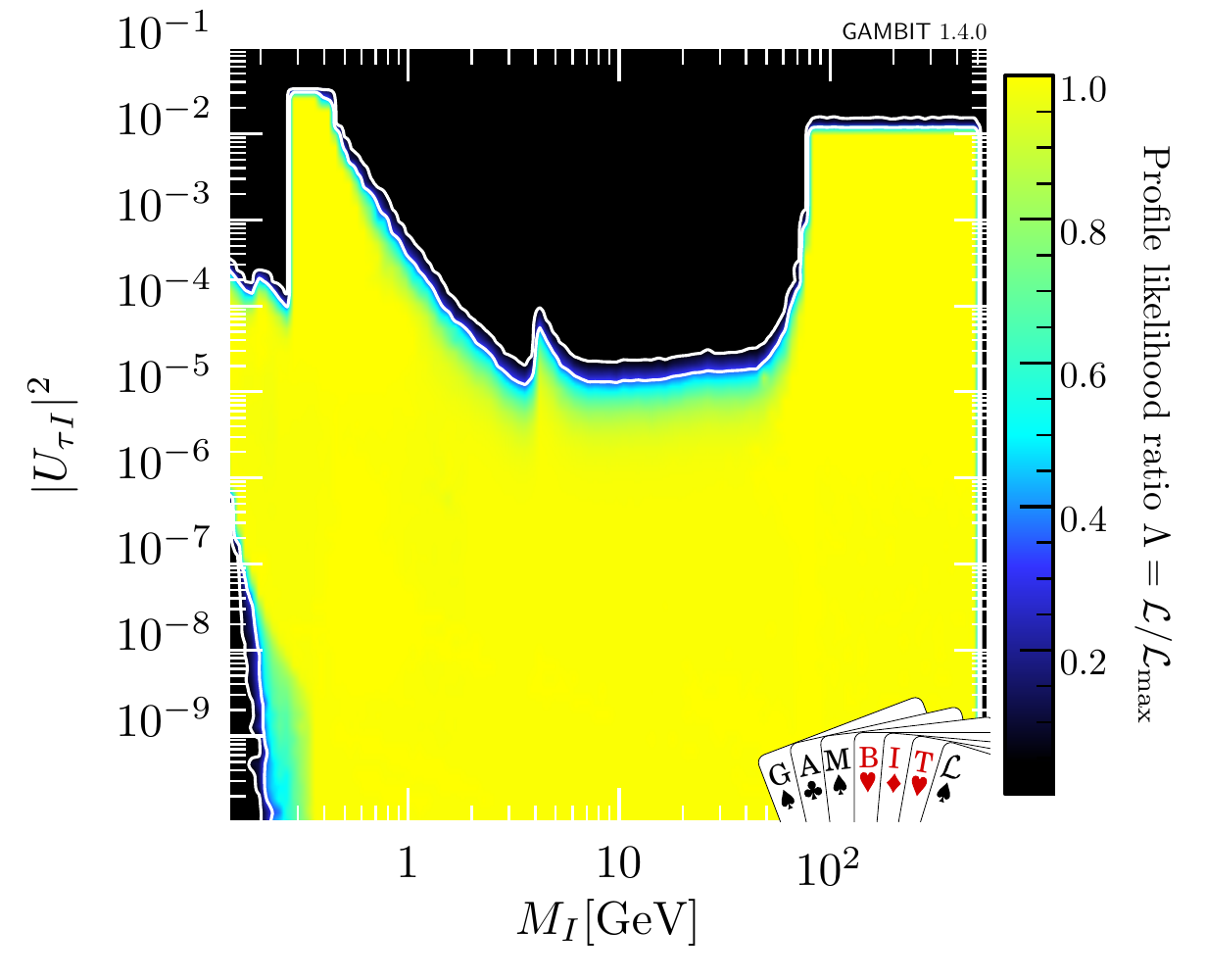}
  \includegraphics[width=0.45\linewidth]{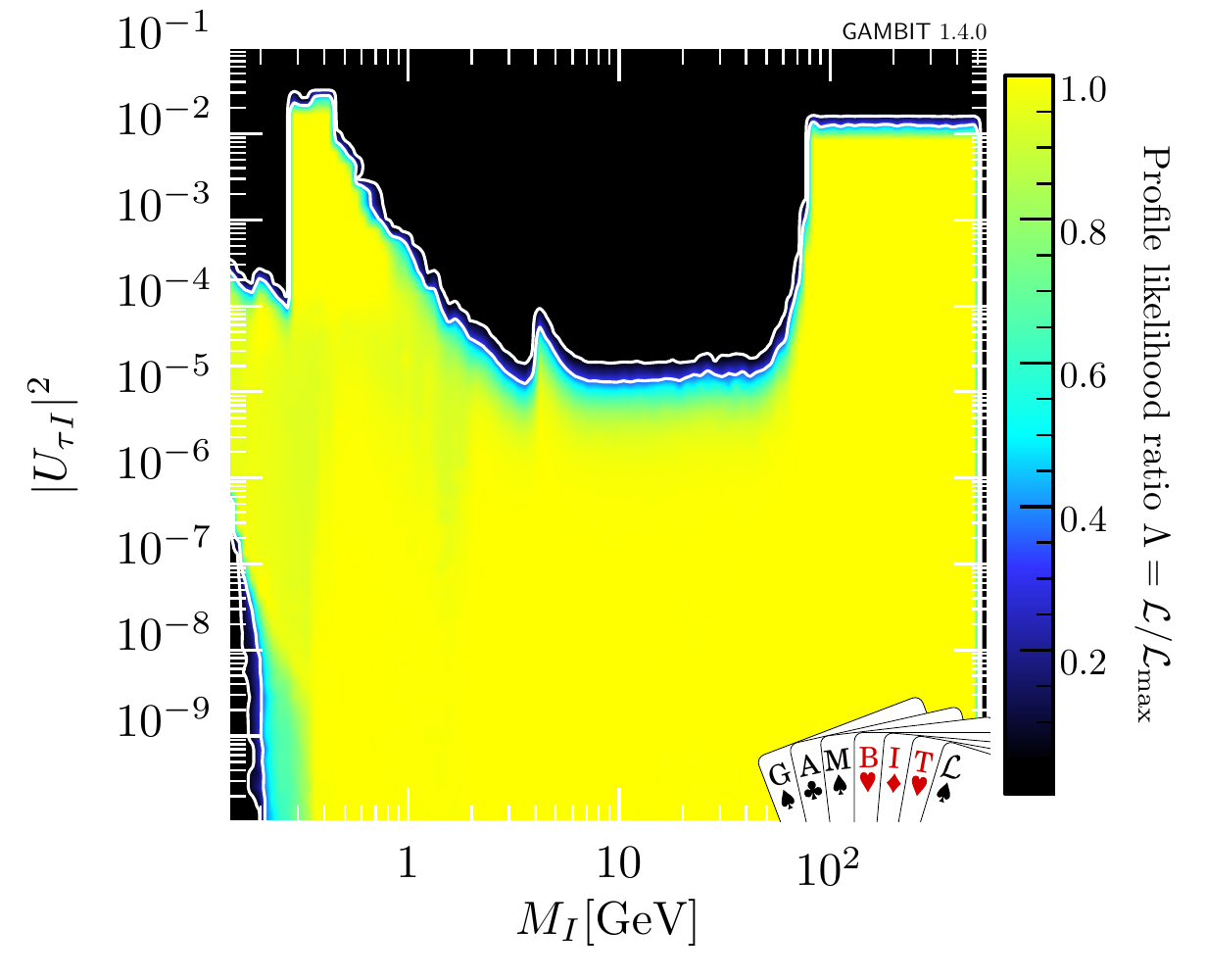}
  \caption{Profile likelihoods of right-handed neutrino models in the $M_I$ vs $U_{eI}^2$ (top), $M_I$ vs $U_{\mu I}^2$ (middle) and $M_I$ vs $U_{\tau I}^2$ (bottom) planes.  Results are shown for normal (left) and inverted (right) neutrino mass ordering. From \cite{RHN}.}
  \label{fig:M_Ue_capped}
\end{figure}

\begin{figure}[tbp]
  \centering
  \includegraphics[width=0.45\linewidth]{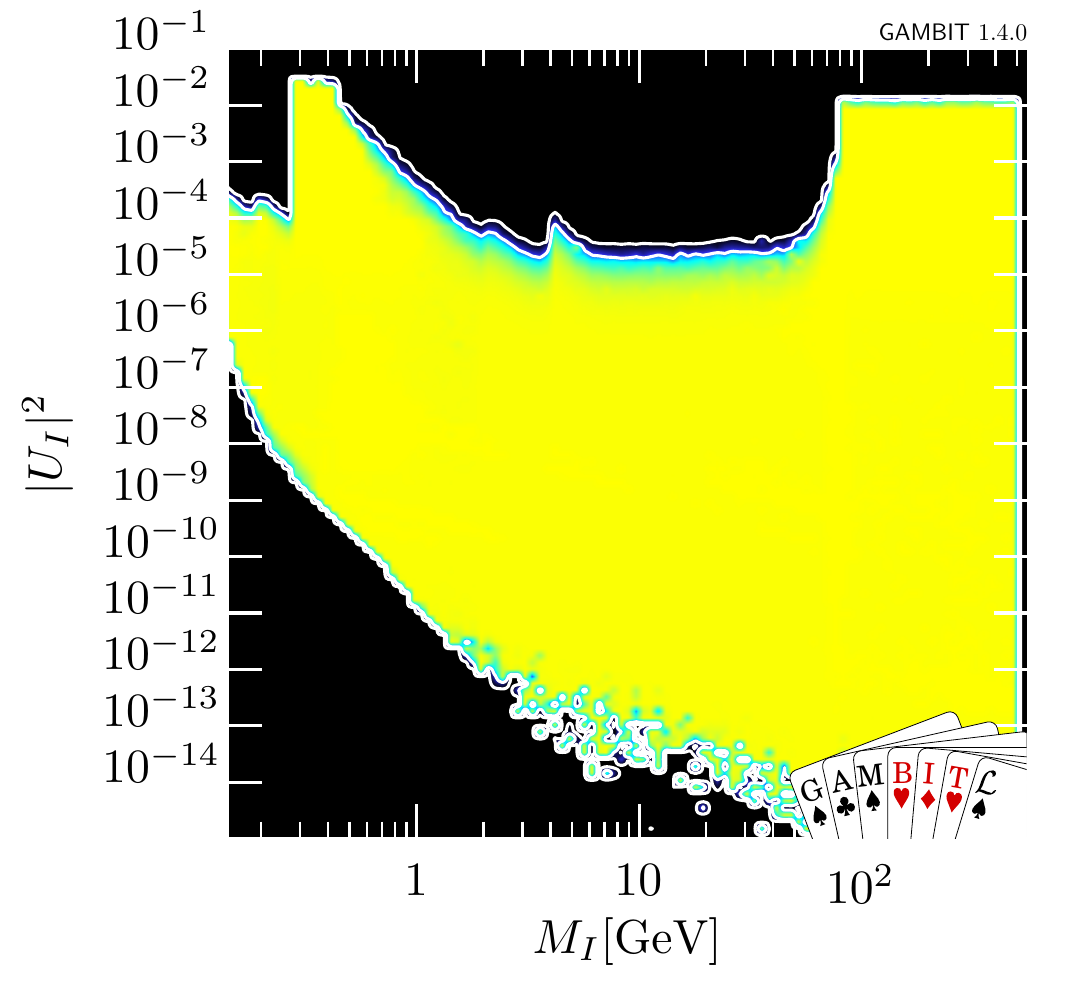}
  \includegraphics[width=0.45\linewidth]{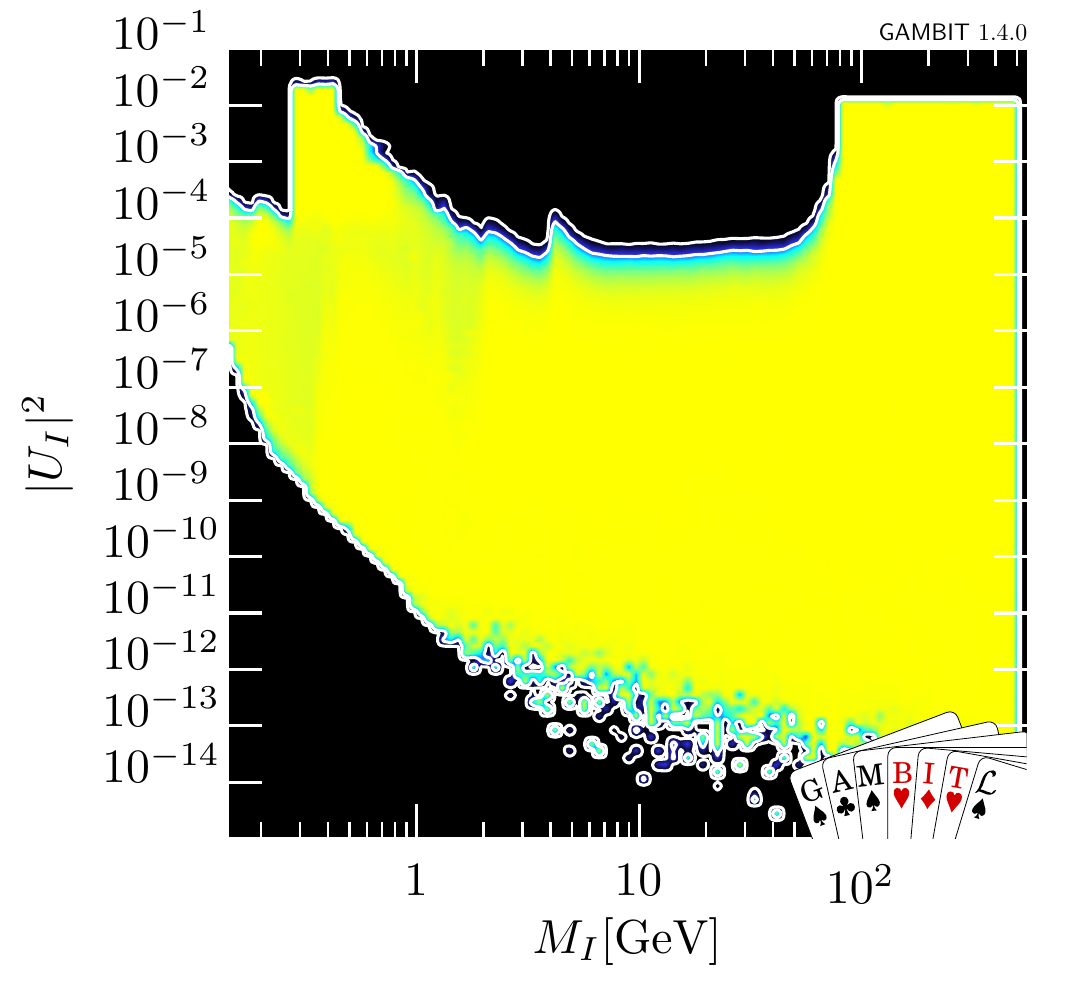}
  \caption{Profile likelihood of right-handed neutrino models in the $M_I$ vs $U_I^2$ plane for normal (left) and inverted (right) neutrino mass ordering. From \cite{RHN}.}
  \label{fig:M_U_capped}
\end{figure}

A comprehensive, frequentist \gambit study of this scenario has recently been completed \cite{RHN}.  The full list of parameters considered in the \gambit RHN study is given in Table~\ref{tab:scanpars}.  Separate scans were done for the cases of a normal and an inverted mass hierarchy, and the scanning strategy used a number of carefully targeted scans to ensure that the regions near the various experimental bounds were convergently sampled. The scans used the full range of likelihoods implemented in \neutrinobit, documented in Section~\ref{sec:neutrinobit}, in addition to the extra routines described in \flavbit, \decaybit and \precisionbit. The main analysis used a capped likelihood, in which each point is forced to have a likelihood which is equal to or worse than the SM ($\mathcal{L} = \min[\mathcal{L}_{\rm{SM}}, \mathcal{L}_{\rm{RHN}}]$). This is due to a number of excesses in individual experimental observations that --- although combining to give a small overall significance --- would bias the presentation of exclusion limits on RHN parameters. In this review, we concentrate on the resulting limits on RHNs, and direct the reader to the original study~\cite{RHN} for a detailed discussion of the excesses.

The 1-loop Casas-Ibarra parameterisation used in the GAMBIT analysis \cite{RHN} is valid for seesaw scenarios where the active-sterile mixing, $|\Theta|^2$, is small. In principle, additional $|\Theta|^4$ corrections could be expected in low-scale seesaw scenarios, and could only be captured by an exact expansion (e.g.\ Schechter-Valle \cite{Schechter:1980gr}). Nevertheless, the loss of generality in the Casa-Ibarra approximation is outweighed by its numerical and computational benefits. This parameterisation allows one to explicitly choose the masses of both active and sterile neutrinos, and is automatically consistent with oscillation data, allowing oscillation parameters to be easily treated as Gaussian nuisances.  Alternative parameterisations would constitute a different effective prior on both the sterile and active neutrino parameters. Because the GAMBIT analysis is based on profile likelihoods, which are by construction prior-independent, switching parameterisation would only have the impact of making sampling less efficient, rather than causing any physical or statistical effect.

\subsubsection{RHN results}

In Fig.\ \ref{fig:M_Ue_capped}, we show, as functions of the heavy neutrino masses $M_I$, the constraints on the couplings $U^2_{\alpha I}$ to the active neutrino flavours $\alpha = (e,\mu,\tau)$; in Fig.\ \ref{fig:M_U_capped} we also show the overall constraints on their combination $U^2_I=\sum_\alpha U^2_{\alpha I}$. The index $I$ can refer to any of the heavy neutrino flavours $I=(1,2,3)$, as their labelling has no physical significance. The profile likelihood is mostly flat at low values of the couplings, but exhibits characteristic drop-offs at higher values that result from specific experimental observations. The most dominant constraint varies with the RHN mass.

Above the masses of the weak gauge bosons, direct searches at colliders are not relevant, and the leading constraints on the RHN properties come from electroweak precision observables, CKM measurements and searches for lepton flavour violation (LFV). The upper limits on the $\tau$ couplings are much larger than on the $e$ and $\mu$ couplings, due to the fact that the EWPO and LFV limits are stronger for the $e$ and $\mu$ flavours.

When $M_I$ is between the $D$ meson and $W$ boson masses, direct search experiments dominate, as RHNs are efficiently produced via the $s$-channel exchange of on-shell $W$ bosons. The DELPHI and CMS results compete to impose the strongest bound in this region.

Below the $D$ meson mass, the dominant constraints come from direct searches at beam-dump experiments, in particular CHARM and NuTeV (above the kaon mass), PS-191 and E949 (between the pion and the kaon mass), and pion decay experiments at even lower mass. In the case of the $\tau$ couplings, the direct search constraints are much weaker, and the most significant constraint instead comes from DELPHI searches for long-lived particles.

For $M_I$ values below 0.3\,GeV, the global constraints are stronger than the sum of the individual contributions, due to an interplay between the lower bound from BBN, the upper bounds from direct searches and the constraints on RHN mixing from neutrino oscillation data (which disfavour large hierarchies amongst the couplings to individual SM flavours). The BBN lifetime constraint does not have an observable effect on the individual couplings, but it does force their combination to be greater than a certain value (as seen in Fig.\ \ref{fig:M_U_capped}).

Finally, we point out that although this analysis included the active neutrino oscillation likelihoods contained in \textsf{NeutrinoBit}, based on the results of \textsf{NuFit} \cite{NuFit15}, it would not change the results even slightly if one were to replace these with nuisance likelihoods from either of the other main 3-flavour neutrino fitting groups \cite{deSalas:2017kay,Capozzi:2018ubv}.  This is because the results of all three groups are highly consistent, and the preferred parameter region is where the approximate $B-L$ symmetry holds and the oscillation constraints are essentially irrelevant.  As the fits allow $m_{\nu_0}\to 0$, there is no lower limit implied on $M_I$ from oscillation data, but rather only from BBN (the effects of which were modelled under the massless neutrino approximation).

\section{Summary}
\label{sec:summary}

\gambit is an open-source software framework for combining all relevant experimental constraints on theories for physics Beyond the Standard Model.  It includes extensive libraries of theory and likelihood routines for dark matter, flavour, collider, neutrino and precision observables, along with spectrum and decay calculations, all for a range of popular and highly plausible theories of new physics.  It features an array of different statistical samplers, a hierarchical model database, an automated engine for building calculations based on graph theory, and the ability to connect to external physics calculators with ease.

In the two years since its release, \gambit has produced seven global analyses of leading theories for physics beyond the Standard Model.  In supersymmetry, this includes analyses of the CMSSM, NUHM1, NUHM2, a 7-parameter weak-scale MSSM, and an electroweakino effective field theory known as the EWMSSSM.  Results indicate a $3.3\sigma$ combined preference in LHC searches for weak production of light charginos and neutralinos.  In Higgs portal dark matter, \gambit results cover $\mathbb{Z}_2$ and $\mathbb{Z}_3$-symmetric scalar singlet models, as well as $\mathbb{Z}_2$-symmetric vector and fermion models.  All can provide good fits to experimental constraints, but fermionic models strongly prefer to violate $CP$.  Scalar models can not only solve the dark matter problem, but can also stabilise the vacuum of the standard model if and only if they possess TeV-scale masses and respect a $\mathbb{Z}_2$ symmetry.  GAMBIT studies indicate that QCD axions are most likely to constitute a fraction of dark matter rather than the entire amount, and to reside in a mass window between about $10^{-7}$ and $10^{-3}$\,eV. Right-handed neutrinos are constrained by a wide array of different searches at different masses; interactions with electrons and muons are the most strongly constrained, with constraints on couplings to tau leptons somewhat weaker.

\gambit is a powerful tool for testing theories of new physics.  The code can be obtained from \url{https://gambit.hepforge.org}.  All samples, input files and benchmark points resulting from the \gambit physics studies discussed in this review can also be obtained from \textsf{Zenodo}, by visiting \url{https://zenodo.org/communities/gambit-official}.

\section*{Acknowledgements}
We thank our collaborators within the \gambit Community for their essential and extensive contributions to the work reviewed in this article, and Tom\'as Gonzalo in particular for comments on right-handed neutrinos.  The majority of plots presented in the review were produced with \pippi \cite{pippi}.  We acknowledge PRACE for awarding us access to Marconi at CINECA, Italy.

\bibliographystyle{JHEP_pat}
\bibliography{R2}

\end{document}